\documentclass[apj,numberedappendix,iop]{emulateapj}
\pdfoutput=1
\usepackage%
[
pagebackref, % or backref
colorlinks=true,
linkcolor=red,filecolor=red,
citecolor=blue,
urlcolor=cyan,
pdftitle={Three-Dimensional Envelopes of Young Planets},
pdfauthor={D'Angelo et al.},
pdfsubject={Three-Dimensional Envelopes of Young Planets},
pdfcreator={KAZEditor},
pdfproducer={Kazzimiei},
pdfkeywords={Accretion, accretion disks; Hydrodynamics; Methods: numerical; Planet-disk interactions; Planets and satellites: formation; Protoplanetary disks},
bookmarks=true,
breaklinks=true,
hypertexnames=false,
pdfstartview=FitV,
hyperfootnotes=true,
bookmarksopen=false,
pdfpagemode=None]{hyperref}
\usepackage{breakurl}
\usepackage{amsmath}
\newcommand{\ddt}[1]{\frac{\partial #1}{\partial t}} % 
\newcommand{\dd}[2]{\frac{\partial #1}{\partial #2}} %
\newcommand{\gvec}[1]{\boldsymbol{#1}}                            % vector
\newcommand{\gdiv}{\nabla\!\boldsymbol{\cdot}\!}                    % Div()
\newcommand{\Mearth}{\mbox{$M_{\mathrm{E}}$}}        % Mearth
\newcommand{\Mp}{\mbox{$M_{p}$}}                     % Mp
\newcommand{\Mc}{\mbox{$M_{c}$}}                     % Mc
\newcommand{\Me}{\mbox{$M_{e}$}}                     % Me
\newcommand{\Ms}{\mbox{$M_{\star}$}}                     % Ms
\newcommand{\Msun}{\mbox{$M_{\odot}$}}                     % Msun
\newcommand{\Rhill}{\mbox{$R_\mathrm{H}$}}           % Rhill
\newcommand{\Rbondi}{\mbox{$R_\mathrm{B}$}}          % Rbondi
\newcommand{\K}{\mbox{$\mathrm{K}$}}                    % K
\newcommand{\AU}{\mbox{AU}}                          % AU
\newcommand{\figlen}{0.9\linewidth}
\newcommand{\afiglen}{0.95\linewidth}
\slugcomment{Published in The Astrophysical Journal, 778:77, 2013}

\shorttitle{Three-Dimensional Envelopes of Young Planets}
\shortauthors{D'Angelo and Bodenheimer}

\begin{document}

\title{Three-Dimensional Radiation-Hydrodynamics Calculations of the Envelopes
of Young Planets Embedded in Protoplanetary Disks}

\author{Gennaro D'Angelo\altaffilmark{1,2,4} and Peter Bodenheimer\altaffilmark{3}}
\altaffiltext{1}{NASA Ames Research Center, MS 245-3, Moffett Field, CA 94035, USA 
(\href{mailto:gennaro.dangelo@nasa.gov}{gennaro.dangelo@nasa.gov})}
\altaffiltext{2}{SETI Institute, 189 Bernardo Avenue, Mountain View, CA 94043, USA}
\altaffiltext{3}{UCO/Lick Observatory, University of California, Santa Cruz, CA 95064, USA
(\href{mailto:peter@ucolick.org}{peter@ucolick.org})}
\altaffiltext{4}{Visiting Research Scientist, Los Alamos National Laboratory, Los Alamos, NM 87545, USA}

\begin{abstract}
We perform global three-dimensional (3D) radiation-hydrodynamics calculations 
of the envelopes surrounding young planetary cores of $5$, $10$, and $15$ 
Earth masses, located in a protoplanetary disk at $5$ and $10\,\AU$ 
from a solar-mass star. 
We apply a nested-grid technique to resolve the thermodynamics of the 
disk at the orbital-radius length scale and that of the envelope 
at the core-radius length scale. 
The gas is modeled as a solar mixture of molecular and atomic hydrogen, 
helium, and their ions. 
The equation of state accounts for both gas and radiation, and gas energy 
includes contributions from rotational and vibrational states of molecular 
hydrogen and from ionization of atomic species.
Dust opacities are computed from first principles, applying the full Mie theory.
One-dimensional (1D) calculations of planet formation are used to supplement 
the 3D calculations by providing energy deposition rates in the envelope due to solids
accretion.
We compare 1D and 3D envelopes and find that masses and gas accretion 
rates agree within factors of $2$, and so do envelope temperatures. 
The trajectories of passive tracers are used to define the size of 3D envelopes,
resulting in radii much smaller than the Hill radius and smaller than the Bondi 
radius.
The moments of inertia and angular momentum of the envelopes are determined
and the rotation rates are derived from the rigid-body approximation, resulting
in slow bulk rotation.
We find that the polar flattening is $\lesssim 0.05$.
The dynamics of the accretion flow is examined by tracking the motion of tracers
that move into the envelope. The anisotropy of this flow is characterized 
in terms of both its origin and impact site at the envelope surface.  
Gas merges with the envelope preferentially at mid- to high latitudes.
\end{abstract}


%--------------------------------------------------------------------------
\section{Introduction}
\label{sec:Intro}
%--------------------------------------------------------------------------

\defcitealias{black1975}{BB75}
\defcitealias{stone1992a}{SN92}

The formation of a gaseous envelope around a planetary core has
been studied, for over thirty years, by means of spherically symmetric 
one-dimensional (1D) calculations \citep[e.g., ][]{pollack1996}. 
Most of the current knowledge
about the process relies on such calculations. 
At early stages, the growth rate of the envelope is controlled by its own 
cooling rate, up the point where envelope and core mass are about equal.
Thereafter, the growth is much more rapid and proceeds through a phase 
of hydrodynamical collapse, limited exclusively by disk supply.
To date, the million-year evolution that is required to grow a planetary embryo
into a giant planet can only be modeled via 1D calculations.

A number of properties and physical effects, however, cannot be accounted
for or described by invoking spherical symmetry and are therefore 
approximated, imposed, or simply neglected in 1D models.
The envelope radius, for example, can only be introduced as a parameter, 
since it depends on the thermal and gravitational energy of the gas 
outside the envelope \citep{bodenheimer1986}. 
The density and temperature at the outer boundary are approximated to 
average values in the unperturbed disk, whereas the actual (perturbed) disk 
values depend also on the planet mass.
Rotation, often neglected, can be included to some extent, but not constrained.
Although the gas dynamics of a disk away from a planet can be 
characterized as plane-parallel reasonably well, the flow becomes inherently 
three-dimensional (3D) as it approaches the planet 
\citep{gennaro2003a,bate2003,masset2006a}.
Part of this flow eventually feeds the envelope and breaks the spherical symmetry 
of the outer layers through transfer of angular and radial momentum, which 
may induce rotation and radial mixing.

3D models of envelopes that account for the interactions
of the planet with the disk can in principle overcome all the limitations of 1D 
models. However, the computational overhead is so large that it is not yet 
feasible to go beyond evolution times of order $10^{2}$--$10^{3}$ years,
depending on the orbital distance.
At the moment, 3D calculations can only be used to investigate particular 
epochs of the core and envelope growth. Nonetheless, they can provide 
a wealth of information, otherwise not accessible through 1D models,
which can help adjust and/or refine 1D calculations.
The scope of this paper is to study one such epoch, early during the planet
evolution when the accretion of solids is still relatively large and the envelope 
mass is much smaller than the core mass.
We show that there is general consistency between 3D and 1D
calculations and that 3D calculations can indeed address the physics
missing in spherically symmetric models.

Local 3D calculations of planetary envelopes with
radiation hydrodynamics were carried out by \citet{ayliffe2009a,ayliffe2012}.
There are several similarities to our study but also important differences. 
For example,
the local approach cannot capture the thermodynamics of the gas 
in and around the horse-shoe orbit region, which supplies the planet with gas
and affects the dynamics of the accretion flow.
They used an interstellar dust opacity, scaled down by numerical 
factors, whereas we compute the dust opacity based on a grain
size distribution that may be more appropriate for circumstellar disks.
They concentrated on somewhat larger core masses, and for the
$10$ and $15$ Earth-mass ($\Mearth$) cases they used core radii 
about $10$ times larger than the physical radii. They did not account 
for energy delivery by solids accretion.
Recently,
\citet{anelson2013} performed high-resolution, local calculations 
of the envelope region surrounding planetary cores of $10\,\Mearth$, 
applying an isothermal and an adiabatic equation of state. The differences
between the above-cited simulations and the calculations presented here 
are such as to render any comparison unfeasible.
Until now, global 3D calculations of planets embedded in disks with
radiation hydrodynamics have been used to investigate tidal torques 
exerted on the planet 
\citep[e.g.,][]{klahr2006b,paardekooper2006,kley2009}, but never 
the details of the planet envelope and of the envelope-disk interactions.

The rest of the paper is organized as follows.
The physical model, including equation of state and opacity, is described
in Section~\ref{sec:DET}, and various aspects of the numerical solution 
are outlined in Section~\ref{sec:NP}. The thermodynamics of the equilibrium 
disk structures is discussed in Section~\ref{sec:DS}, while the comparison
with the 1D envelopes and the properties of the 3D envelopes are examined 
in Section~\ref{sec:PCE}. The conclusions are given in Section~\ref{sec:summ}.

%%--------------------------------------------------------------------------
\section{Disk and Envelope Thermodynamics}
\label{sec:DET}
%%--------------------------------------------------------------------------

%%--------------------------------------------------------------------------
\subsection{Gas Dynamics}
\label{sec:GD}
%%--------------------------------------------------------------------------

Consider a frame of reference with its origin fixed 
to the star and rotating about the origin at a rate $\Omega$, 
the angular velocity of the planetary core around the star,
\begin{equation}
\Omega=\sqrt{\frac{G(\Ms+\Mc)}{a^{3}}},
\label{eq:omega}
\end{equation}
where $\Ms$ is the stellar mass, $\Mc$ indicates the core mass,
and $a$ is the core's semimajor axis.
Equation~(\ref{eq:omega}) implicitly assumes that the envelope
mass is small compared to $\Mc$.

Consider now a spherical polar coordinate system $\{r,\theta,\phi\}$,
where 
$r\in [a/2, 2 a]$ is the radial distance from the origin, 
$\theta\in [23\pi/50,\pi/2]$ is the meridional angle measured from 
the north pole (co-latitude), and $\phi\in [0,2\pi]$ is the azimuthal angle. 
Since the planet's orbit lies in the disk's equatorial plane ($\theta=\pi/2$),
we assume that the disk is symmetric relative to this plane. 
The geometrical opening angle of the disk (above and below the equatorial plane) 
is $2\pi/25$, or about $14^{\circ}$. 

As customary in many astrophysical applications, 
the disk's gas is approximated as a viscous fluid with kinematic
viscosity $\nu$, volume density $\rho$, and velocity $\gvec{u}$.
The dynamics of the gas is described via the mass continuity equation
\begin {equation}
  \label{eq:mce}
  \ddt{\rho} + \gdiv(\gvec{u}\rho)  = 0
\end{equation}
and the Navier-Stokes equations \citep[see, e.g.,][]{m&m}.
Let us denote with $u_{r}$, $u_{\theta}$, and $u_{\phi}$, the
spherical polar components of the velocity vector $\gvec{u}$, 
and let the quantities
$H_{r}=\rho u_{r}$,
$H_{\theta}=\rho u_{\theta} r$, and
$H_{\phi}=\rho (u_{\phi}+\Omega r\sin{\theta})r\sin{\theta}$,
be the absolute linear ($H_{r}$) and angular
($H_{\theta}$ and $H_{\phi}$) momenta of the gas per unit volume.

By transformation and substitution, the Navier-Stokes 
equations can be re-written in the following conservative form
in terms of absolute linear and angular momenta
\begin {eqnarray}
  \label{eq:H_r}
  \ddt{H_{r}} + \gdiv(\gvec{u}H_{r}) & = & 
  \frac{\rho}{r}\left[\left(\frac{H_{\theta}}{r}\right)^{2}% 
   +\left(\frac{H_{\phi}}{r\sin{\theta}}\right)^{2}\right] \nonumber\\
   & - &  \dd{P_{\mathrm{gas}}}{r} - \rho\dd{\Phi}{r} \nonumber \\
   & + &\mathcal{V}_{r} + \frac{\rho\kappa}{c}F_{r},
\end{eqnarray}
\begin{eqnarray}
 \label{eq:H_theta}
 \ddt{H_{\theta}} + \gdiv(\gvec{u}H_{\theta}) & = &
  \rho\left(\frac{\cos{\theta}}{\sin{\theta}}\right)\left(\frac{H_{\phi}}{r\sin{\theta}}\right)^{2} \nonumber \\
  & - & \dd{P_{\mathrm{gas}}}{\theta} - \rho\dd{\Phi}{\theta} \nonumber \\
  & + & r\mathcal{V}_{\theta}
     +  \frac{\rho\kappa}{c} r F_{\theta},
\end{eqnarray}
\begin {eqnarray}
\label{eq:H_phi}
 \ddt{H_{\phi}} + \gdiv(\gvec{u}H_{\phi}) & = & 
  -  \dd{P_{\mathrm{gas}}}{\phi}- \rho\dd{\Phi}{\phi} \nonumber \\
 & + & r\sin{\theta}\mathcal{V}_{\phi} 
    +  \frac{\rho\kappa}{c}r\sin{\theta} F_{\phi}\,.
\end{eqnarray}

In the above equations, $P_{\mathrm{gas}}$ is the gas pressure,
which we shall discuss in more detail below,
and $\Phi$ is the gravitational potential in the disk
 \begin{equation}
   \label{eq:phi}
   \Phi=%
          \Phi_{c}
        - \frac{G\,\Ms}{r}%
        + \frac{G\,\Mc}{r_{c}^{3}}%
        \,\gvec{r}\!\boldsymbol{\cdot}\!\gvec{r}_{c},
 \end{equation}
in which $\Phi_{c}$ is the potential of the planetary core and
$\gvec{r}_{c}$ is the vector position of its center.
We use a piecewise polynomial representation of the core's potential, 
explicitly given in Appendix~\ref{sec:phi_c}, and assume that any gas
bound to the core has a mass small compared to $\Mc$.
The third term on the right hand side
of Equation~(\ref{eq:phi}) accounts for the fact that the reference
frame is non-inertial since its origin is attached to the star.

The quantities $\mathcal{V}_{r}$, $\mathcal{V}_{\theta}$, and 
$\mathcal{V}_{\phi}$ in Equations~(\ref{eq:H_r}), (\ref{eq:H_theta}), 
and (\ref{eq:H_phi}) represent the viscous force acting on a unit volume 
of gas. They depend on the components $\mathrm{S}_{ij}$ of 
the viscous stress tensor $\mathbf{S}$, which is assumed to be
that of a Newtonian fluid without bulk viscosity.
Explicit expressions for $\mathcal{V}_{r}$, $\mathcal{V}_{\theta}$, 
and $\mathcal{V}_{\phi}$ in spherical polar coordinates, along with 
those for the components $\mathrm{S}_{ij}$, are given in \citet{m&m}.
Notice that Equations~(\ref{eq:H_theta}) and (\ref{eq:H_phi}), 
which evolve angular momenta per unit volume, involve torques 
rather than forces.
The last terms on the right-hand side of 
Equations~(\ref{eq:H_r}), (\ref{eq:H_theta}), and (\ref{eq:H_phi})
are the forces/torques, per unit volume, imparted to the gas by 
the absorbed/scattered photons in the radiation field, of which 
$\gvec{F}$ represents a frequency-integrated energy flux, 
$\kappa$ is a frequency-integrated opacity coefficient, 
and $c$ is the speed of light.
As discussed below, we will identify $\kappa$ with the Rosseland mean opacity.

%%--------------------------------------------------------------------------
\subsection{Gas Thermodynamics}
\label{sec:GT}
%%--------------------------------------------------------------------------

Gas thermodynamics is considered under the approximation 
of local thermodynamic equilibrium and 
a single temperature, $T$, for the gas and the radiation. 
The radiation energy density is then written as 
$E_{\mathrm{rad}}=(4\pi/c)B(T)$,
where $B(T)=(\sigma_{\mathrm{SB}}/\pi)T^{4}$  
is the frequency-integrated Planck function and
$\sigma_{\mathrm{SB}}$ is the Stefan-Boltzmann constant. 
Indicating with $E_{\mathrm{gas}}$ the gas energy density, 
the evolution of the gas and radiation energies are governed 
by \citep[e.g.,][]{turner2001}
\begin {equation}
\label{eq:dEgas}
 \ddt{E_{\mathrm{gas}}} + \gdiv(\gvec{u}E_{\mathrm{gas}}) =%
  - P_{\mathrm{gas}}\gdiv{\gvec{u}} + \Psi + \varepsilon,
\end{equation}
and
\begin {equation}
\label{eq:dErad}
 \ddt{E_{\mathrm{rad}}} + \gdiv(\gvec{u}E_{\mathrm{rad}}) =%
  - \gdiv{\gvec{F}} - P_{\mathrm{rad}}\gdiv{\gvec{u}},
\end{equation}
which here assume that the radiation pressure tensor is represented 
by a scalar matrix with scalar $P_{\mathrm{rad}}$.
In a non-equilibrium situation, the right-hand sides of the above
equations contain, respectively, the terms 
$\mp[4\pi\rho\kappa B(T)-c\rho\kappa E_{\mathrm{rad}}]$,
which describe matter-radiation interaction
but which vanish here on account of the assumed relation between 
$E_{\mathrm{rad}}$ and $B(T)$.
Adding up Equations~(\ref{eq:dEgas}) and (\ref{eq:dErad}), 
the evolution equation of the total internal energy per unit volume,
$E=E_{\mathrm{gas}}+E_{\mathrm{rad}}$, can be written as 
\citep[e.g.,][]{yorke1995}
\begin {equation}
\label{eq:dEtot}
 \ddt{E} + \gdiv(\gvec{u}E) = - \gdiv{\gvec{F}} - P\gdiv{\gvec{u}}%
                                            + \Psi + \varepsilon.
\end{equation}

In Equations~(\ref{eq:dEgas}) and (\ref{eq:dEtot}), 
$\varepsilon V_{c}$ is the gravitational energy 
per unit time released by planetesimals penetrating
the planet's envelope, defined through the volume integral
\begin {equation}
\label{eq:eps}
\varepsilon V_{c}=\int{\frac{G\Mc\dot{M}_{c}}{R_{c}}}%
                                   \delta(\gvec{r}-\gvec{r}_{c}) dV,
\end{equation}
in which $V_{c}\sim (4\pi/3)R^{3}_{c}$ and $R_{c}$ is the core radius. 
The $\delta$-function is meant to signify that gravitational energy
carried by planetesimals is released at the core surface. 
The accretion rate of the core, $\dot{M}_{c}$, and the core radius are 
input parameters discussed in Section~\ref{sec:1D}.

The function $\Psi$, in Equations~(\ref{eq:dEgas}) and (\ref{eq:dEtot}),
accounts for viscous energy dissipation. In terms of the components 
of the viscous stress tensor, the dissipation function is \citep{m&m}.
\begin {eqnarray}
\Psi     =   \frac{1}{2\nu\rho}\mathrm{S}_{ij}\mathrm{S}^{ij} %
         =   \frac{1}{2\nu\rho}&&\left(%
             \mathrm{S}^{2}_{rr}+\mathrm{S}^{2}_{\theta\theta}
            +\mathrm{S}^{2}_{\phi\phi} + \right.         \nonumber \\
                               &&\left.
            2\mathrm{S}^{2}_{r\theta}+2\mathrm{S}^{2}_{r\phi}
           +2\mathrm{S}^{2}_{\theta\phi} \right).        \label{eq:Psi}
\end{eqnarray}

The gas and radiation pressures are given by, respectively,
$P_{\mathrm{gas}}=\rho k_{\mathrm{B}}T/(\mu m_{\mathrm{H}})$ 
and $P_{\mathrm{rad}}=E_{\mathrm{rad}}/3$.
The mean molecular
weight, $\mu$, accounts for the presence of molecules, atoms,
and ions, and will be discussed in detail below ($k_{\mathrm{B}}$ 
is the Boltzmann constant and $m_{\mathrm{H}}$ is the atomic
hydrogen mass).
As mentioned earlier, the radiation pressure is a tensor whose
components depend on the Eddington factor \citep[e.g.,][]{turner2001}, 
although here we retain only the diagonal elements (assumed all equal) 
and use the general property that the trace of the tensor is 
equal to $E_{\mathrm{rad}}$ \citep{castor2007}.
This approximation works best for optically thick gas.
In Equation~(\ref{eq:dEtot}), $P=P_{\mathrm{gas}}+P_{\mathrm{rad}}$
refers to the total pressure.

Energy transport via radiation is taken into account in the so-called
flux-limited diffusion approximation \citep{levermore1981,castor2007}.
The radiation energy flux is written as
\begin {equation}
\label{eq:F}
 \gvec{F} = - \mathcal{D}\nabla{E_{\mathrm{rad}}}.
\end{equation}
The flux-limited diffusion coefficient is
\begin {equation}
\label{eq:D}
 \mathcal{D} = \frac{c\lambda}{\rho\kappa},
\end{equation}
in which the so-called flux-limiter, $\lambda$, is
a function of the ratio 
$\mathcal{R}=|\nabla{E_{\mathrm{rad}}}|/(\rho\kappa E_{\mathrm{rad}})$.
The choice of the flux-limiter is problem dependent 
\citep[see discussion in][]{turner2001}. In fact, there are only constraints
in limiting cases.
In the diffusion limit, i.e., for $\mathcal{R}\rightarrow 0$, $\lambda$ must
tend to $1/3$, so that 
$\gvec{F}=-c/(3\rho\kappa) \nabla{E_{\mathrm{rad}}}=%
-(16/3)\sigma_{\mathrm{SB}}/(\rho\kappa) T^{3}\nabla{T}$.
In the streaming limit,  i.e., for $\mathcal{R}\rightarrow \infty$,
the asymptotic behavior must be 
$\lambda\rightarrow 1/\mathcal{R}$, 
so that 
$\gvec{F}\rightarrow -\gvec{n} c E_{\mathrm{rad}}$ with
$\gvec{n}=\nabla{E_{\mathrm{rad}}}/|\nabla{E_{\mathrm{rad}}}|$.
Here, we adopt the rational approximation to the flux-limiter 
of \citet{levermore1981}
\begin {equation}
\label{eq:lambda}
 \lambda=\frac{2+\mathcal{R}}{6+3\mathcal{R}+\mathcal{R}^{2}}.
\end{equation}
It should be mentioned that, 
as discussed by \citet{castor2007}, regardless of the choice of the function
$\lambda$, the flux-limited diffusion approximation can hardly describe the 
angular distribution of the radiation field to an accuracy better than
$\sim 10$\%.

The radiation flux in Equation~(\ref{eq:dEtot}) includes only radiation 
generated internally by the gas. However,
the radiation flux from additional sources, such as irradiation by the star 
or other external sources, can be simply added to the flux in 
Equation~(\ref{eq:F}), without any further modifications.

%%--------------------------------------------------------------------------
\subsection{Equation of State}
\label{sec:EoS}
%%--------------------------------------------------------------------------

As anticipated above, we apply an equation of state for an ideal gas 
that accounts for the effects due to the dissociation of molecular
hydrogen and of the ionization of atomic hydrogen and helium.
Contributions from radiation are also taken into account.
The mass fractions of hydrogen and helium are set equal, 
respectively, to $X=0.7$ and $Y=0.28$. These numbers deviate 
somewhat from current estimates of protosolar values, principally 
due to the availability of gas opacity tables, as discussed below. 
\citet{asplund2009} and \citet{lodders2010}
reported protosolar composition values of $X=0.71$ and $Y=0.27$.
We neglect heavy elements in constructing the equation of state, 
although their contribution to gas opacity is taken into account 
(see Section~\ref{sec:opa}).

Following \citet{black1975}, 
let us introduce the degree of dissociation of molecular hydrogen,
$y=\rho_{\mathrm{H}}/(\rho_{\mathrm{H}}+\rho_{\mathrm{H}_{2}})$,
the degree of ionization of atomic hydrogen,
$x=\rho_{\mathrm{H}^{+}}/(\rho_{\mathrm{H}^{+}}+\rho_{\mathrm{H}})$,
and the degrees of single and double ionization of helium,
$z_{1}=\rho_{\mathrm{He}^{+}}/(\rho_{\mathrm{He}^{+}}+\rho_{\mathrm{He}})$
and
$z_{2}=\rho_{\mathrm{He}^{2+}}/(\rho_{\mathrm{He}^{2+}}+\rho_{\mathrm{He}^{+}})$,
respectively. Applying the Saha equation \citep[see][]{kippenhahn2013,kowalski2006}, 
the dissociation and ionization degrees can be derived from the following 
relations
\begin {eqnarray}
\label{eq:Saha}
\frac{y^{2}}{1-y}&=&\frac{m_{\mathrm{H}}}{2X\rho}\!%
                                \left(\frac{m_{\mathrm{H}} k_{\mathrm{B}}T}{4\pi\hbar^{2}}\right)^{3/2}%
                                e^{-4.48\mathrm{eV}/(k_{\mathrm{B}}T)} \label{eq:y}\\
\frac{x^{2}}{1-x}&=&\frac{m_{\mathrm{H}}}{X\rho}\!%
                                \left(\frac{m_{\mathrm{e}} k_{\mathrm{B}}T}{2\pi\hbar^{2}}\right)^{3/2}%
                                e^{-13.60\mathrm{eV}/(k_{\mathrm{B}}T)} \label{eq:x}\\                            
\frac{z_{1}}{1-z_{1}}&=&\frac{4m_{\mathrm{H}}}{\rho}\!%
                                \left(\frac{m_{\mathrm{e}} k_{\mathrm{B}}T}{2\pi\hbar^{2}}\right)^{3/2}%
                                \frac{e^{-24.59\mathrm{eV}/(k_{\mathrm{B}}T)}}{X+z_{1}Y/4} \label{eq:z1}\\
\frac{z_{2}}{1-z_{2}}&=&\frac{m_{\mathrm{H}}}{\rho}\!%
                                \left(\frac{m_{\mathrm{e}} k_{\mathrm{B}}T}{2\pi\hbar^{2}}\right)^{3/2}%
                                \frac{e^{-54.42\mathrm{eV}/(k_{\mathrm{B}}T)}}{X+(z_{2}+1)Y/4} \label{eq:z2},
\end{eqnarray}
where $m_{\mathrm{e}}$ is the electron mass and $\hbar$ is Planck's constant
divided by $2\pi$.
The mean molecular weight, $\mu$, of the mixture is such that
\citep[e.g.,][]{black1975,kippenhahn2013}
\begin {equation}
\label{eq:mue}
\frac{\mu}{4}=[2 X (1+y+2 x y)+Y (1+z_{1}+z_{1} z_{2})]^{-1}.
\end{equation}
The internal energy density of the mixture can be written as
\begin {eqnarray}
\label{eq:Emix}
E_{\mathrm{gas}} 
          & = & \left(E_{\mathrm{H}_{2}}+E_{\mathrm{H}}+E_{\mathrm{He}}
                       + E_{\mathrm{H+H}}+E_{\mathrm{H}^{+}}\right. \nonumber \\%
          & + & \left.E_{\mathrm{He}^{+}}+E_{\mathrm{He}^{2+}}\right)%
                         k_{\mathrm{B}}T\rho/m_{\mathrm{H}},
\end{eqnarray}
where all contributions in the parenthesis are dimensionless and all, 
except for the contribution of $\mathrm{H}_{2}$, are straightforward
\citep[see, e.g.,][]{black1975}:
$E_{\mathrm{H}}=3 X (1+x) y/2$,
$E_{\mathrm{He}}=3 Y (1+z_{1}+z_{1} z_{2})/8$,
$E_{\mathrm{H+H}}=4.48\mathrm{eV}\, X y /(2k_{\mathrm{B}}T)$,
$E_{\mathrm{H}^{+}}=13.60\mathrm{eV}\, X x y/(k_{\mathrm{B}}T)$,
$E_{\mathrm{He}^{+}}=24.59\mathrm{eV}\, Y z_{1} (1-z_{2}) /(4k_{\mathrm{B}}T)$,
and
$E_{\mathrm{He}^{2+}}=54.42\mathrm{eV}\, Y z_{1} z_{2} /(4k_{\mathrm{B}}T)$.
The second and third terms in Equation~(\ref{eq:Emix}) are the translational 
energies of hydrogen and helium atoms.
The last four terms represent contributions due to dissociation 
of molecular hydrogen and ionization of atomic hydrogen and helium.

The energy of molecular hydrogen in Equation~(\ref{eq:Emix}) takes
into account, along with translational, also rotational and vibrational 
degrees of freedom \citep[e.g.,][]{pathria2011}
\begin{equation}
\label{eq:EH2_0}
E_{\mathrm{H}_{2}}=\frac{X (1-y)}{2}%
     \left[\frac{3}{2}+\frac{T}{\zeta_{v}}\dd{\zeta_{v}}{T}%
                            +\frac{T}{\zeta_{r}}\dd{\zeta_{r}}{T}\right],
\end{equation}
in which
$\zeta_{v}$ and $\zeta_{r}$ are the vibrational
and rotational partition functions of the molecule, respectively.

The vibrational energy levels of a diatomic molecule can be described
by the partition function of a quantum harmonic oscillator 
\begin{equation}
\label{eq:zeta_v}
     \zeta_{v}=\left(1-e^{-\Theta_{v}/T}\right)^{-1},
\end{equation}
with $\Theta_{v}=6140\,\K$ for the $\mathrm{H}_{2}$ molecule.
The rotational energy levels must take into consideration the relative 
spin states of the two nuclei.
Parahydrogen (anti-parallel spins) forms a singlet state, while
orthohydrogen (parallel spins) forms an excited triplet state.
At equilibrium, the percentage of the two forms depends on temperature.
At temperatures $T\lesssim 50\,\K$, the parahydrogen singlet 
is the most populated energy state and more than $80$\% of the molecules 
are in para-form. As the temperature rises, the orthohydrogen triplet state 
starts to be occupied.
At temperatures $T\gtrsim 300\,\K$, all energy levels are equally populated, 
yielding an ortho-to-para ratio of $3/1$.

Approximating the rotational energy levels of $\mathrm{H}_{2}$ to those 
of a quantum rigid rotor, the rotational partition function of 
para/orthohydrogen can be expressed as
\begin{equation}
\label{eq:zeta_PO}
\zeta_{\mathrm{P,O}}=\sum_{j} \left(2j+1\right) e^{-j(j+1)\Theta_{r}/T}
\end{equation}
with $\Theta_{r}=85.5\,\K$. The sum is performed over \textit{even}
integers for parahydrogen and over \textit{odd} integers for orthohydrogen.
Assuming equilibrium of the two forms at all temperatures and because 
of the spin degeneracy of the orthohydrogen triplet state, 
the total partition function is 
$\zeta_{r}=\zeta_{\mathrm{P}} + 3\zeta_{\mathrm{O}}$ 
\citep[e.g.,][]{kittel2004,pathria2011}.
However, conversion from one form to the other is quite inefficient 
in absence of a catalyst \citep[e.g.,][]{schmauch1964}, due to weak 
magnetic interaction of the nucleus spin with the outside world
\citep[e.g.,][]{pathria2011,draine2011}.

Therefore, the two forms of $\mathrm{H}_{2}$ may be regarded as independent 
species (different molecules) with a given occurrence ratio 
$f_{\mathrm{O}}/f_{\mathrm{P}}$.
In this case, the total partition function is the product of the single partition
functions \citep[e.g.,][]{pathria2011}
\begin{equation}
\label{eq:zeta_r}
\zeta_{r}=\zeta_{\mathrm{P}}^{f_{\mathrm{P}}}%
\left(\zeta_{\mathrm{O}} e^{2\Theta_{r}/T}\right)^{f_{\mathrm{O}}},
\end{equation}
where $f_{\mathrm{O}}+f_{\mathrm{P}}=1$.
Here we apply Equation~(\ref{eq:zeta_r}) and assume a fixed 
number ratio $f_{\mathrm{O}}/f_{\mathrm{P}}=3$.
As mentioned by \citet{boley2007b}, the exponential is meant to
regularize $\zeta_{r}$ in the limit $T\rightarrow 0$. 
In this limit, $\zeta_{\mathrm{P}} \rightarrow 1$ and
$\zeta_{\mathrm{O}} \rightarrow 0$ 
(recall that the triplet is an excited state), 
so that the requirement of a fixed number ratio would be violated.

By using Equations~(\ref{eq:zeta_r}), (\ref{eq:zeta_PO}) and 
(\ref{eq:zeta_v}), Equation~(\ref{eq:EH2_0}) becomes
\begin{eqnarray}
\label{eq:EH2}
E_{\mathrm{H}_{2}}& = &\frac{X (1-y)}{2}%
     \left[\frac{3}{2}+\left(\frac{\Theta_{v}}{T}\right)
                              \frac{e^{-\Theta_{v}/T}}{1-e^{-\Theta_{v}/T}}\right. \nonumber\\%
                            & + & \left. 
                            f_{\mathrm{P}}\frac{d\ln{\zeta_{\mathrm{P}}}}{d\ln{T}}%
                          +f_{\mathrm{O}}\left(%
                          \frac{d\ln{\zeta_{\mathrm{O}}}}{d\ln{T}}-\frac{2\Theta_{r}}{T}\right)\right].
\end{eqnarray}

As explained later, a stability condition for the numerical calculations
requires an estimate of the adiabatic sound speed of the gas, 
$c_{\mathrm{gas}}=\sqrt{\Gamma_{1} P_{\mathrm{gas}}/\rho}$. 
The first adiabatic exponent, defined as
$\Gamma_{1}=(\partial\ln{P_{\mathrm{gas}}}/\partial\ln{\rho})$ 
at constant entropy \citep{kippenhahn2013}, 
can also be expressed as \citep{cox2006}
\begin{equation}
\label{eq:Gamma_1}
\Gamma_{1}=%
\frac{1}{c_{V}}\left(\frac{P_{\mathrm{gas}}}{T\rho}\right)\chi^{2}_{T}+\chi_{\rho},
\end{equation}
where the specific heat at constant volume, $c_{V}$, is calculated
by taking the derivative with respect to $T$ of the specific energy of 
the gas $E_{\mathrm{gas}}/\rho$ (in Equation~(\ref{eq:Emix}),
$y$, $x$, $z_{1}$, and $z_{2}$ are all functions of $T$ and $\rho$),
and the so-called temperature and density exponents, 
$\chi_{T}$ and $\chi_{\rho}$, 
are defined by
\begin{equation}
\label{eq:chi_T}
\chi_{T}=\left(\dd{\ln{P_{\mathrm{gas}}}}{\ln{T}}\right)_{\rho}%
=1-\dd{\ln{\mu}}{\ln{T}},
\end{equation}
and
\begin{equation}
\label{eq:chi_rho}
\chi_{\rho}=\left(\dd{\ln{P_{\mathrm{gas}}}}{\ln{\rho}}\right)_{T}%
=1-\dd{\ln{\mu}}{\ln{\rho}}.
\end{equation}
All the summations required in the calculation of $\zeta_{\mathrm{P}}$,
$\zeta_{\mathrm{O}}$, and their first and second derivatives with
respect to $T$, use a sufficiently large number of terms so that the
magnitude of the relative difference between the approximated and 
true sum is $\le 10^{-8}$.

%%%%%%%
\begin{figure*}[]
\centering%
\resizebox{\figlen}{!}{%
\includegraphics{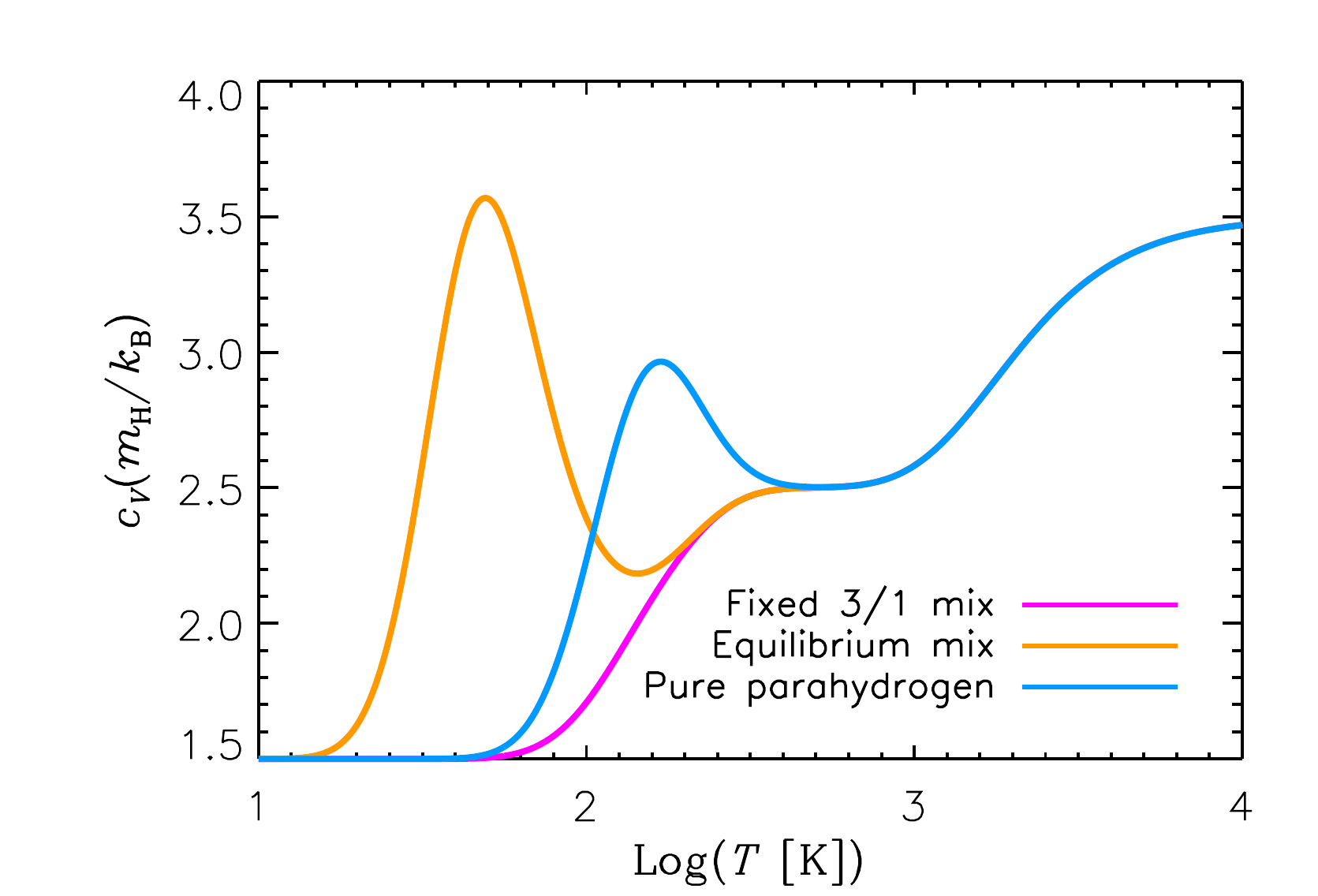}%
\includegraphics{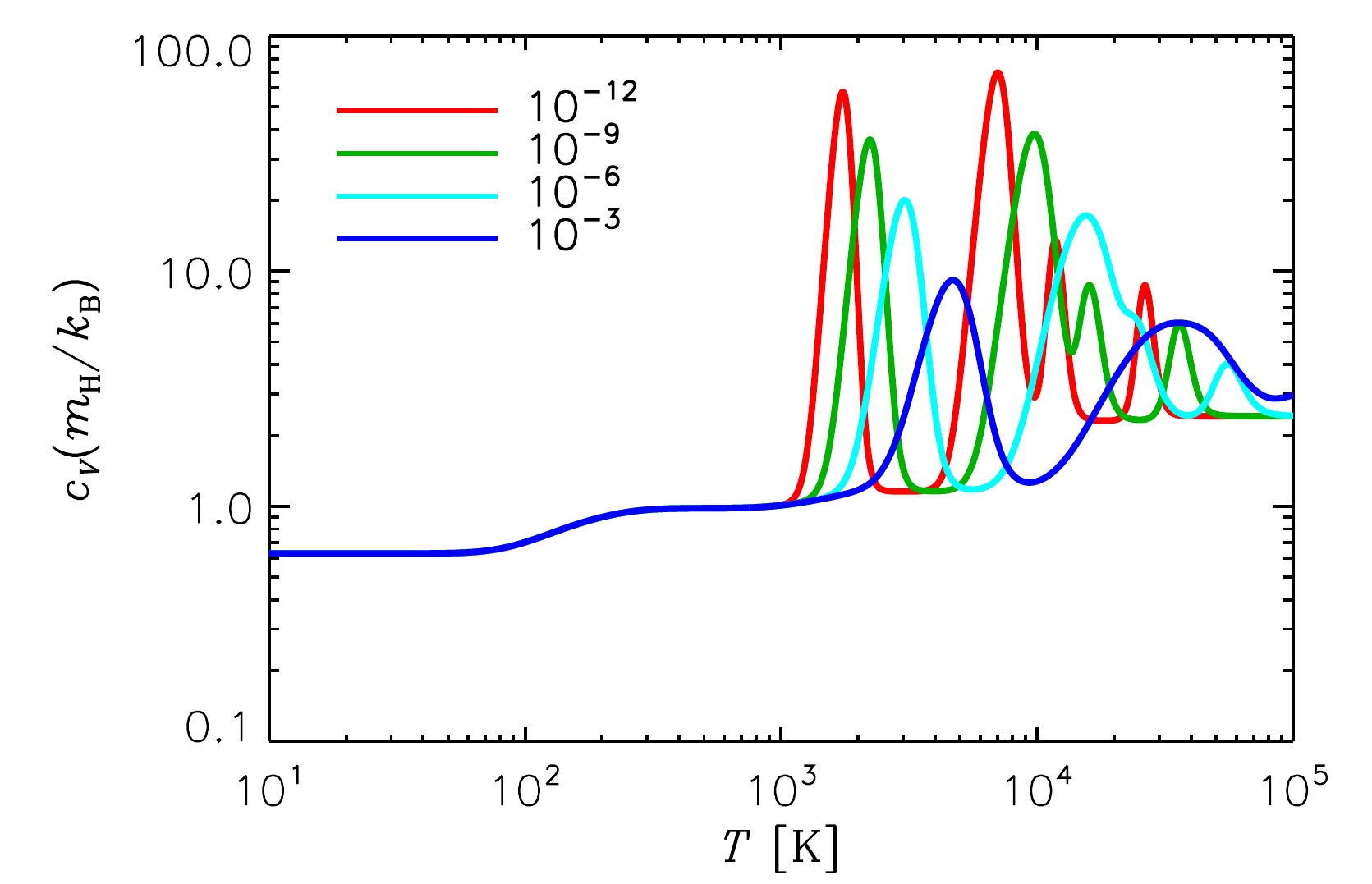}}
\resizebox{\figlen}{!}{%
\includegraphics{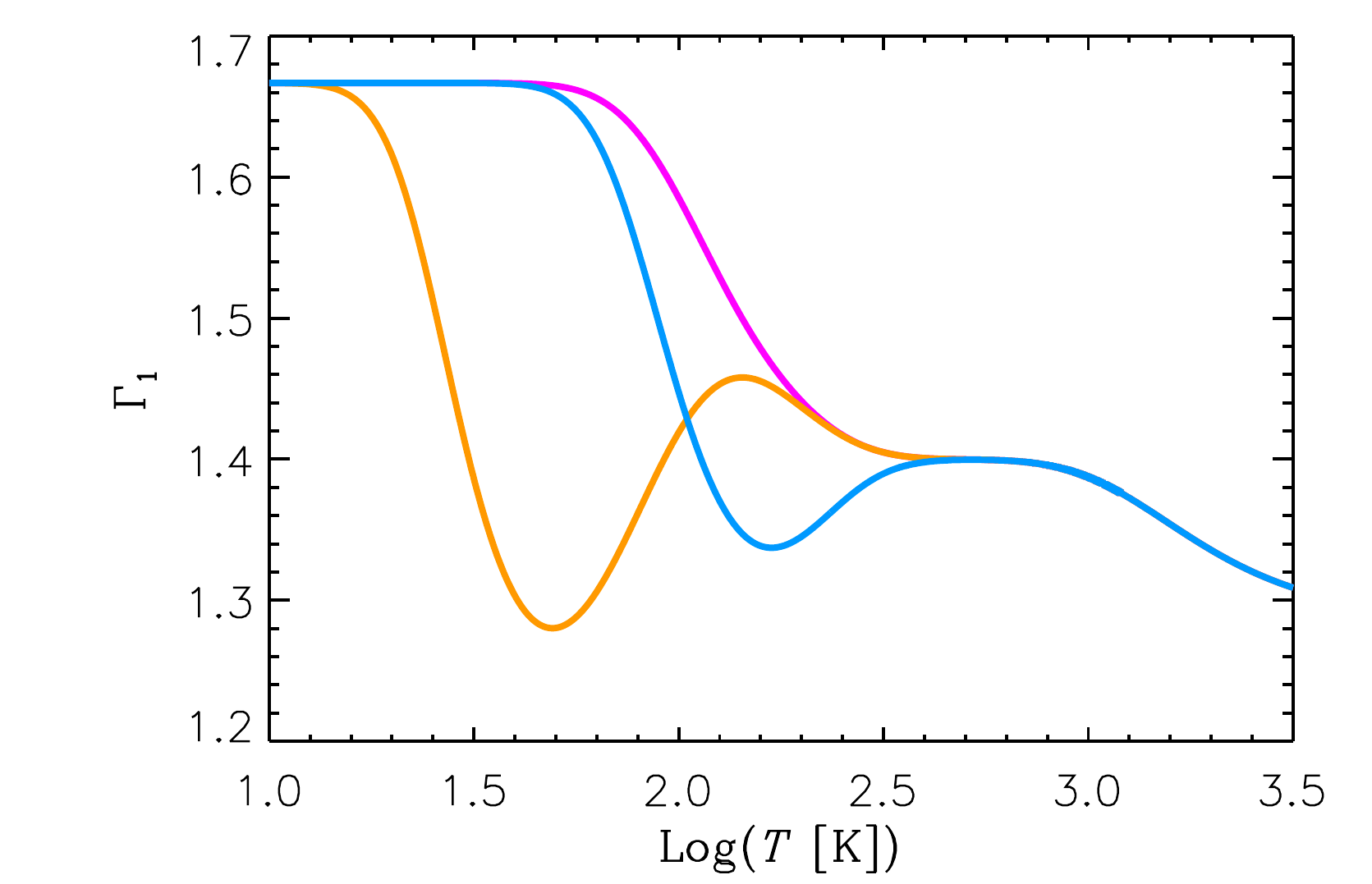}%
\includegraphics{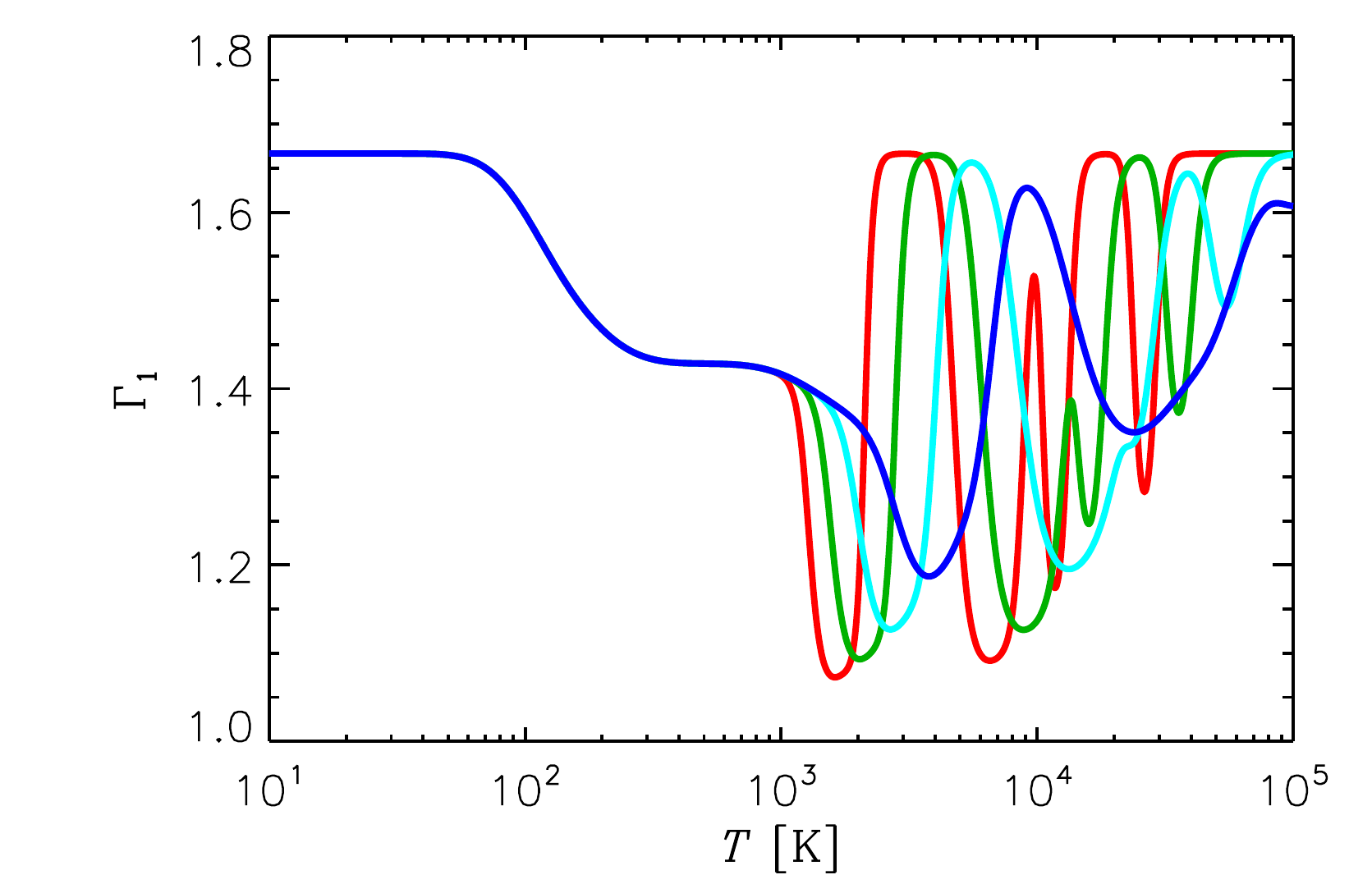}}
\resizebox{\figlen}{!}{%
\includegraphics{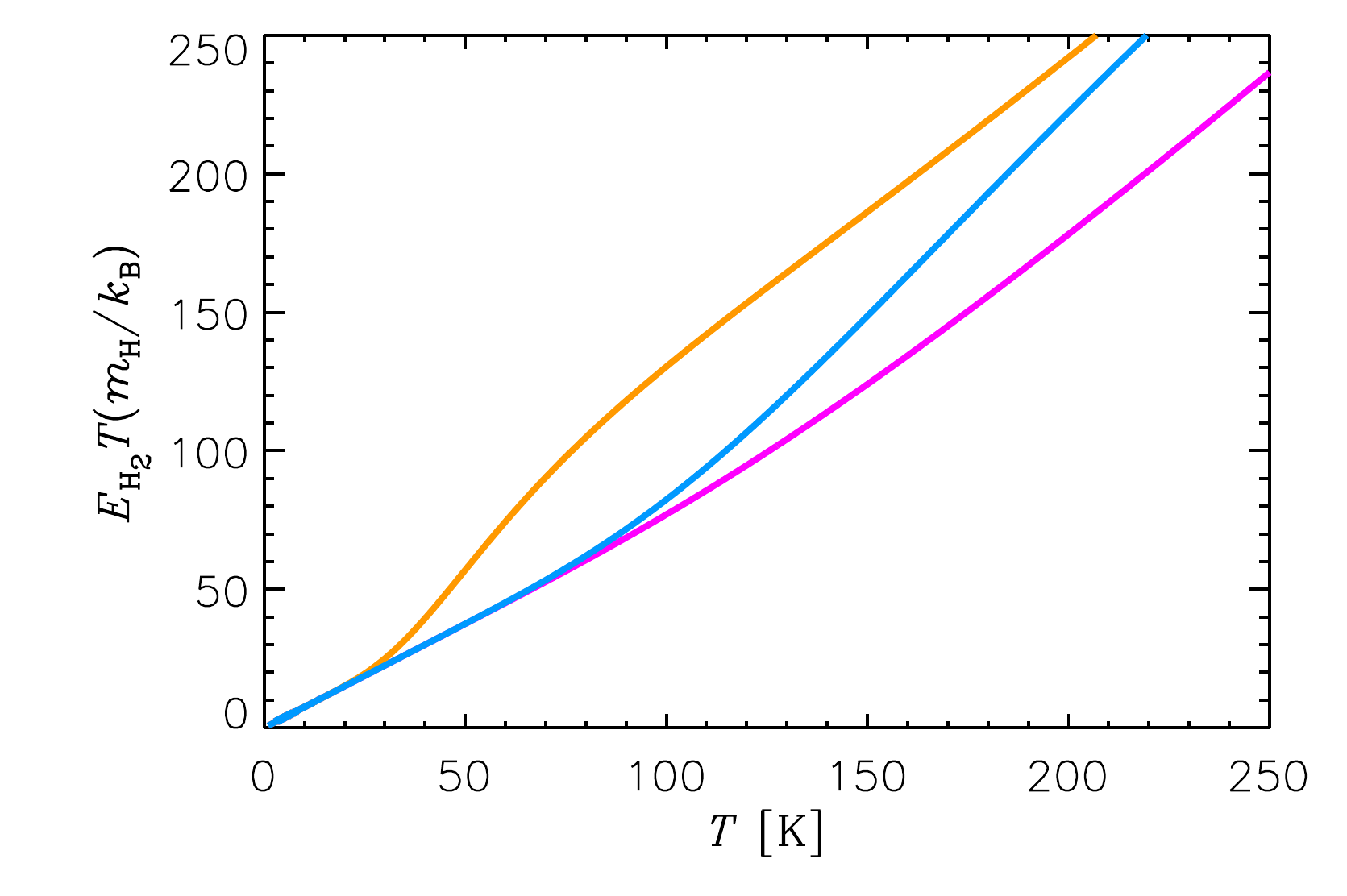}%
\includegraphics{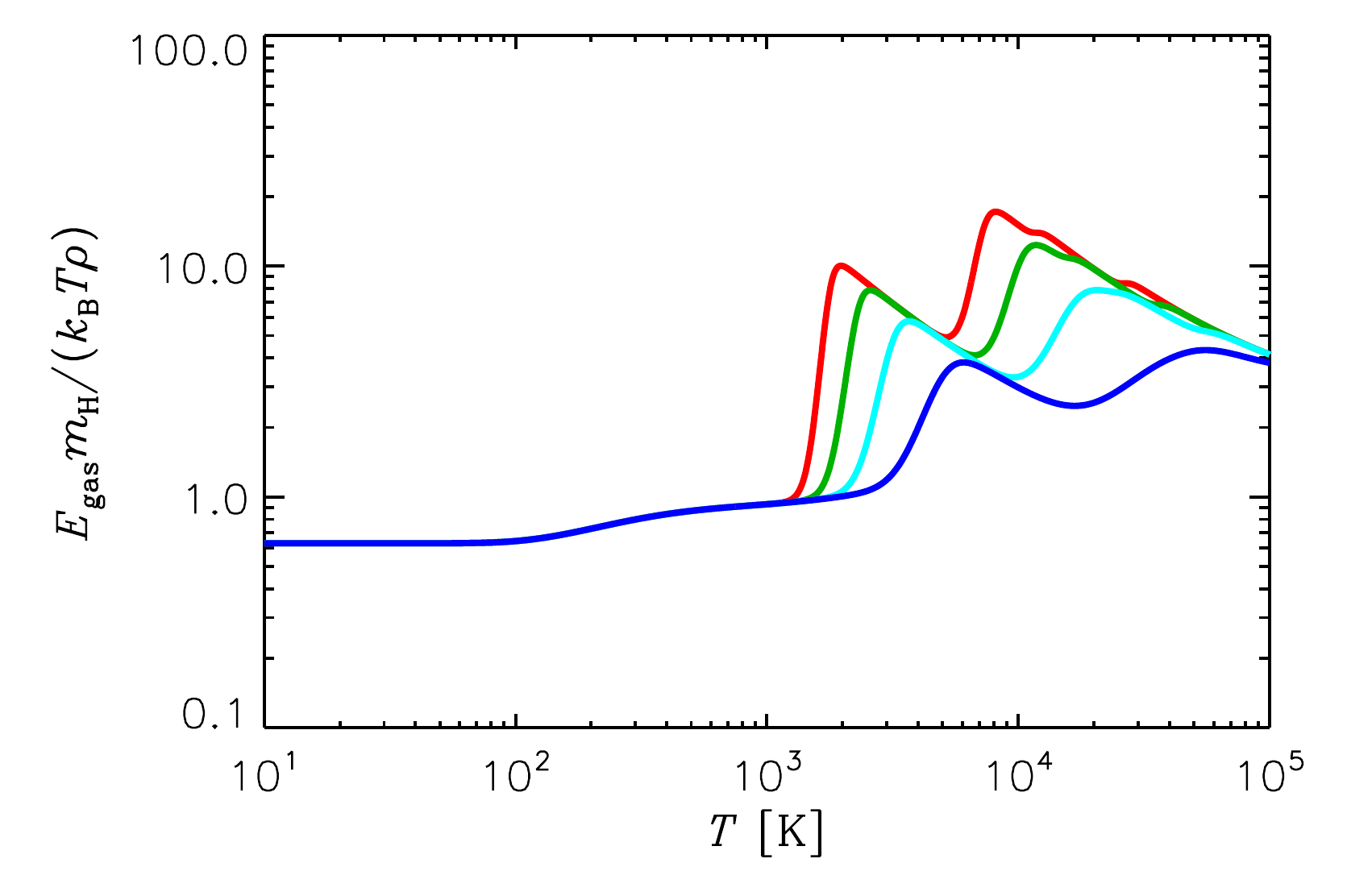}}
\caption{%
              Left: specific heat at constant volume (top), first adiabatic
              exponent (center) and specific energy times temperature (bottom) 
              of molecular hydrogen. 
              Note that both specific heat and specific energy are 
              divided by $k_{\mathrm{B}}/m_{\mathrm{H}}$. 
              The top and center panels are intended to reproduce 
              Figures~1 and 2 of
              \citet{decampli1978}, while the bottom panel replicates Figure~2
              of \citet{boley2007b}. The different line styles refer, as indicated
              in the top panel,
              to a gas mixture with a fixed $1/3$ ratio between parahydrogen 
              and orthohydrogen, a normal equilibrium mixture, and a
              parahydrogen gas.
              Right: same as for in the left panels, but for the actual gas mixture
              used in the calculations and four reference gas densities, as indicated
              in the top panel in units of $\mathrm{g\,cm^{-3}}$. The bottom panel shows the quantity
              in parenthesis on the right-hand side of Equation~(\ref{eq:Emix}).
              }
\label{fig:EoS}
\vspace{2mm}
\end{figure*}
%%%%%%%
%%%%%%%
\begin{figure}[]
\centering%
\resizebox{\linewidth}{!}{%
\includegraphics{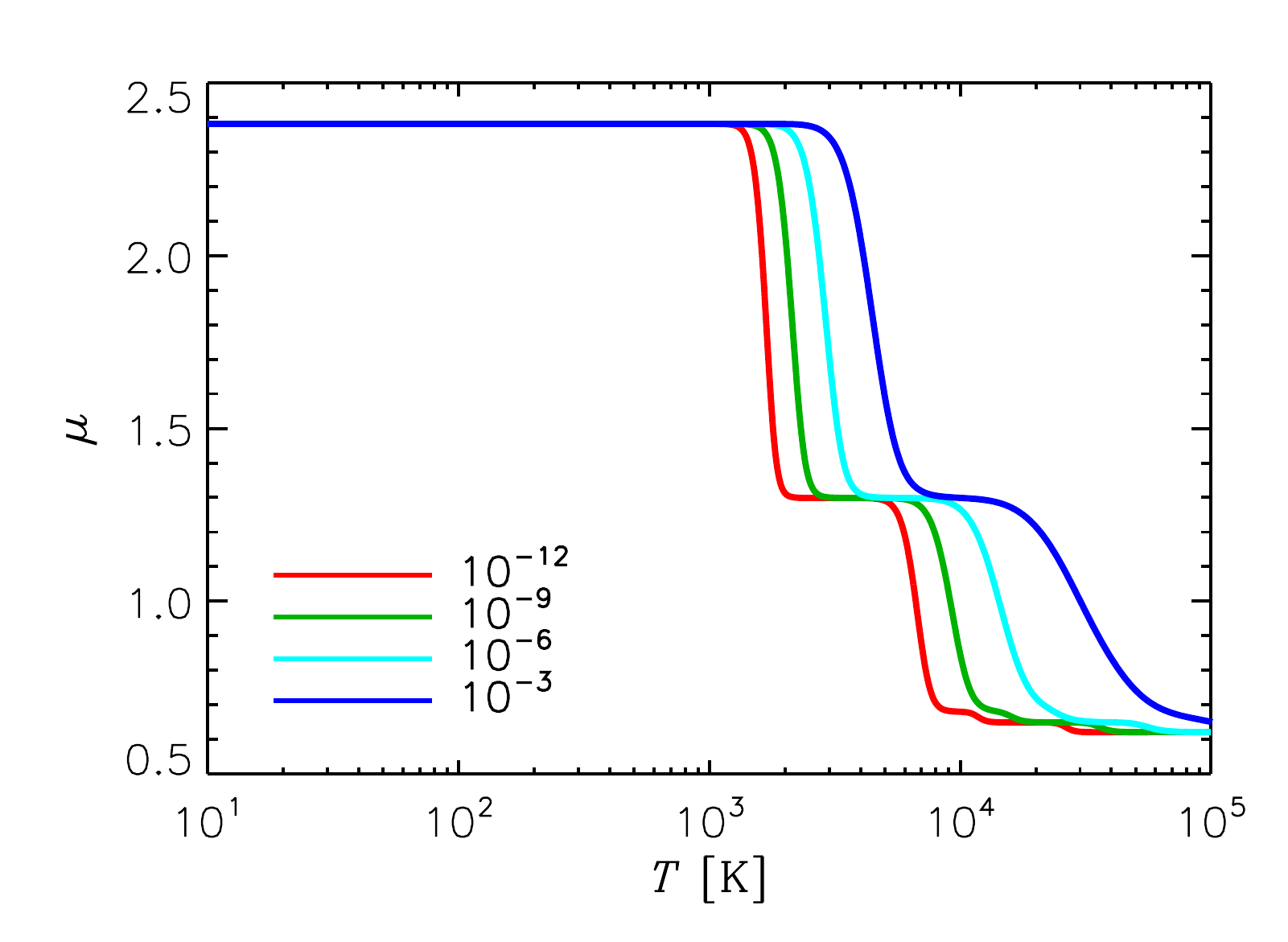}}
\caption{%
              Mean molecular weight, $\mu$, in Equation~(\ref{eq:mue}) 
              for the gas mixture used in the calculations. Four reference
              values of the gas density $\rho$ are used, as indicated in the 
              legend in units of $\mathrm{g\,cm^{-3}}$. 
              }
\label{fig:mue_sim}
\vspace{3mm}
\end{figure}
%%%%%%%
The left panels of Figure~\ref{fig:EoS} show the specific heat (top),
the first adiabatic exponent (center), and the specific energy (bottom)
of $\mathrm{H}_{2}$ (see the figure caption for further details). 
The curves in the top and center panels should be compared to those 
in Figures~1 and 2 of \citet{decampli1978}\footnote{There 
is a typo in Equation~(1) of \citet{decampli1978}, in which $\chi_{T}$ 
should be squared, as in Equation~(\ref{eq:Gamma_1})
\citep[see also][]{wuchterl1991a,hansen2004}.}.
The top panel also reproduces Figure~1 of 
\citet[there is a typo in their Equation~(11), as the leading squared 
parenthesis of the second term should not be there]{black1975}.
The curves in the bottom panel should be compared to the corresponding 
curves in Figure~2 of \citet{boley2007b}. 
The right panels of the figure show, for the actual gas mixture
used in this work, the variation with temperature 
of $c_{V}=(\partial E_{\mathrm{gas}}/\partial T)_{\rho}$ divided by 
$k_{\mathrm{B}}/m_{\mathrm{H}}$ (top), 
$\Gamma_{1}$ in Equation~(\ref{eq:Gamma_1})
(center), and $E_{\mathrm{gas}}$ in Equation~(\ref{eq:Emix})
divided by $k_{\mathrm{B}}T\rho/m_{\mathrm{H}}$ (bottom).
The first adiabatic exponent of the actual gas mixture is basically 
constant below $\approx 50\,\K$, but undergoes substantial variations 
at higher temperatures.
A detailed description of the features visible in the plot of 
$\Gamma_{1}$ is given by \citet{wuchterl1990}. 
The mean molecular weight of the gas mixture is plotted
in Figure~\ref{fig:mue_sim}. 
At densities $\rho\gtrsim 10^{-14}\,\mathrm{g\,cm^{-3}}$,
the gas mixture has $\mu=2.381$ for $T\lesssim 1000\,\K$.
At the reference densities used in the figure, full
dissociation of molecular hydrogen occurs between 
$\sim 2000$ and $\sim 10^{4}\,\K$.

%%--------------------------------------------------------------------------
\subsection{Opacity Coefficient}
\label{sec:opa}
%%--------------------------------------------------------------------------

Absorption and scattering of radiation are contributed to by both
gas and dust grains. As gas opacity, we use the Rosseland mean 
opacity tables provided by \citet{ferguson2005}, based on the protosolar 
elemental composition of \citet{grevesse1998}.
The opacity calculations of \citet{ferguson2005} include, along with
continuous opacity sources, the line opacities of $40$ atomic species 
and their ions, and of $40$ molecules.

At temperatures below $1600$--$1800\,\K$ (depending on $\rho$), 
dust grains start to dominate the opacity.
Monochromatic dust opacities are calculated from the basic scattering 
and absorption properties of (spherical) grains, following the procedures of 
\citet{pollack1985} and \citet{pollack1994}, and using the full Mie theory. 
We consider the contributions of seven different species of grains.
Details on the calculation of the dust opacity are given in 
Appendix~\ref{sec:opa_calc}.
We use a dust size distribution such that the number of grains,
as function of size, is a power-law of the grain radius with 
exponent equal to $-3$. 
The minimum and maximum radii of the size distribution
are $0.005\,\mu\mathrm{m}$ and $1\,\mathrm{mm}$, respectively. 
These values are within ranges derived from models of the spectral 
energy distributions of T~Tauri disks \citep{dalessio2001}.
For comparison purposes, a dust opacity based on the size distribution
of interstellar grains \citep{draine1984} is also presented 
in Appendix~\ref{sec:opa_calc}.
In general, the interstellar dust opacity divided by some numerical factor 
does not replicate the opacity produced by a size distribution with larger 
grains.

Dust and gas opacities are blended, using a linear interpolation, 
over a temperature interval around the highest vaporization temperature 
of the various grain species. The width of the interval is approximately
$20$\% of said temperature.

%%--------------------------------------------------------------------------
\section{Numerical Procedures}
\label{sec:NP}
%%--------------------------------------------------------------------------

Equations~(\ref{eq:H_r}) through (\ref{eq:H_phi}) are solved
by means of a finite-difference code 
\citep{gennaro2002,gennaro2003a,gennaro2005}.
The solution is obtained in a stepwise fashion 
\citep[see, e.g.,][]{stone1992a}.
The advection part of the equations is solved by using
an operator-splitting technique and then by applying the 
second-order monotonic transport of \citet{vanleer1977} 
to the split operators.
The solution is subsequently updated by taking into account the
terms on the right-hand side of the equations. 
The terms involving the forces/torques per unit volume
imparted to the gas by the radiation field are applied after
updating the radiation energy density, as explained below.

Equations~(\ref{eq:dEgas}) and (\ref{eq:dErad}) are also
integrated in a stepwise fashion.
Instead of advecting separately $E_{\mathrm{gas}}$ and
$E_{\mathrm{rad}}$, the code performs the advection
of the total energy density $E$, that is, it integrates the
left-hand side of Equation~(\ref{eq:dEtot}), using the same 
technique as for the advection of the linear and angular momenta. 
The equations
\begin {equation}
\label{eq:dEgasnum}
 \ddt{E_{\mathrm{gas}}} =%
  - P_{\mathrm{gas}}\gdiv{\gvec{u}} + \Psi + \varepsilon,
\end{equation}
and
\begin {equation}
\label{eq:dEradnum}
 \ddt{E_{\mathrm{rad}}} =%
  - \gdiv{\gvec{F}} - P_{\mathrm{rad}}\gdiv{\gvec{u}},
\end{equation}
are then integrated separately in multiple steps. 
In order to do so, however, the energy densities
$E_{\mathrm{gas}}$ and $E_{\mathrm{rad}}$ must be
obtained from the total energy density, $E$. 
For this purpose, we introduce the quantity $\Upsilon$, defined by
\begin {equation}
\label{eq:Ups}
\left(\frac{1}{\Upsilon-1}\right)=%
                 \left(\frac{\mu m_{\mathrm{H}}}{\rho k_{\mathrm{B}}T}\right)%
                 E_{\mathrm{gas}},
\end{equation}
where $E_{\mathrm{gas}}$ is given by Equation~(\ref{eq:Emix})
and the mean molecular weight $\mu$ by Equation~(\ref{eq:mue}).
We then express the total internal energy density as the following sum
\begin {equation}
\label{eq:Etot}
 E=\left(\frac{1}{\Upsilon^{*}-1}\right)\frac{\rho k_{\mathrm{B}}}{\mu m_{\mathrm{H}}}T
   +\left(\frac{4\sigma_{\mathrm{SB}}}{c}\right)T^{4},
\end{equation}
in which $\Upsilon^{*}$ is the quantity $\Upsilon$ 
computed during the previous time step at any point in space. 

Since $E$ is known after the advection step in 
Equation~(\ref{eq:dEtot}), Equation~(\ref{eq:Etot})
represents a fourth-order polynomial (sometimes referred 
to as a quartic) in $T$, whose roots can be found analytically.
The solution of this equation proceeds first by transforming 
the quartic into the so-called \textit{auxiliary cubic},
using Ferrari's formulae, and then by solving the cubic 
equation using the formulae of Cardano-Tartaglia.
Procedures to find the only physically acceptable root
are given in Appendix~\ref{sec:QS}.
Once the temperature is determined, the energy densities
$E_{\mathrm{gas}}$ and $E_{\mathrm{rad}}$ are also known,
and so are the pressures 
$P_{\mathrm{gas}}$ and $P_{\mathrm{rad}}$. Thus, one
can solve separately Equations~(\ref{eq:dEgasnum}) and 
(\ref{eq:dEradnum}), and eventually compute the updated
total energy density, $E$.
Equation~(\ref{eq:Etot}) is also solved to find the total pressure,
$P=P_{\mathrm{gas}}+P_{\mathrm{rad}}$, 
for the evaluation of the right-hand side of Equation~(\ref{eq:Dtlim})
below.

Momenta and energy equations are written in a covariant
form \citep{stone1992a}. This formalism allows for the solution
of these equations is cartesian, cylindrical, and spherical polar 
coordinates.

A numerical stability analysis \citep[e.g.,][]{pressF1992} shows
that any explicit solution of Equations~(\ref{eq:H_r}), (\ref{eq:H_theta})
(\ref{eq:H_phi}), (\ref{eq:dEgas}), and (\ref{eq:dErad})
is only conditionally stable, and as such is subject to a restriction 
on the size of the marching time step
\citep[the Courant-Friedrichs-Lewy condition, see e.g.,][]{stone1992a}. 

Let us indicate with $\Delta S$ the minimum of the lengths
$\Delta r$, $r\Delta\theta$, and $r \sin{\theta}\Delta\phi$, 
over the grid. 
The limiting time step $(\Delta t)_{\mathrm{CFL}}$ 
that assures stability is such that
\begin {eqnarray}
\left(\frac{1}{\Delta t}\right)^{2}_{\mathrm{CFL}}&=&%
\max\left[ \left(\frac{u_{r}}{\Delta r}\right)^{2},%
                 \left(\frac{u_{\theta}}{r\Delta\theta}\right)^{2},%
                 \left(\frac{u_{\phi}}{r\sin{\theta}\Delta\phi}\right)^{2},%
         \right. \nonumber\\%
         &  &\left. %
                 \frac{\bar{\Gamma}_{1}(P/\rho)}{(\Delta S)^{2}},%
                 \frac{36\nu^{2}}{(\Delta S)^{4}} \label{eq:Dtlim}
         \right],
\end{eqnarray}
where $\bar{\Gamma}_{1}=\max{(4/3,\Gamma_{1})}$.
The first three terms on the right-hand side are imposed by advection,
the fourth term by the propagation of acoustic waves \citep{turner2001},
and the last by the (physical) viscous diffusion
\citep[artificial viscosity, not applied here, 
would add another term, see][]{stone1992a}.
The ratio of $(\Delta t)_{\mathrm{CFL}}$ to the actual time step,
the Courant number, varies between $2$ and $5$, and is typically set 
to $2$ in these calculations.

%%--------------------------------------------------------------------------
\subsection{Radiation Diffusion Solver}
\label{sec:RDS}
%%--------------------------------------------------------------------------

The radiation diffusion part of Equation~(\ref{eq:dErad}) would
impose a term in Equation~(\ref{eq:Dtlim}) of order 
$\mathcal{D}^{2}/(\Delta S)^{4}$, which would be much larger than 
all other terms in many practical situations. 
In fact, if we consider a typical accretion disk at a few \AU\ from the star,
$u^{2}_{\phi}\sim 10^{12}\,\mathrm{cm^{2}\,s^{-2}}\gg u^{2}_{r}, u^{2}_{\theta}$,
$\nu\sim 10^{15}\,\mathrm{cm^{2}\,s^{-1}}$,
$\rho\sim 10^{-10}\,\mathrm{g\,cm^{-3}}$,
and 
$\mathcal{D}\approx c/(\rho\kappa)\gtrsim 10^{20}\,\mathrm{cm^{2}\,s^{-1}}$
if $\kappa \lesssim 1\,\mathrm{cm^{2}\,g^{-1}}$.
Therefore, $\mathcal{D}\gg \nu$ and
$\mathcal{D}^{2}/(\Delta S)^{2}\gg u^{2}_{\phi}$ when
$\Delta S\ll 10^{14}$, or $\Delta\phi\ll 1$, which is typically the case.
Only in a very dense and opaque gas, radiation transfer in 
the flux-limited diffusion approximation may be treated explicitly.

We approximate Equation~(\ref{eq:dEradnum}) as
\begin {eqnarray}
 \frac{E_{\mathrm{rad}}-E^{*}_{\mathrm{rad}}}{\Delta t} & = &%
  \mathcal{D}^{*}\nabla^{2}E_{\mathrm{rad}}+%
  (\nabla\mathcal{D}^{*})\!\boldsymbol{\cdot}\!(\nabla{E_{\mathrm{rad}}})\nonumber\\
 & - & E_{\mathrm{rad}}\frac{\gdiv{\gvec{u}^{*}}}{3} \label{eq:dEradimp},
\end{eqnarray}
and solve it implicitly. The first and second spatial derivative 
operators, $\nabla$ and $\nabla^{2}$, here are intended as 
centered differencing operators, written in covariant form for
integration in cartesian, cylindrical, and spherical polar coordinates.
Explicit expressions for these two operators can be found in \citet{stone1992a}.
Quantities marked with an asterisk represent known values
(from a previous step).
Notice that the spatial discretization can also be applied directly 
to the divergence of the flux in Equation~(\ref{eq:dEradnum}) or,
alternatively, this term can be discretized by exploiting the divergence theorem.

The time differentiation in Equation~(\ref{eq:dEradimp}) follows the 
backward Euler method, which is first-order accurate in time. 
Second-order accuracy can be obtained by performing the time 
differentiation according to the Crank-Nicolson method 
\citep[e.g.,][]{pressF1992}, that is, by replacing $E_{\mathrm{rad}}$
on the right-hand side with the time average
$(E_{\mathrm{rad}}+E^{*}_{\mathrm{rad}})/2$.
Both implicit methods are unconditionally stable, but 
the Crank-Nicolson differentiation can be prone to oscillations
in the presence of rapid transients, whereas the backward Euler 
differentiation is not.
One strategy to retain the second-order accuracy in time, 
but mitigate possible spurious oscillations, is to alternate between
these two differentiation schemes \citep{britz2003}. 
We typically perform
a ``backward Euler'' time step every five ``Crank-Nicolson'' time steps.

Regardless of the  time differentiation scheme, 
Equation~(\ref{eq:dEradimp}) can be expressed 
through the linear system
\begin{equation}
\label{eq:Axb}
\mathbf{A}\gvec{x}=\gvec{b}
\end{equation}
of $N$ equations in $N$ unknowns, where each unknown is 
the value of $E_{\mathrm{rad}}$ at a grid point and $N$ is the
total number of grid points. Note that Equation~(\ref{eq:Axb})
bears no recollection of the number of dimensions in the physical 
problem, but in a 3D problem, $N$ can very easily 
reach beyond $10^{6}$!
Applying the backward Euler or Crank-Nicolson differentiation 
changes the form of the right-hand side $\gvec{b}$, 
but it alters the elements of the matrix $\mathbf{A}$ only 
by numerical factors of $1/2$.

The matrix of the linear system coefficients, $\mathbf{A}$, is 
sparse. In fact, it has at most seven non-zero elements per
row (using a second-order accurate differentiation in space). 
There are various strategies to invert the matrix $\mathbf{A}$
and solve Equation~(\ref{eq:Axb}), including direct and
iterative solvers.
Direct solvers for sparse linear systems, which typically use 
some version of Gaussian elimination, 
have become quite competitive over the past decade and are 
known for their robustness and accuracy (they should deliver 
an exact solution within round-off errors). However, they still 
suffer from large memory storage requirements and lack of 
performance when applied to large (e.g., 3D) 
problems \citep{gutknecht2006}.
In fact, the direct solution of a linear system is generally a 
$\mathcal{O}(N^{3})$ process for dense matrices.
The occurrence of sparse matrices may not improve
performance significantly, as efficient handling of sparse matrices 
involves complex algorithms, which entail a substantial  
computational overhead \citep{demmel2000}.

We apply two classes of iterative solvers for sparse and non-symmetric 
linear systems, referred to as Krylov subspace solvers, which provide 
an approximation $\gvec{\tilde{x}}$ to the solution $\gvec{x}$.
The first class is a generalization of the 
Bi-Conjugate Gradient Stabilized method 
\citep{sleijpen1993,vandervorst2003}, abbreviated as BiCGStab($l$),
where $l$ is the degree of the Minimal Residual Polynomials
\citep[see][]{sleijpen1993}. This solver does not suffer from some of 
the breakdowns of the BiCGStab algorithm and typically delivers 
better convergence performance \citep{vandervorst2003}.
The second class is a variant of the Generalized Minimal Residual 
method \citep[known as GMRES, see][]{saad2003} 
introduced by \cite{vandervorst1994} and abbreviated as GMRESR.
This is actually a family of recursive schemes, which may provide 
a considerable improvement over other variants of GMRES methods 
in terms of memory requirements and computing efficiency.
Both methods are widely used to solve large sparse linear systems, 
such as those arising from the discretization of partial differential 
equations (variants of these methods are also available in 
commercial computational softwares, such as \textit{Mathematica}
and \textit{MATLAB}).
We refer to the cited literature, and references therein, for an in-depth 
description of the mathematical properties and implementation aspects 
of these solvers.

The reason for using two different classes of solvers is that,
depending on the mathematical and structural properties 
of $\mathbf{A}$, it is known that one type of solver may succeed 
where the other may fail.
Both solvers perform an educated search of characteristic vector 
spaces (the Krylov subspaces) of increasing dimension 
in an attempt to minimize the residual $\gvec{b}-\mathbf{A}\gvec{x}$.
There are local and global criteria to establish whether or not 
convergence is achieved. We choose a global relative criterion
based on the $L^{2}$-norm, so that the approximate solution
satisfies the inequality
\begin{equation}
\label{eq:L2norm}
\|\gvec{b}-\mathbf{A}\gvec{\tilde{x}}\|\le \eta\|\gvec{b}\|,
\end{equation}
where the relative tolerance $\eta$ has a minimum value of 
$10^{-5}$ and a maximum of $0.01$.
These numbers are a result from direct numerical experiments
on the actual problems dealt with here and are a compromise 
between accuracy and computational effort. 
At each time step, a solution of Equation~(\ref{eq:Axb}) is
attempted with the BiCGStab($2$) solver. 
If convergence within the minimum tolerance is not reached
and the approximate solution achieved within the maximum 
number of iterations (typically $\sim 1000$) returns a relative
tolerance $\eta >0.01$, a solution is attempted with the BiCGStab($4$) 
solver. If again the solution does not satisfy the imposed 
requirements, a solution is attempted with the GMRESR
solver. If also the last attempt fails, the maximum number of iterations 
is raised until it is no longer convenient to continue the calculation. 
We find that the GMRESR solver is typically very robust\footnote{%
The full Generalized Minimal Residual method
\citep[i.e., the one not re-started after each cycle of a fixed number 
of iterations, see][]{saad2003},
is guaranteed to deliver the exact solution, within round-off errors, 
in a maximum of $N$ iterations.}, 
but it is also the slowest of our Krylov subspace solvers.

Since both classes of iterative solvers are very general, neither can
take advantage of the structural properties of $\mathbf{A}$ 
to improve robustness and expedite convergence.
A way around this drawback is to apply a \textit{preconditioner}, that
is, a matrix $\mathbf{P}$ such that $\mathbf{P}^{-1}$ is a ``good''
approximation to $\mathbf{A}^{-1}$ and so that the structural
properties of the product matrix $\mathbf{P}^{-1}\mathbf{A}$ 
allow for an easier solution (in terms of computational effort) 
of the linear system 
$\mathbf{P}^{-1}\mathbf{A}\gvec{x}=\mathbf{P}^{-1}\gvec{b}$,
which clearly admits the same solution as Equation~(\ref{eq:Axb}).
In this case, $\mathbf{P}$ is referred to as a left-preconditioner.

In the words of Yousef Saad\footnote{Yousef Saad and Martin Schultz
introduced the Generalized Minimal Residual method in 1986.}
(\citeyear{saad2003}):
``Finding a good preconditioner to solve a given sparse linear system 
is often viewed as a combination of art and science.''
We implemented and tested a Jacobi preconditioner, in which
$\mathbf{P}$ is a diagonal matrix whose elements are the diagonal
elements of $\mathbf{A}$. This is among the simplest of all 
preconditioners, it is inexpensive to build 
and it can work effectively as long as $N$ is ``small'' 
(this limitation can actually be shown mathematically).
Another, more complex and efficient preconditioner 
we implemented is the Incomplete LU
(ILU) factorization\footnote{The letters
``L'' and ``U'' stand for lower-triangular and upper-triangular matrices.} 
\citep{saad2003,vandervorst2003}.
This preconditioner was constructed starting from the
properties of the discretized Equation~(\ref{eq:dEradimp}),
which generates a coefficient matrix with a regular structure, 
using the strategy outlined by \citet{saad2003}. 
The ILU preconditioner proves to be very effective, leading to convergence in
a number of iterations considerably smaller than that
necessary for the convergence of the non-conditioned system
(see comments in Appendix~\ref{sec:diff_test}).
However, the construction of an ILU preconditioner requires
a substantial computational overhead. 
It is also important to bear in mind that the solution of
the preconditioned system, $\gvec{\hat{x}}$, minimizes the 
residual $\mathbf{P}^{-1}(\gvec{b}-\mathbf{A}\gvec{x})$.
Therefore, the solution satisfies the inequality 
$\|\mathbf{P}^{-1}(\gvec{b}-\mathbf{A}\gvec{\hat{x}})\|<%
\eta\|\mathbf{P}^{-1}\gvec{b}\|$, 
but non necessarily the inequality
$\|\gvec{b}-\mathbf{A}\gvec{\hat{x}}\|<\eta\|\gvec{b}\|$.
Therefore, some precautions are needed when applying
left-preconditioners, and preconditioners in general  
\citep[see discussion in][]{vandervorst2003}. 
We typically solve a preconditioned system while also gathering 
information on the solution of the non-conditioned system.
In a production run that uses the preconditioner, some portions of the 
calculation are performed without preconditioner so that solutions can 
be compared and convergence monitored.
Except for testing purposes, the Jacobi preconditioner is
rarely used in production runs.

In Appendix~\ref{sec:diff_test}, we present some tests
of the iterative linear solvers mentioned here, applied to 
diffusion and radiative transfer problems. 
The BiCGStab($l$) solver, with $l=2$ and $4$, and the
GMRESR solver are tested separately on the
same problems. The Jacobi and ILU preconditioners are 
also tested in these numerical experiments.
The tests we perform include the the standard 
diffusion of pulses, the stationary problems proposed
by \citet{boss2009}, and the ``relaxation'' problems
proposed by \citet{boley2007a}. We also derive solutions
to the diffusion equation and test the solvers against
these solutions. Furthermore, we perform tests of the
streaming limit by calculating the propagation of 
fronts at the speed of light \citep[e.g.,][]{turner2001}.

We opt here to remove the pressure work term from 
Equation~(\ref{eq:dEradimp}). Once $E_{\mathrm{rad}}$
is updated through the solution of Equation~(\ref{eq:Axb}),
the updated radiation flux $\gvec{F}$ is used to compute the new
momenta in Equations~(\ref{eq:H_r}), (\ref{eq:H_theta}), and 
(\ref{eq:H_phi}). The updated value of the divergence 
$\gdiv{\gvec{u}}$ is
then used to correct $E_{\mathrm{rad}}$ (and $E_{\mathrm{gas}}$)
by integrating the radiation (and gas) pressure work \citep{gennaro2003b}.

%%--------------------------------------------------------------------------
\subsection{Nested-grid Structure}
\label{sec:NGS}
%%--------------------------------------------------------------------------

All equations are discretized over a spherical polar grid with 
constant spacing in the three coordinate directions.
We apply a nested-grid refinement technique 
\citep{yorke1995,gennaro2002,gennaro2003a} that increases
the volume resolution by a factor of $2^{3}$ for any grid level
added to the grid structure.
The integration cycle requires the equations in Section~\ref{sec:DET}
to be solved independently on any grid level and information
to be exchanged between neighboring grids.
In this study, we employ either $9$ ($a=5\,\AU$) or $10$ 
($a=10\,\AU$) grid levels. The basic
level contains $103\times 44\times 423$ zones, in $r$, $\theta$,
and $\phi$, respectively, whereas refinement levels contain all
$64\times 64\times 64$ zones. 
Any such level needs to be integrated twice for each integration 
of the next coarser level, which means that level $10$ is integrated 
$512$ as many times as level $1$.

The grid spacings on the basic grid level are such that
$\Delta r/a= a\,\Delta \phi/(r\sin{\theta})= 0.015$
and $a\,\Delta \theta/r= 0.003$.
The linear resolution at the top-most level is a factor 
$256$ or $512$ as high. In physical units, the average grid spacing 
varies between $1.3\,R_{c}$ and $1.9\,R_{c}$,
where the core radius $R_{c}$ varies in the range from 
$\approx 1.34\times10^{9}$ ($\Mc=5\,\Mearth$) to 
$\approx 1.9\times10^{9}\,\mathrm{cm}$ ($\Mc=15\,\Mearth$).

The boundary conditions at the inner and outer radius of the disk 
are handled using the procedure of \citet{devalborro2006}, extended
to gas and radiation energy densities. Reflective boundary conditions 
are applied at the disk surface and mirror symmetry is imposed at the 
equatorial plane. 
For the solution of the linear system in Equation~(\ref{eq:Axb}),
periodicity in the azimuthal direction and symmetry at the equatorial plane 
are directly imposed through the definition of the elements of the coefficient 
matrix $\mathbf{A}$. The boundary conditions on refinement levels are 
interpolated from coarser grids 
\citep[see ][and references therein]{gennaro2003a}

%%--------------------------------------------------------------------------
\section{Protoplanetary Disk Structure}
\label{sec:DS}
%%--------------------------------------------------------------------------

%%%%%%
\begin{figure*}[]
\centering%
\resizebox{\figlen}{!}{%
\includegraphics[clip]{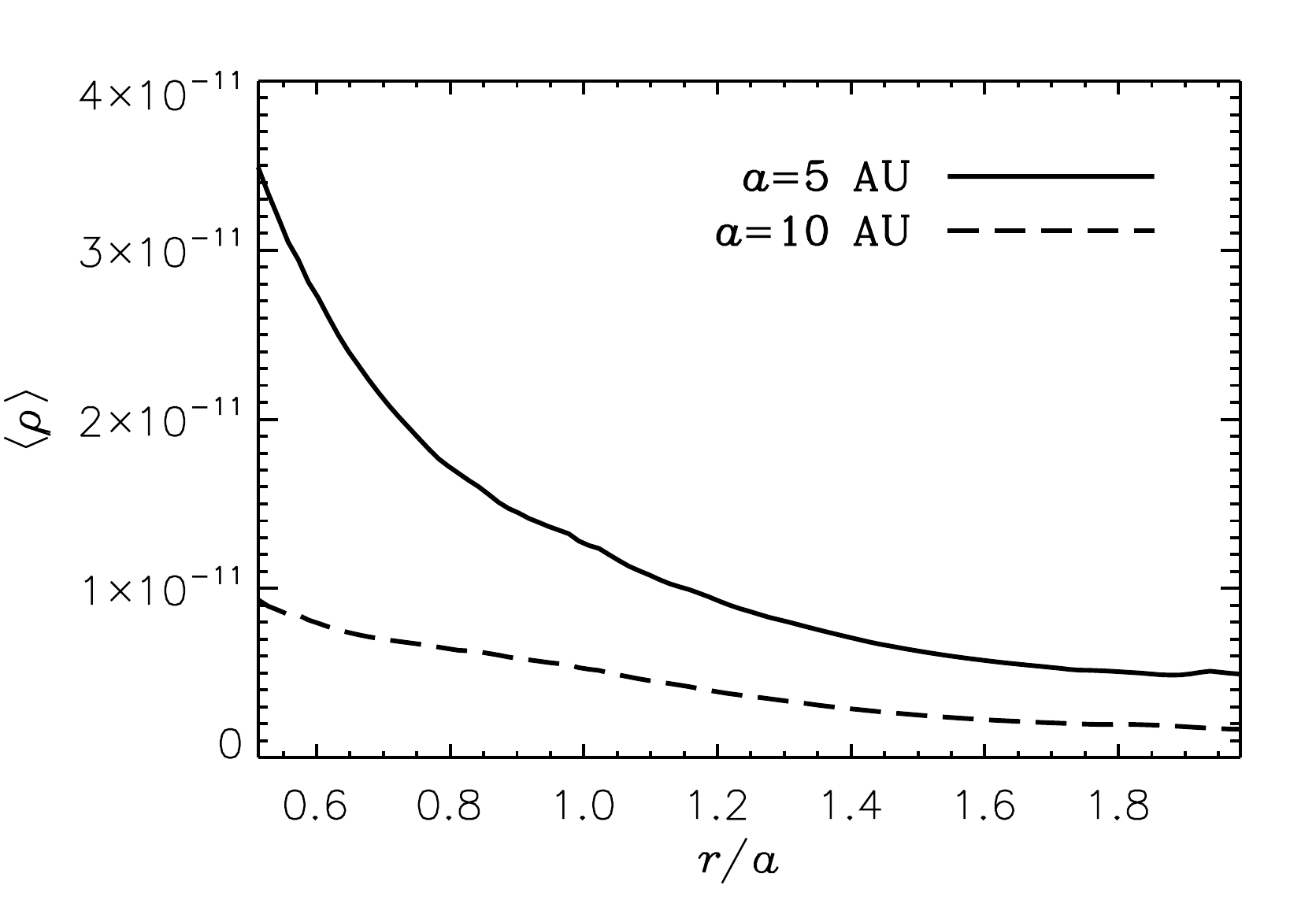}%
\includegraphics[clip]{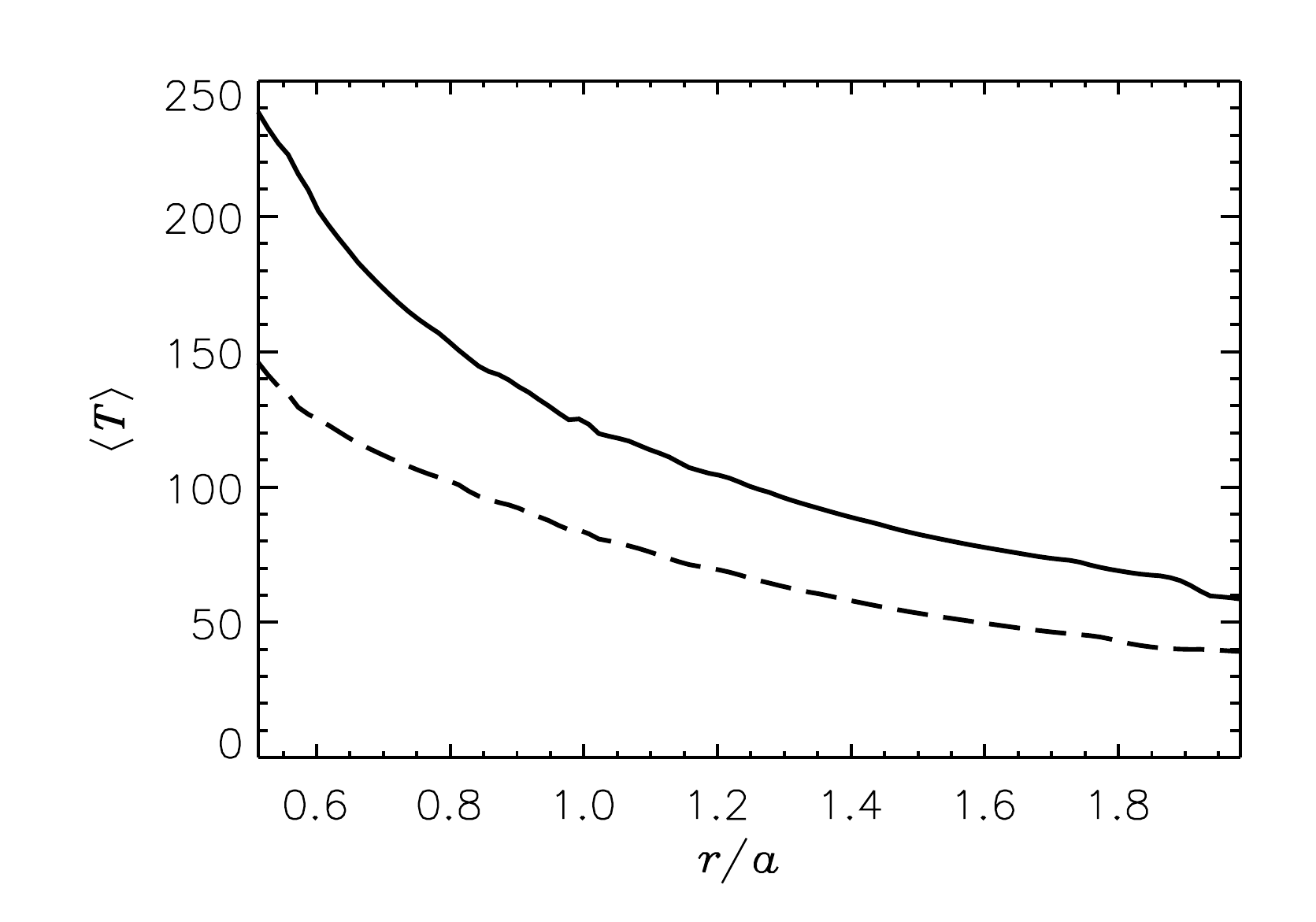}}
\caption{%
             Azimuthally averaged density in units of $\mathrm{g\,cm}^{-3}$
             (left) and temperature in units of $\K$ (right), in the disk's 
             equatorial plane ($\theta=\pi/2$), for planetary cores located 
             at $a=5\,\AU$ (solid line) and $10\,\AU$ (dashed line). 
             These results refer to the thermodynamics state 
             of quasi-equilibrium reached by each disk.
             }
\label{fig:av_disk}
\vspace{2mm}
\end{figure*}
%%%%%%%
%%%%%%%%%
\begin{deluxetable}{cccccc}
\tablecolumns{6}
\tablewidth{0pc}
\tablecaption{Azimuthally Averaged Disk Properties\label{table:dgc}}
\tablehead{
\colhead{$a$\tablenotemark{a}}&\colhead{$\Sigma$\tablenotemark{b}}&
\colhead{$\rho$\tablenotemark{c}}&%
\colhead{$T$\tablenotemark{c}}&\colhead{$\mu$\tablenotemark{c}}&%
\colhead{$\Gamma_{1}$\tablenotemark{c}}
}
\startdata
 $5$ & $120$ & $1.3\times 10^{-11}$ & $124$ & $2.38$ & $1.55$\\
$10$ & $130$ & $5.0\times 10^{-12}$ &  $80$ &  $2.38$ & $1.64$
\enddata
\tablenotetext{a}{Core's orbital radius in \AU.}
\tablenotetext{b}{Gas surface density in $\mathrm{g\,cm^{-2}}$.}
\tablenotetext{c}{Mid-plane quantities, in cgs units where applicable.}
\end{deluxetable}
\vspace{2mm}
%%%%%%%%%

We use two sets of initial conditions for the disks embedding the
planetary cores at $a=5$ and $10\,\AU$.
In both, the initial surface density is of the type $\Sigma\propto 1/\sqrt{r}$
\citep[][]{davis2005}
and the initial values are $142$ and $170\,\mathrm{g\,cm^{-2}}$
at $a=5$ and $10\,\AU$, respectively. The initial temperatures
at those orbital distances are, respectively, $110$ and $95\,\K$.
The kinematic viscosity, in units of $a^{2}\Omega$, 
is given by $\nu=4\times 10^{-6}\,\sqrt{r/a}$. For a local isothermal 
disk with no radial velocity stratification (i.e., $d u_{r}/d \theta=0$), 
this condition implies an initial steady-state with respect to the viscous
evolution since $\nu\Sigma$ is constant in radius.

The disk-planet systems are evolved until they settle into a thermodynamics
state of quasi-equilibrium.
The evolved disk mass in the models extending from $2.5$ to $10\,\AU$ 
($a=5\,\AU$) is $\approx 0.0035\,\Ms$, with $\Ms=1\,\Msun$, and 
the azimuthally averaged surface density at $5\,\AU$ is 
$\approx 120\,\mathrm{g\,cm^{-2}}$.
The evolved disk mass in the models extending from $5$ to $20\,\AU$ 
($a=10\,\AU$) is $\approx 0.015\,\Ms$, and the averaged surface 
density at $10\,\AU$ is $\approx 130\,\mathrm{g\,cm^{-2}}$.
These density values would correspond to those of a
$\sim 5\times 10^{5}$ to $\sim 10^{6}$ years old disk, whose initial 
mass (within $\sim 100\,\AU$ of the star) was between $\sim 0.04$ and 
$\sim 0.08\,\Msun$ and whose initial density at $1\,\AU$ was 
between $\sim 2000$ and $\sim 4000\,\mathrm{g\,cm^{-2}}$ 
\citep{gennaro2012}.
The formation of a planetary core with a mass between $5$ and 
$15\,\Mearth$ requires a surface density of solids between 
$\approx 6$ and $\approx 13\,\mathrm{g\,cm^{-2}}$ at $5\,\AU$
and on the order of a few to several $\mathrm{g\,cm^{-2}}$ at $10\,\AU$ 
\citep{lissauer1987,pollack1996}, consistent with the gas-augmented 
initial surface density in these disks.
%%%%%%
\begin{figure*}[]
\centering%
\resizebox{\afiglen}{!}{%
\includegraphics[clip]{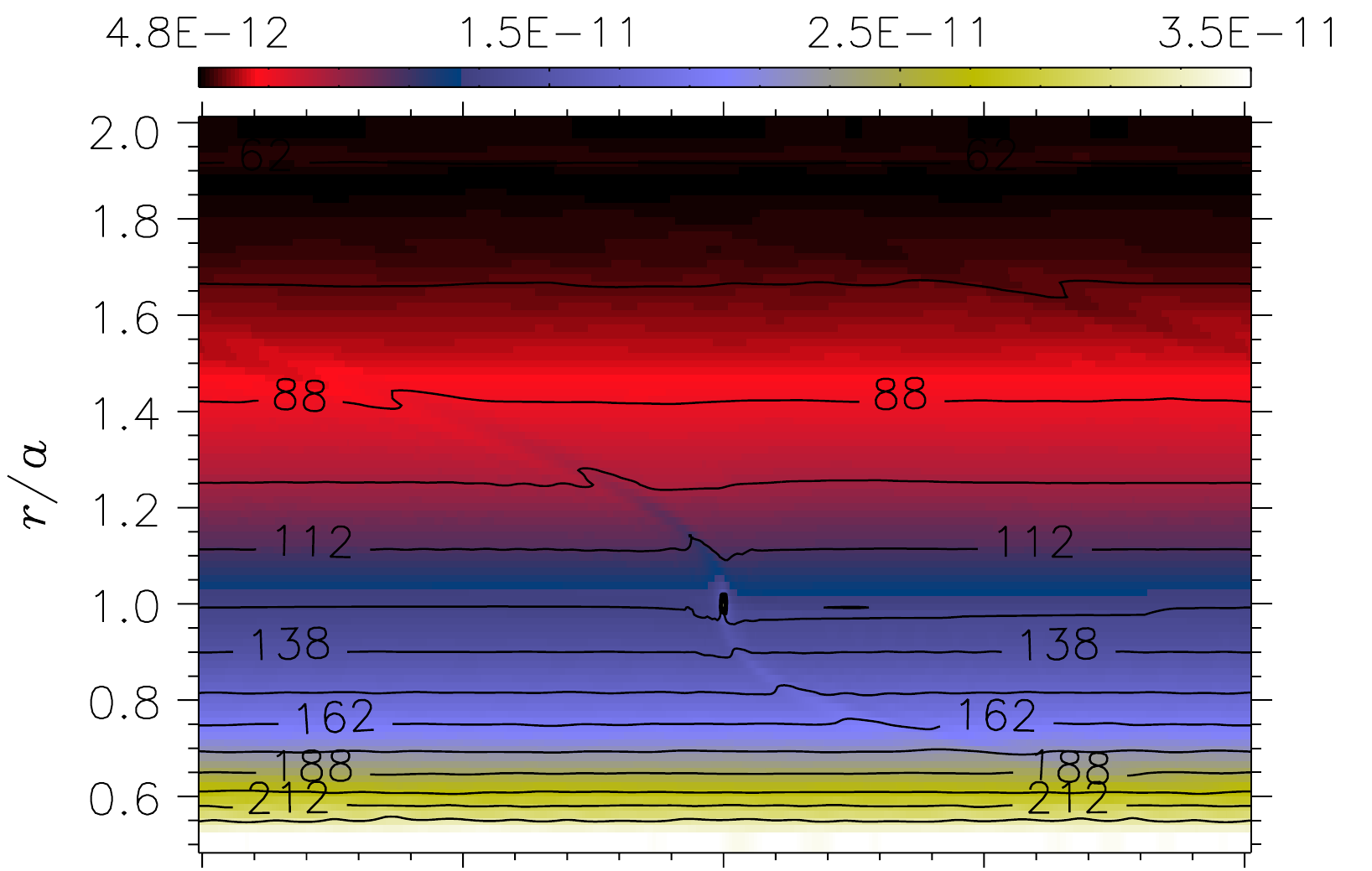}%
\includegraphics[clip]{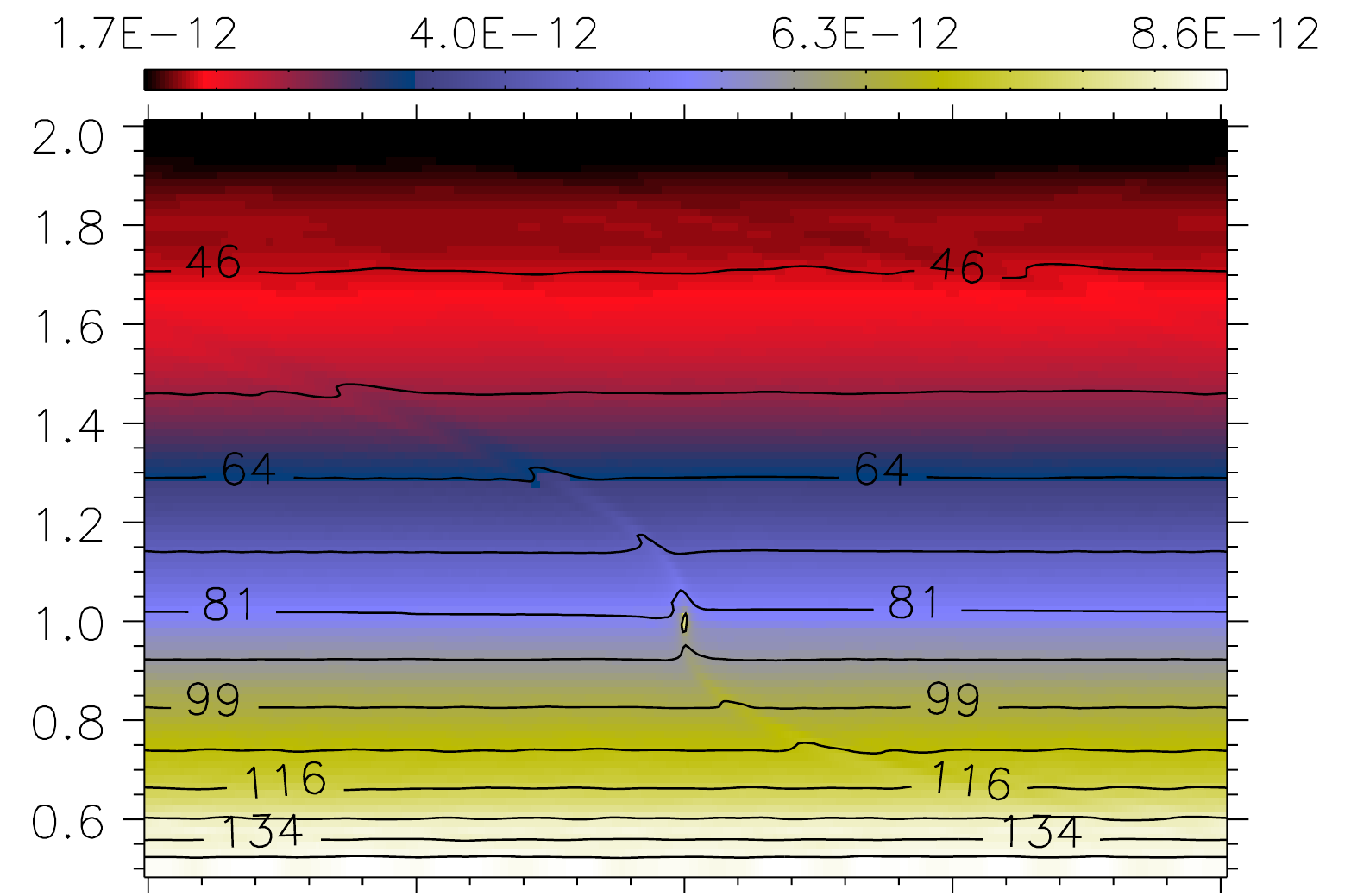}}
\resizebox{\afiglen}{!}{%
\includegraphics[clip]{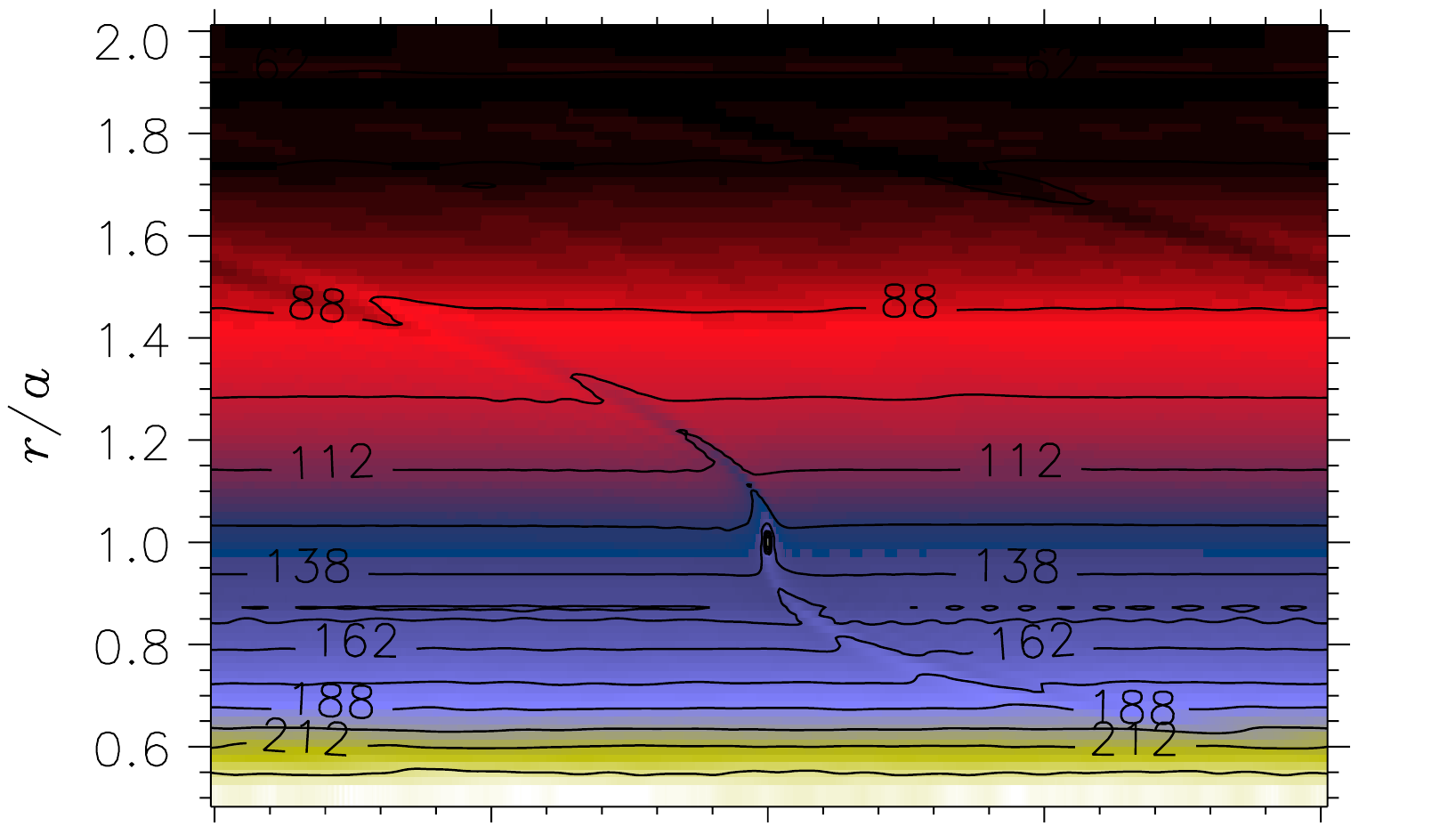}%
\includegraphics[clip]{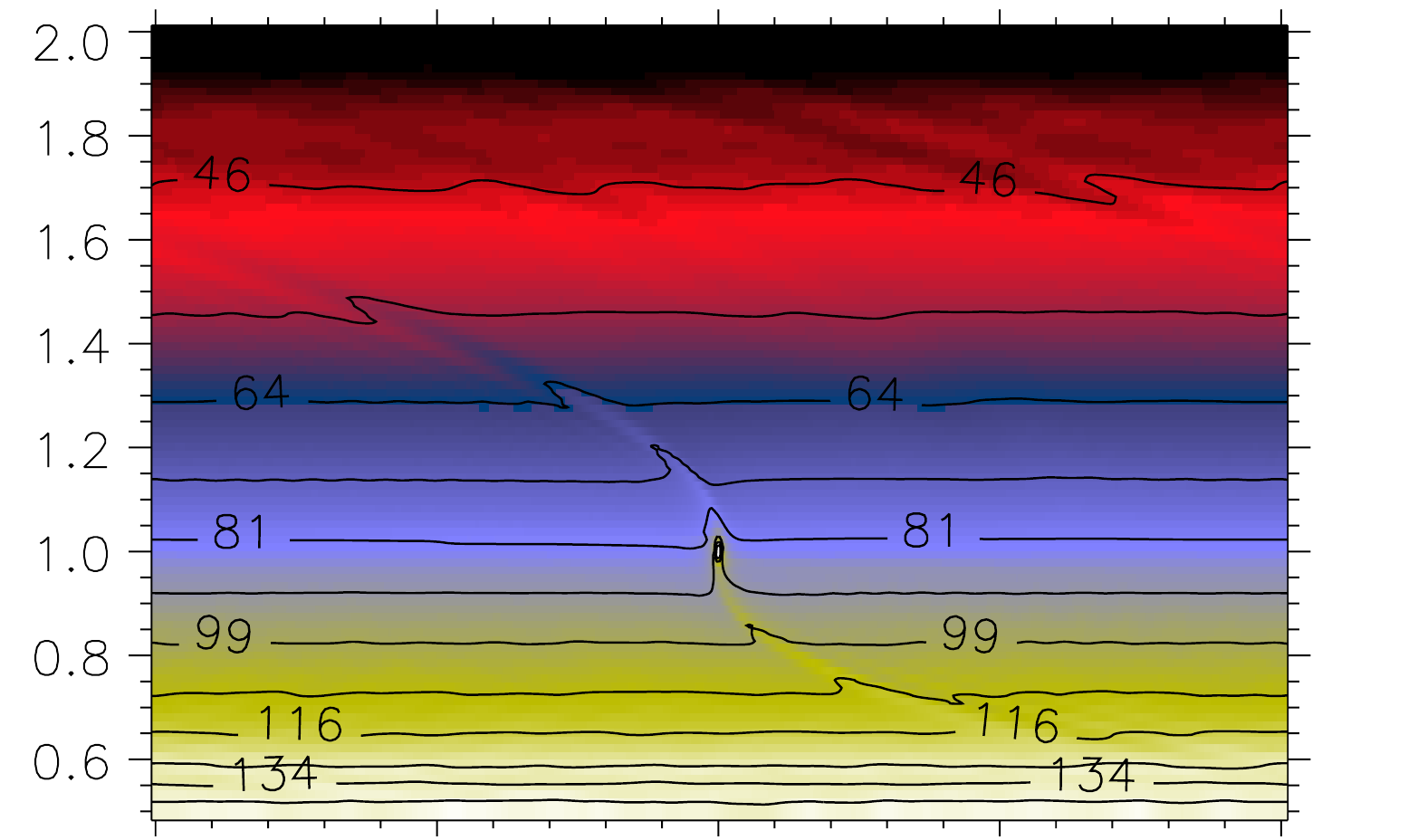}}
\resizebox{\afiglen}{!}{%
\includegraphics[clip]{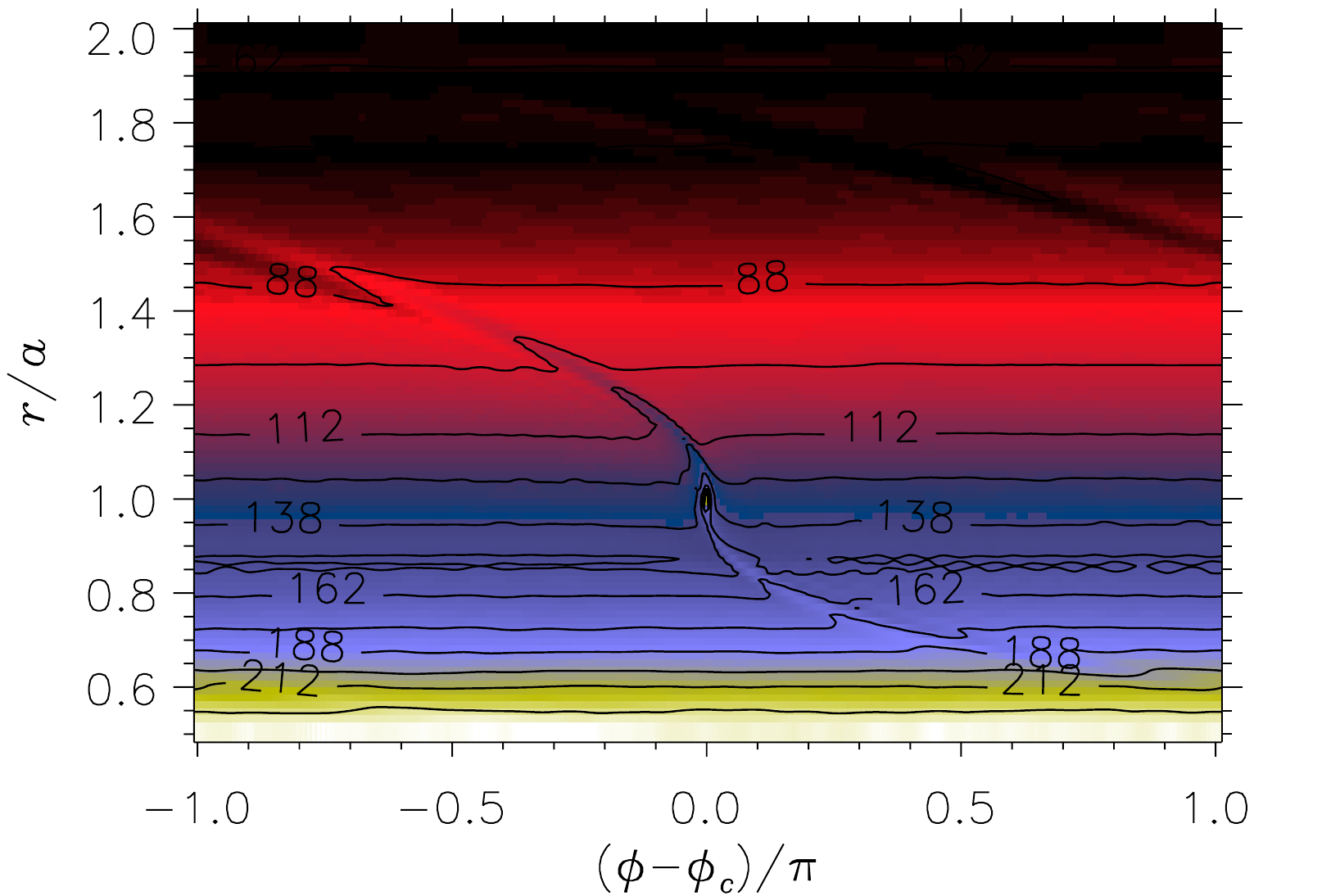}%
\includegraphics[clip]{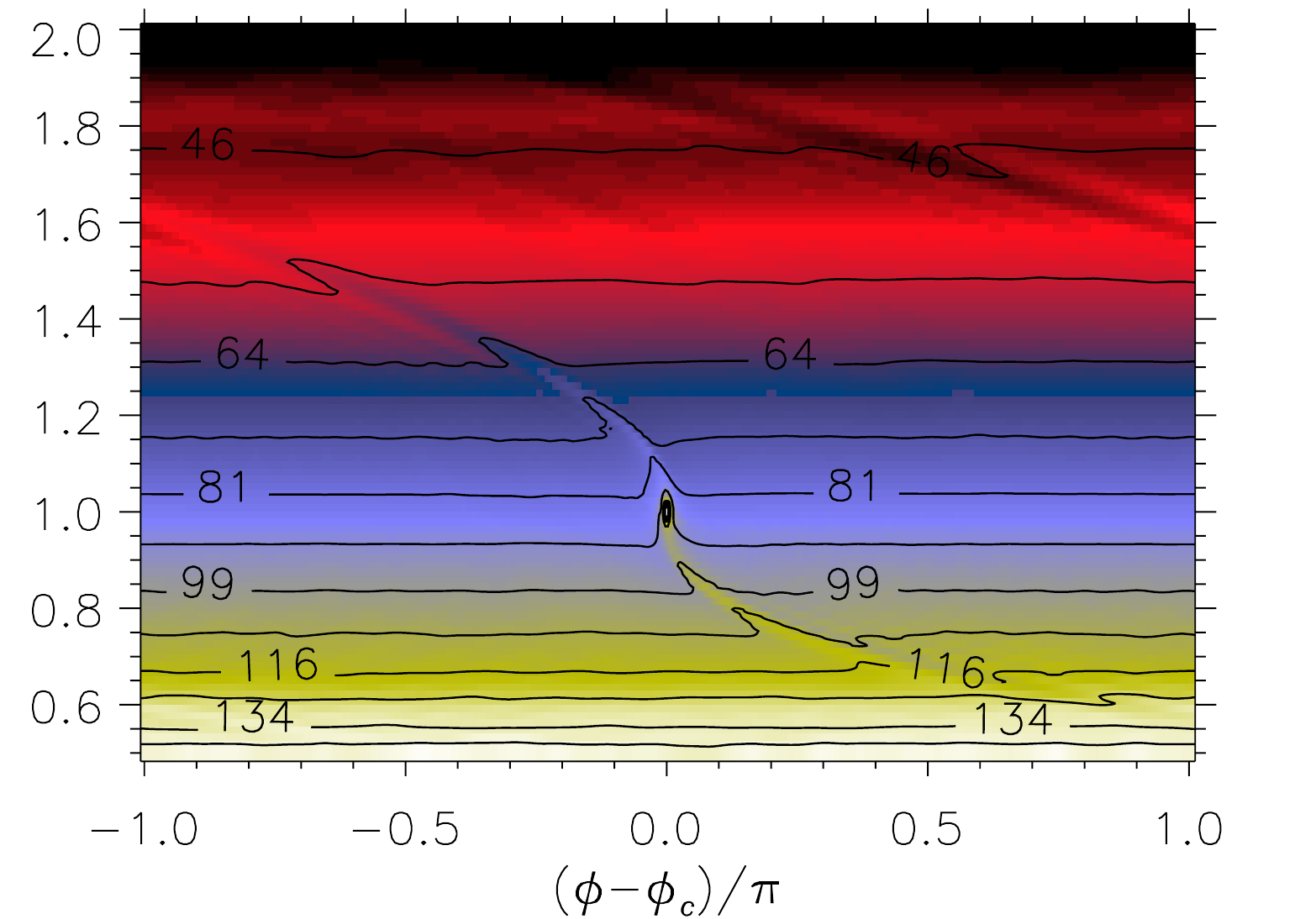}}
\caption{%
             Disk density (in units of $\mathrm{g\,cm}^{-3}$, color scale) 
             and temperature (in units of $\K$, contours) in the 
             disk's equatorial plane. The planetary core has azimuthal 
             angle $\phi_{c}$ and semimajor axis $a$. 
             From top to bottom, the core has a mass
             $\Mc=5$, $10$, and $15\,\Mearth$ and is
             located at $5\,\AU$ on the left and at $10\,\AU$ on the right.
             }
\label{fig:xz_disk}
\end{figure*}
%%%%%%%
%%%%%%
\begin{figure*}[]
\centering%
\resizebox{\afiglen}{!}{%
\includegraphics[clip]{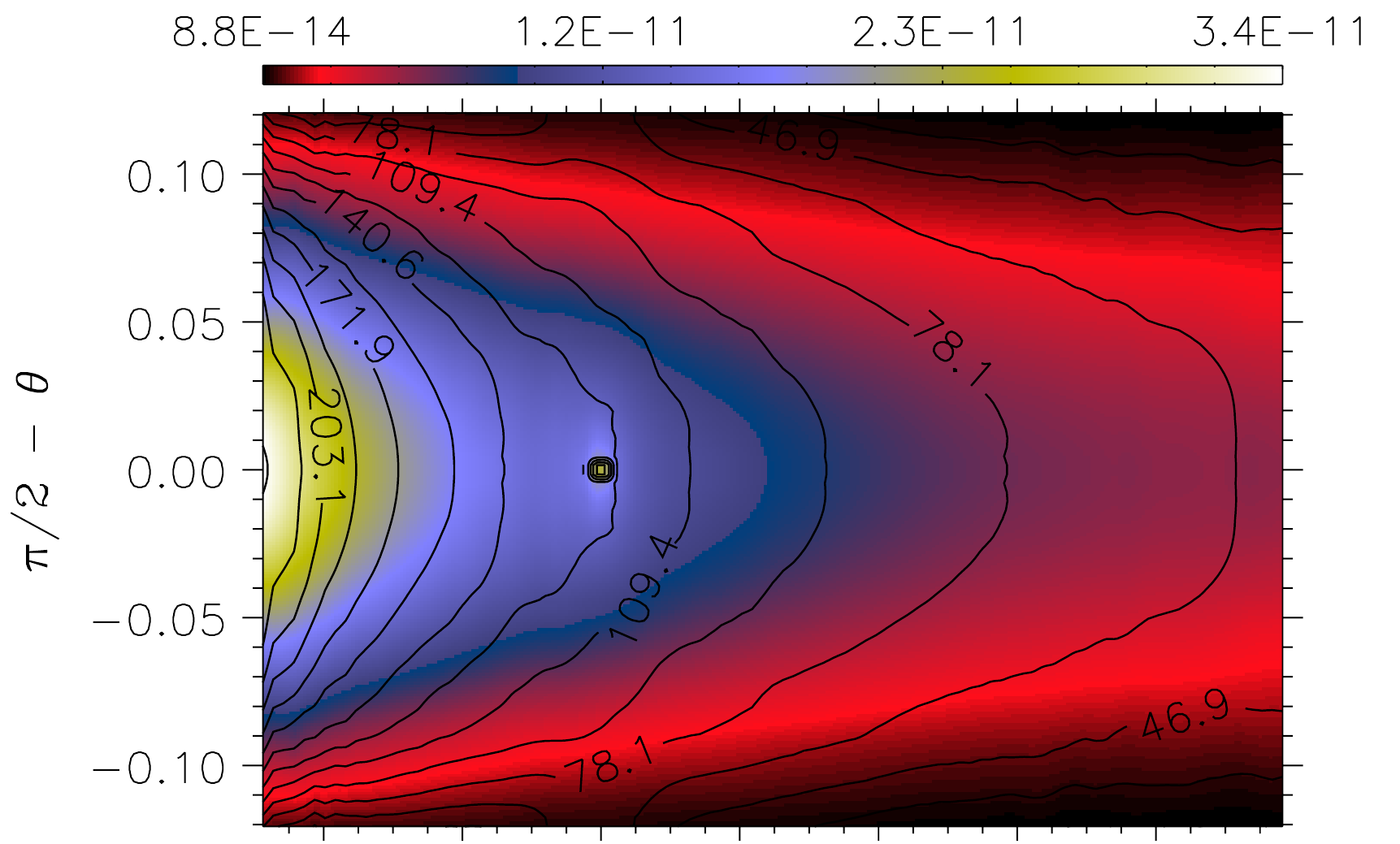}%
\includegraphics[clip]{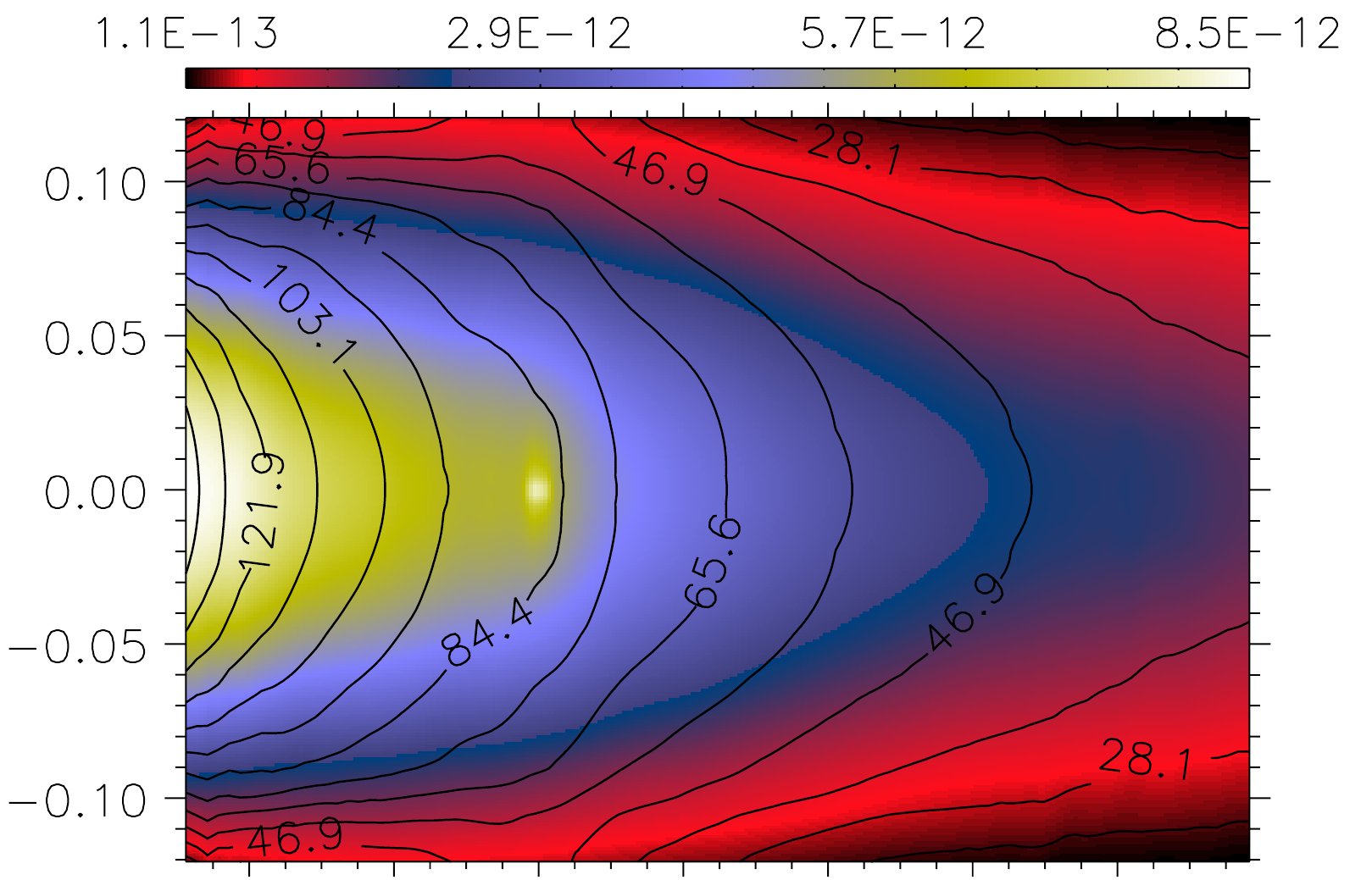}}
\resizebox{\afiglen}{!}{%
\includegraphics[clip]{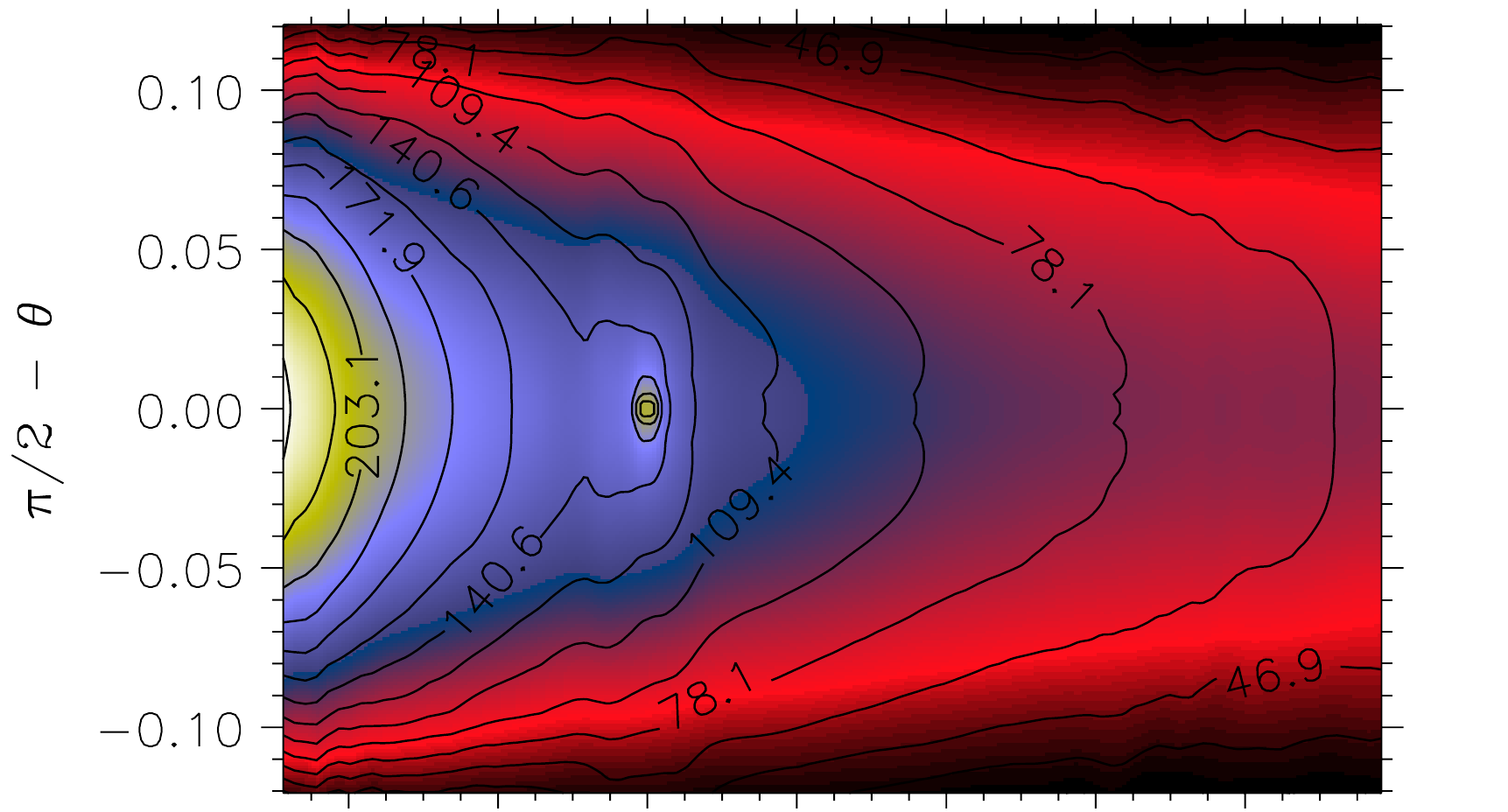}%
\includegraphics[clip]{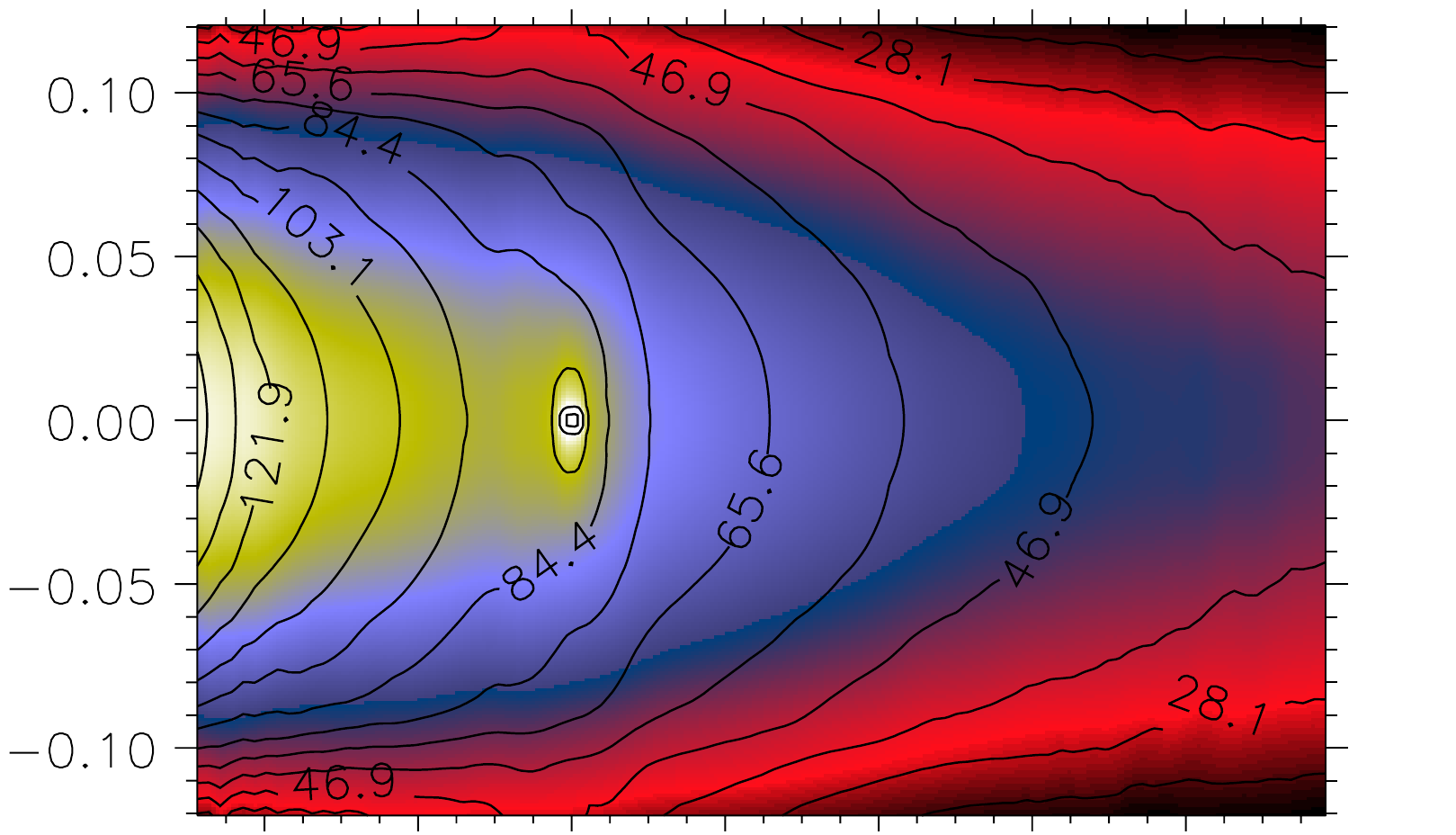}}
\resizebox{\afiglen}{!}{%
\includegraphics[clip]{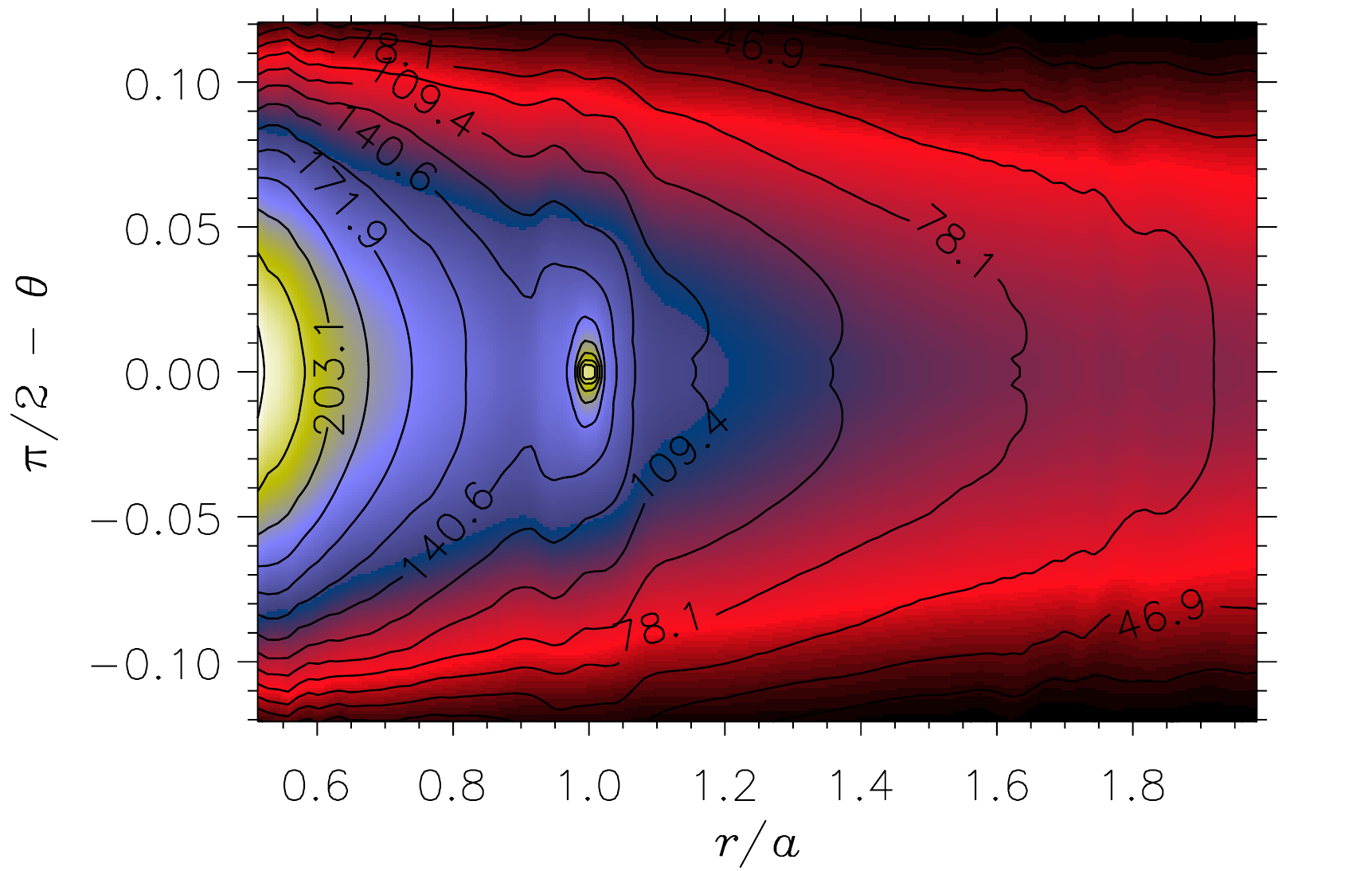}%
\includegraphics[clip]{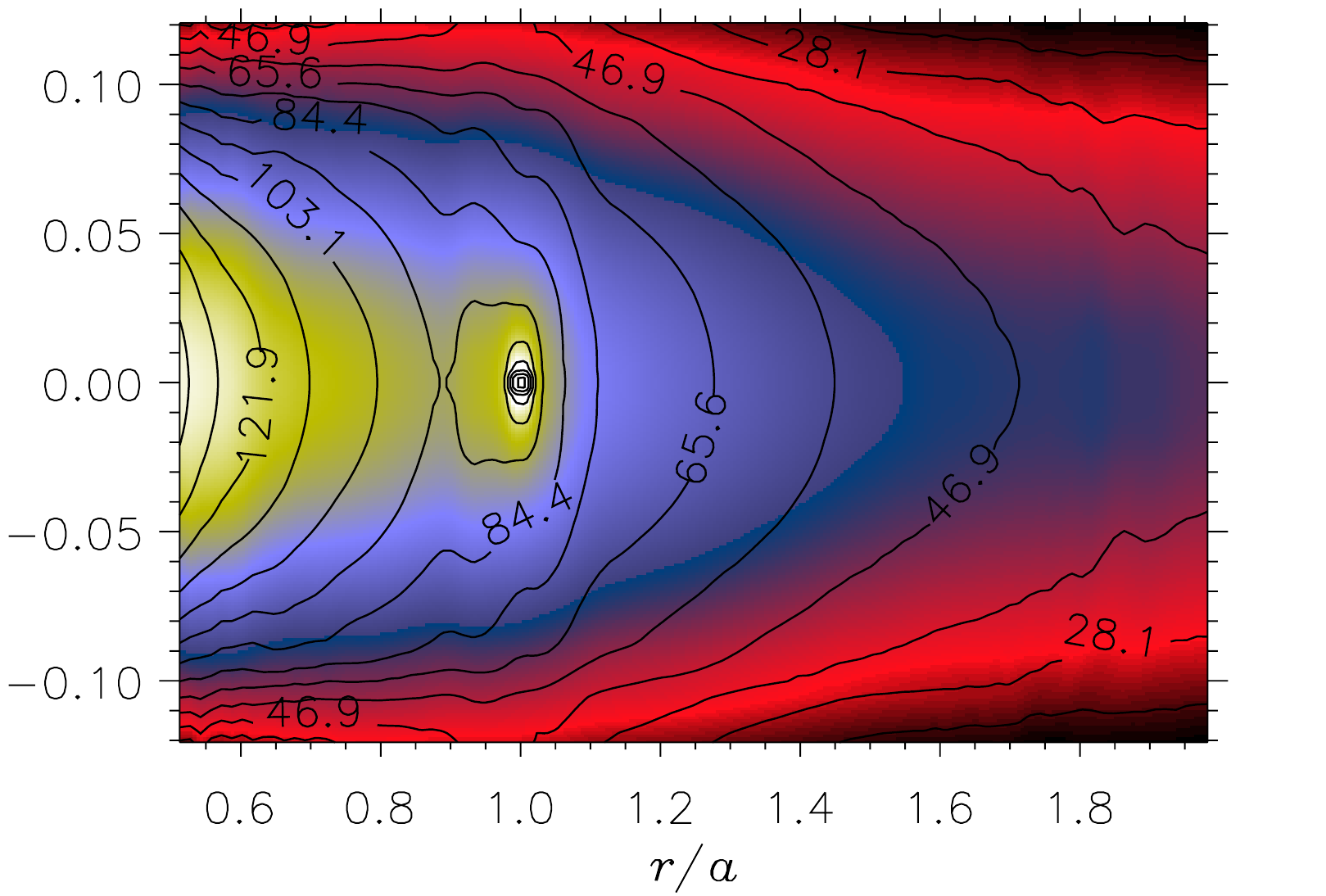}}
\caption{%
             As in Figure~\ref{fig:xz_disk}, but for the density and temperature
             in $r$--$\theta$ plane passing through the azimuthal position 
             of the planetary core
             (at coordinates $r=a$ and $\theta=\pi/2$ in these maps).
             Embedded cores have masses
             $\Mc=5$, $10$, and $15\,\Mearth$ (from top to bottom), and orbital 
             radii of $5\,\AU$ (left) and $10\,\AU$ (right).             
             }
\label{fig:xy_disk}
\end{figure*}
%%%%%%%

The quasi-equilibrium density and temperature distributions, averaged in
the azimuthal direction around the star in the disk mid-plane, 
is plotted in Figure~\ref{fig:av_disk}. The solid lines refer to models 
with $a=5\,\AU$ and the dashed lines to models with $a=10\,\AU$. 
The mean radial slope of the density is such that $\langle\rho\rangle$ 
is roughly proportional to $r^{-3/2}$ around $5\,\AU$, with a somewhat 
shallower slope around $10\,\AU$. For the mid-plane temperature,
the mean slope is roughly such that $\langle T\rangle\propto 1/r$.
This slope is consistent with an approximate balance between 
viscous heating and vertical radiative cooling \citep[e.g.,][]{gennaro2003b}, 
considering that $\kappa_{\mathrm{R}}$ is either roughly proportional 
to $T^{0.1}$ or about constant at temperatures $40<T<250\,\K$
(Figure~\ref{fig:k}, lower-left panel).
A summary of the azimuthally averaged disk's properties,
at the core's orbital radius, is given in Table~\ref{table:dgc}.
Note that the results presented in Figure~\ref{fig:av_disk}, 
and in the rest of this section, are plotted for the first grid level
but calculated on the entire nested-grid structure 
(see Section~\ref{sec:NGS}).

%%%%%%
\begin{figure*}[]
\centering%
\resizebox{\figlen}{!}{%
\includegraphics[clip]{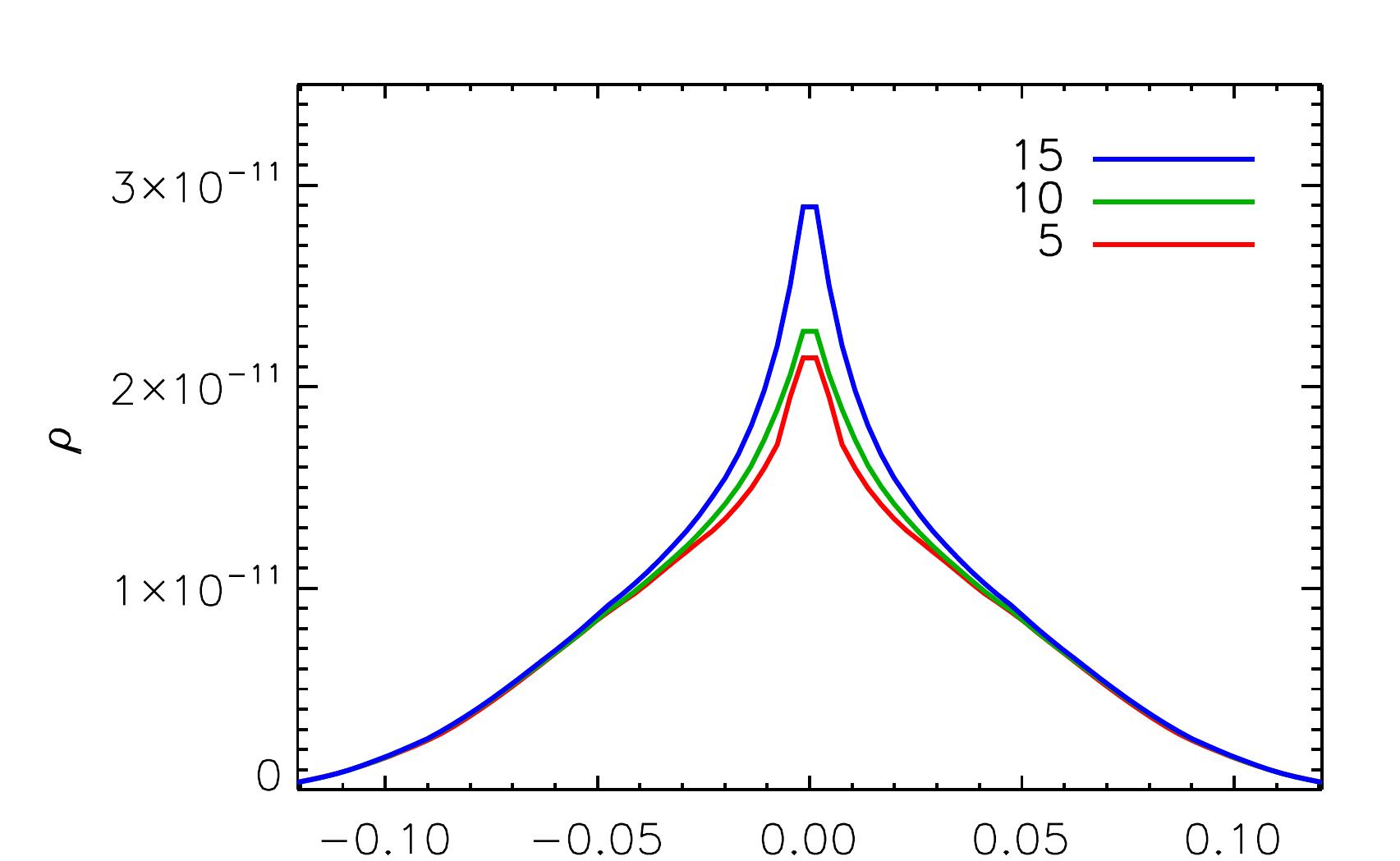}%
\includegraphics[clip]{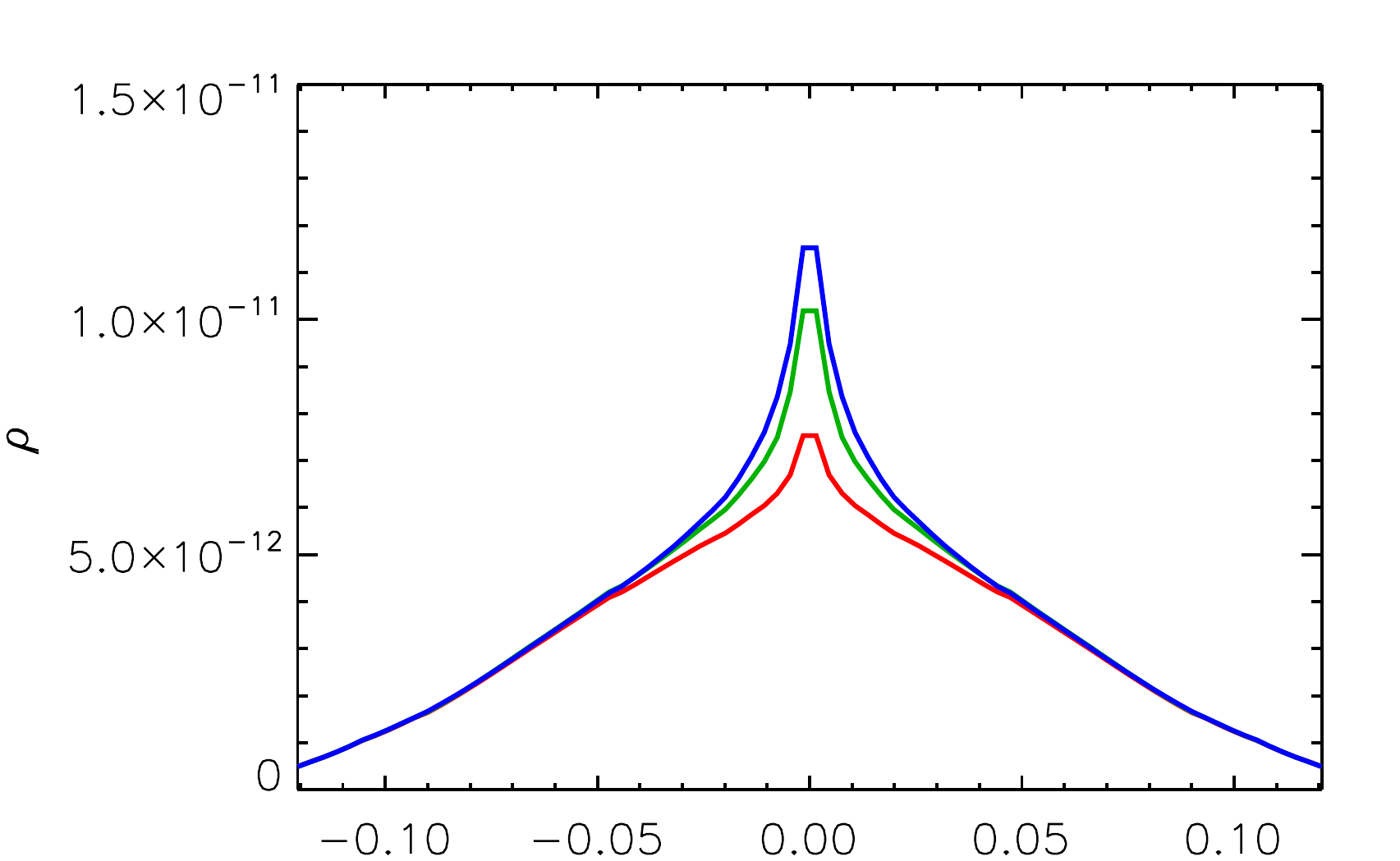}}
\resizebox{\figlen}{!}{%
\includegraphics[clip]{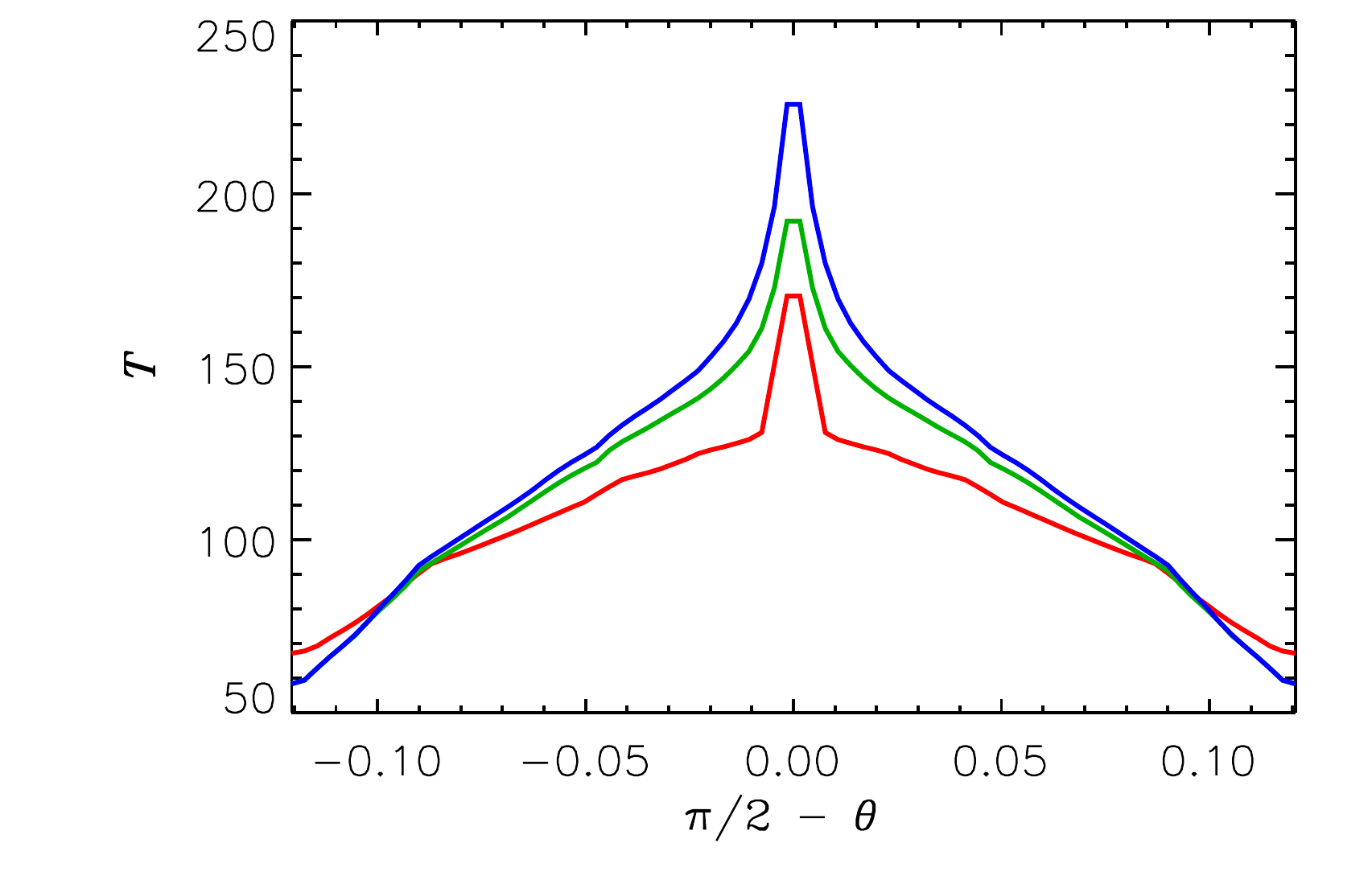}%
\includegraphics[clip]{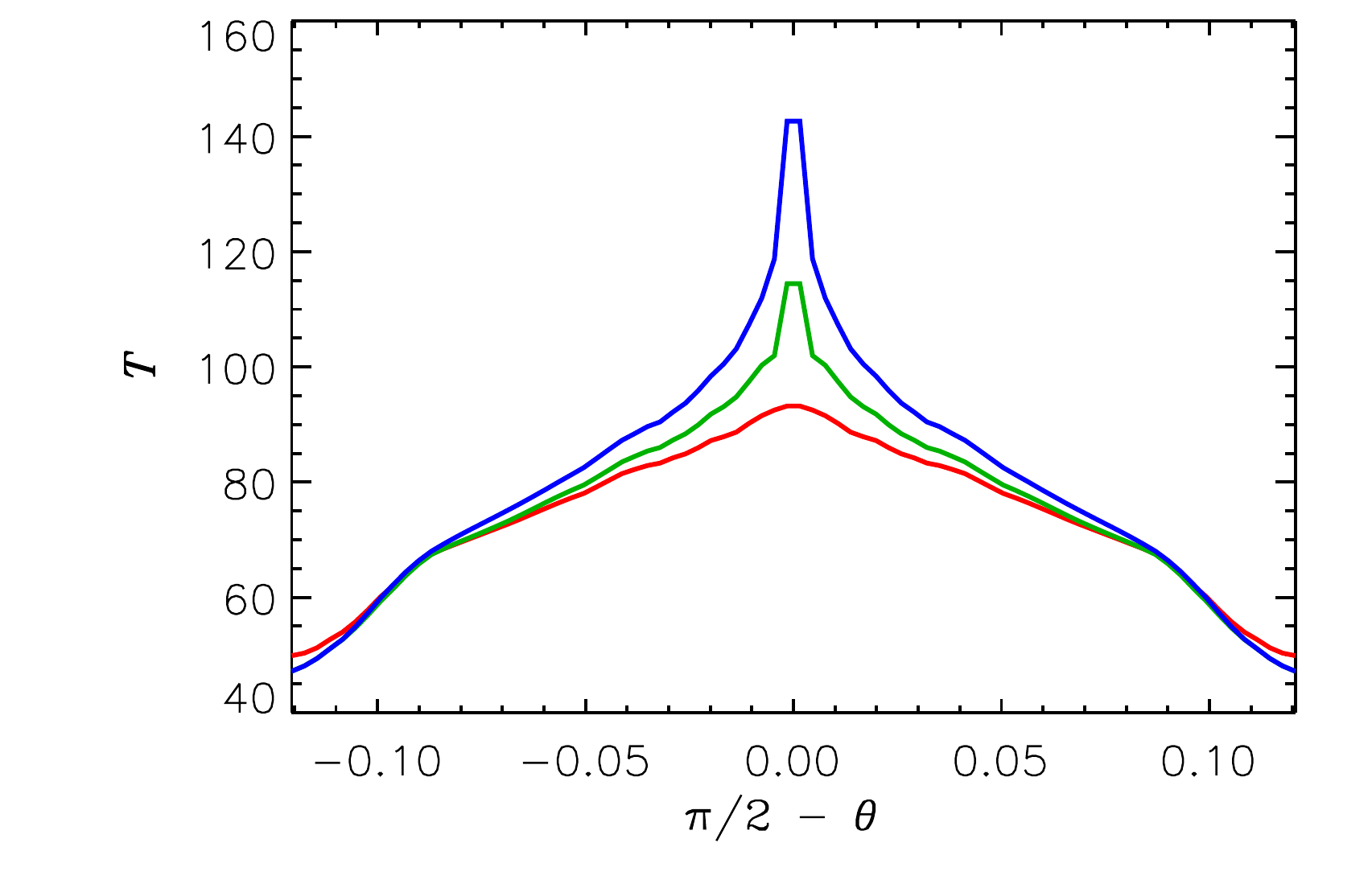}}
\caption{%
             Density in units of $\mathrm{g\,cm}^{-3}$ (top) and temperature in units 
             of $\K$ (bottom), along the disk's vertical (co-latitude) direction
             ($\theta=\pi/2$ indicates
             the mid-plane of the disk), at the radial and azimuthal position of the planetary core.
             The semimajor axis is $a=5\,\AU$ in the left panels and $10\,\AU$
             in the right panels. The curves refer to cores of different mass,
             as listed in the top-left panel in units of $\Mearth$.
             }
\label{fig:z_disk}
\end{figure*}
The distributions of density and temperature in the
disk's equatorial plane are illustrated in Figure~\ref{fig:xz_disk},
where the images refer to the density and contour levels to the temperature 
(see the figure caption for further details).
Similarly, Figure~\ref{fig:xy_disk} shows the vertical stratification of density 
and temperature, at the azimuthal position of the planet.
The effects of the core's perturbation on the temperature in 
the equatorial plane are mostly confined to regions where compression 
occurs due to the propagation of spiral density waves. More local effects 
can be seen in Figure~\ref{fig:xy_disk}, where the isothermal (contour) lines 
indicate a temperature increase in the region around the radial 
position of the planet, an effect that becomes larger as the core mass
increases. Density and temperature profiles in the vertical (i.e., $\theta$) 
direction, at the radial and azimuthal position of the core, are plotted in
Figure~\ref{fig:z_disk}. The figure shows the extent to which both 
density and temperature in the disk are enhanced, approaching
the disk mid-plane ($\theta=\pi/2$), by gas compression due to the gravity 
of the planet. 
Effectively, these curves represent quantities averaged over the minimum 
spacing of the basic grid (see Section~\ref{sec:NGS}). In reality, as discussed
in the next sections, density and temperature can be larger by orders of
magnitudes in close proximity of the planet, but at distances from the core 
not resolved in these plots.

An estimate of disk aspect ratio can be obtained from the mid-plane 
temperature as 
$h\sim c_{\mathrm{gas}}/u_{\mathrm{K}}$, where
$u_{\mathrm{K}}$ is the Keplerian velocity of the gas,
so that
$h\sim \sqrt{\Gamma_{1}k_{\mathrm{B}}T r/(\mu m_{\mathrm{H}}G\Ms)}$.
The resulting aspect ratio is $\approx 0.06$ and $\approx 0.07$,
respectively, for the disk models with $a=5$ and $10\,\AU$. 
Since these values are computed using thermodynamical quantities
at the mid-plane, they are likely to overestimate the value of $h$.
Alternatively, the vertical density distribution can be approximated as
hydrostatic, i.e., as a gaussian at any given radius. 
In a spherical geometry, said approximation corresponds to
the profile along the $\theta$-direction
$\exp{[(\sin\theta -1)/h^{2}]}$ \citep{masset2006a}. 
This procedure results in typical values of $h$, averaged over one scale 
height from the equatorial plane, of $\approx 0.055$ and $\approx 0.06$, 
respectively.

The azimuthally averaged density is not much affected by disk-planet tidal 
interactions, for any of the core masses considered, at both $5$ and $10\,\AU$
(see left panel of Figure~\ref{fig:av_disk}). Tidal perturbations are confined to
the excitation of spiral density waves (see Figure~\ref{fig:xz_disk}).
The absence of gap formation is in accord with 
simple arguments based on the balance of viscous and tidal torques exerted 
on the disk.
In fact, when $\Rhill/a<h$, the condition for significant tidal interactions
(leading to gap formation) is approximately
$(\Mc/\Ms)^{2}\gtrsim 3\pi\alpha h^{5}$ \citep[e.g.,][]{gennaro2010},
assuming that the envelope mass $\Me$ is negligible.
The equivalent $\alpha$-viscosity \citep{S&S1973} in these disks at 
the planet's orbital radius is $\alpha=\nu/(h^{2}a^{2}\Omega)\sim 0.001$,
hence the condition above requires a core (plus envelope) mass
$\gtrsim 30\,\Mearth$ for significant tidal perturbation of the disk's density.

%%--------------------------------------------------------------------------
\section{Envelopes of Planetary Cores}
\label{sec:PCE}
%%--------------------------------------------------------------------------

There are two length scales that are relevant to the formation
of a gaseous envelope around a solid core, both dictated by
energy arguments. The first is determined by thermodynamics
and the second by gravity.

The mean thermal velocity of the gas is
$u_{\mathrm{th}}=\sqrt{(8/\pi)k_{\mathrm{B}}T/(\mu m_{\mathrm{H}})}$
\citep[e.g.,][]{m&m}.
Disk gas moving within a maximum distance, $\Rbondi$, 
of a planetary core may become bound to the core 
if $u_{\mathrm{th}}$ is smaller than the escape velocity from 
the core at that distance, $u_{\mathrm{esc}}=\sqrt{2G\Mc/\Rbondi}$
\citep[e.g.,][]{bodenheimer1986}, where
\begin{equation}
\label{eq:RB}
\Rbondi=\left(\frac{\pi}{4}\right)\frac{G\Mc \mu m_{\mathrm{H}}}{k_{\mathrm{B}}T}
\end{equation}
is the Bondi radius and is defined through the condition 
$u_{\mathrm{th}}=u_{\mathrm{esc}}$. 
The disk region where the gravity of 
the core dominates that of the star is set by the (circular) restricted
three-body problem dynamics and is a fraction of the Hill radius,
$\Rhill=a\left[\Mc/(3\Ms)\right]^{1/3}$.
In fact, the radius of the sphere having the same volume as the 
Roche lobe is $\approx 2\Rhill/3$ \citep{paczynski1971,kopal1978}
and the radius of a sphere entirely contained in the Roche lobe
is $\lesssim 3\Rhill/5$, as can be calculated from the equations
describing the equipotential surfaces of the three-body problem 
\citep[e.g.,][]{murray2000}.

Therefore, a gaseous envelope may form around a core 
within the smaller of $\Rbondi$ and $3\Rhill/5$. 
These two characteristic lengths become equal for a core 
mass
\begin{equation}
\label{eq:RBeqRH}
\Mc\approx\left[\frac{9}{2\Ms}%
           \left(\frac{a}{\pi}\frac{k_{\mathrm{B}}T}{G\mu m_{\mathrm{H}}}\right)^{3}\right]^{1/2},
\end{equation}
or
$\Mc/\Mearth\approx 3.6 \mu^{-3/2}(T/100\,\K)^{3/2}(a/1\,\AU)^{3/2}$
for a solar-mass star
(in the equation above, $(5/4)^{3}$ is approximated to $2$).
In our disk models, at $5\,\AU$, the ratio of the two lengths,
$(5/3)\Rbondi/\Rhill$, varies
from $\approx 0.5$ ($\Mc=5\,\Mearth$) to $\approx 1$ ($\Mc=15\,\Mearth$).
At $10\,\AU$, the Bondi radius is larger by a factor of about $3/2$ 
(due to the lower disk temperature, see Figure~\ref{fig:av_disk}), 
but the Hill radius is twice as large. 
Therefore, the ratio of the characteristic lengths is reduced by a factor 
of $4/3$. In  all cases considered here, the envelope radius should be 
generally set by thermal arguments ($\Rbondi\lesssim 3\Rhill/5$).
The envelope is therefore expected to be confined within the
Bondi sphere and the Bondi radius is expected to be
a hard limit for the envelope radius.
Note that additional energy sources, such as kinetic energy of
the background flow, may facilitate gas escape from the core at
even shorter distances.

It should be stressed that while $\Rhill$ has a non-ambiguous definition
(if the mass contributed by a planet's envelope is small compared
to $\Mc$, as in these calculations), there is an ambiguity 
in the definition of $\Rbondi$, since it relies on an average temperature
of the background flow. In the estimates given above, this temperature
is taken as the azimuthal average (around the star) at the planet's 
orbital radius. In reality, the temperature should be some \emph{local} 
mean calculated outside, but in the vicinity (i.e., on the length scale),
of the envelope radius.
As a local mean around the planet, such temperature is expected
to be somewhat larger than the disk azimuthal average and also
to depend on the core mass.  
Therefore, the Bondi radius may be somewhat smaller than
the estimates presented above and Equation~(\ref{eq:RB}) should
represent an upper bound. In the following, to make the definition
less ambiguous and more workable for our purposes, we shall refer 
to this upper bound as the nominal length of the Bondi radius.

%%--------------------------------------------------------------------------
\subsection{1D Calculations of Envelopes}
\label{sec:1D}
%%--------------------------------------------------------------------------

We perform 1D calculations of the accumulation
of gaseous envelopes around planetary cores using the
planet evolution code of
\citet{pollack1996,hubickyj2005,lissauer2009}, and references therein. 
We also apply the procedures and approximations detailed in those papers.
The purpose of the 1D calculations is to produce reference models
for the envelope stratification (e.g., of temperature and density) around
cores of $5$, $10$, and $15\,\Mearth$, at both $5$ and $10\,\AU$. 
Additionally, they provide the core accretion rate $\dot{M}_{c}$ 
that is needed for the energy source term in Equation~(\ref{eq:eps}),
which represents the gravitational energy released at the base of the
envelope by incoming solid material.

In these models, a $0.1\,\Mearth$ core accretes solids 
(planetesimals of $100\,\mathrm{km}$ in radius) and gas. 
The accretion rate of  planetesimals is proportional 
to the local surface density of solids. We use values of $10$ and
$13\,\mathrm{g\,cm^{-2}}$ for cores forming at $5\,\AU$, and
$6\,\mathrm{g\,cm^{-2}}$ for cores forming at $10\,\AU$.
Given the  gas-to-dust mass ratio of $\approx 70$ assumed here 
(see Appendix~\ref{sec:opa_calc} for details), such values are 
consistent with the expected gas-augmented initial surface density 
($\approx 2000\,\mathrm{g\,cm^{-2}}$ or less at $1\,\AU$) 
in the disks described in Section~\ref{sec:DS}.

In standard core accretion calculations, the cross-over mass, 
$\Mc=\Me$, is about equal to $\sqrt{2}$ times the isolation mass
\citep{pollack1996}. 
Since the 3D radiation hydrodynamics calculations
performed for this study neglect the effects of gas self-gravity, we should 
restrict the discussion to earlier phases of the planet evolution when the envelope is
still much less massive than the core.
At $5\,\AU$, the cross-over mass is about $15\,\Mearth$ when 
the surface density of solids is $10\,\mathrm{g\,cm^{-2}}$.
For a solids' surface density of $13\,\mathrm{g\,cm^{-2}}$,
the cross-over mass is instead $\approx 22\,\Mearth$, and 
$\Me\ll \Mc$ when $\Mc=15\,\Mearth$ (see Section~\ref{sec:1D-3D}). 
At $10\,\AU$, the cross-over mass ($\propto a^{3}$) is larger 
than $50\,\Mearth$ in the 1D models considered here.

The accretion rate of gas is dictated by the contraction rate of the
envelope. At these early stages of formation, it mainly depends on 
the ability of the outer envelope to cool by radiating away the gravitational 
energy produced by contraction and by the accretion of planetesimals.
Since dust grains represent the main source of opacity in the outer
envelope, their optical properties, abundance, and depth distribution
are critical to the determination of the gas accretion rate
\citep{naor2010}.
Our 1D models use interstellar dust opacities \citep{pollack1985}
reduced by a factor $50$, to mimic the reduction caused by 
grain growth and settling in the envelope \citep{podolak2003}.
As explained in the next section (see also Appendix~\ref{sec:opa_calc}), 
the 3D calculations use different dust opacities, applying a size
distribution of grains whose presence in T~Tauri disks is
suggested by observations. These opacities also fall well below
interstellar values in the relevant temperature range.

The exterior boundary of the envelope is defined as in \citet{lissauer2009},
so that the inverse of the envelope radius is equal to $1/\Rbondi + 4/\Rhill$.
Note that for $\Rbondi\approx\Rhill$, the envelope radius becomes
$\approx \Rbondi/5$. 
At the exterior boundary, densities and temperatures are matched to 
the disk values, azimuthally averaged around the star, obtained from 
the 3D calculations and given in Table~\ref{table:dgc}. 
Therefore, we set 
$\rho=1.3\times 10^{-11}\,\mathrm{g\,cm^{-3}}$ and $T=124\,\K$ 
at $5\,\AU$, and $\rho=5\times 10^{-12}\,\mathrm{g\,cm^{-3}}$ 
and $T=80\,\K$ at $10\,\AU$ (see also Figure~\ref{fig:av_disk}).

%%--------------------------------------------------------------------------
\subsection{Comparisons between 1D and 3D Envelopes}
\label{sec:1D-3D}
%%--------------------------------------------------------------------------

%%%%%%%%%
\begin{deluxetable*}{ccccccccccccc}
%\rotate
\tablecolumns{13}
\tablewidth{0pc}
\tablecaption{Envelope Masses and Accretion Rates in 1D and 3D Models\label{table:13D}}
\tablehead{
\colhead{}&\colhead{}&%
\multicolumn{3}{c}{$5\,\Mearth$}&\colhead{}&%
\multicolumn{3}{c}{$10\,\Mearth$}&\colhead{}&%
\multicolumn{3}{c}{$15\,\Mearth$}\\%
\cline{3-5} \cline{7-9} \cline{11-13}\\[-5pt]
\colhead{$a$\tablenotemark{a}}&\colhead{}&%
\colhead{$\dot{M}_{c}$\tablenotemark{b}}&\colhead{$\dot{M}_{e}$\tablenotemark{b}}&\colhead{$\Me$\tablenotemark{c}}&%
\colhead{}&%
\colhead{$\dot{M}_{c}$}&\colhead{$\dot{M}_{e}$}&\colhead{$\Me$}&%
\colhead{}&%
\colhead{$\dot{M}_{c}$}&\colhead{$\dot{M}_{e}$}&\colhead{$\Me$}%
}
\startdata
$5$ & & $8.8\times 10^{-5}$ & $9.5\times 10^{-8}$ & $1.8\times 10^{-3}$%
       & & $5.9\times 10^{-5}$ & $1.4\times 10^{-6}$ & $3.2\times 10^{-2}$%
       & & $1.2\times 10^{-4}$ & $4.0\times 10^{-6}$ & $8.1\times 10^{-2}$\\
$10$ & & $5.6\times 10^{-5}$ & $8.4\times 10^{-8}$ & $2.8\times 10^{-3}$%
       & & $1.2\times 10^{-4}$ & $6.9\times 10^{-7}$ & $1.9\times 10^{-2}$%
       & & $1.8\times 10^{-4}$ & $2.0\times 10^{-6}$ & $6.5\times 10^{-2}$\\
\cline{1-13}\\[-2pt]
       & & $\dot{M}_{c}$[3D]   & $\dot{M}_{e}$[3D]  & $\Me$[3D]%
       & & $\dot{M}_{c}$[3D]   & $\dot{M}_{e}$[3D]  & $\Me$[3D]%
       & & $\dot{M}_{c}$[3D]   & $\dot{M}_{e}$[3D]  & $\Me$[3D]\\[2pt]
\cline{1-13}\\[-5pt]
$5$ & & $8.8\times 10^{-5}$ & $1.5\times 10^{-7}$ & $1.3\times 10^{-3}$%
       & & $5.9\times 10^{-5}$ & $8.7\times 10^{-7}$ & $1.8\times 10^{-2}$%
       & & $1.2\times 10^{-4}$ & $2.1\times 10^{-6}$ & $3.5\times 10^{-2}$\\
$10$ & & $5.6\times 10^{-5}$ & $9.2\times 10^{-8}$ & $1.6\times 10^{-3}$%
       & & $1.2\times 10^{-4}$ & $5.1\times 10^{-7}$ & $8.7\times 10^{-3}$%
       & & $1.8\times 10^{-4}$ & $1.4\times 10^{-6}$ & $3.3\times 10^{-2}$
\enddata
\tablenotetext{a}{Core's orbital radius in \AU.}
\tablenotetext{b}{Accretion rate in Earth masses per year.}
\tablenotetext{c}{Envelope mass in Earth masses.}
\end{deluxetable*}
%%%%%%%%%
%%%%%%
\begin{figure*}[]
\centering%
\resizebox{\figlen}{!}{%
\includegraphics[clip]{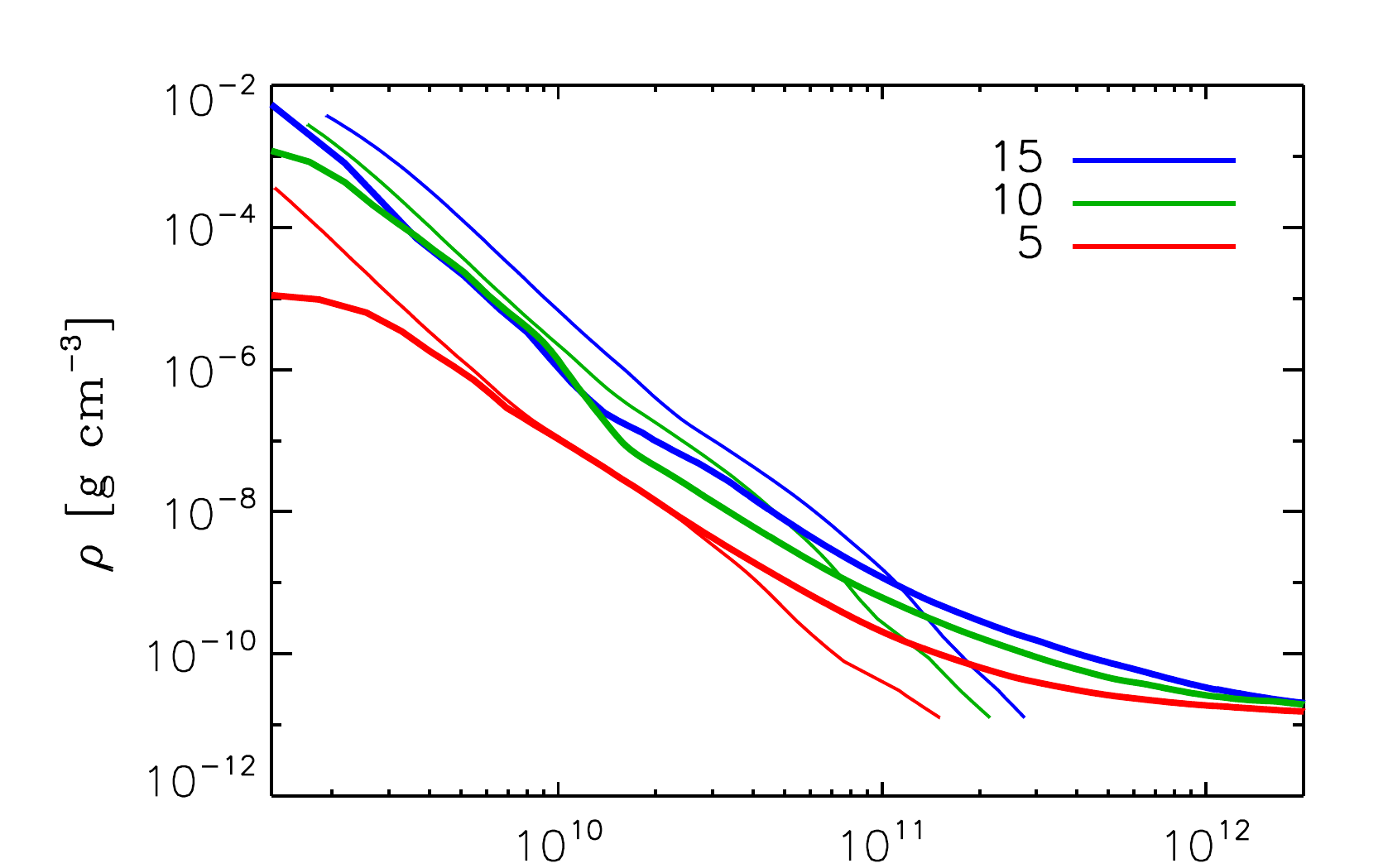}%
\includegraphics[clip]{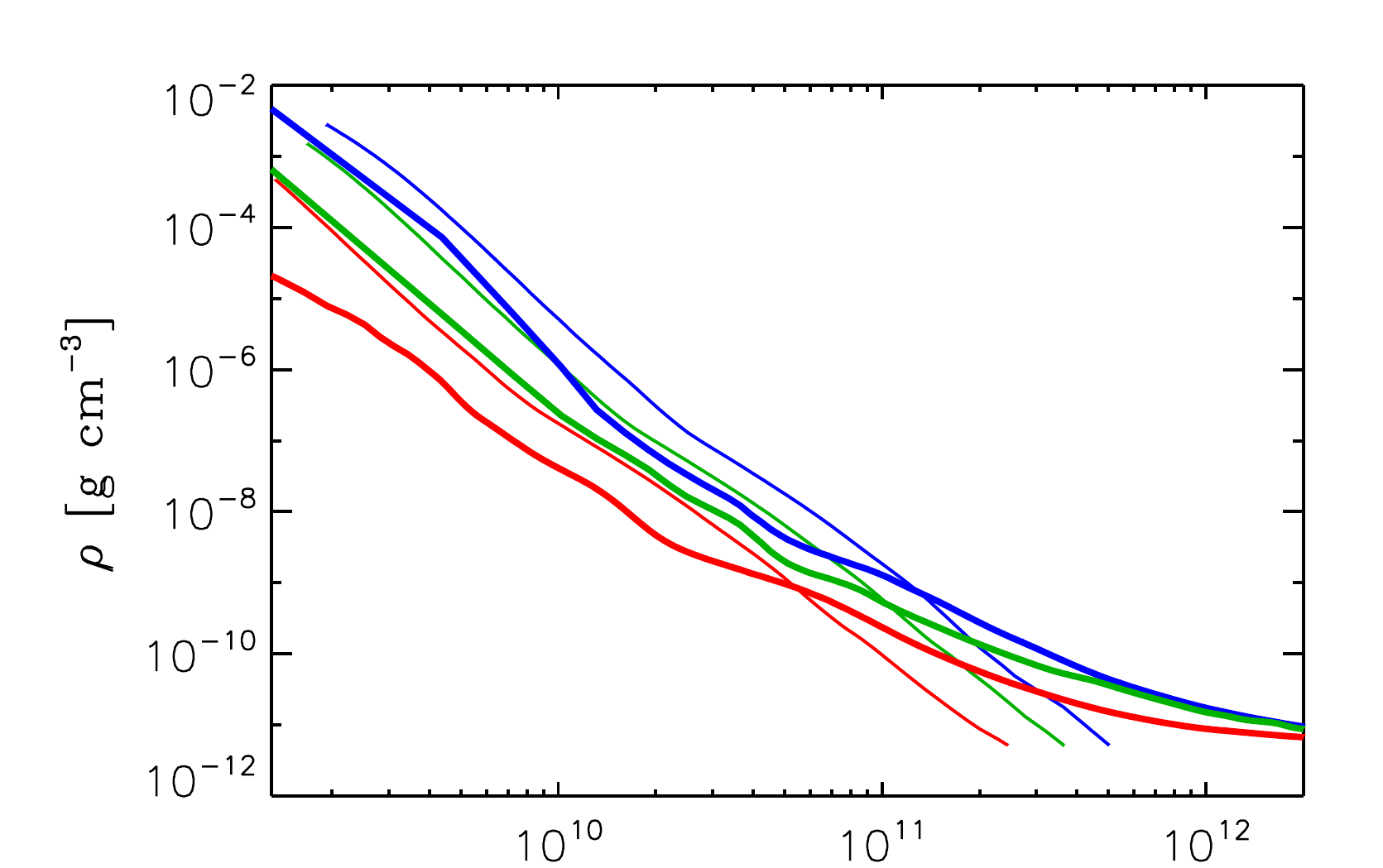}}
\resizebox{\figlen}{!}{%
\includegraphics[clip]{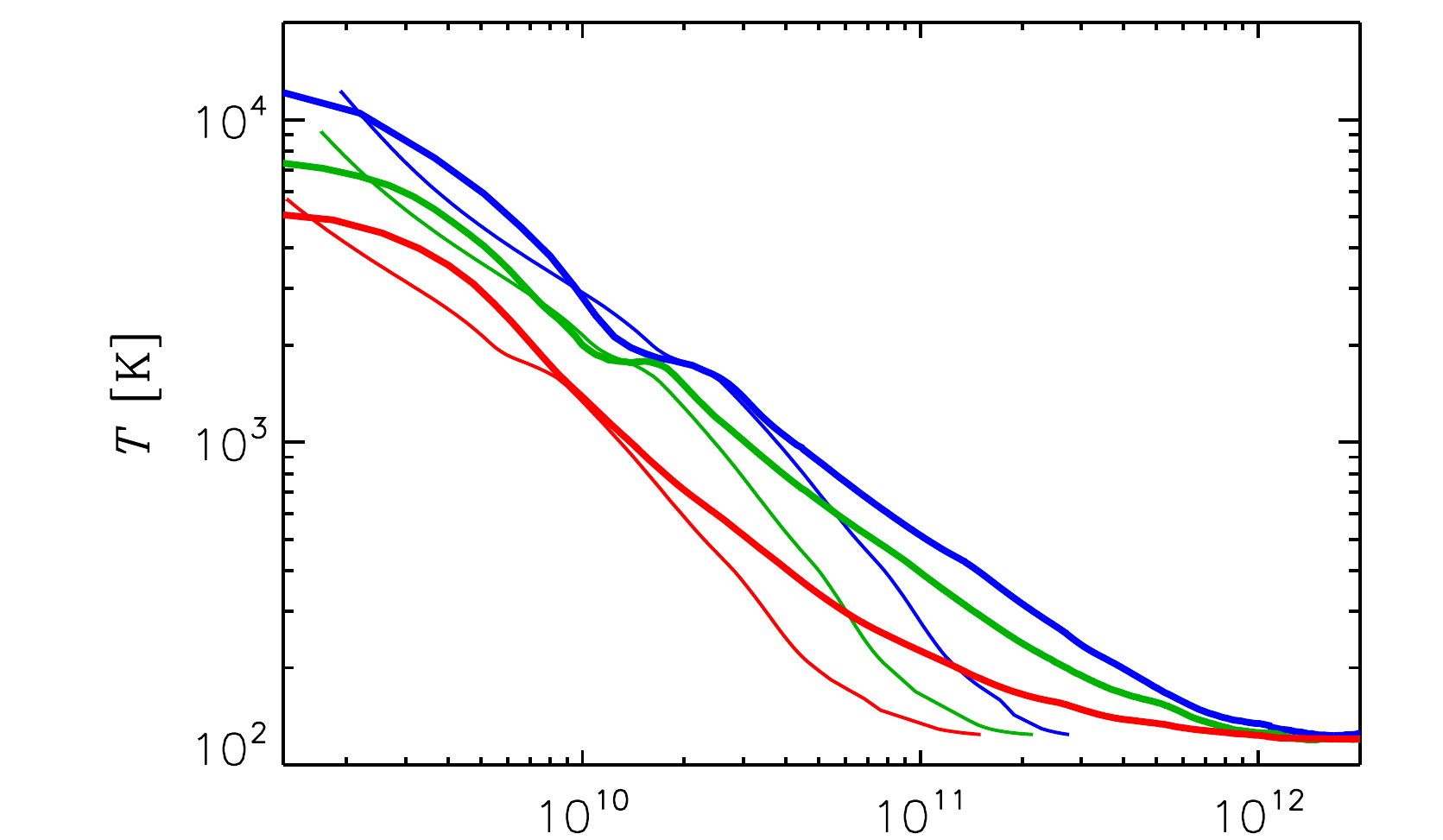}%
\includegraphics[clip]{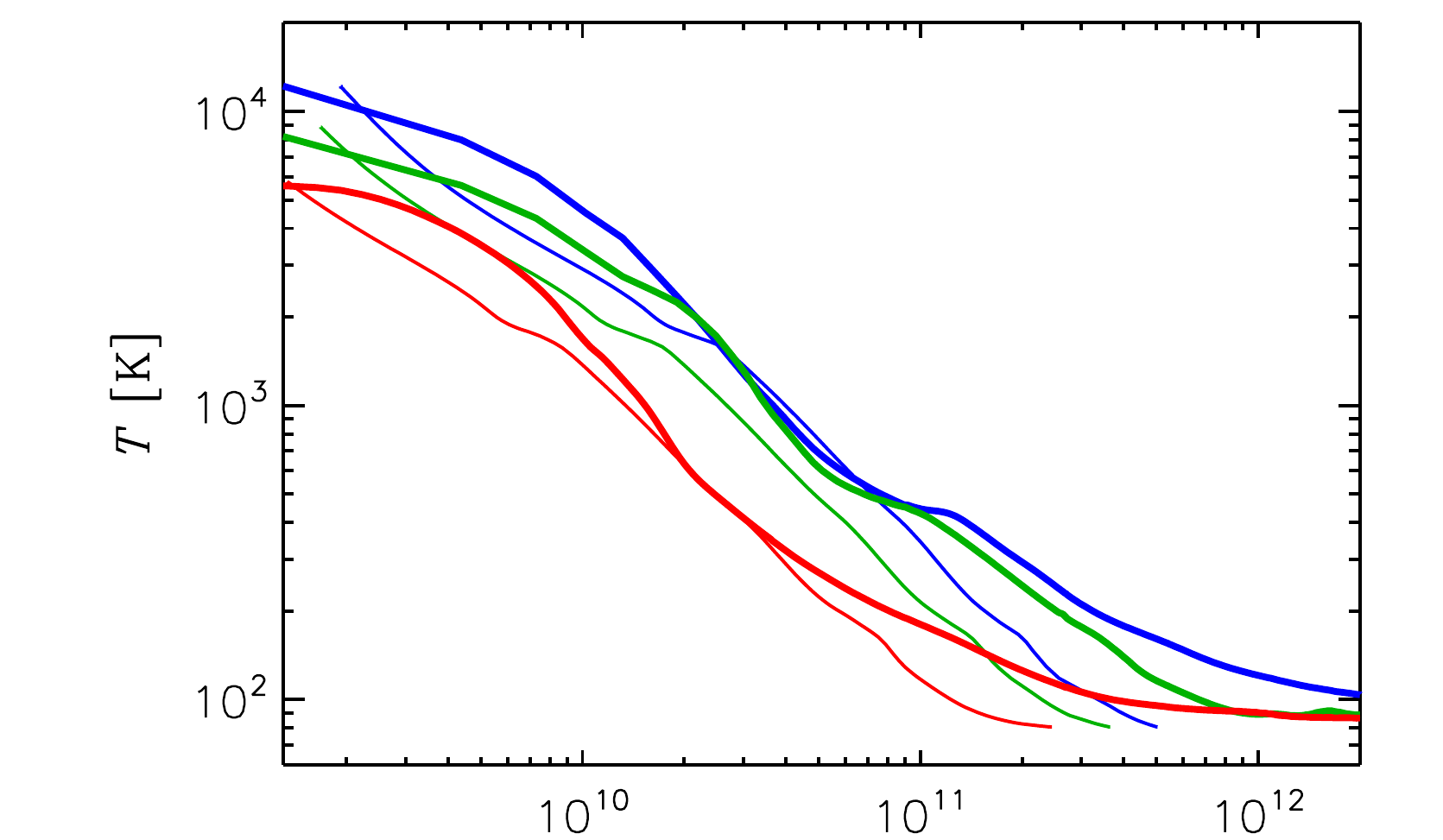}}
\resizebox{\figlen}{!}{%
\includegraphics[clip]{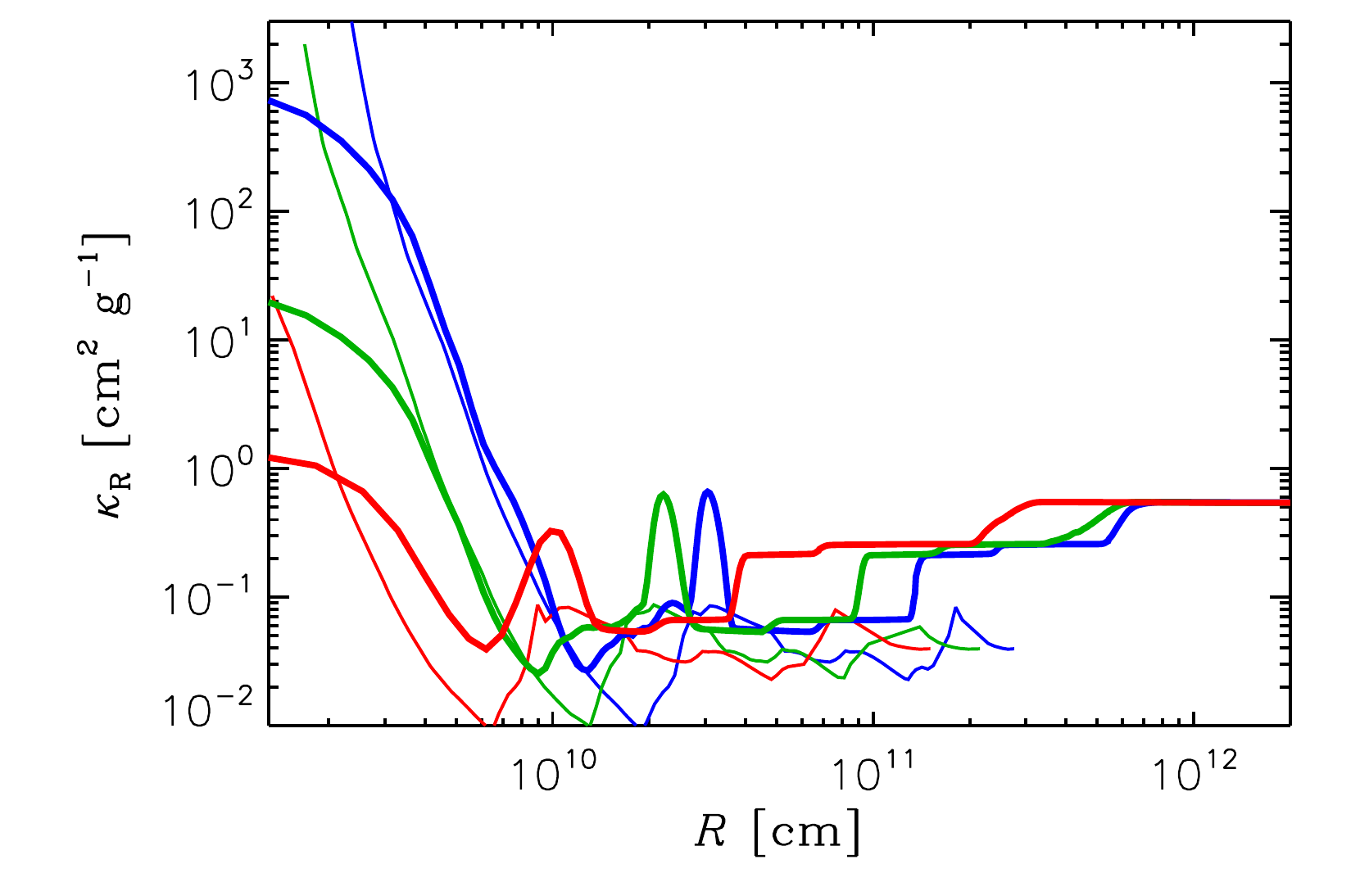}%
\includegraphics[clip]{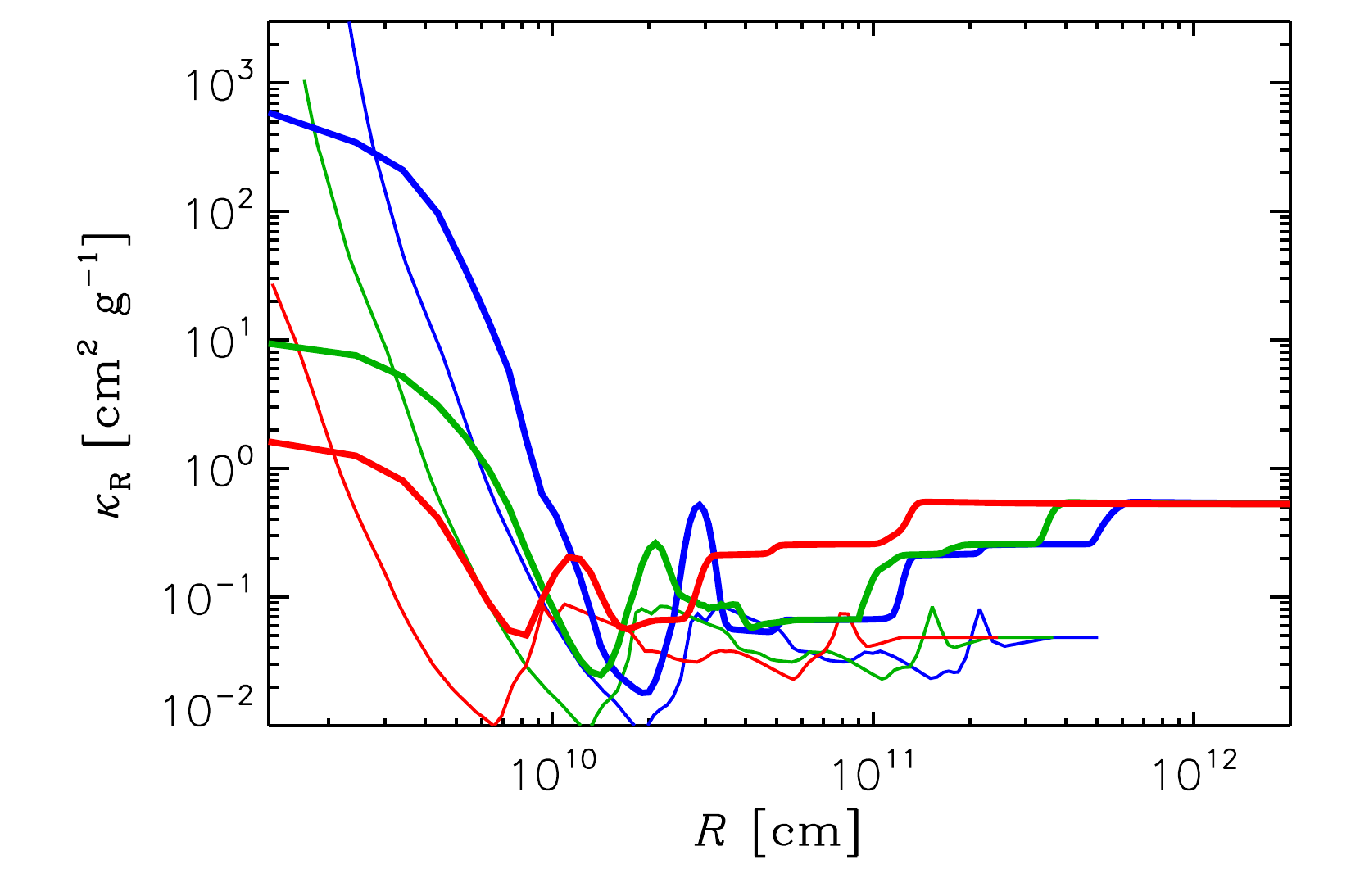}}
\caption{%
             Averaged density (top), temperature (center), and  
             Rosseland mean opacity (bottom), versus distance from
             the core center, of 3D envelopes around
             planetary cores of different masses, as indicated in the legend
             of the top-left panel in units of $\Mearth$. 
             The cores are located at $5$ (left) and $10\,\AU$ (right).
             The thinner lines, which start at the core radius $R_{c}$
             (between $\approx 1.3\times10^{9}$ and 
             $\approx1.9\times10^{9}\,\mathrm{cm}$), 
             represent the same quantities
             obtained from the 1D models discussed in Section~\ref{sec:1D}.
             }
\label{fig:env1D3D}
\end{figure*}
%%%%%%%
Some bulk quantities of the 1D and 3D envelope models are reported 
in Table~\ref{table:13D}.
As explained above, $\dot{M}_{c}$ is calculated in the 1D models and 
applied to the 3D calculations (hence the same entries in the table), 
in order to modify the energy budget of the gas on a length scale 
$\approx R_{c}$ around the core. 
For consistency, the envelope masses $\Me$[3D] are computed using
the same envelope radii as in the 1D models.
The gas accretion rate of the 3D envelopes, $\dot{M}_{e}$[3D], 
is evaluated from the change of the envelope mass over a time 
of roughly $50$ orbits of the core.

The envelope masses reported in Table~\ref{table:13D} also allow us 
to evaluate possible effects caused by the envelope gravity, which are
unaccounted for in the 3D calculations. In fact, the ratio of the radial
component (i.e., toward the core center) of the gravitational force due 
to the gas and that due to the core is at most $\Me/\Mc$.  
According to 1D models, this ratio ranges from $10^{-4}$ to $10^{-3}$,
and similar ratios result from the 3D models. Hence, only a relatively small 
contribution is expected to arise from the envelope gravity at these stages
of evolution.

There are several physical differences between the 1D and the 3D
calculations. In particular, the 1D models are computed as 
a sequence of hydrostatic envelope structures whereas the 3D models 
are intrinsically hydrodynamical, with a complex velocity field.
Diffusion and convection of energy are mutually exclusive in 
the 1D calculations (and the adiabatic temperature gradient
is applied to the convective layers) whereas they always occur 
together in the 3D calculations, through an advection-diffusion 
equation (see Section~\ref{sec:GT}), regardless of which transport 
mechanism is dominant. 
Conversion of mechanical energy into thermal energy, determined by
viscosity through Equation~(\ref{eq:Psi}), is taken into account in 
the 3D, but not in the 1D, calculations.
The release of energy by planetesimals penetrating the envelope
occurs gradually and is depth-dependent in the 1D models, whereas
all the energy is released at the bottom of the envelope in the 3D models, 
effectively as if no ablation took place and planetesimals were intact upon 
hitting the core.
The dust opacity of the outer layers of the envelope is different, 
typically lower in the 1D models.

Considering all these differences and the fact that some properties
of 1D envelopes are related to their history, we should only seek 
for consistency between 1D and 3D calculations. 
In this sense, the bulk quantities reported in Table~\ref{table:13D} 
do show a general agreement: both envelope masses and gas 
accretion rates, $\dot{M}_{e}$, differ by factors of $2$ or less. 
In particular, as gas accretion is a consequence of contraction, 
the numbers in the Table suggest that the contraction 
time scales of the 1D and 3D envelopes are comparable. 
The relative increase of $\dot{M}_{e}$, for increasing core mass, 
is also comparable. 
It is important to notice that gas accretion rates differ by factors of oder 
unity, between calculations with $a=5$ and $10\,\AU$ (in both 1D and 
3D), an indication that they are dictated by the internal envelope 
properties, as just mentioned, rather than imposed by the external 
disk thermodynamics (which is different at the two orbital locations,
see Section~\ref{sec:DS}).

Figure~\ref{fig:env1D3D} shows a more detailed comparison, between
the 1D and 3D calculations, of the density (top), temperature (center), 
and Rosseland mean opacity (bottom), versus distance from the core 
center, $R$, for values of the ratio $\Mc/\Mearth$ indicated in the legend 
of the top-left panel. Left panels refer to planets with a semimajor axis 
of $5\,\AU$ and right panels refer to planets with $a=10\,\AU$.
The envelope properties of the 1D calculations (thin lines) are similar
at the two orbital radii, except in the outer parts because of the different
boundary conditions (see Section~\ref{sec:1D}).
The results from the 3D calculations (thick lines), computed as averages 
around the core at $\theta=\pi/2$ and plotted up to a distance $R>\Rhill/2$,
show a somewhat larger contrast between cases at $5$ and $10\,\AU$.
Overall, the 3D envelope models are less dense in the interiors, 
denser in the outer parts, and generally hotter than the 1D envelope models.

Differences should be expected at length scales of order $R_{c}$ from
the core 
($1.3\times10^{9} \lesssim R_{c} \lesssim 1.9\times10^{9}\,\mathrm{cm}$), 
as the 3D models have a linear resolution between about $R_{c}$ 
and $2\,R_{c}$, while the resolution of the 1D models is better by two (or more) 
orders of magnitude! 
The gravitational potential is also different at $R_{c}$, since the 3D calculations 
use a softened potential (see  Appendix~\ref{sec:phi_c}), producing a shallower 
gravity field. 
Despite these limitations, thermodynamical quantities at the base of 
the envelope are comparable in most cases. 
In the 1D models, the density at $R_{c}$ ranges from
$\approx 4\times 10^{-4}$ to $\approx 4\times 10^{-3}\,\mathrm{g\,cm^{-3}}$,
the temperature varies from $\approx 5700$ to $\approx 12000\,\K$, and
the pressure is between
$\approx 10^{8}$ and $\approx 3\times 10^{9}\,\mathrm{dyne\,cm^{-2}}$.
In the 3D models, the density at the core ranges from
$\approx 10^{-5}$ to $\approx 10^{-3}\,\mathrm{g\,cm^{-3}}$,
the temperature varies from $\approx 5000$ to $\approx 11000\,\K$, and
the gas pressure is between
$\approx 10^{6}$ and $\approx 4\times 10^{9}\,\mathrm{dyne\,cm^{-2}}$.
The largest discrepancies at the base of the envelope occur for the density 
(and hence pressure) around the $5\,\Mearth$ cores, as visible in the top 
panels of Figure~\ref{fig:env1D3D}, although the case at $5\,\AU$ is also 
the one that displays the best agreement with the 1D calculation beyond 
a few core radii.

Differences in density, temperature, and pressure are also expected at 
the outer radius of the 1D envelopes, as values there are affected 
by the boundary conditions (see discussion in Section~\ref{sec:1D}).
At that distance from the core, between $10^{11}$ and $5\times 10^{11}\,\mathrm{cm}$,
the density of the 3D envelopes is larger by factors between 
$7$  and $12$, the temperature is higher by factors between 
$1.5$  and $2$, and the gas pressure is greater by factors between 
$10$ and $25$.
These factors also represent the largest relative differences, between 
the 1D and 3D calculations, in the density and temperature 
distributions throughout most of the envelope ($R\gtrsim 3\,R_{c}$). 

The Rosseland mean opacity in the envelope, of both gas and dust,
is illustrated in the bottom panels of Figure~\ref{fig:env1D3D}
(see the lower-left panel of Figure~\ref{fig:k} for a plot of $\kappa_{\mathrm{R}}$ 
as a function of temperature).
Distinct transitions can be seen, corresponding to the sublimation/formation 
of the various grain species included in the opacity calculation 
(see Appendix~\ref{sec:opa_calc}). The most prominent transitions are
those associated with the vaporization of water ice grains at $T\approx 160\,\K$
and of refractory organics grains at $T\approx 420\,\K$ 
(the average temperatures at the outer radius of the 3D envelopes, defined 
in the next section, are $\lesssim160\,\K$).
Minor transitions can also be identified, such as the one corresponding to the 
sublimation of troilite (FeS) at $T\approx 680\,\K$.
The reduction of opacity due to the vaporization deeper in the envelope 
of more refractory species, such as silicates at $T\gtrsim 1400\,\K$ when 
$\rho\gtrsim 10^{-8}\,\mathrm{g\,cm^{-3}}$, is compensated for by 
the increase of molecular opacity, which peaks around $2000\,\K$
\citep[see][]{ferguson2005}. 
At temperatures below the sublimation temperature of refractory 
organics, the opacity of the 3D envelopes is larger than that of 
the 1D envelopes, on average by factors of $5$--$7$.
Above such temperature, and up to $\sim 1500\,\K$, the opacities
differ by a factor $2$, or less.
As can be seen in the lower-left panel of Figure~\ref{fig:k}, 
between $\sim 100$ and $\sim 1000\,\K$ the opacity of the 3D models 
is a factor $\sim 10$ lower than the interstellar dust opacity (due 
to the presence of larger grains). The grain opacity of 1D models 
is interstellar, but reduced by a factor $50$.
In the envelope interiors, differences in (gas) opacity are likely less 
relevant as energy transport is expected to occur mostly via convection.

%%%%%%
\begin{figure*}[]
\centering%
\resizebox{\figlen}{!}{%
\includegraphics[clip]{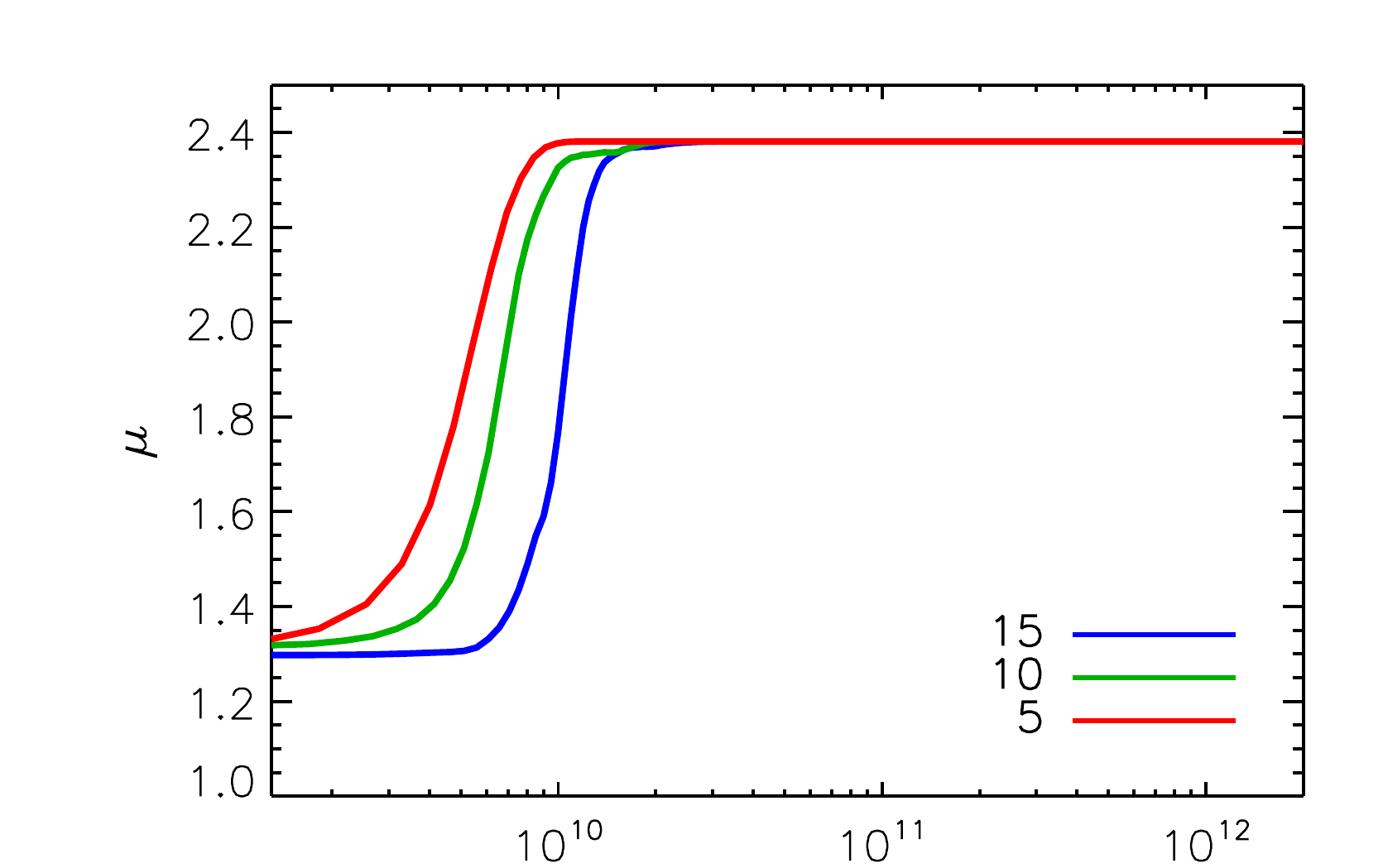}%
\includegraphics[clip]{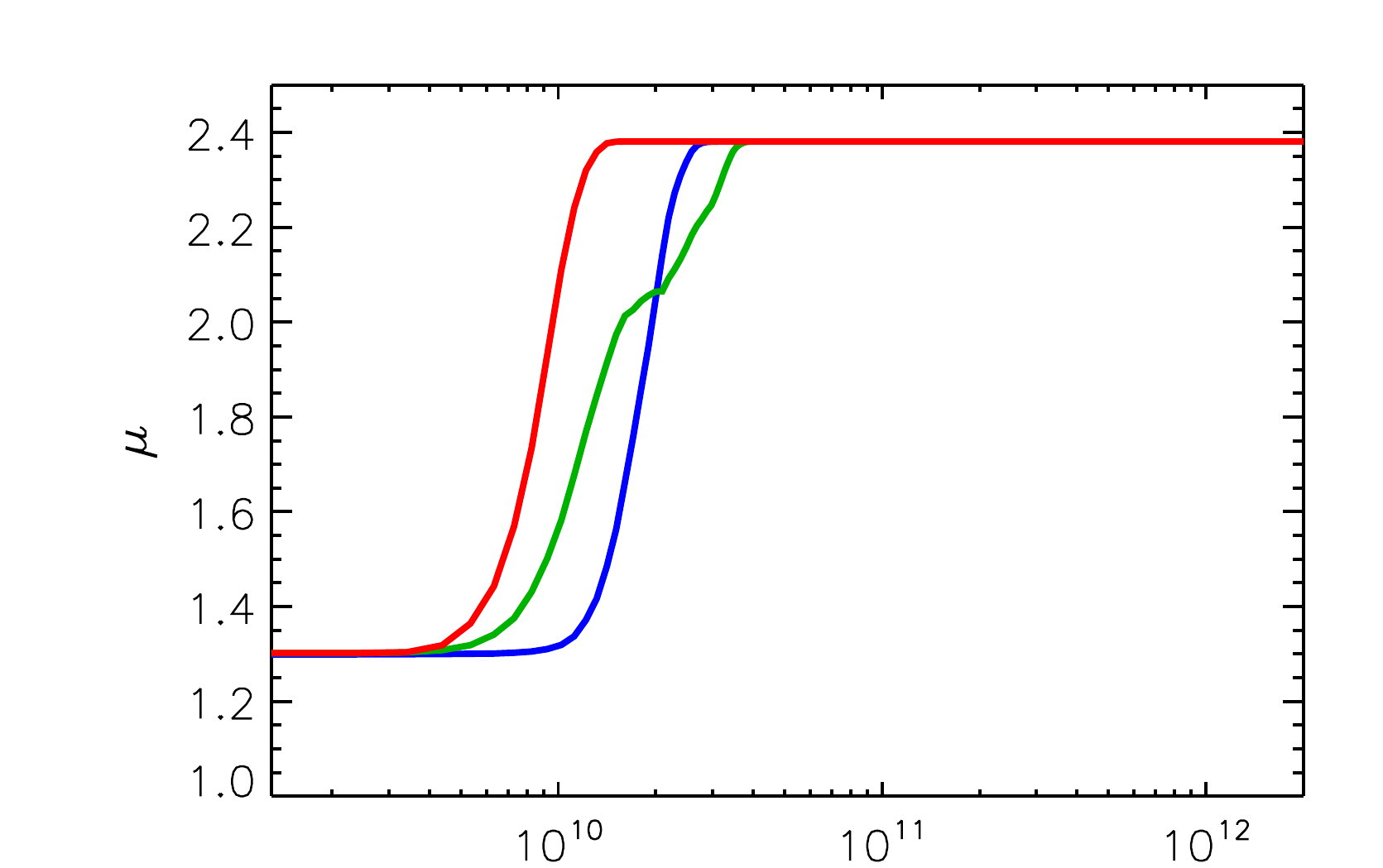}}
\resizebox{\figlen}{!}{%
\includegraphics[clip]{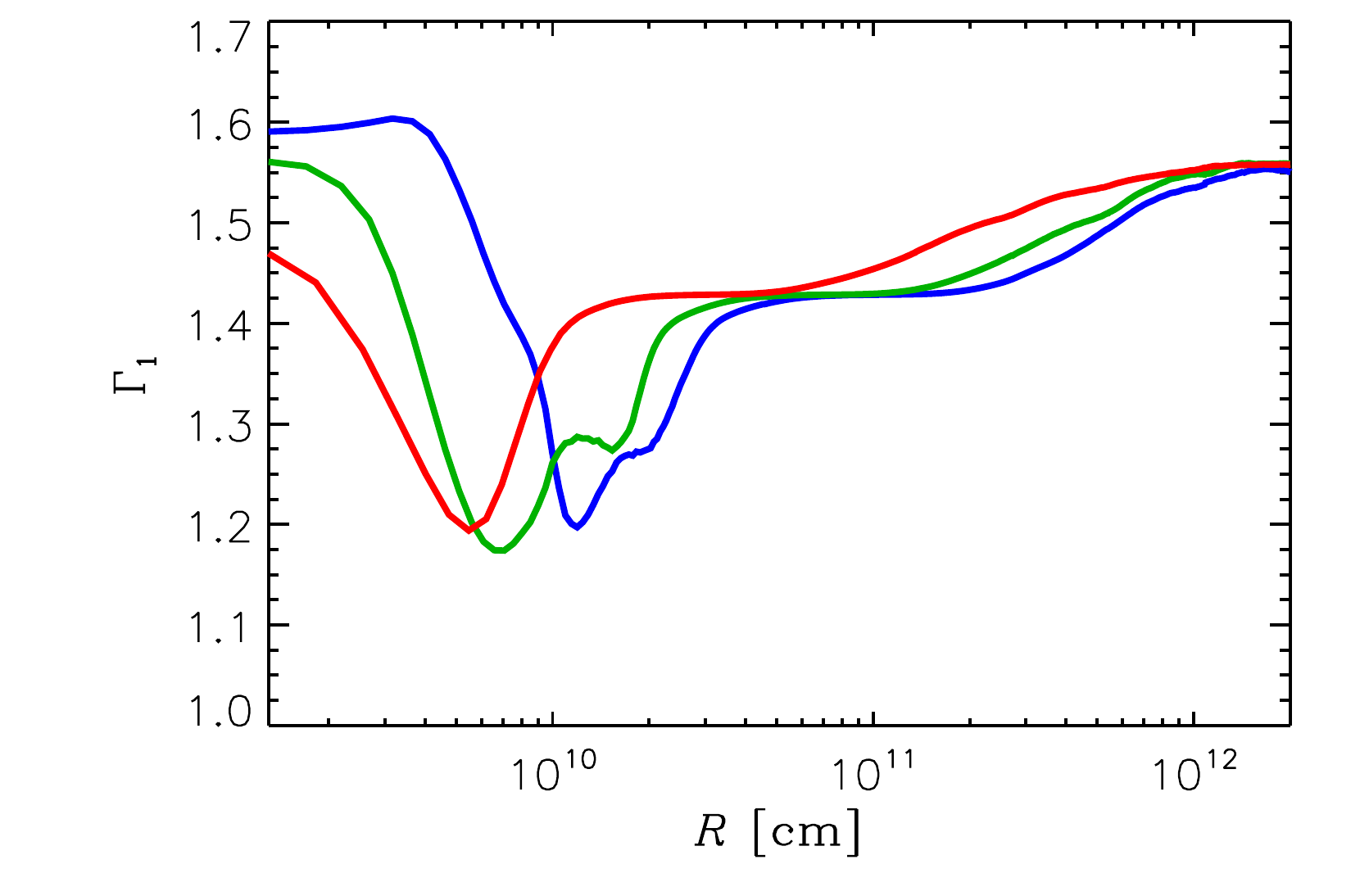}%
\includegraphics[clip]{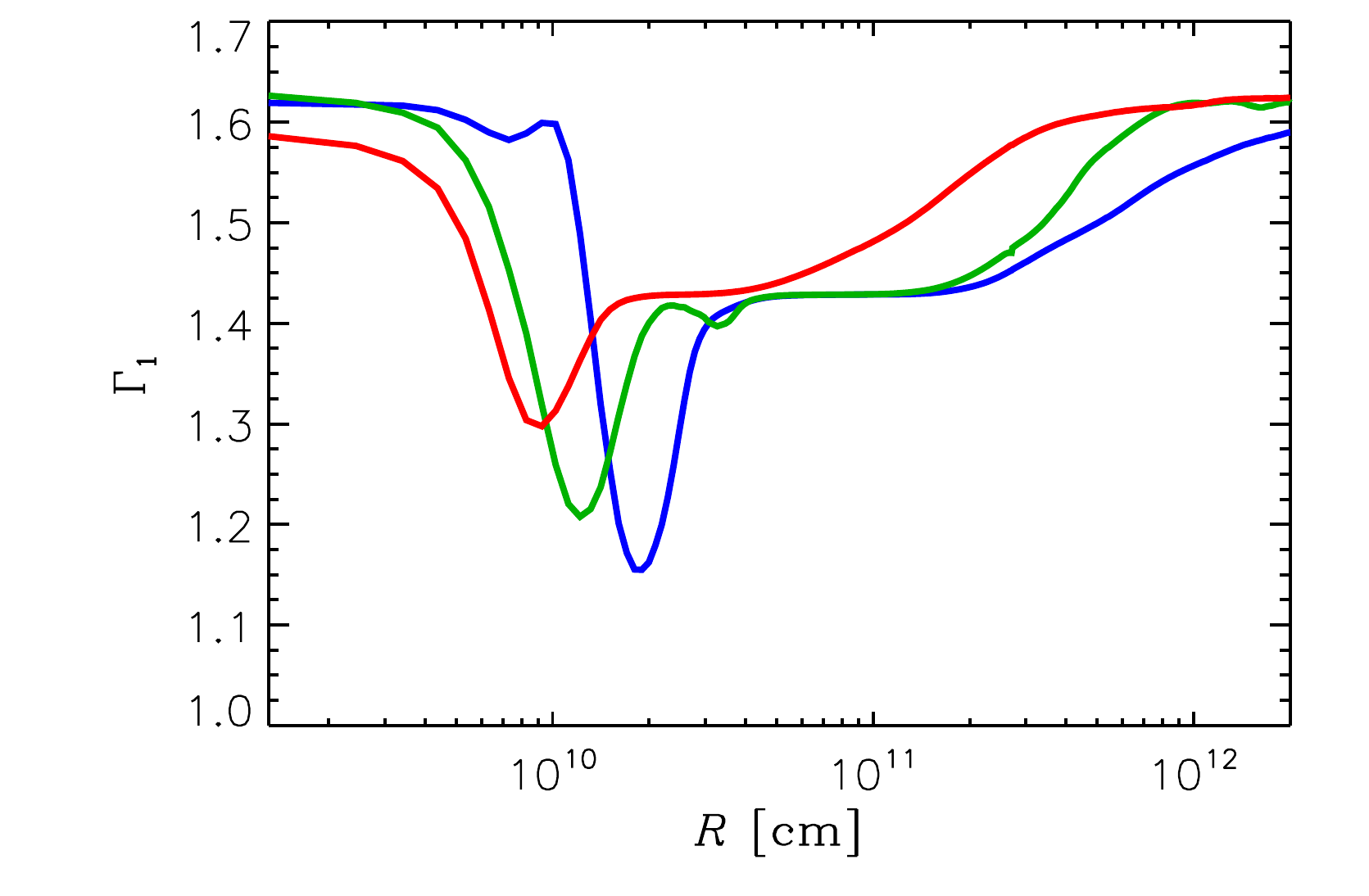}}
\caption{%
             Average values of the mean molecular weight, $\mu$, in Equation~(\ref{eq:mue})
             (top) and of the first adiabatic exponent,$\Gamma_{1}$, in Equation~(\ref{eq:Gamma_1}) 
             (bottom), versus distance from the core center, in 3D envelopes around
             cores of various masses, as indicated in the top-left panel in units of $\Mearth$. 
             Left and right panels refer to the core's orbital radius of $5$ and $10\,\AU$, 
             respectively.
             }
\label{fig:envGM}
\end{figure*}
%%%%%%%
The mean molecular weight (Equation~(\ref{eq:mue})) and the first adiabatic 
exponent (Equation~(\ref{eq:Gamma_1})) of the gas in the envelope are
plotted in Figure~\ref{fig:envGM}, for all core masses. 
The left and right panels refer, respectively, to cases with semimajor axis 
of $5$ and $10\,\AU$.
Significant dissociation of $\mathrm{H}_{2}$ begins at $T\sim 1500\,\K$
($\rho\sim 10^{-8}\,\mathrm{g\,cm^{-3}}$) and is nearly complete
(when $\mu\simeq 1.3$) at $T\gtrsim 4000\,\K$, depending on the local 
gas density (see Figure~\ref{fig:mue_sim}). 
The dissociation begins farther away from the core, at a distance
about $25$\% greater, in the model with $\Mc=10\,\Mearth$ at $10\,\AU$ 
than in the $15\,\Mearth$ case at the same orbital distance 
(see top-right panel) because of the similar temperatures but lower densities. 
Otherwise, the volume of atomic hydrogen increases as the core mass
becomes larger. 
No significant ionization is observed deep in the envelope.
The first adiabatic exponent, in the bottom panels, dips to a minimum
during the dissociation of $\mathrm{H}_{2}$. The curves also show, 
to the right of the minimum, the reduction of $\Gamma_{1}$ caused
by the excitation of vibrational and rotational states of $\mathrm{H}_{2}$ 
(see Figure~\ref{fig:EoS}).

%%--------------------------------------------------------------------------
\subsection{Size, Shape, and Rotation of 3D Envelopes}
\label{sec:ESI}
%%--------------------------------------------------------------------------

%%%%%%
\begin{figure*}[]
\centering%
\resizebox{0.85\linewidth}{!}{%
\includegraphics[clip]{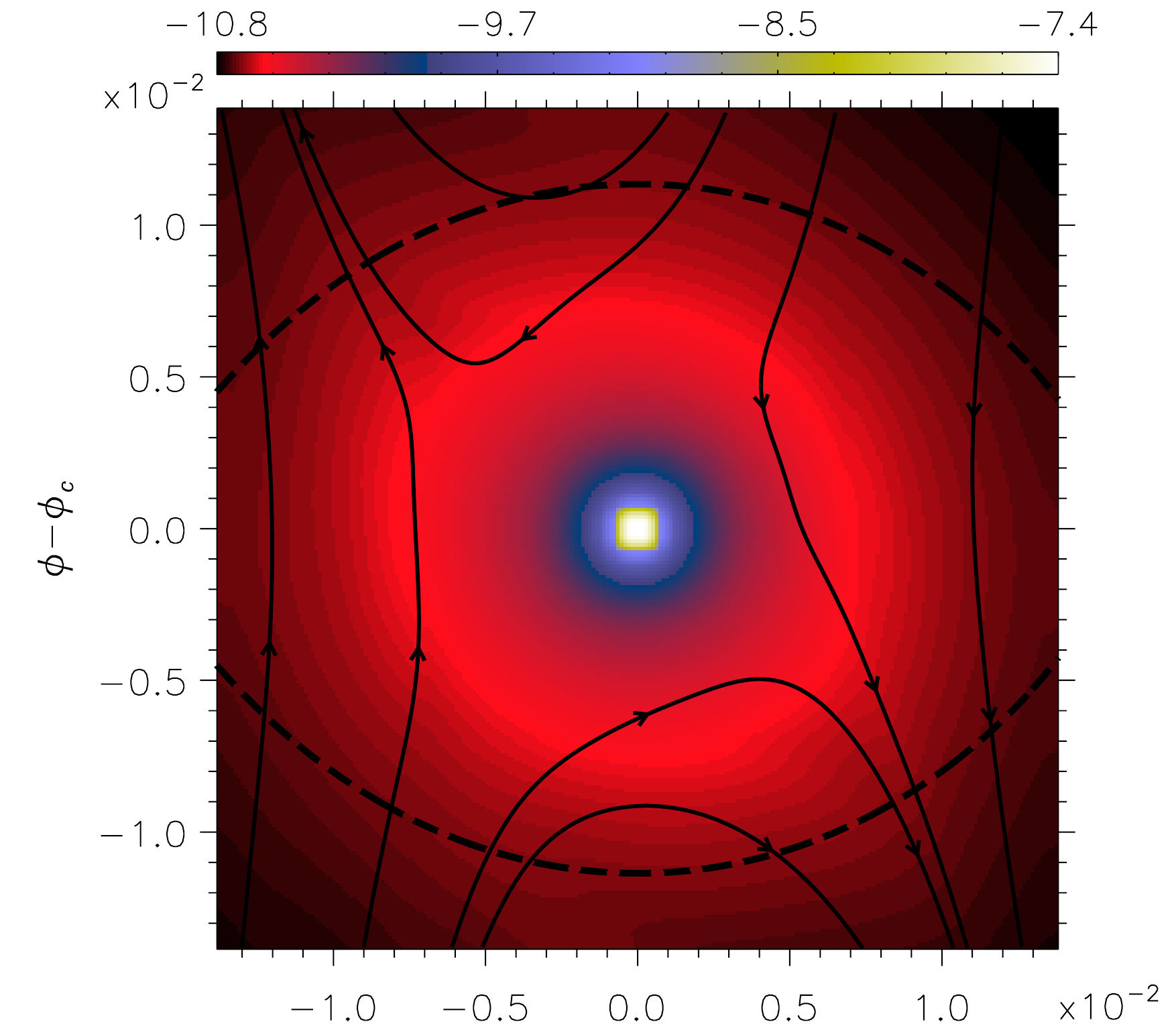}%
\includegraphics[clip]{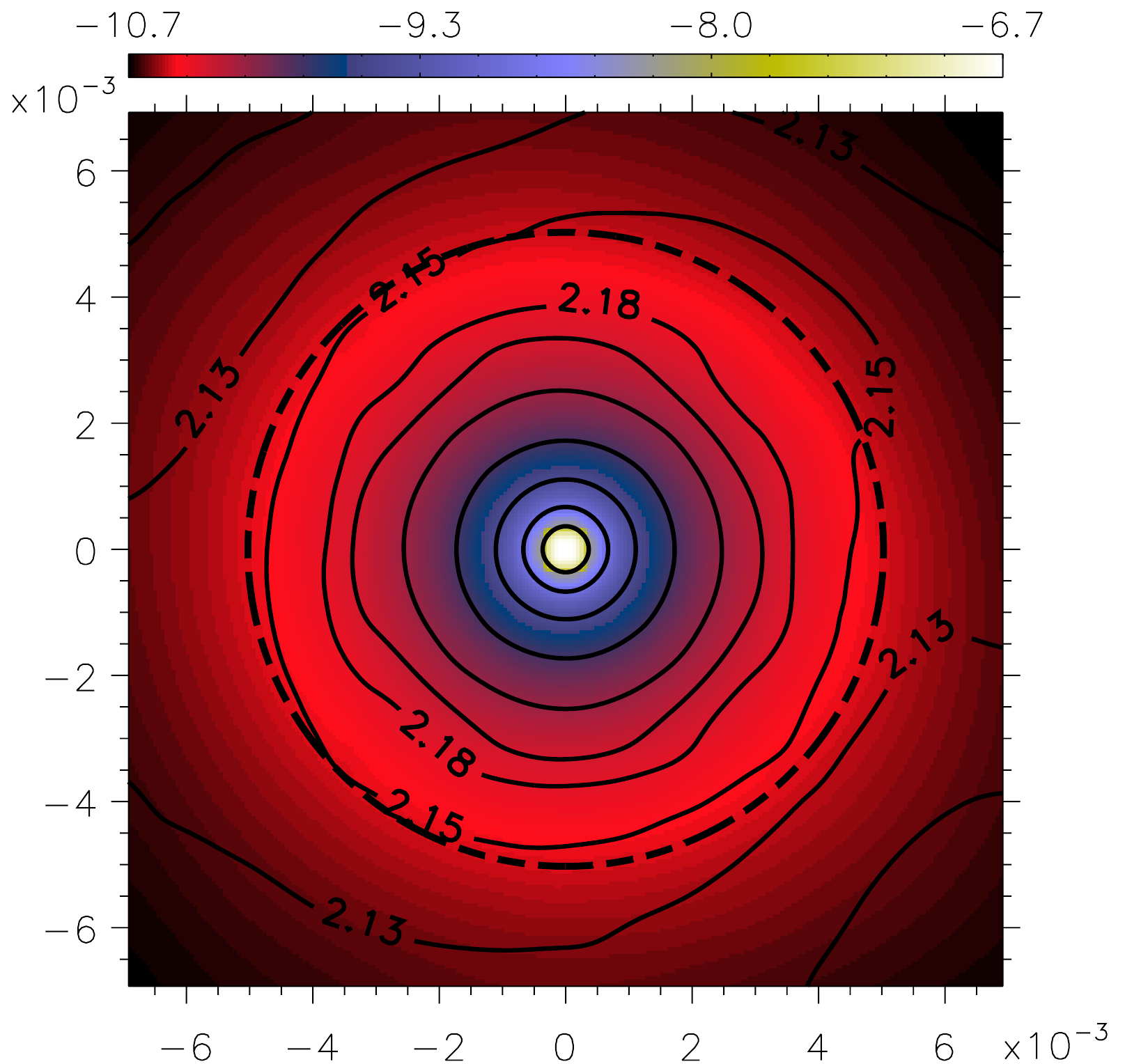}}
\resizebox{0.85\linewidth}{!}{%
\includegraphics[clip]{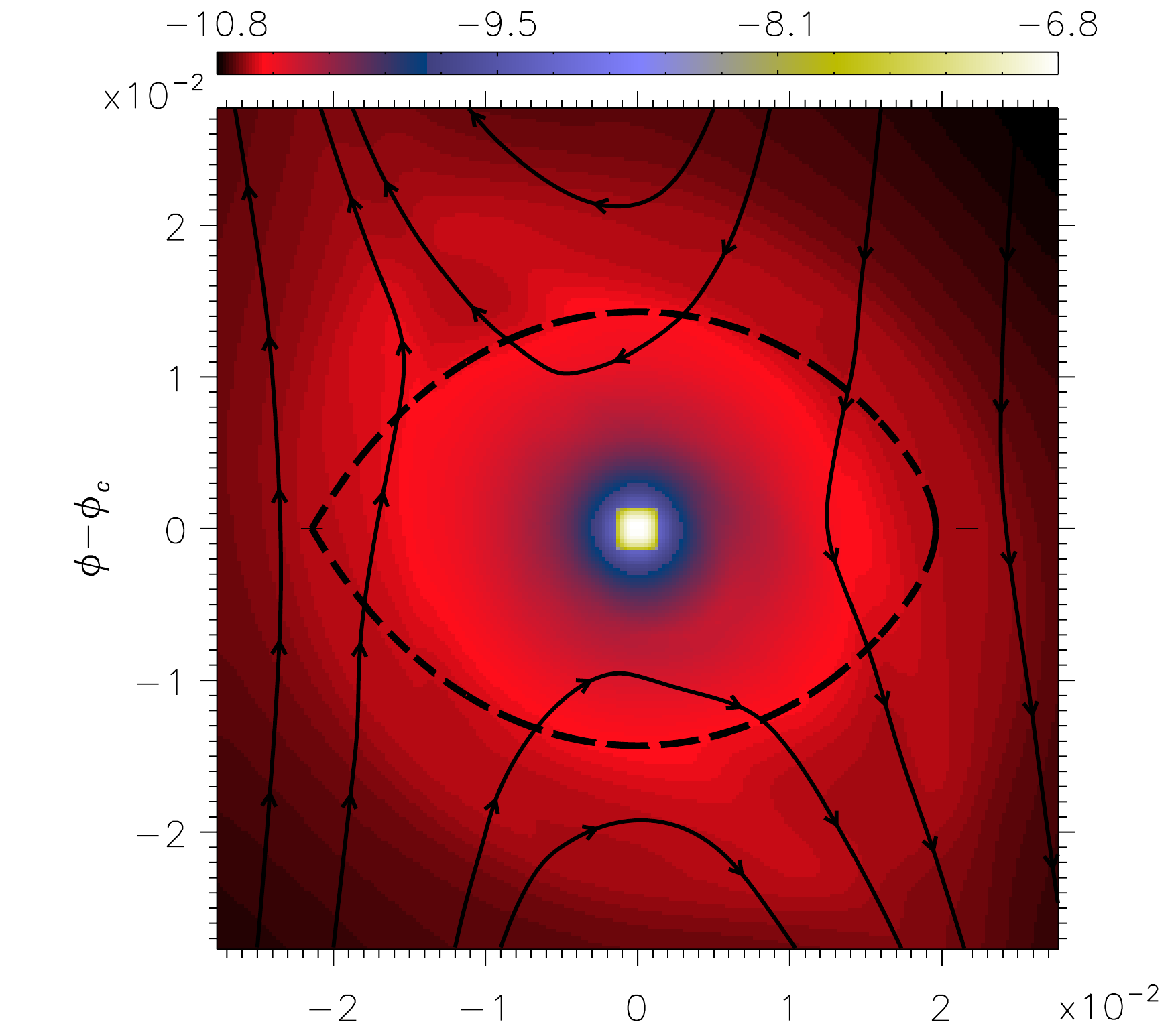}%
\includegraphics[clip]{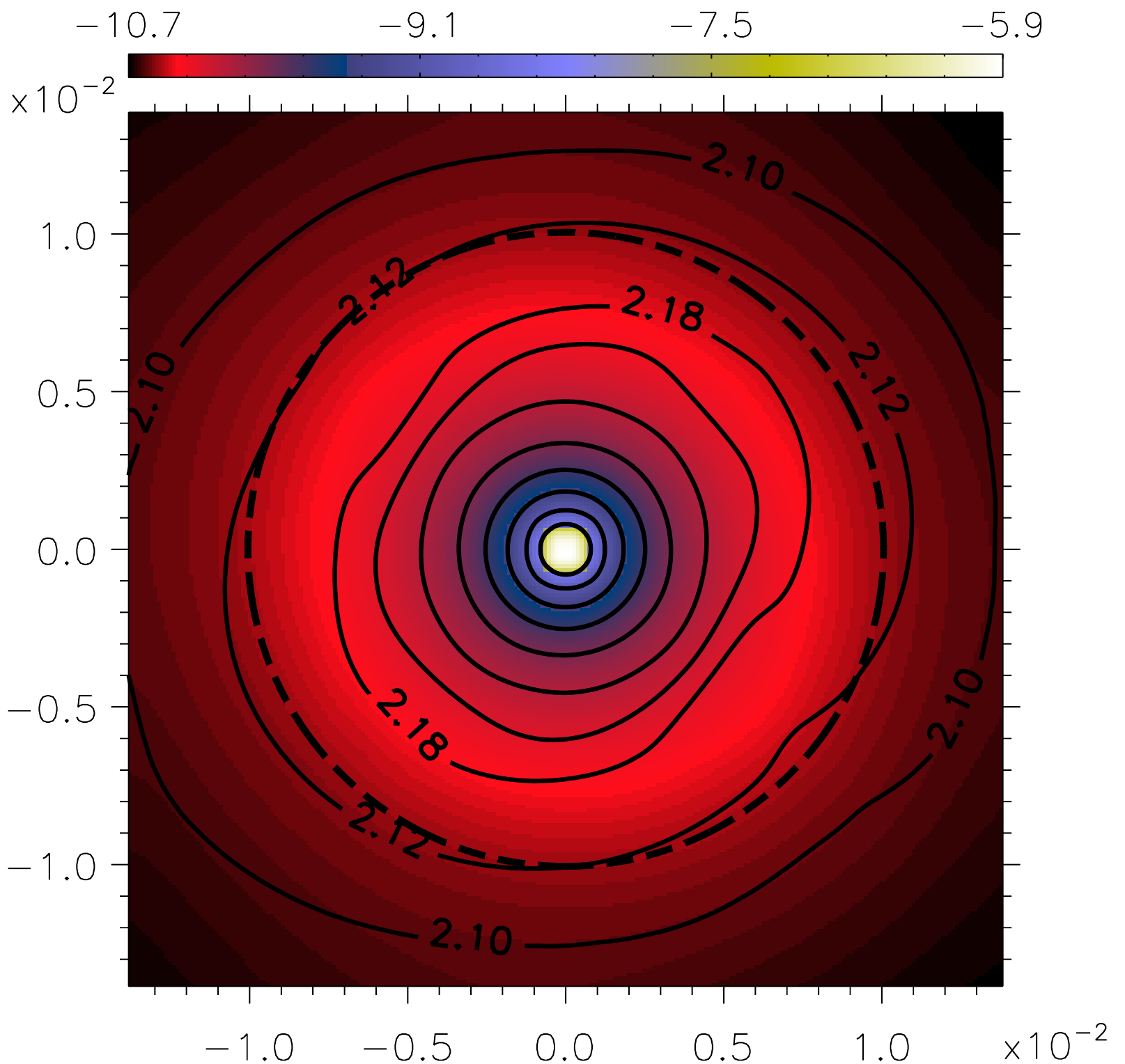}}
\resizebox{0.85\linewidth}{!}{%
\includegraphics[clip]{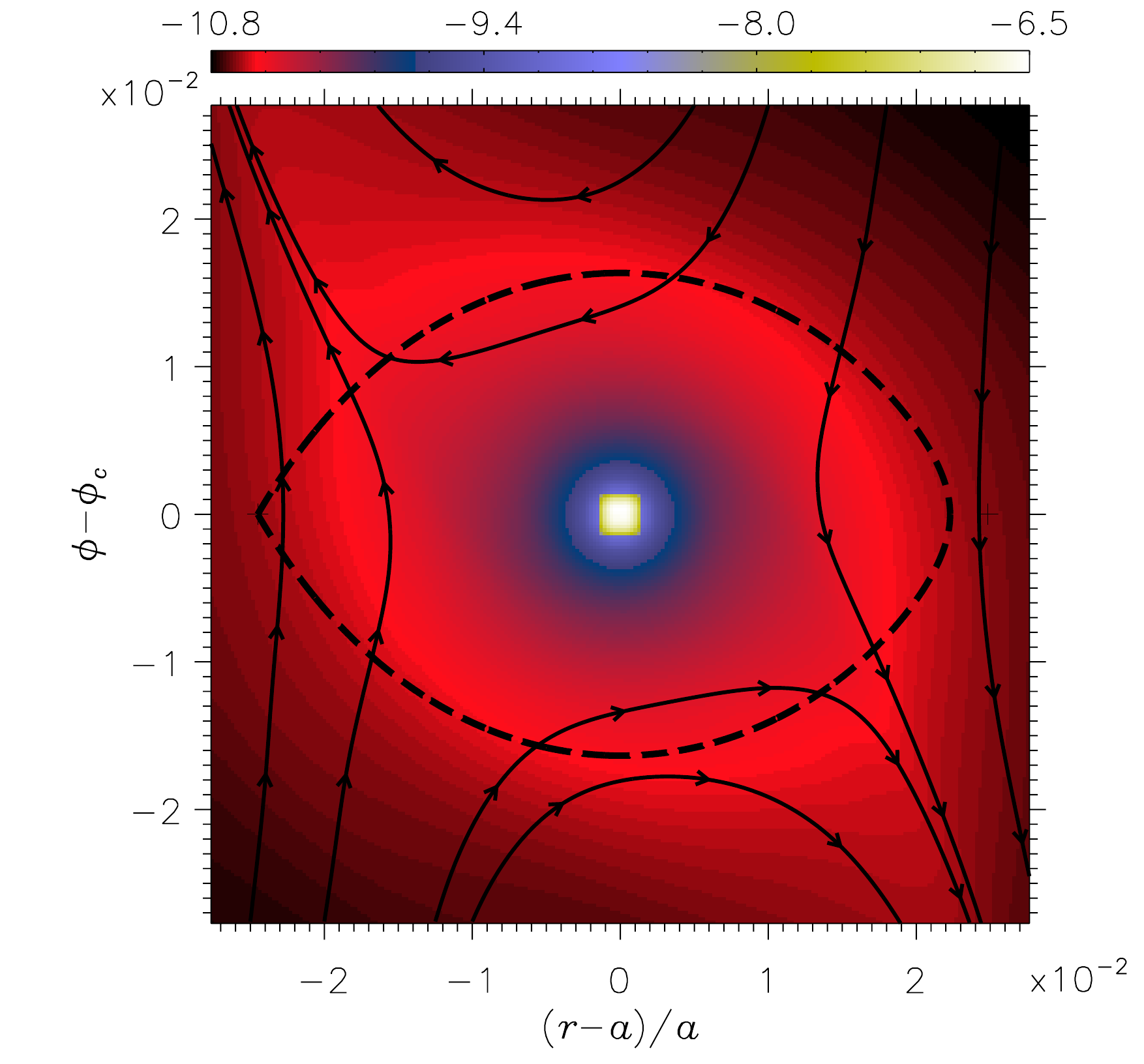}%
\includegraphics[clip]{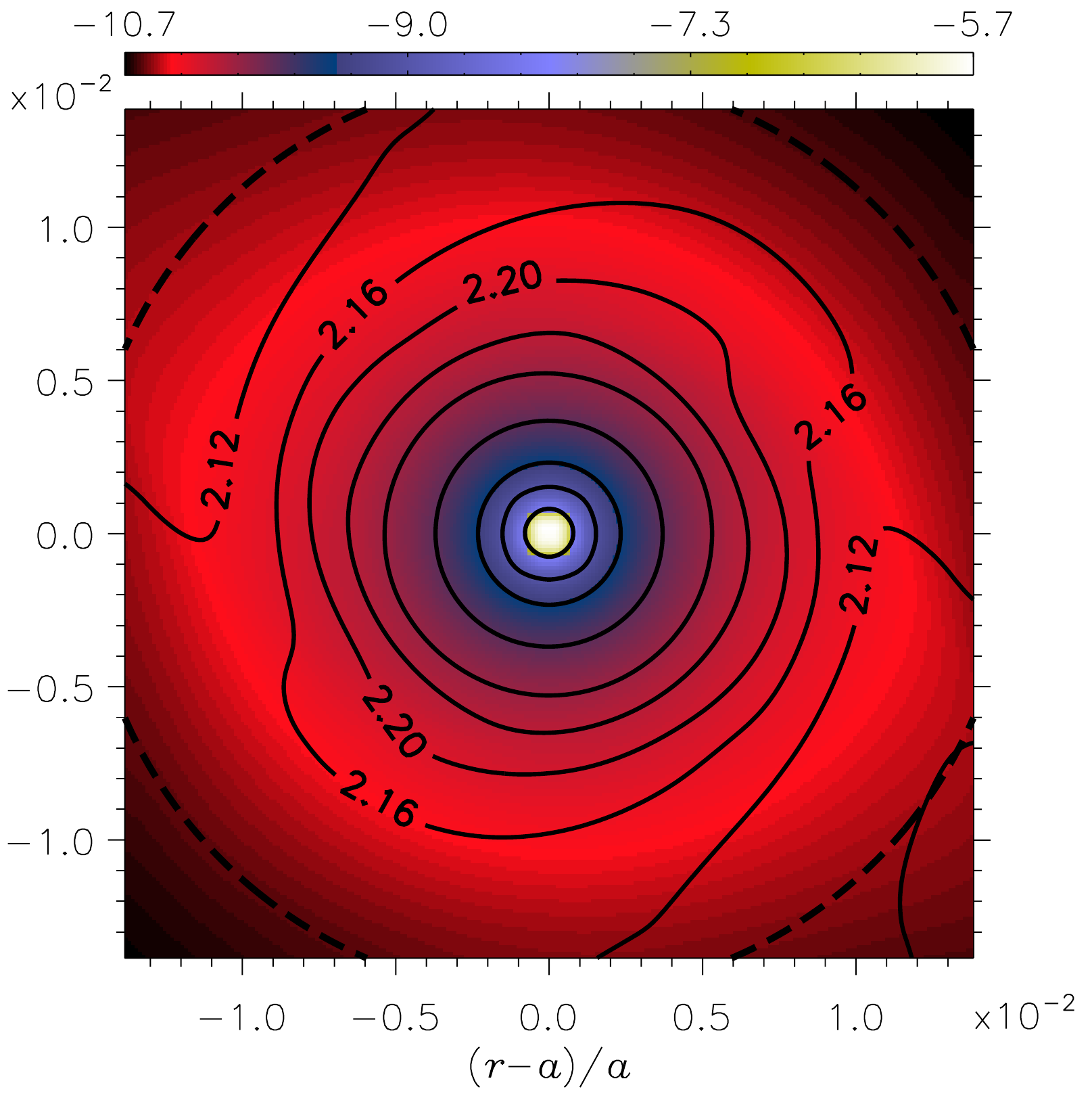}}
\caption{%
             Density around the envelope region (on a logarithmic scale 
             in units of $\mathrm{g\,cm}^{-3}$) in the disk mid-plane,
             for planetary cores located at $5\,\AU$. From top to bottom, the core mass is 
             $5$, $10$, and $15\,\Mearth$. The density is saturated around
             the core position to improve the scale contrast.
             The left panels also show the flow streamlines at the mid-plane, 
             along with the intersection with the  Roche lobe (dashed line).
             The right panels show the temperature contours (on a logarithmic scale 
             in units of $\K$). The dashed circle represents the intersection with the Bondi 
             sphere (see Section~\ref{sec:PCE}). 
             }
\label{fig:xz_env_5}
\end{figure*}
%%%%%%%
%%%%%%
\begin{figure*}[]
\centering%
\resizebox{0.85\linewidth}{!}{%
\includegraphics[clip]{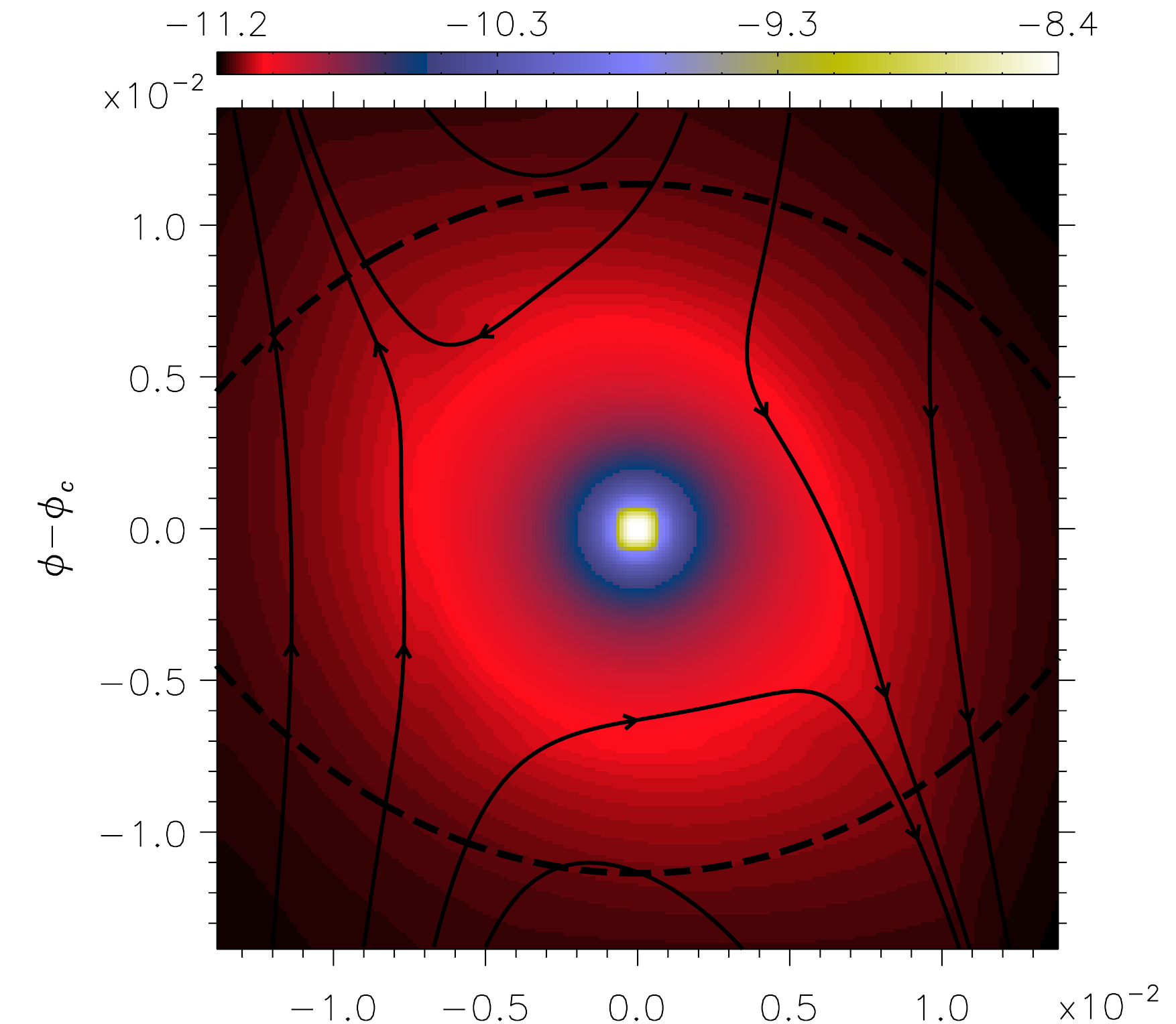}%
\includegraphics[clip]{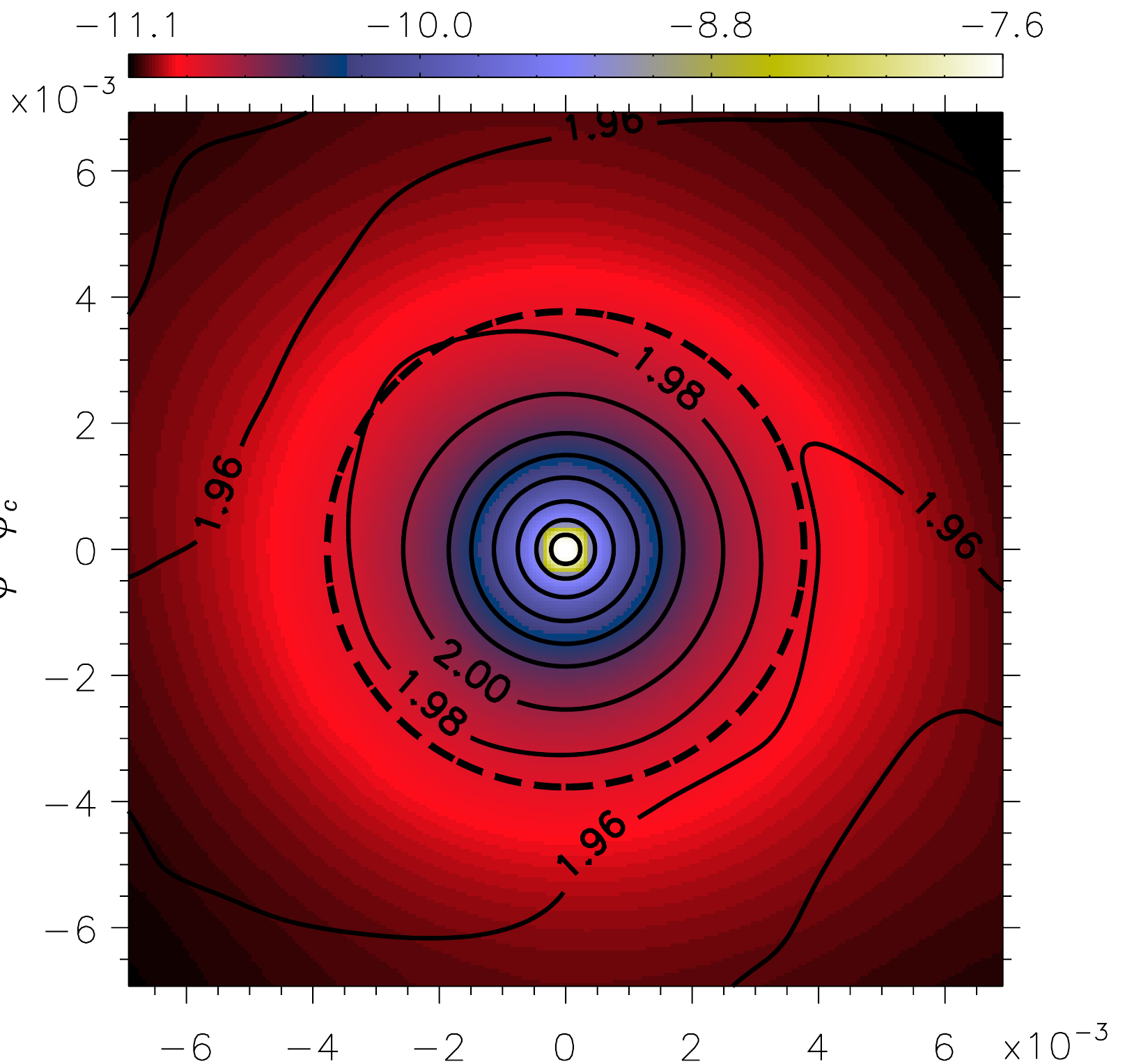}}
\resizebox{0.85\linewidth}{!}{%
\includegraphics[clip]{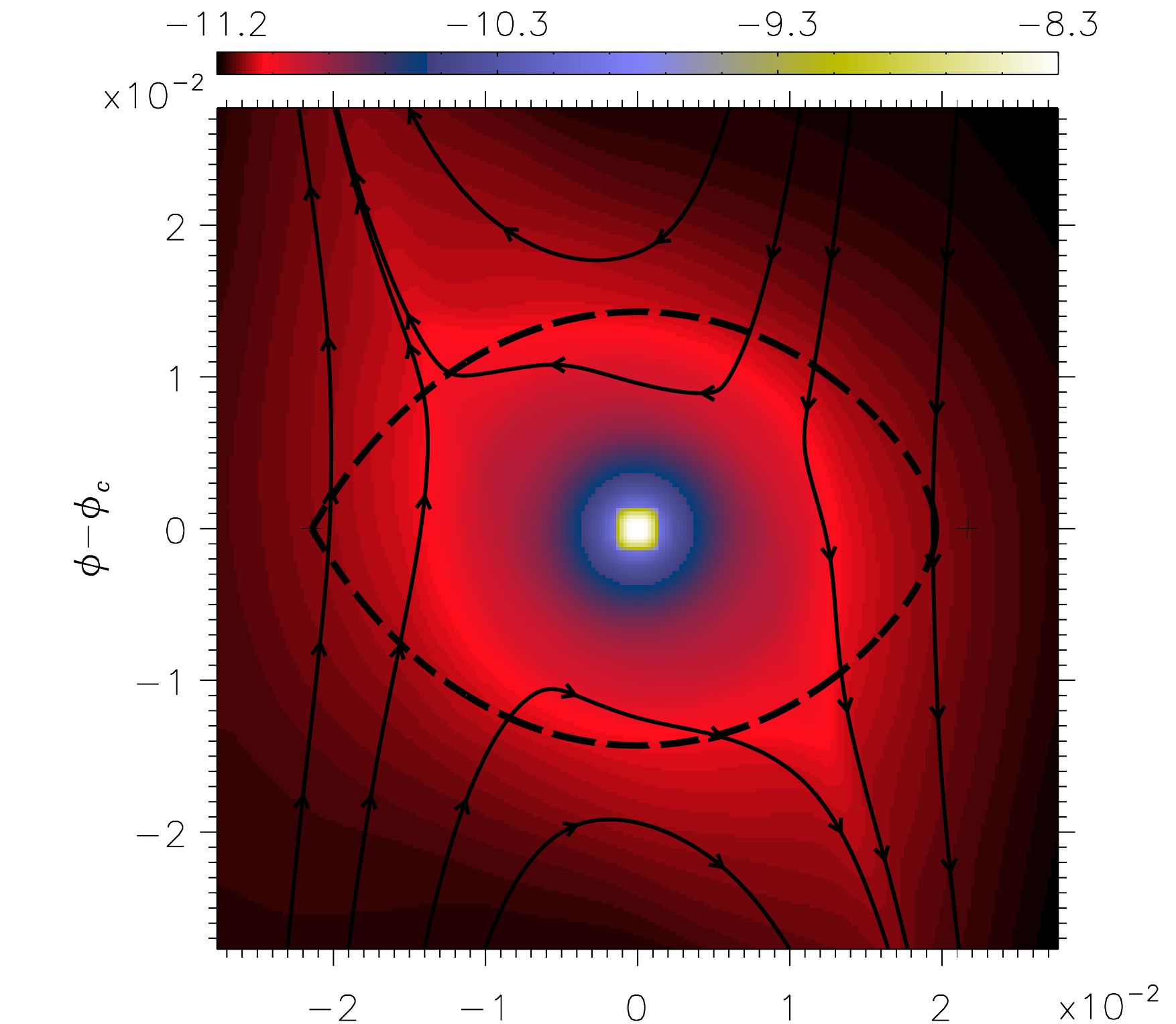}%
\includegraphics[clip]{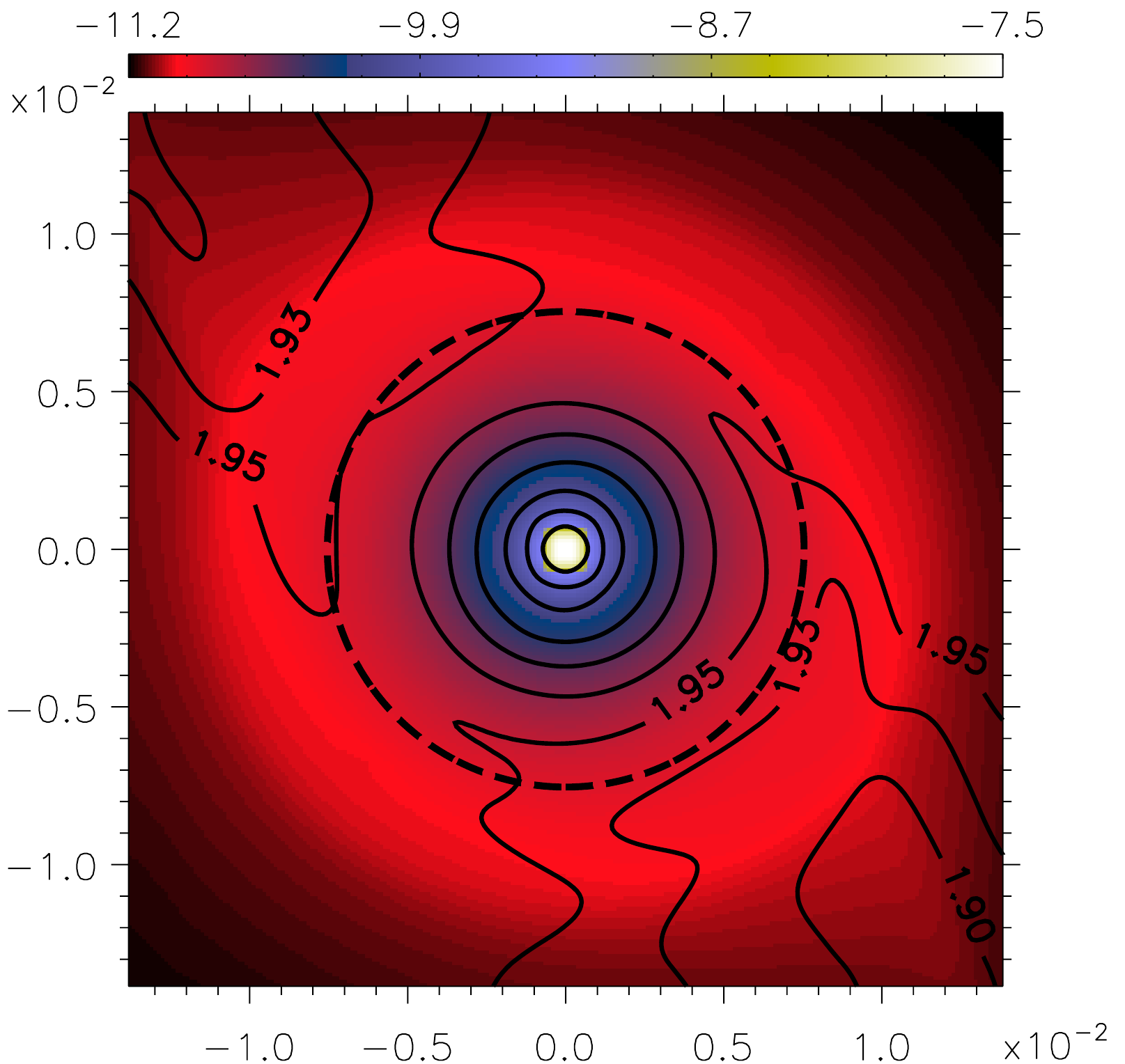}}
\resizebox{0.85\linewidth}{!}{%
\includegraphics[clip]{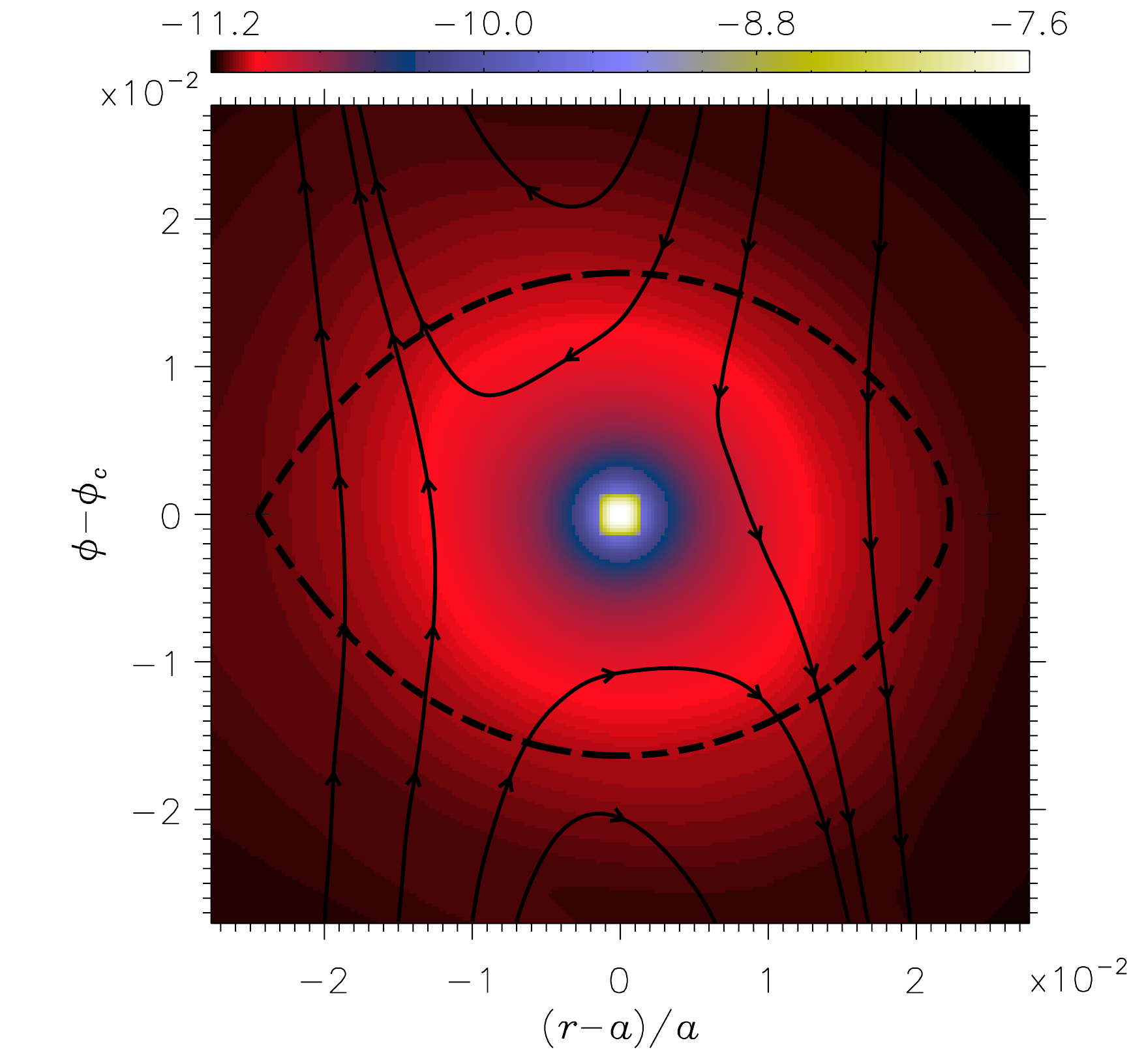}%
\includegraphics[clip]{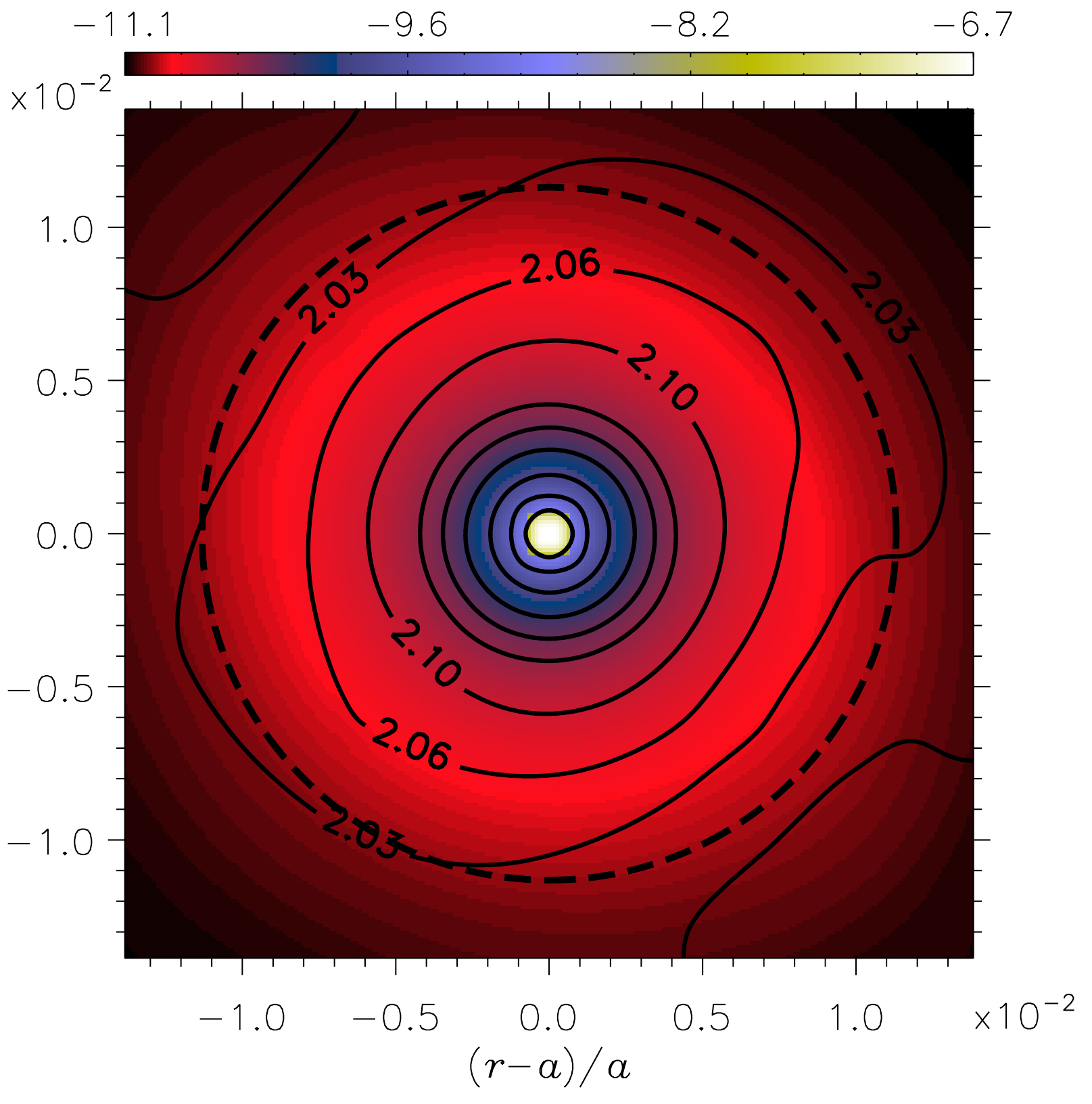}}
\caption{%
             Same quantities as in Figure~\ref{fig:xz_env_5}, 
             but around cores located at $10\,\AU$.
             }
\label{fig:xz_env_10}
\end{figure*}
%%%%%%%
We wish to identify a volume around the core that can be
defined as its ``envelope''. Gas inside this volume does not participate
in the disk circulation any longer. If the gas behaved as a collision-less
system only subjected to gravity, the Roche lobe would define such a volume.
For simplicity, we assume that the envelope is a sphere.

Figures~\ref{fig:xz_env_5} and \ref{fig:xz_env_10} show density maps,
flow streamlines at the mid-plane (left panels), and temperature contours 
(right panels) in regions around cores located at $5$ and $10\,\AU$, respectively 
(see the caption of Figure~\ref{fig:xz_env_5} for further details).
Also plotted in the panels are the intersection of the core's Roche lobe (left) 
and Bondi sphere (right) with the disk mid-plane ($\theta=\pi/2$).
The streamlines clearly indicate that the volume of each envelope must 
be significantly smaller than that of the corresponding Roche lobe.
In fact, horse-shoe or circulating orbits reach as close to the core as 
$0.4\,\Rhill$.
Symmetry properties of the gas may also help to identify the envelope region. 
Contours of equal temperature suggest that each envelope is confined 
within the corresponding Bondi sphere (Equation~(\ref{eq:RB})), as also 
argued in Section~\ref{sec:PCE}. 
Based on the shape of these contours illustrated in the right panels 
of Figures~\ref{fig:xz_env_5} and \ref{fig:xz_env_10}, the envelopes would
appear to extend over radii between $\approx 0.4$ and $\approx 0.6\,\Rbondi$.
These contours also suggest that, in units of $\Rbondi$, 
the envelope radius tends to be smaller for larger cores.

There is no obvious reason, though, that justifies the spherical symmetry
assumption of thermodynamical quantities for a non-isolated planet, 
particularly in the outer envelope layers, which are likely affected 
by interactions with the flow circulation exterior to the envelope.
Layers whose density and pressure are comparable to those of the external 
flow are the most affected, and their dynamics must bear some similarities 
to that of the unbound gas in contact with the envelope.
In other words, the properties of these envelope layers must be affected 
by the \textit{accretion flow}, which need not be (and is not!) spherically 
symmetric around the core (see Section~\ref{sec:ANI}).
Consequently, symmetry arguments may not be appropriate to define 
the outer envelope regions, and tracking of the actual motion of the gas 
is then necessary to determine the envelope volume.

%%%%%%%%%
\begin{deluxetable*}{cccccccccccc}
\tablecolumns{12}
\tablewidth{0pc}
\tablewidth{0.9\linewidth}
\tablecaption{Envelope Properties from the 3D Models\label{table:mom}}
\tablehead{
\colhead{$a$}&
\colhead{$\Mc$}&
\colhead{$\Me$}&
\colhead{$R_{e}$}&
\colhead{$I_{11}$}&\colhead{$I_{22}$}&\colhead{$I_{33}$}&\colhead{$I_{12}$}&
\colhead{$L_{3}$}&\colhead{$\omega_{3}$}&
\colhead{$b_{3}/b_{1}$}&\colhead{$b_{3}/b_{2}$}
\\
\colhead{[$\AU$]}&
\colhead{[$\Mearth$]}&
\colhead{[$\Mearth$]}&
\colhead{[$a$]}&
\colhead{[$\Me R^{2}_{e}$]}&\colhead{[$\Me R^{2}_{e}$]}&\colhead{[$\Me R^{2}_{e}$]}&
\colhead{[$\Me R^{2}_{e}$]}&
\colhead{[$\Me R^{2}_{e}\Omega$]}&\colhead{[$\Omega$]}&\colhead{}&\colhead{}
}
\startdata
$5$&  $\phantom{1}5$  & $2.5\times 10^{-3}$ & $0.0046$ 
                  & $0.207$ & $0.208$ & $0.209$ & $\phantom{-}9.4\times 10^{-5}$ 
                  & $1.07\times 10^{-2}$ & $0.05$                   & $0.988$ & $0.991$ \\
$5$& $10$ &$2.2\times 10^{-2}$ &$0.0065$
                  & $0.096$ & $0.096$ & $0.101$ &                  $-3.3\times 10^{-5}$ 
                 & $1.37\times 10^{-1}$ & $1.36$                    & $0.949$ & $0.951$ \\
$5$&$15$ &$4.9\times 10^{-2}$ &$0.0096$
                  & $0.131$ & $0.132$ & $0.138$ &                  $-7.0\times 10^{-5}$ 
                  & $4.41\times 10^{-1}$ & $3.19$                    & $0.944$ & $0.953$ \\[-5pt] \\
$10$&$\phantom{1}5$&$2.2\times 10^{-3}$ &$0.0026$
                  & $0.220$ & $0.220$ & $0.221$ &                  $-1.7\times 10^{-5}$ 
                  & $1.62\times 10^{-2}$ & $0.07$                    & $0.996$ & $0.997$ \\
$10$&$10$&$1.0\times 10^{-2}$ &$0.0036$
                   & $0.192$ & $0.193$ & $0.194$&                  $-1.0\times 10^{-4}$ 
                  & $2.08\times 10^{-1}$ & $1.07$                    & $0.990$ & $0.992$ \\
$10$&$15$&$3.5\times 10^{-2}$ &$0.0045$
                  & $0.106$ & $0.106$ & $0.107$ &                  $-2.5\times 10^{-5}$ 
                  & $1.47\times 10^{-1}$ & $1.38$                    & $0.988$ & $0.989$
\enddata
\end{deluxetable*}
%%%%%%%%%
We follow the approach of \citet{lissauer2009} and use passive tracers 
to characterize the motion of gas around a core and determine the volume
where gas is bound to the core. 
The tracer particles are advected by the flow, and thus follow the trajectory 
of gas parcels. 
The position of the tracers is advanced in time according to the method
described in Appendix~D of \citet{gennaro2008}. The trajectories
are second-order accurate in both space and time, and use velocity fields 
at the highest resolution available.
The particles are deployed on concentric spherical surfaces centered 
at the core center.
The spheres have radii ranging from $\approx R_{c}$ to $3\Rhill/5$, 
the largest possible radius of an envelope when $\Rbondi$ exceeds
this distance (see discussion in Section~\ref{sec:PCE}). 
In total, $60000$ tracers are deployed on the northern hemisphere 
of $50$ spherical shells.
Mirror symmetry conditions are applied to positions and velocities 
of tracers that cross the equatorial plane toward the southern hemisphere.
Denoting with $s(t)$ the distance from the center of the core ($R=0$) 
along the trajectory of a particle, the envelope radius is taken as the radius 
$R_{e}$ of the largest spherical surface for which $s_{i}(t)\le 3\Rhill/5$, 
for all tracers $i$ initially deployed on that surface.
For sensitivity purposes, one calculation also uses $360000$ tracers
distributed on $100$ hemispheres, but no significant difference is observed.

Particles that travel beyond $3\Rhill/5$ are either on horse-shoe or circulating
orbits (see left panels of Figures~\ref{fig:xz_env_5} and \ref{fig:xz_env_10}), 
and rapidly leave the region.
At the end of the trajectory integrations, tracers are either located inside
the envelope ($R\le R_{e}$) or in the disk, and no particle is left between
$R_{e}$ and $3\Rhill/5$.
A fraction of the tracers deployed outside of the envelope do move inside
the envelope. These tracers define the accretion flow that will be discussed
in the next section.

The estimates of the envelope radius are listed in the fourth column 
of Table~\ref{table:mom}, preceded by the envelope mass, $\Me$, contained
within this radius. (Notice that the masses $\Me$[3D] in Table~\ref{table:13D} 
are those inside the envelope radii of the 1D calculations.)
The radius $R_{e}$ increases with core mass and, for a given $\Mc$, varies 
by $\lesssim 10$\% with respect to the orbital distances. 
The ratio $R_{e}/\Rbondi$ decreases with increasing core mass. 
The cores located at $5\,\AU$ have radii between $0.64$ and $0.92\,\Rbondi$, 
and between $0.40$ and $0.69\,\Rbondi$ at $10\,\AU$.
In terms of Hill radius, the cores at $5\,\AU$ have envelope radii smaller
than $\approx \Rhill/3$ and smaller than $\Rhill/5$ at $10\,\AU$.
Overall, these results are consistent with the energy-based arguments
of Section~\ref{sec:PCE}, according to which 
$R_{e}\lesssim \Rbondi$ if $\Rbondi<\Rhill$.
They are also in agreement with the analysis based on the streamlines, 
discussed above, according to which $R_{e}\lesssim 0.4\,\Rhill$. 
Moreover, as argued above, spherical symmetry of, e.g., isothermal surfaces
should not be expected in the outer envelope layers of non-isolated planets.

The densities at $R=R_{e}$ are between $3.5\times 10^{-11}$
and $\approx 5\times 10^{-11}\,\mathrm{g\,cm^{-3}}$ at $a=5\,\AU$, 
and between $2\times 10^{-11}$ and 
$\approx 3\times 10^{-11}\,\mathrm{g\,cm^{-3}}$ at $a=10\,\AU$,
$3$ to $6$ times larger than the azimuthally averaged densities around
the star.
The temperatures range from $\approx 140$ to $\approx 160\,\K$ and
from $\approx 100$ to $\approx 140\,\K$ at the smaller and larger orbital 
distance, respectively. The perturbed temperatures are $\lesssim 30$\%
($a=5\,\AU$) and $\lesssim 75$\% ($a=10\,\AU$)
higher than the corresponding azimuthally averaged temperatures
around the star (see Table~\ref{table:dgc}).

Let us introduce a cartesian reference frame, whose origin is attached 
to the center of the core and in which $\{x_{i}\}$, with $i=1$, $2$, and $3$,
indicate the coordinate axes. The axis $x_{1}$ is parallel to the
core-star direction, pointing toward the star; axis $x_{2}$ is
tangent to the orbit, pointing in the opposite direction of the orbital 
motion; axis $x_{3}$ is perpendicular to the orbit, so that
$\boldsymbol{\hat{x}}_{1} \boldsymbol{\times} \boldsymbol{\hat{x}}_{2}%
=\boldsymbol{\hat{x}}_{3}$.

In order to determine some bulk properties, such as rotation and flattening,
we shall assume that the envelope can be approximated as a rigid body 
and introduce the inertia tensor \citep{llvI1976}
\begin{equation}
I_{ij}=\int_{V_{e}} \left(\delta_{ij}\sum_{k}x^{2}_{k}-x_{i}x_{j} \right)\rho dV,
\label{eq:Iij}
\end{equation}
where $V_{e}=(4\pi/3) R^{3}_{e}$ is the volume comprising the envelope. 
The diagonal components $I_{ii}$ are the moments of inertia of the envelope 
about the corresponding axes, and the negative of the off-diagonal components
are the products of inertia. Equation~(\ref{eq:Iij}) implies that $I_{ij}=I_{ji}$,
hence only six elements of the tensor matrix are independent.
Furthermore, by construction the envelope is symmetric with respect to
the $x_{1}$--$x_{2}$ plane (see Section~\ref{sec:GD}), 
therefore $I_{13}=I_{23}=0$.

The angular momentum with respect to the origin is
\begin{equation}
\boldsymbol{L}=\int_{V_{e}}%
                          \boldsymbol{x} \boldsymbol{\times} \boldsymbol{\dot{x}}\rho dV,
\label{eq:Lvec}
\end{equation}
whose components for a rigid body can be expressed as
\begin{equation}
L_{i}=\sum_{j}I_{ij} \omega_{j},
\label{eq:Li}
\end{equation}
where $\omega_{i}$ indicates the rotation rate about the axis $x_{i}$.
From Equation~(\ref{eq:Lvec}) and from the assumed symmetry
relative to the $x_{1}$--$x_{2}$ plane, we have that $L_{1}=L_{2}=0$.
Thus, from Equation~(\ref{eq:Li}), it follows that $\omega_{1}=\omega_{2}=0$,
whereas the rotation rate about axis $x_{3}$ is $\omega_{3}=L_{3}/I_{33}$.
Note that $\omega_{3}>0$ corresponds a counter-clockwise rotation, in the
same direction as the orbital revolution.

Due to symmetry, axis $x_{3}$ is a principal axis of inertia while axes
$x_{1}$ and $x_{2}$ are not, since the inertia tensor is not diagonal. 
In fact, as reported in Table~\ref{table:mom}, $I_{12}\neq 0$. 
However, the table shows than $|I_{12}| \ll I_{ii}$ in all cases 
and therefore axes  $x_{1}$ and $x_{2}$ can be considered as 
principal axes to a very good approximation.
Comparing $I_{33}$ of the envelopes at $a=5$ and $10\,\AU$, a significant 
difference (a factor $2$) is found only for $15\,\Mearth$ case. Otherwise, 
they agree to a level better than $20$\%.

%%%%%%%
\begin{figure}[]
\centering%
\resizebox{\linewidth}{!}{\includegraphics[clip]{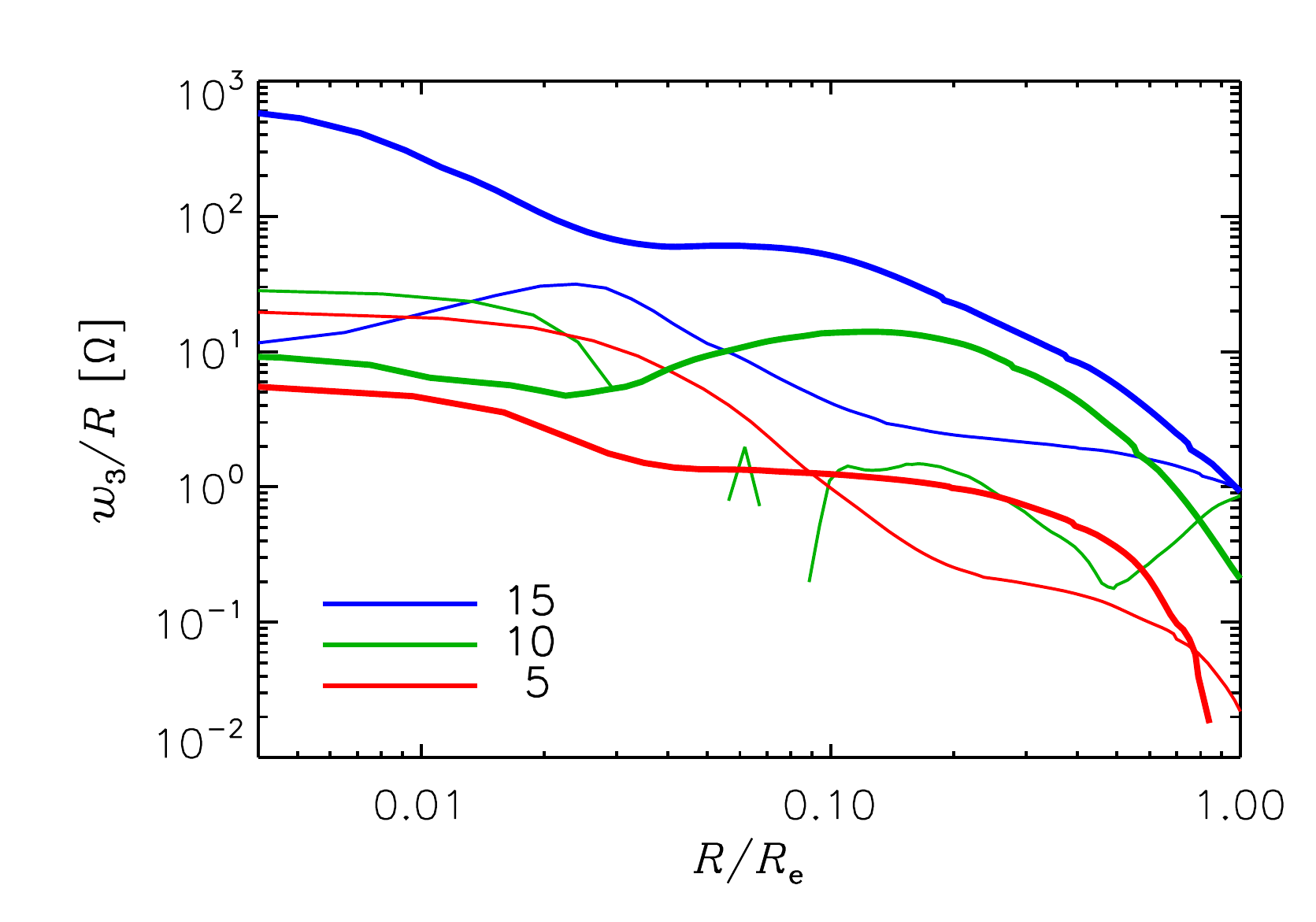}}
\caption{%
             Average angular velocity of the gas about the polar axis $x_{3}$, 
             in the equatorial plane of the envelope ($\theta=\pi/2$). 
             The quantity $w_{3}$ denotes the rotational velocity about the axis. 
             Positive values indicate a counter-clockwise rotation. 
             Gaps in the curves represent negative values of $w_{3}$.
             The thick (thin) lines refer to the core's orbital radius $a=5\,\AU$ ($10\,\AU$). 
             The core mass in units of $\Mearth$ is indicated in the legend.
             }
\label{fig:om3}
\end{figure}
%%%%%%%
Table~\ref{table:mom} also includes the values of $L_{3}$ and $\omega_{3}$,
which imply that the envelopes are slow rotators, unless the rigid-body
approximation is grossly inapplicable.
We compute the average rotational velocity of the gas, $w_{3}$, about 
the axis $x_{3}$ in the equatorial plane. In Figure~\ref{fig:om3}, we plot 
the angular velocity $w_{3}/R$ as a function of the distance from the axis.
There is differential rotation at the equator, and the normalized derivative 
$d\ln{(w_{3}/R)}/d\ln{R}$ averages out to values between 
$\approx 0$ and $\approx -2$ for $0.1\lesssim R/R_{e}\lesssim 1$.
Although the physical nature of $w_{3}/R$ and $\omega_{3}$ are quite 
different, it is plausible that $\omega_{3}$ samples the rotation of 
the outer envelope (carrying most of the angular momentum $L_{3}$) 
rather than that of the interior. If this is indeed the case, 
the slow rotation predicted by $\omega_{3}$ is consistent with the actual 
gas rotation rates at the equator: averages of $w_{3}/R$ in Figure~\ref{fig:om3},
between $R_{e}/2$ and $R_{e}$, result in values comparable to 
$\omega_{3}$ in Table~\ref{table:mom}.
Although the envelopes appear to rotate slowly, their specific angular momentum,
$L_{3}/\Me$, ranges from 
$2\times 10^{13}$ to $4\times 10^{15}\,\mathrm{cm^{2}\,s^{-1}}$. 
For comparison, the giant planets of the solar system have specific angular
momenta between $\sim 10^{14}$ and $\sim 10^{15}\,\mathrm{cm^{2}\,s^{-1}}$
(supposing uniform rotation).

Let us approximate the shape of the envelope to that of a triaxial 
ellipsoid
\begin{equation}
\sum_{j}\left(\frac{x_{j}}{b_{j}}\right)^{2}=1,
\label{eq:esoid}
\end{equation}
of semimajor axes $b_{j}$. In case of a uniform density and mass 
$\Me$, the moments of inertia of this solid figure are
\begin{equation}
I_{ii}=\frac{1}{5}\Me\left(\sum_{k}b^{2}_{k}-b^{2}_{i}\right).
\label{eq:Iesoid}
\end{equation}
By inverting this system of equations, one can express the ratios $b_{i}/b_{j}$
in terms of moments of inertia
\begin{equation}
\left(\frac{b_{i}}{b_{j}}\right)^{2}=\frac{\sum_{k}I_{kk}-2I_{ii}}{\sum_{k}I_{kk}-2I_{jj}}.
\label{eq:bibj}
\end{equation}
The flattening (or oblateness), $f_{S}=1-2 b_{3}/(b_{1}+b_{2})$, of gas giants 
in the solar system is $0.065$, $0.098$, $0.023$, and $0.017$, respectively, 
for Jupiter, Saturn, Uranus and Neptune.
The flattening of these 3D envelopes, evaluated via Equation~(\ref{eq:bibj}),
ranges from $\approx 0.004$ to $\approx 0.05$ (see Table~\ref{table:mom}), 
typically smaller than those of the solar system giants by a factor of $\lesssim 20$.
Yet, the spin rate of these envelopes is smaller by a factor of $\lesssim 10^{-4}$.
Thus, one may wonder 
whether the oblateness is caused entirely by rotation.

Let $q=\omega^{2}_{3} R^{3}_{e}/(G \Mp)$ be the ratio of the centrifugal 
to the gravitational accelerations at the equator and $\Mp=\Mc+\Me$. 
To first order in $q$, the rotational flattening $f_{S}$ of a self-gravitating 
(isolated) spheroid in hydrostatic equilibrium is given by the 
Radau-Darwin relation \citep[see][]{cook2009}
\begin{equation}
\frac{q}{f_{S}} = \frac{2}{5}+\frac{5}{2}\left(1-\frac{3}{2} \frac{I_{33}}{\Me R^{2}_{e}}\right)^{2},
\label{eq:fS}
\end{equation}
in which the core is assumed to be a point mass.
For a homogeneous body ($I_{33}=2\Me  R^{2}_{e}/5$), 
the classical expression $f_{S}/q= 5/4$ is recovered \citep{chandra1967}.
The oblateness predicted by Equation~(\ref{eq:fS}), in accord with that
obtained from Equation~(\ref{eq:bibj}), is an increasing function of $\Mp$.
But Equation~(\ref{eq:fS}) typically provides smaller numbers.
This may suggest that the flattening is not due to rotation alone, 
as the planets are not isolated. Alternatively, the homogeneous ellipsoid 
approximation does not produce accurate enough estimates. 

The quadrupole moment of the envelope density, $J_{2}$,  is related to 
the principal moments of inertia via the McCullagh's theorem 
\citep[][]{cook2009}
\begin{equation}
J_{2} = \frac{1}{\Me R^{2}_{e}}\left[I_{33} -\frac{1}{2}\left(I_{11}+I_{22}\right)\right].
\label{eq:J2}
\end{equation}
The envelope models provide values of $J_{2}$ that are larger for
increasing core mass ($J_{2}$ is similar at $\Mc=10$ and $15\,\Mearth$), 
ranging from $\approx 2\times 10^{-3}$ to
$\approx 7\times 10^{-3}$ at $a=5\,\AU$ and from $\approx 8\times 10^{-4}$ to
$\approx 2\times 10^{-3}$ at $a=10\,\AU$.
As $J_{2}$ can be related to $f_{S}$ and $q$, two quantities that can be 
measured from observations, Equation~(\ref{eq:J2}) is used to estimate 
the difference between the polar and equatorial moment of inertia of planets. 
This procedure would be somewhat less useful here, since it is not clear 
to what extent rotation contributes to the flattening of non-isolated planets.

%%--------------------------------------------------------------------------
\subsection{Anisotropy of Envelope Accretion}
\label{sec:ANI}
%%--------------------------------------------------------------------------

%%%%%%
\begin{figure*}[]
\centering%
\resizebox{\afiglen}{!}{%
\includegraphics[clip]{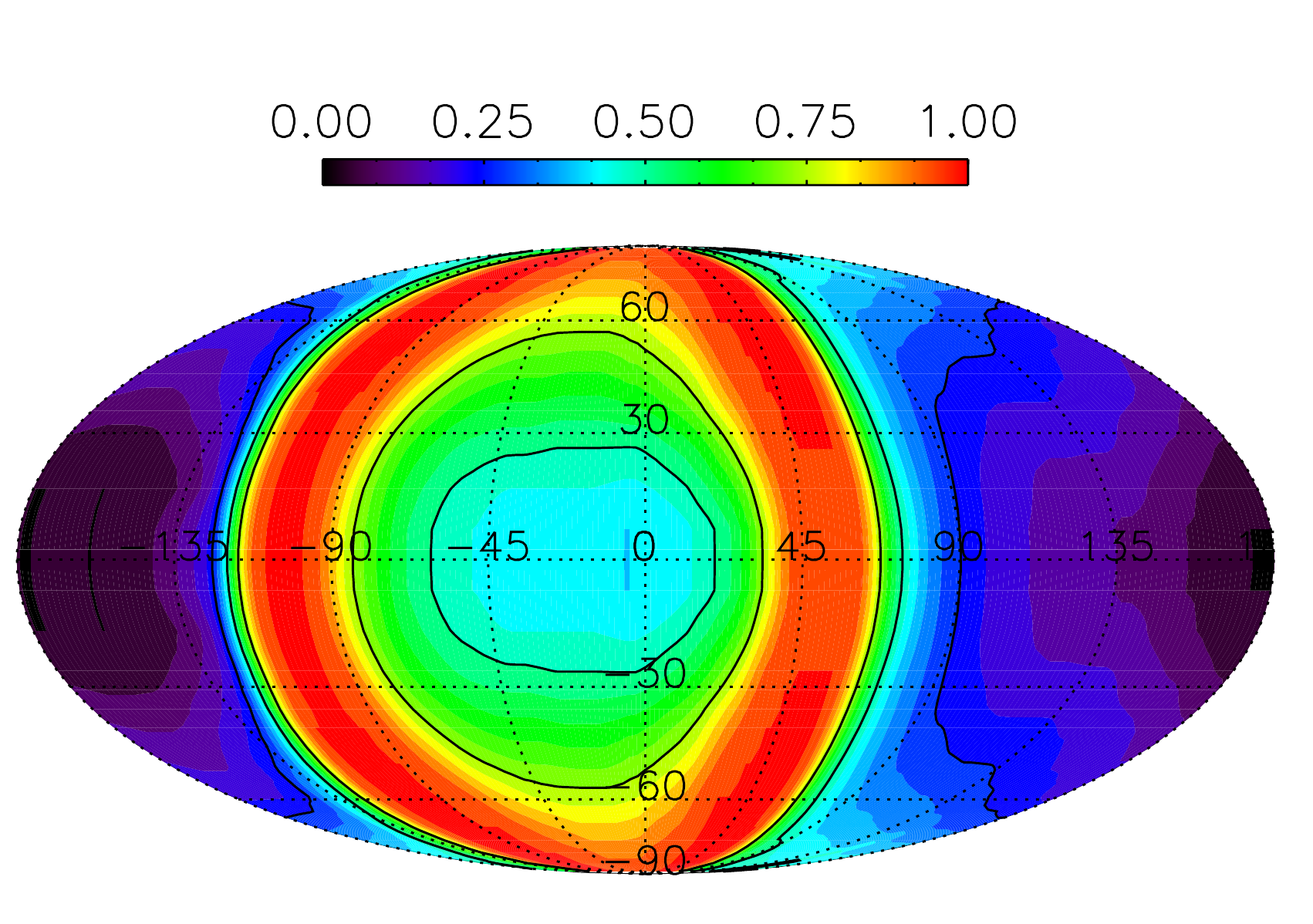}%
\includegraphics[clip]{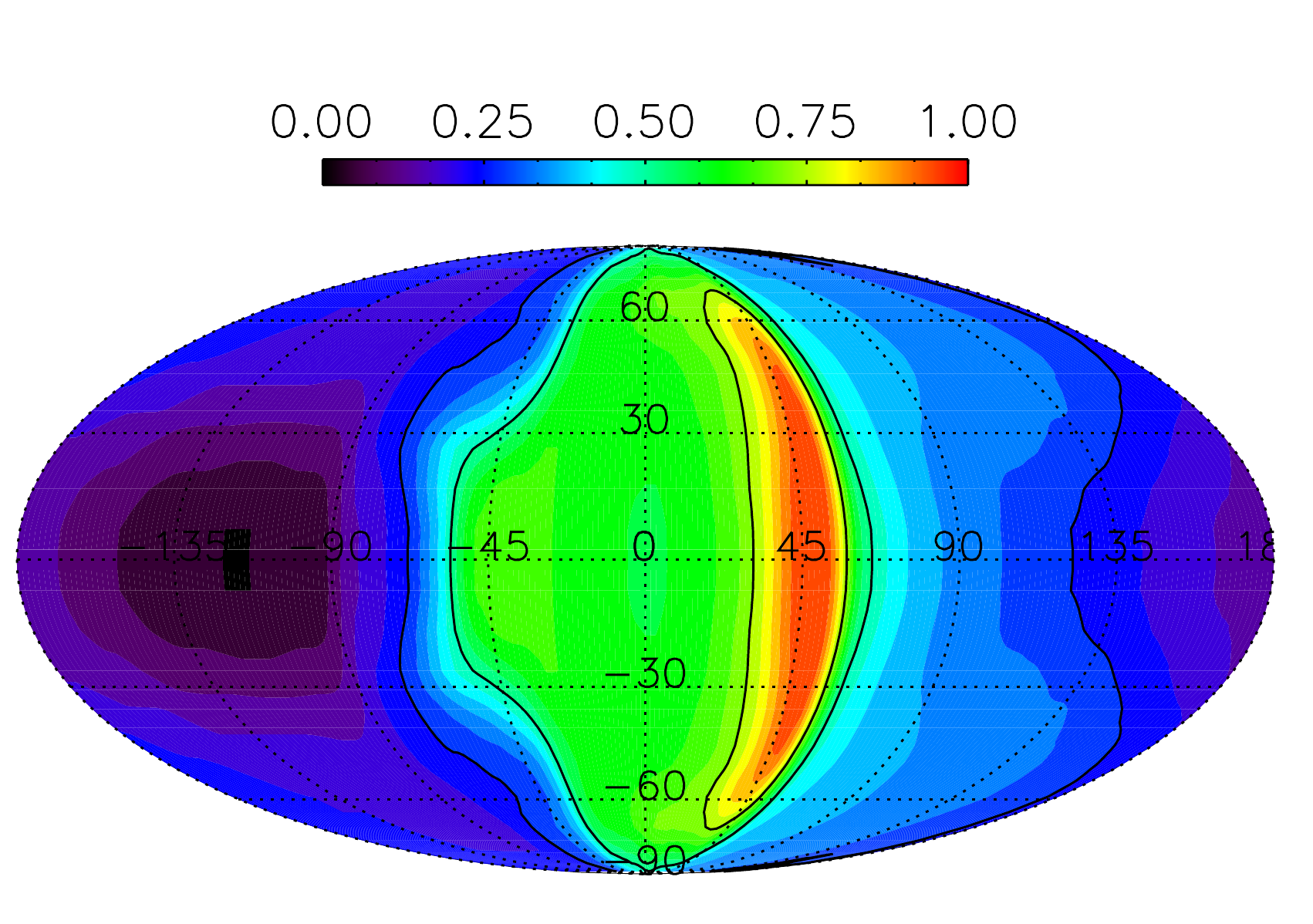}}
\resizebox{\afiglen}{!}{%
\includegraphics[clip]{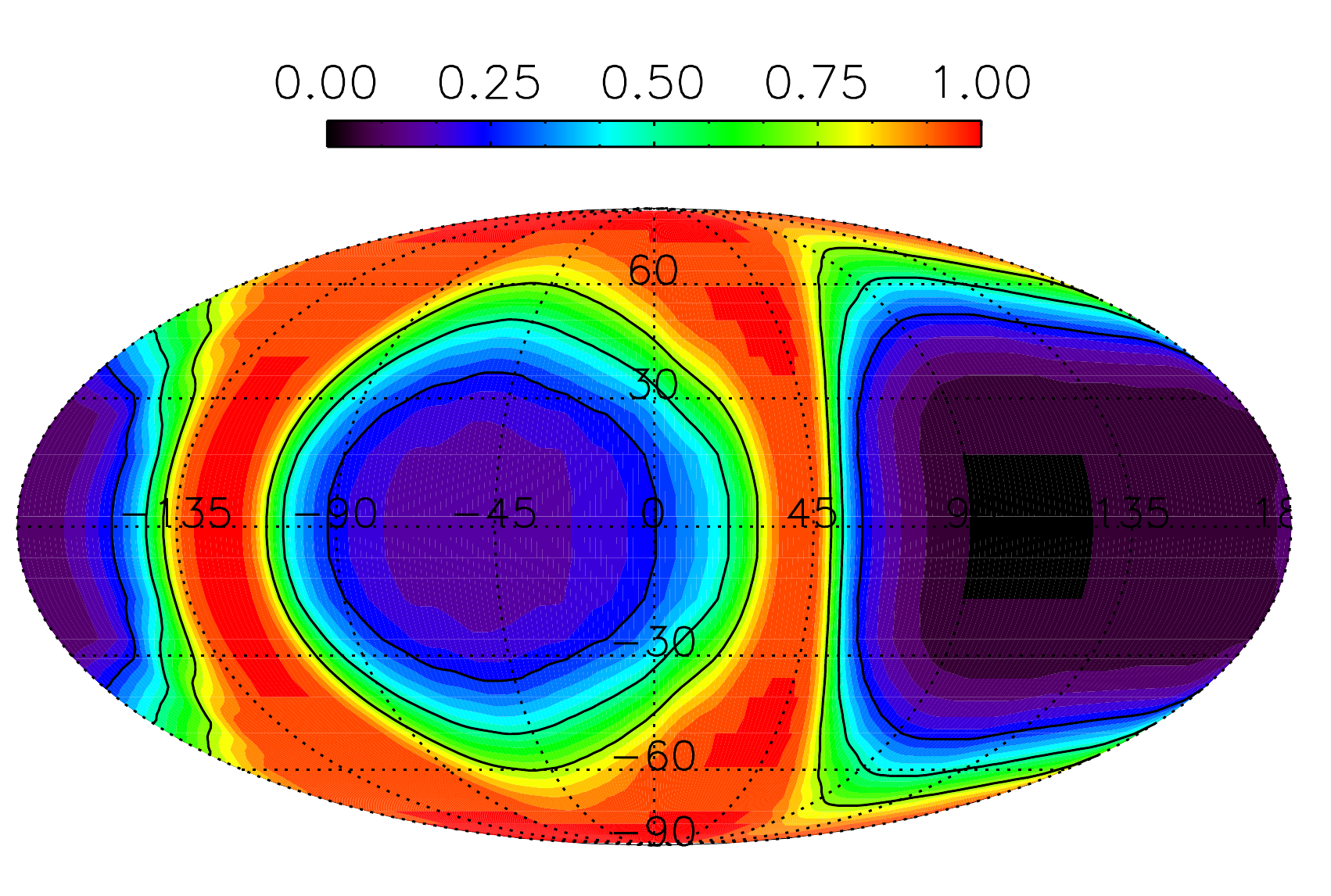}%
\includegraphics[clip]{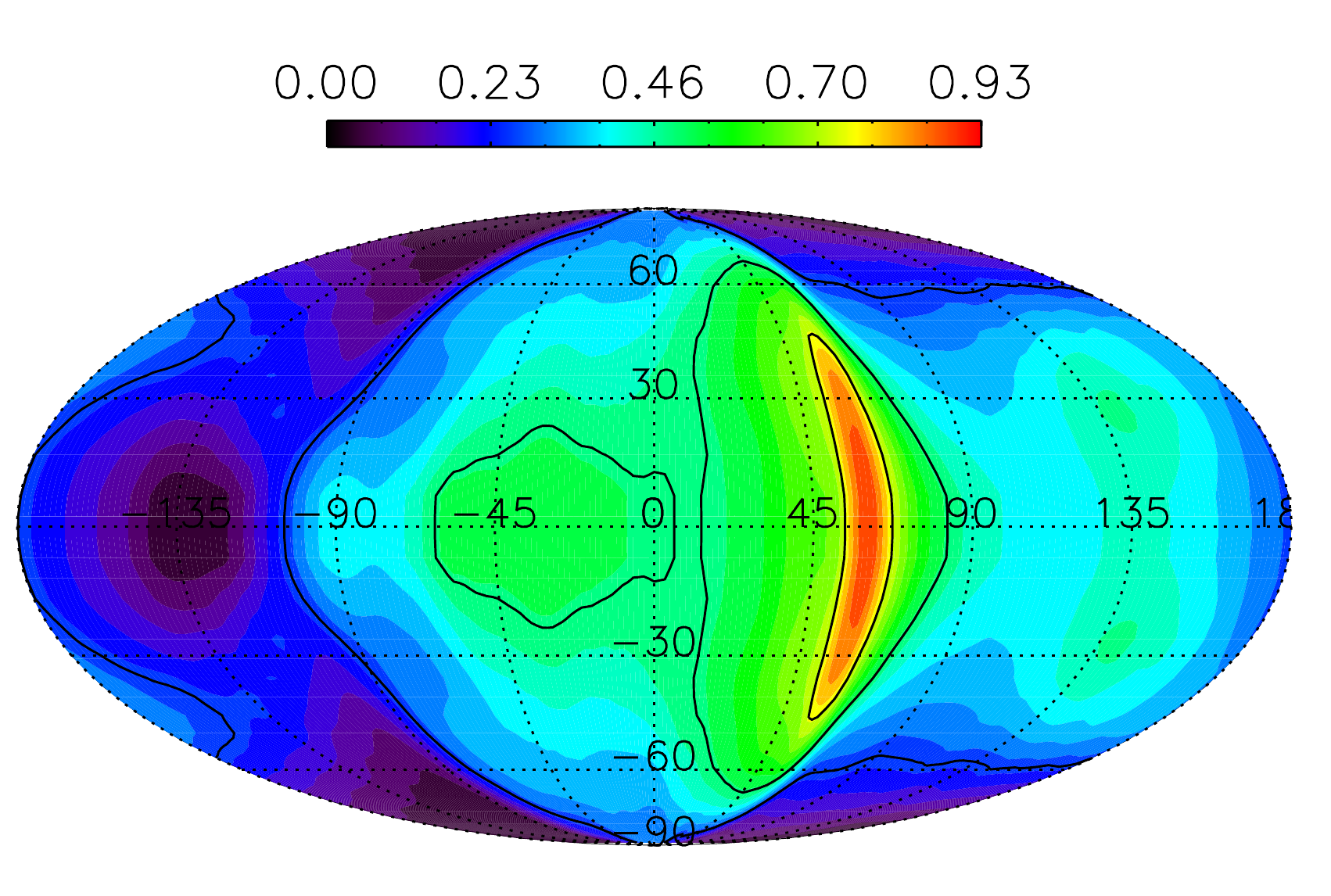}}
\resizebox{\afiglen}{!}{%
\includegraphics[clip]{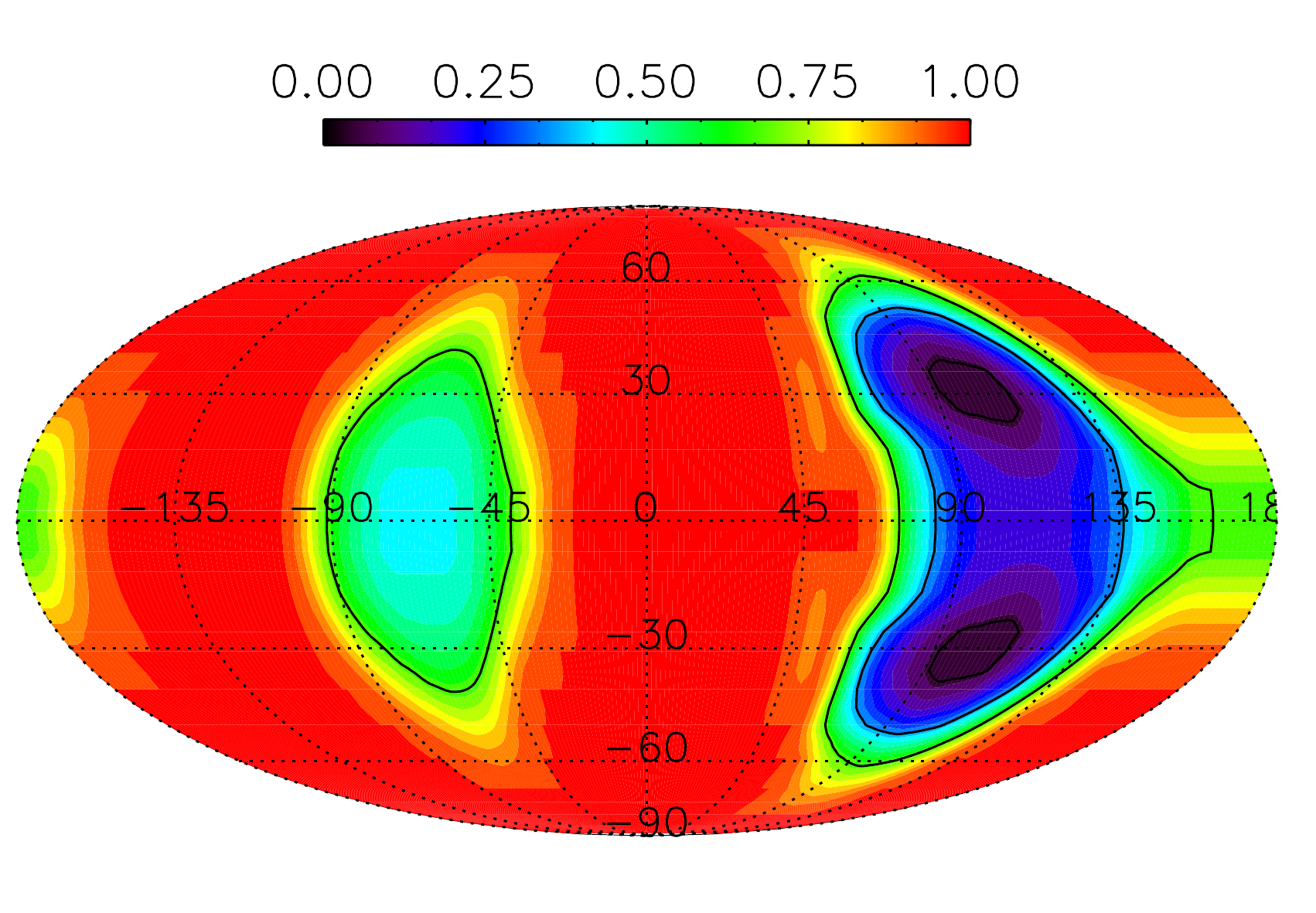}%
\includegraphics[clip]{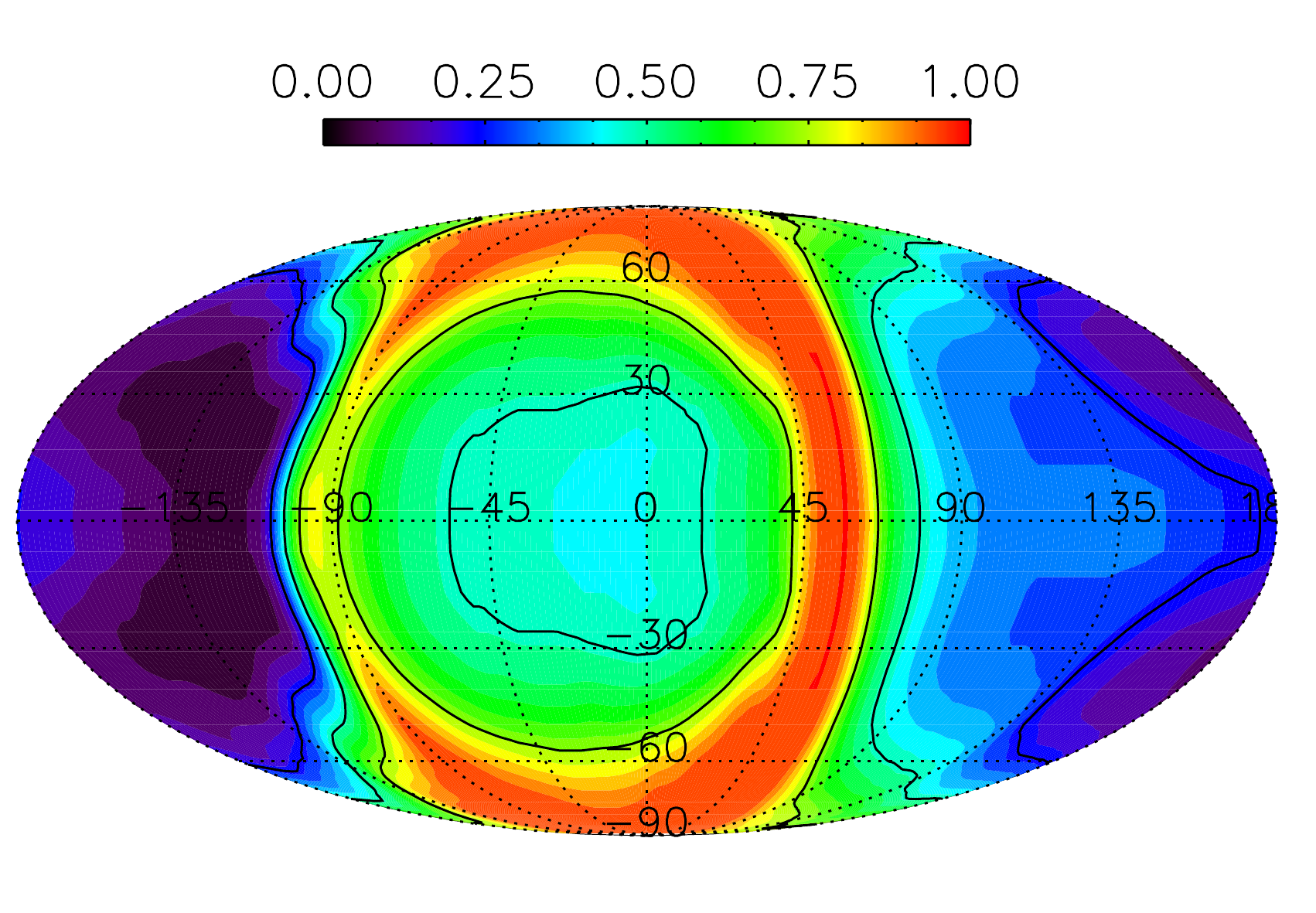}}
\caption{%
             Anisotropy of the mean accretion intensity (see text), as seen from 
             the center of the planet, projected on Mollweide maps. The longitude 
             $\mp 180^{\circ}$ indicates the direction toward the star and 
             $+90^{\circ}$ is the direction along the orbital motion.
             An intensity equal to $1$ implies that gas from that direction accretes
             on the planet, regardless of its place of origin along that line of sight
             (but within the accretion flow).
             From top to bottom, the core mass is $5$, $10$, and $15\,\Mearth$. 
             The core is located at $5\,\AU$ on the left and at $10\,\AU$ on the right.
             }\vspace*{2mm}
\label{fig:mollweide}
\end{figure*}
%%%%%%%
As anticipated in the previous section, the accretion flow is defined
as the gas that crosses the envelope surface and originates from within 
a distance of $3\Rhill/5$ from the core center.
Tracer particles deployed between $R_{e}$ and $3\Rhill/5$ are used
to track the motion of the accretion flow. These tracers are either carried 
inside the envelope or to horse-shoe/circulating orbits of the disk.
The accretion flow itself is fed by gas coming from the disk, as clearly shown 
by the streamlines in the left panels of Figures~\ref{fig:xz_env_5} and 
\ref{fig:xz_env_10}.
It is thus expected that the accretion flow, as seen from the core center, 
is directionally dependent (as opposed to a strictly spherical accretion).

In order to quantify the anisotropy of the gas accreting on the envelope,
we introduce an ``accretion intensity'' along a given direction, as seen 
from the center of the planet. 
Each tracer initially released on a spherical shell exterior to the envelope 
is assigned an ``intensity'' of $1$ if it moves inside the envelope. 
Otherwise, the accretion intensity of the tracer is $0$. 
For  a given direction, we take the mean of the accretion intensities 
defined on all these spherical shells. 
This quantity represents a measure of the anisotropy of the
accretion flow, but it does not bear information on the mass flux delivered 
to the planet.

The mean accretion intensity is projected on Mollweide maps in
Figure~\ref{fig:mollweide}. The contours levels on the maps indicate
locations where the intensity is $0.25$, $0.5$, and $0.75$.
The longitude $0^{\circ}$ is the direction along the core-star
line, pointing away from the star.
Longitudes $\mp 90^{\circ}$ are, respectively, the directions
opposite and along the orbital motion.
Each map refers to a core of different mass and semimajor axis
(see the figure caption for further details).
The scale on these maps is absolute in the sense that
an intensity of $1$ ($0$) implies that gas originating from that
direction (but inside the accretion flow!) always (never) accretes
on the planet, whatever its distance. 
An intensity of $0.5$ implies that, along that line of sight, gas is equally
both accreting and non-accreting.
Consequently, the scale on the maps should not necessarily start from $0$
(although it does in Figure~\ref{fig:mollweide}) or reach $1$,
as is the case for the mean accretion intensity around the $10\,\Mearth$
core located at $10\,\AU$.

There are some common features on the accretion intensity maps as, 
for example, the relatively low tendency for gas to accrete along the equator 
at longitude $\sim 180^{\circ}$ (star-ward direction) and the relatively
high tendency to accrete around longitude $+45^{\circ}$.
In general, for a given meridian, there is higher tendency for gas to accrete 
away from the equator, although this trend is less clear around the 
$10\,\Mearth$ core at $10\,\AU$.

%%%%%%%
\begin{figure}[]
\centering%
\resizebox{\linewidth}{!}{\includegraphics[clip]{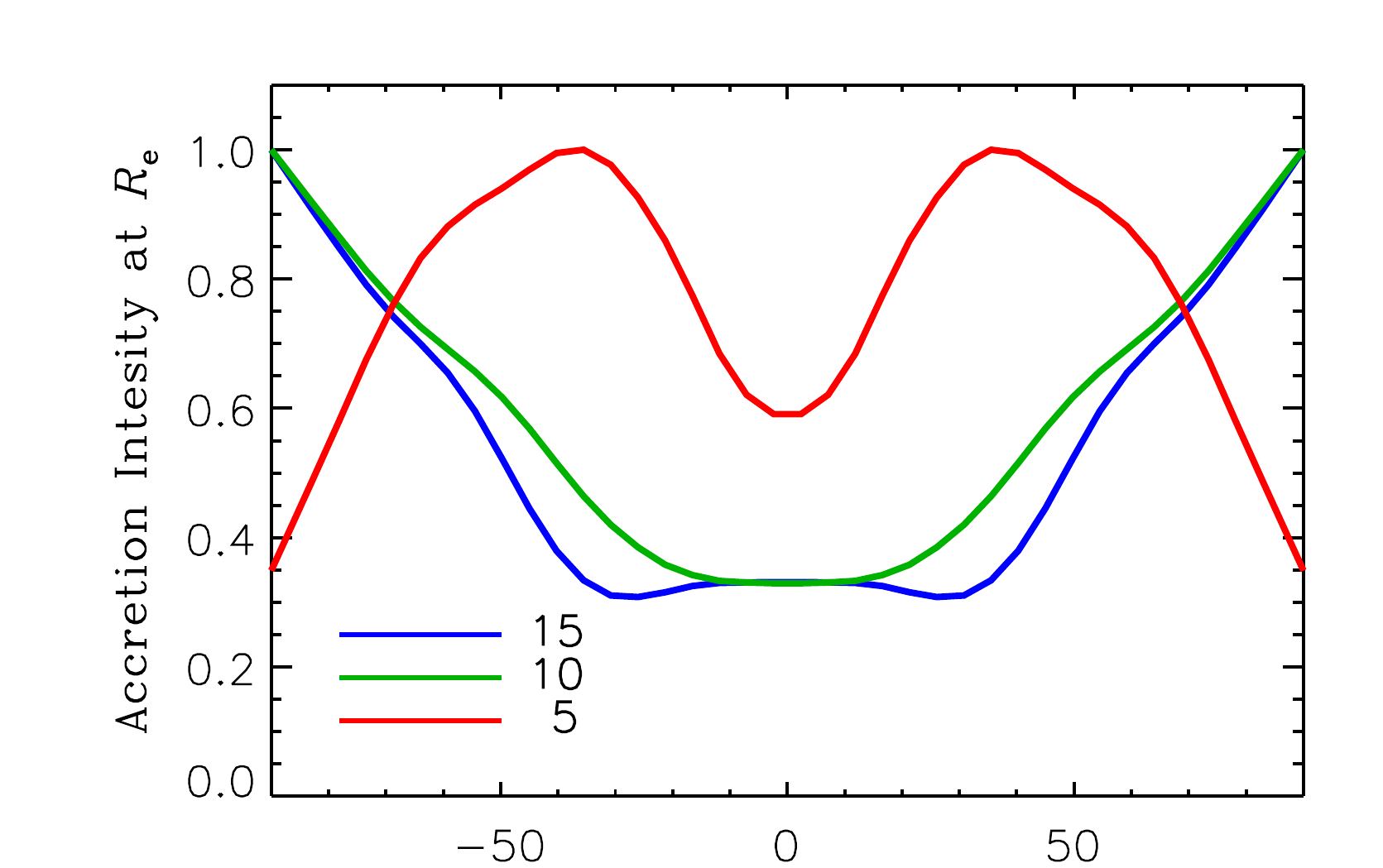}}
\resizebox{\linewidth}{!}{\includegraphics[clip]{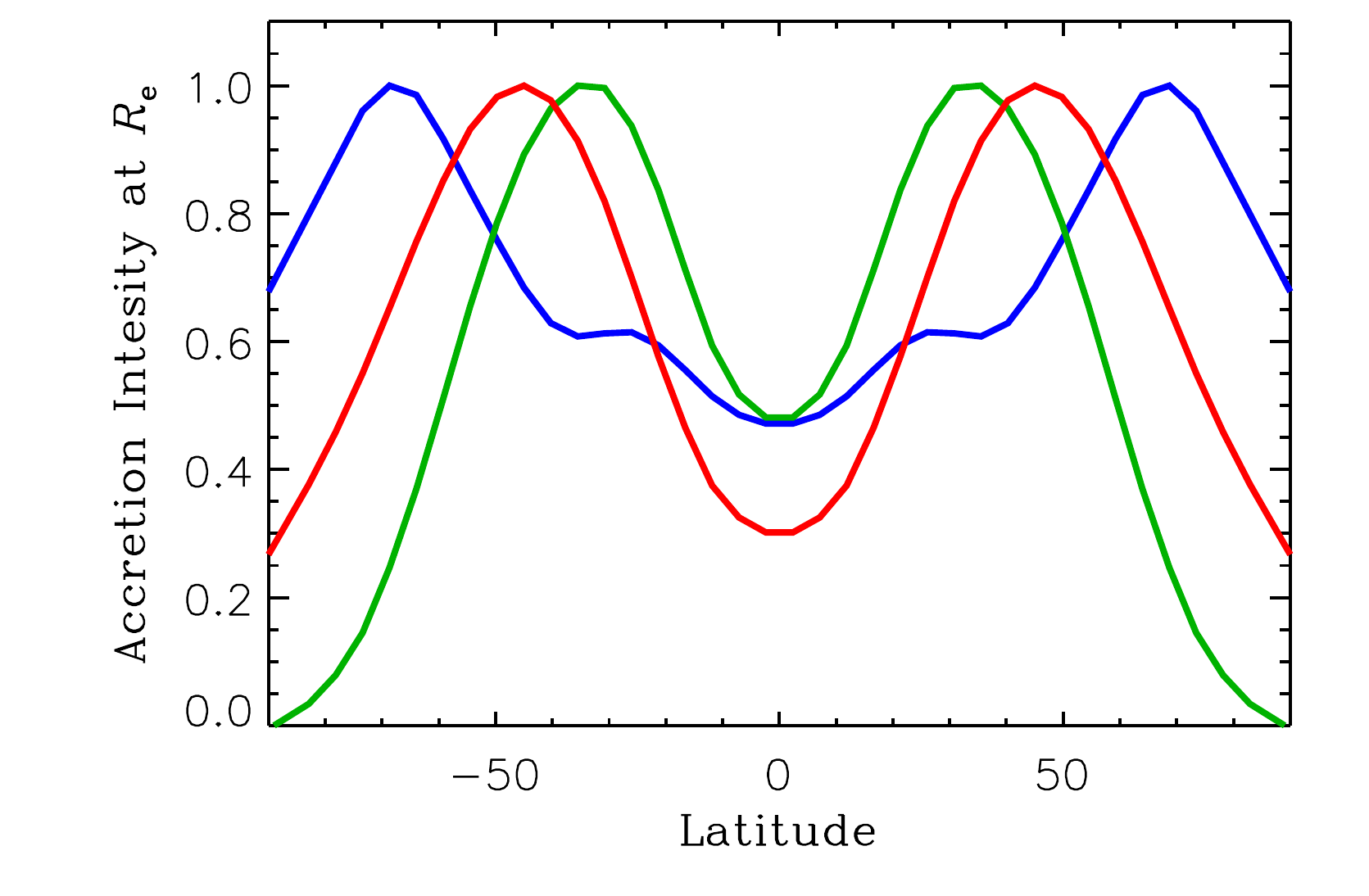}}
\caption{%
             Normalized intensity of accretion at the envelope surface, integrated 
             in longitude, as a function of the latitude. The curves represent the
             relative tendency for accreting gas to penetrate the envelope surface. 
             The top and bottom panels refer to planets located at $5$ and $10\,\AU$,
             respectively. The core mass is indicated in the top panel in units of $\Mearth$.
              }
\label{fig:accin}
\end{figure}
%%%%%%%
The maps in Figure~\ref{fig:mollweide} provide information about the 
anisotropy of the accretion flow integrated along the line of sight, that is, 
the direction from which accreting gas originates. 
However, they convey no direct information about the angular distribution 
of the locations where accreting gas actually enters the envelope. 
Such a distribution can be obtained from the trajectories of the tracers 
in the accretion flow, as they intersect the sphere of radius $R_{e}$.
We count the number of intersections as a function of the latitude, 
defined as in Figure~\ref{fig:mollweide}, and constructed an equivalent
of the accretion intensity at $R=R_{e}$.
The resulting distributions, normalized to their maxima, are plotted in 
Figure~\ref{fig:accin}. The curves clearly show that gas accreting 
on the envelope does so preferentially at mid- to high latitudes.

%%--------------------------------------------------------------------------
\section{Summary and Conclusions}
\label{sec:summ}
%%--------------------------------------------------------------------------

We present detailed and global radiation-hydro\-dynamics calculations
of 3D envelopes around planetary cores embedded in protoplanetary
disks. 
The global approach allows us to fully take into account the circulation
of gas as it orbits the star and moves toward the planet.
We consider cores of $5$, $10$, and $15\,\Mearth$ at $5$ and $10\,\AU$
from a solar-mass star.
The equation of state includes both gas and radiation, and gas is treated as
a solar mixture of H$_{2}$, H, He, and their ions (see Section~\ref{sec:EoS}). 
Molecular hydrogen is modeled as a mixture of  parahydrogen and orthohydrogen
with a fixed $1/3$ ratio. 
Detailed calculations of dust opacity are also performed, assuming the presence
of multiple grain species (see Section~\ref{sec:opa}).
Nested grids are used to resolve the flow at various length scales, from the orbital
to the core radius (see Section~\ref{sec:NGS}).
Some average properties of equilibrium disk structures (see Section~\ref{sec:DS})
are summarized in Table~\ref{table:dgc}.
The energy budget of the envelopes accounts for deposition of energy due to
solids accretion, obtained from 1D calculations (see Section~\ref{sec:1D}).

The masses and gas accretion rates of 1D and 3D envelopes differ by a factor
$2$ or less
(see Table~\ref{table:13D}). The density (for $R>3\,R_{c}$) and temperature 
in the envelope differ by factors smaller than $10$ and $2$, respectively
(see Figure~\ref{fig:env1D3D}). 
The largest differences typically occur at the boundaries of 1D envelopes, 
whose density and temperature are matched to the azimuthally averaged 
values of the disk at the corresponding orbital distance (see Figure~\ref{fig:av_disk}).
We find that perturbations induced by the core can raise the local density 
and temperature above these average disk values.
The interior structure of 1D envelopes is not very sensitive to boundary
conditions (compare left and right panels in Figure~\ref{fig:env1D3D}).
Despite the approximate gravitational potential at $R_{c}$ and limited 
linear resolution ($\approx R_{c}$) of the 3D calculations, density, 
temperature, and pressure at $R\approx R_{c}$ are comparable to those
of the 1D envelopes in most cases.
The general consistency of two very different physical approximations and
entirely different numerical solutions represents an important, two-way 
validation of the 1D and 3D models.

Energy-based arguments (see Section~\ref{sec:PCE}) suggest that when
$\Rbondi < \Rhill$ (as in all the cases studied here), the envelope
does not extend beyond the Bondi sphere. The behavior of the gas streamlines 
around the cores agrees with such a conclusion (see Figures~\ref{fig:xz_env_5}
and \ref{fig:xz_env_10}). 
By means of passive tracers we identify the volume of the gas bound to the core. 
We obtain envelope radii that increase with core mass, and range from 
$R_{e}\approx 0.4$ to $\approx 0.9\,\Rbondi$ (see Table~\ref{table:mom}). 
Marginal variations of $R_{e}$ ($\lesssim 10$\%) are obtained between 
envelopes around cores located at $5$ and $10\,\AU$.
We find that interactions with the external flow can produce asymmetries
(e.g., in temperature) in the outer layers of an envelope.

We determine the moments of inertia and angular momenta of the
envelopes and estimate bulk rotation rates using the rigid-body
approximation (see Section~\ref{sec:ESI}). The results indicate
slow rotation, consistent with the mean angular velocity of the gas
at the equator, between $R_{e}/2$ and $R_{e}$. 
The oblateness of the envelopes is estimated by using the homogeneous
ellipsoid and the hydrostatic equilibrium approximations. Both solutions
point to a moderate to low flattening ($\lesssim 0.05$), although the latter
method typically provides smaller values than the former. This
may suggest that the oblateness is not caused entirely by rotation
or that one or both approximations are not accurate enough.

We define an accretion flow in the region between $R_{e}$ and $3\Rhill/5$
and study its directional dependence as seen from the planet center
(see Section~\ref{sec:ANI}). The anisotropy clearly shows that the 
accretion is not spherical (see Figure~\ref{fig:mollweide}).
We also identify the angular distribution of the locations where
the accretion flow enters the envelope (see Figure~\ref{fig:accin}), 
which displays a tendency toward merging at mid- to high latitudes.

Estimates of the specific angular momentum of Jupiter and Saturn
are, respectively, $\approx 2\times 10^{15}$ and 
$\approx 10^{15}\,\mathrm{cm^{2}\,s^{-1}}$.
The envelopes surrounding the $10$ and $15\,\Mearth$ cores at
$5\,\AU$ have specific angular momenta of order $10^{15}\,\mathrm{cm^{2}\,s^{-1}}$, 
and $\approx 4\times 10^{14}\,\mathrm{cm^{2}\,s^{-1}}$ for the same cores 
at $10\,\AU$ (see Table~\ref{table:mom}). 
These figures also give the specific angular momentum of the gas 
in the corresponding accretion flows (as defined here).
If such planets evolved into Jupiter and Saturn, via continued accretion
of gas, the specific angular momentum of the accreted gas should be 
comparable to that of the gas accreted during these earlier phases 
of evolution. 

Overall, these 3D calculations, applied to relatively low-mass
gaseous envelopes around protoplanetary cores, provide a firm
basis for the calculation of gas accretion rates onto such cores
and for the study of non-spherically symmetric envelope properties.
These results indicate that the 3D code can now be extended to
the later stages of evolution when the gas accretion rates are high
and are limited by the detailed physics of the disk rather than
by the thermal properties of the planetary envelope. At present,
such accretion rates, which cannot be determined in 1D simulations,
are generally simulated in 3D applying a local isothermal equation of state
\citep{lissauer2009,bodenheimer2013}. Future simulations should
determine the effect on these rates of the radiative feedback of
the planet onto the disk. At the later stages, a subdisk is
expected to form around the planet.  The simulations should be
able to estimate the effects of this subdisk on the 
gas flow onto the planet and whether this flow, as suggested above, 
carries low specific relative angular momentum 
\citep[see also][]{tanigawa2012,ayliffe2012}.
Ultimately, such calculations should
be able to determine the final mass and angular momentum of a giant
planet, given a set of initial conditions.

\acknowledgments

We are grateful to Jack Lissauer, Morris Podolak, and Uma Gorti 
for useful feedback on this work.
We thank the referee for constructive and helpful comments.
Primary support for this project was provided by NASA Outer Planets 
Research Program grant 202844.02.02.01.75;
additional support was provided by NASA Origins of Solar Systems 
grant NNX11AK54G.
Resources supporting this work were provided by the NASA High-End
Computing (HEC) Program through the NASA Advanced Supercomputing
(NAS) Division at Ames Research Center.
G.D.\ thanks Los Alamos National Laboratory for its hospitality.

\appendix

%%--------------------------------------------------------------------------
\section{Dust Opacity Calculation}
\label{sec:opa_calc}
%%--------------------------------------------------------------------------

Let us indicate with $\kappa_{\lambda}$ the monochromatic opacity
(absorption plus scattering) coefficient at the radiation wavelength $\lambda$, 
so that 
$\tau_{\lambda}=\int{\kappa_{\lambda}\rho dl}$ is the optical thickness of the
medium along the direction $l$ at that same wavelength.
(Notations used in this Appendix may be unrelated to the same symbols 
adopted in other parts of the paper.)
The equations of radiation hydrodynamics are often applied in 
a frequency-integrated form, as are Equations~(\ref{eq:dEgas}),
(\ref{eq:dErad}), and (\ref{eq:dEtot}).
Under the local thermodynamic equilibrium, the frequency-integrated 
opacity coefficient involved in the radiation flux in Equation~(\ref{eq:F})
is the Rosseland mean opacity, $\kappa_{\mathrm{R}}$, defined by
\citep[e.g.,][]{castor2007}
\begin{equation}
\frac{1}{\kappa_{\mathrm{R}}}=%
\frac{\int_{0}^{\infty}{(1/\kappa_{\lambda})(\partial B_{\lambda}/\partial T)d\lambda}}%
{\int_{0}^{\infty}{(\partial B_{\lambda}/\partial T)d\lambda}},
\label{eq:kR}
\end{equation}
where $B_{\lambda}=B_{\lambda}(T)$ is the Planck's function
\citep[see][]{gray1992}.
Although, not employed in this study, the Planck mean opacity
\begin{equation}
\kappa_{\mathrm{P}}=%
\frac{\int_{0}^{\infty}{\kappa_{\lambda} B_{\lambda} d\lambda}}%
{\int_{0}^{\infty}{ B_{\lambda} d\lambda}}
\label{eq:kP}
\end{equation}
may also be required if, for example, 
the total radiation flux includes contributions from external irradiation 
sources \citep[e.g.,][]{rafikov2006}.

If we consider protoplanetary disk regions with densities 
$\rho\lesssim 10^{-8}\,\mathrm{g\,cm}^{-3}$, the Rosseland mean 
opacity due to atoms and molecules at temperatures $T\lesssim 1500\,\K$
is less than $10^{-3}\,\mathrm{cm^{2}\,g^{-1}}$,
and less than $10^{-4}\,\mathrm{cm^{2}\,g^{-1}}$ below $1000\,\K$
\citep[e.g.,][]{freedman2008}. At such temperatures, however, dust grains
entrained in the gas also contribute to absorption and scattering of
radiation. In fact, below $1000\,\K$, dust opacity typically dominates 
by a large margin over gas opacity. In the following, we assume that
gas opacity is small compared to dust opacity in the range of 
temperatures that allows for the presence of dust.

Consider a size distribution so that the number of grains, of radius $a$, 
of the dust  species $i$ is $n^{i}(a)$. 
Let us indicate the cross-section for absorption and scattering of photons
by a dust particle as 
$\sigma^{i}(a,\lambda)=\pi a^{2} \mathcal{Q}^{i}(a,\lambda)$,
where $\mathcal{Q}$ is the total extinction 
(absorption plus scattering) efficiency of the particle. 
The opacity coefficient of species $i$, 
$\kappa^{i}_{\lambda}$, is given by
\begin{equation}
\kappa^{i}_{\lambda}=%
\frac{%
f_{s}^{i}\int{n^{i}(a)\sigma^{i}(a,\lambda) d a} }{%
(4\pi/3)\rho_{s}^{i}\int{n^{i}(a)a^{3} d a} },
\label{eq:kli}
\end{equation}
where $\rho_{s}^{i}$ is the density of the solid grain and
$f_{s}^{i}$ is the mass fraction of the solid species, 
so that its mass per unit volume is $\rho f_{s}^{i}$ 
($\rho$ is the gas density). 
Summing over all grain species, 
the total monochromatic opacity of the dust is
\begin{equation}
\kappa_{\lambda}=\sum_{i}\kappa^{i}_{\lambda}.
\label{eq:kl}
\end{equation}
 
The grain extinction efficiency, $\mathcal{Q}$, is the sum
of the absorption efficiency, $\mathcal{Q}_{\mathrm{ab}}$, and the scattering 
efficiency $\mathcal{Q}_{\mathrm{sc}}$. 
\citet{pollack1985} discusses the presence of a modulation factor 
that multiplies the extinction efficiency and that accounts for 
anisotropic scattering, which may be important when the radiation 
wavelength is comparable to the grain size. Here, this factor is 
embedded in $\mathcal{Q}$.
Since we assume that grains are
spherical particles, Mie theory can be applied to compute the 
extinction efficiency. 
Let us introduce the ratio $\xi=2\pi a/\lambda$, and the
real and imaginary parts of the refraction index of the material,
$\mathcal{N}_{r}$ and $\mathcal{N}_{i}$, respectively. 
If $(\mathcal{N}_{r}-1)\xi\ll 1$, i.e., in the limit of small particles, it
can be shown \citep{pollack1985} that
$\mathcal{Q}_{\mathrm{ab}}\propto \xi$ and $\mathcal{Q}_{\mathrm{sc}}\propto \xi^{4}$, 
and hence $\mathcal{Q}$ is dominated by absorption. In particular,
for small particles the asymptotic form of the absorption efficiency is
$\mathcal{Q}_{\mathrm{ab}}\approx 8\mathcal{N}_{r}\mathcal{N}_{i}/%
[(\mathcal{N}^{2}_{r}-\mathcal{N}^{2}_{i}-2)^2 +4\mathcal{N}^{2}_{r}\mathcal{N}^{2}_{i}]$
\citep{pollack1985}.
In the limit of large particles, $\xi\gg 1$, one typically finds that 
$\mathcal{Q}_{\mathrm{ab}} + \mathcal{Q}_{\mathrm{sc}}\approx 1$.

To compute the extinction efficiency of a solid grain we employ the Mie theory
according to the formulation of \citet{bohren1983} for homogeneous spheres.
We follow \citet{pollack1994} and consider the dust as mixture of seven
grain species, made of iron, olivine, orthopyroxene, troilite (FeS), refractory
and volatile organics, and water ice. The material densities, $\rho_{s}$,
the mass fractions, $f_{s}$, as well as the representative vaporization 
temperatures versus gas density are taken from Tables~2 and 3 of
\citet{pollack1994}.
The combined gas-to-dust mass ratio is $71.5$.
The optical properties of the various materials, that is the real and imaginary 
parts of the refraction index $\mathcal{N}_{r}$ and $\mathcal{N}_{i}$ 
as a function of wavelength, 
come from the Database of Optical Constants maintained by 
the Astrophysical Institute of the University of Jena\footnote{%
See the Institute Homepage at \url{http://www.astro.uni-jena.de}.}.

We use the same size distribution for all grain species, given by
a simple power-law of the grain radius, $n(a)\propto a^{-3}$.
The minimum and maximum radii of the distribution, 
$0.005\,\mu\mathrm{m}$ and $1\,\mathrm{mm}$, are also 
the same for all species. Both the power index of the distribution
and its maximum radius compare favorably with observations
of T~Tauri disks, as shown by models of the spectral 
energy distributions \citep{dalessio2001}. In fact, the study of
\citet{dalessio2001} suggests that grain growth in these
disks can lead to distributions containing grains with 
radii possibly larger than $\sim 1\,\mathrm{cm}$.
The size distribution of interstellar grains is typically taken as 
$n(a)\propto a^{-3.5}$, with radii varying between $0.005$ and 
$0.25\,\mu\mathrm{m}$ \citep{draine1984}. 
For comparison purposes, we also compute interstellar opacities 
applying this size distribution and the composition and mass fractions 
adopted by \citet{pollack1994} for molecular clouds.

The integrals in Equation~(\ref{eq:kli}) are performed by means
of a ``cautious extrapolation'' integration based on an adaptive 
Romberg method \citep{davis2007}. 
The convergence of each integration
is achieved within a relative error tolerance of $10^{-7}$ or 
an absolute error tolerance of $10^{-15}$.
The monochromatic opacity is calculated with a relative wavelength 
resolution $d\lambda/\lambda\approx 1.4\times 10^{-4}$.
The integration of the Rosseland and Planck opacities, in
Equations~(\ref{eq:kR}) and (\ref{eq:kP}), is performed
applying the same integration algorithm and requiring the same precision.

Grain opacities depend on gas density only through the vaporization 
temperature \citep{pollack1994}. For temperatures higher than the
highest vaporization temperature $T^{\mathrm{mx}}_{\mathrm{ev}}$, 
grains are considered entirely vaporized
and only gas contributes to the opacity of the medium.
We use a linear interpolation between the dust opacity
discussed here and the gas opacity of \citet{ferguson2005} 
(see Section~\ref{sec:opa}) in the temperature range 
$[0.8\,T^{\mathrm{mx}}_{\mathrm{ev}},T^{\mathrm{mx}}_{\mathrm{ev}}]$.

%%%%%%
\begin{figure}[]
\centering%
\resizebox{\figlen}{!}{%
\includegraphics[clip]{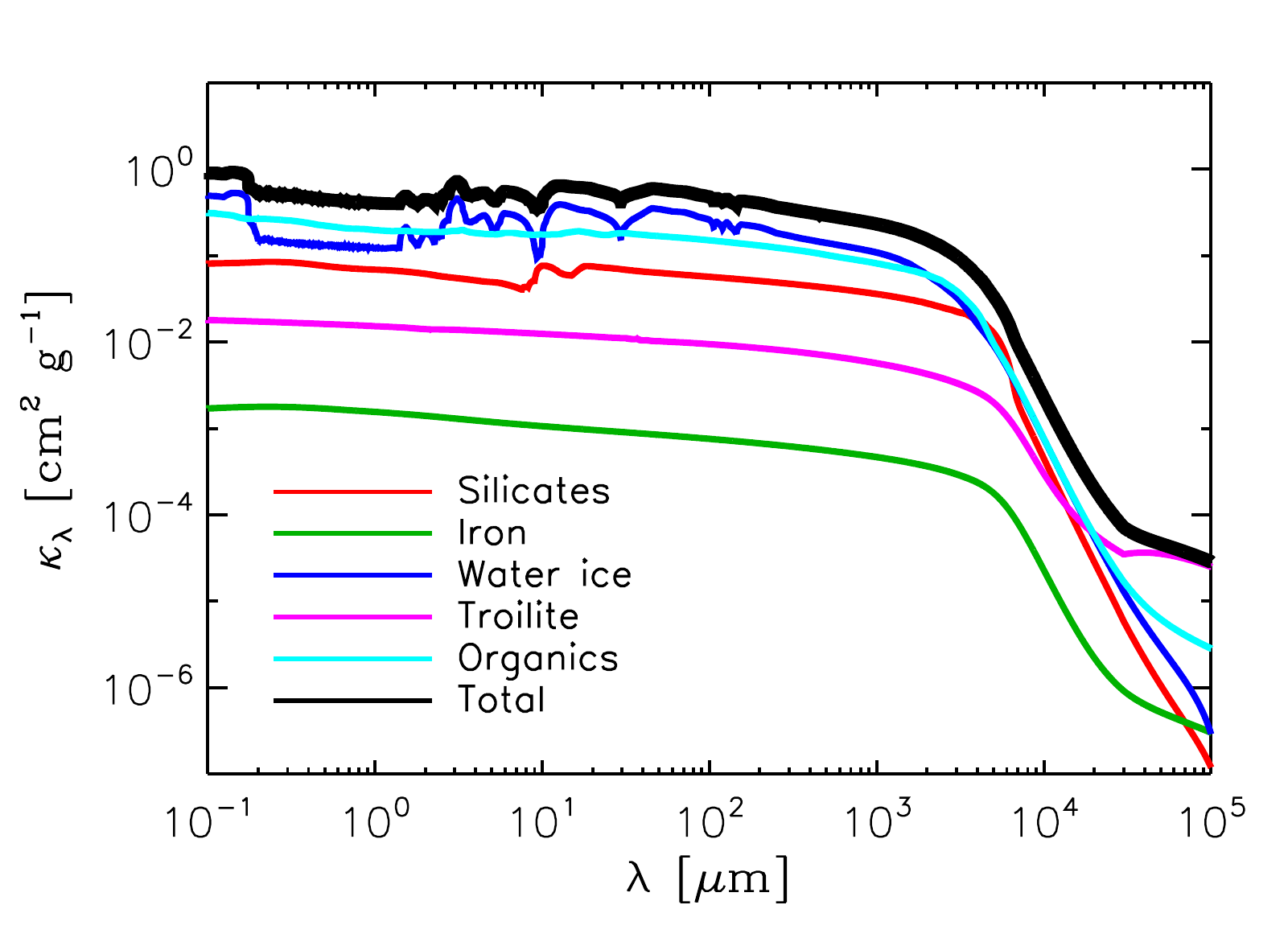}%
\includegraphics[clip]{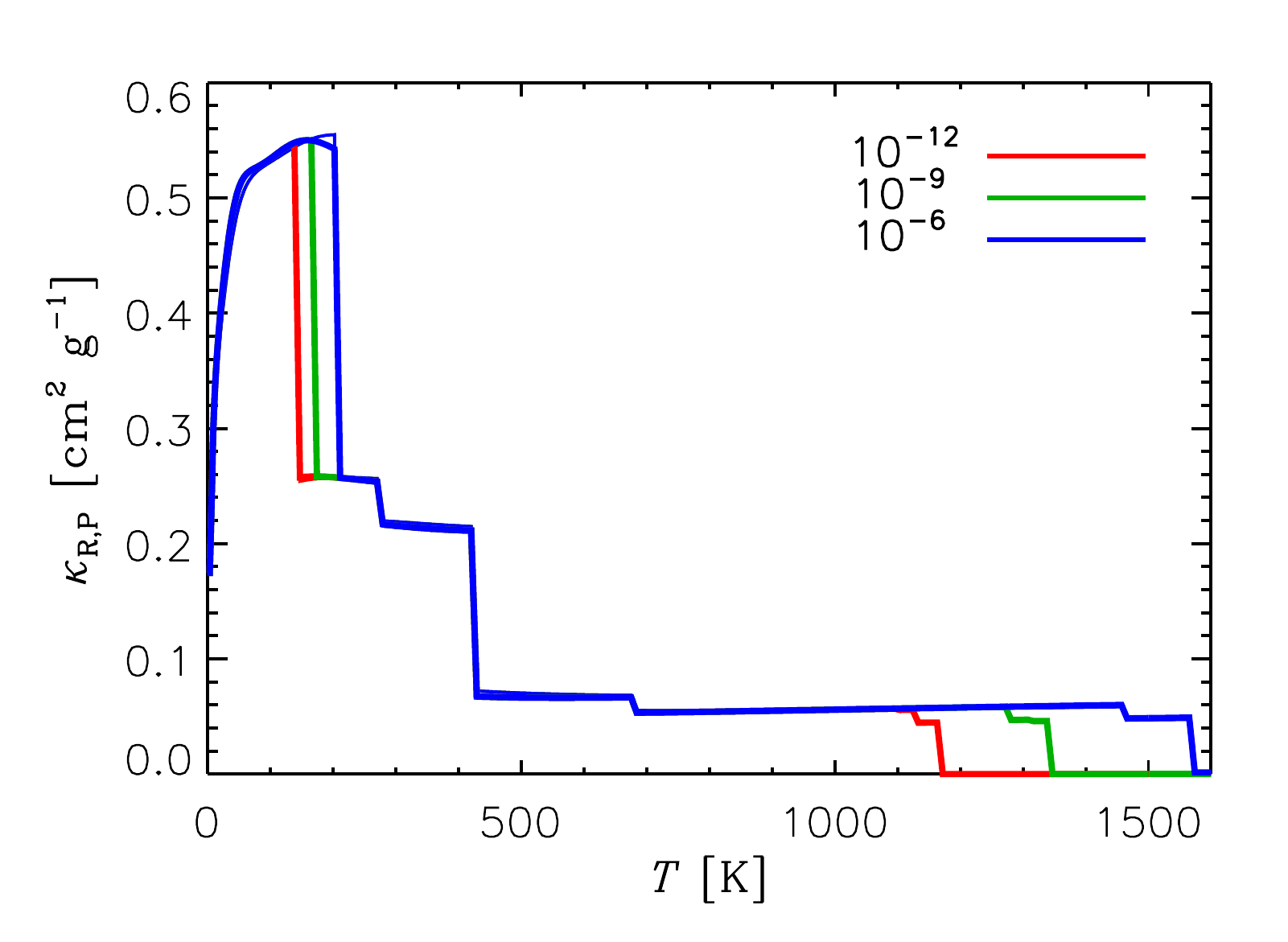}}
\resizebox{\figlen}{!}{%
\includegraphics[clip]{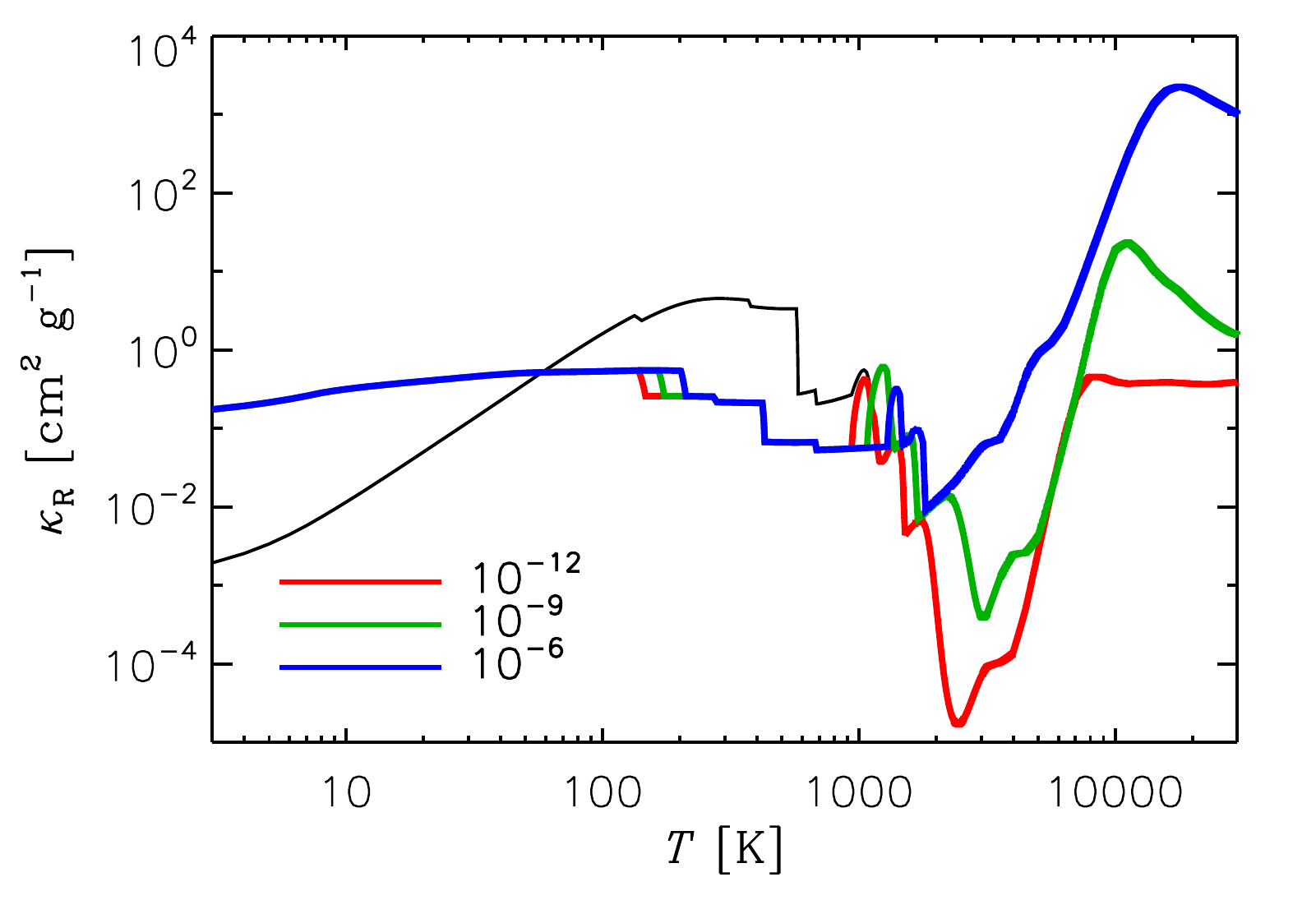}%
\includegraphics[clip]{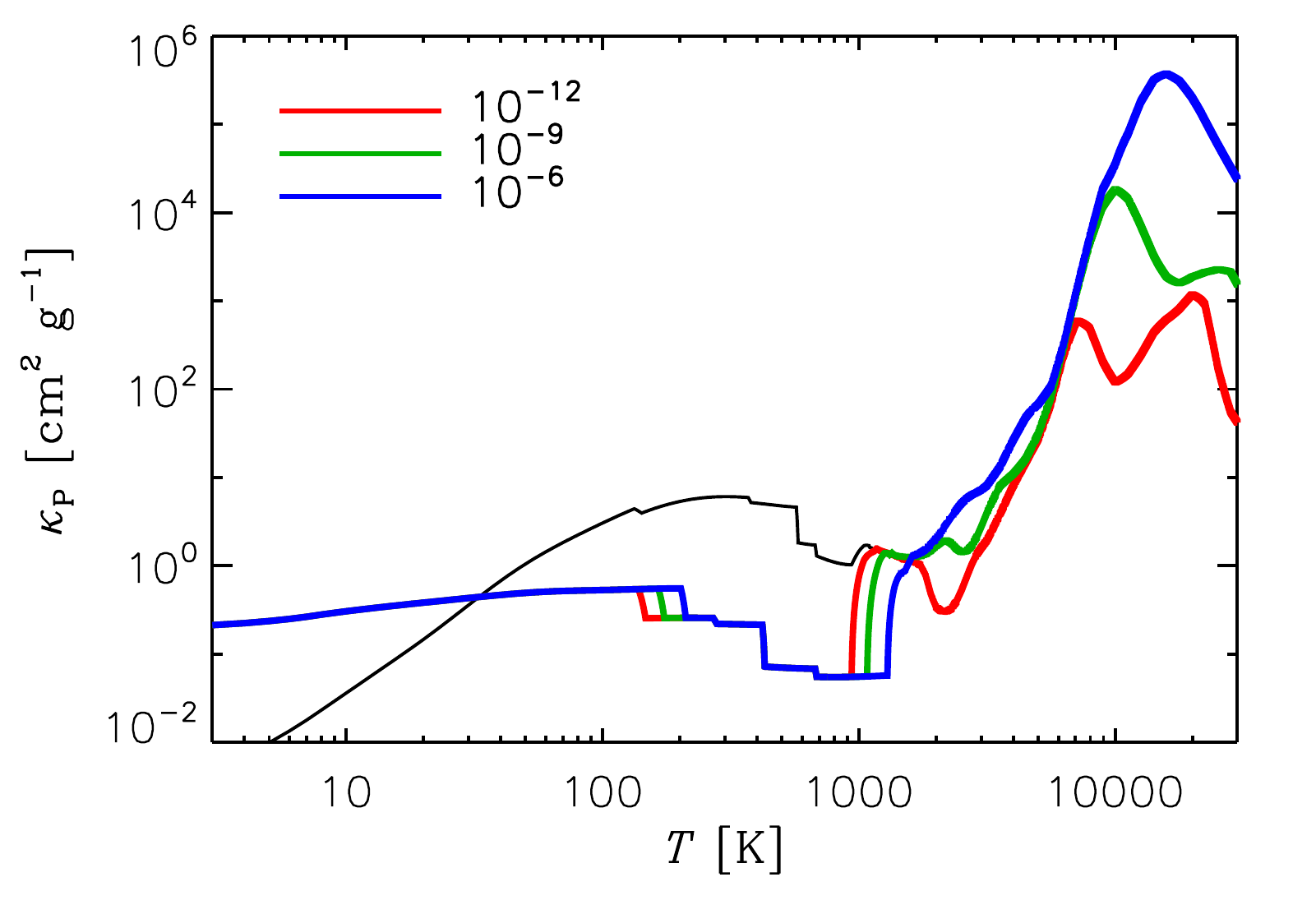}}
\caption{%
             Top-left: monochromatic opacity (Equation~(\ref{eq:kli})) as a function
             of wavelength for the various grain species, as labelled. Olivine 
             and orthopyroxene are grouped under silicates, while organics
             include both refractory and volatile species. The thick solid line
             is the sum of the single contributions (Equation~(\ref{eq:kl})).
             Top-right: Rosseland (Equation~(\ref{eq:kR}), thicker lines) 
             and Planck mean opacity (Equation~(\ref{eq:kP}), thinner lines) of 
             the dust mixture for three reference gas densities (as indicated
             in units of $\mathrm{g\,cm}^{-3}$).
             Bottom: Rosseland (left) and Planck mean opacity (right) of dust 
             and gas, for three reference gas densities (as labelled 
             in units of $\mathrm{g\,cm}^{-3}$). The thin black lines represent
             the interstellar opacity (see text), assuming a density of 
             $10^{-12}\,\mathrm{g\,cm}^{-3}$.
             }
\label{fig:k}
\end{figure}
%%%%%%%
The top-left panel of Figure~\ref{fig:k} shows the monochromatic opacities
versus wavelength, $\kappa^{i}_{\lambda}$, 
of silicates (olivine and orthopyroxene), iron,
water ice, troilite, and (refractory and volatile) organics, as indicated.
The thicker line is the total monochromatic opacity (Equation~(\ref{eq:kl})).
The Rosseland mean opacity (Equation~(\ref{eq:kR})) of the dust versus
temperature is shown in the top-right panel, for three values of $\rho$, 
as indicated in the legend in units of $\mathrm{g\,cm}^{-3}$.  
In the same panel, the Planck mean opacity (Equation~(\ref{eq:kP})) 
of the dust is plotted for comparison, showing only minor differences
over that range of temperatures.
In the bottom panels of the figure, 
the Rosseland and Planck mean opacity of dust and gas are plotted
for the same reference densities (see the figure caption for further details).
The bottom panels also show the interstellar opacities (thin curves) computed 
as explained above. The grain species are the same as in the other opacities, 
but the mass fractions $f_{s}$ are different (the gas-to-dust mass ratio is
$106.5$). 
At temperature $100 \lesssim T \lesssim 1000\,\K$,  the larger grains
used for disk opacity reduce $\kappa_{\mathrm{R}}$ by a factor $\approx 10$
relative to that produced by interstellar grains. Note that the largest
disk grains are such that $\xi\gg 1$, and their contribution to
$\kappa_{\mathrm{R}}$ is approximately proportional to $1/a$. 
For $T\ll1000\,\K$, the interstellar grains are such that $\xi\ll 1$ and 
$\kappa_{\mathrm{R}}$ becomes approximately proportional to some 
power of $T$ \citep{pollack1985}.

%%--------------------------------------------------------------------------
\section{Solution of a Special Quartic Equation}
\label{sec:QS}
%%--------------------------------------------------------------------------

In this Appendix, we first show in a mathematical sense that 
there is one, and one only, temperature that satisfies Equation~(\ref{eq:Etot}),
and then provide the formal solutions to this equation.
We refrain from giving a full derivation of these solutions, 
as such derivation can be found in many textbooks. 
We caution the reader that notations used in this Appendix
have no connection with the same notations used elsewhere in the paper.

Consider a  fourth-order polynomial of the type
\begin{equation}
\label{eq:gz}
g(z)=a z^{4}+b z - E,
\end{equation}
in which the constants $a$, $b$, and $E$ are non-zero, 
positive real numbers and the variable $z$ is a real number.
By taking the derivative $dg/dz$, one finds that $g$ is an
increasing function for $z^{3}>-b/(4a)$, and decreasing otherwise. 
Since $g(0)=-E<0$ and the minimum has a negative abscissa,
the polynomial $g(z)$ has two real roots: one negative and the other positive, 
which proves the existence of a single physically relevant temperature 
that satisfies Equation~(\ref{eq:Etot}). These same conclusions
apply if $b=0$.

In order to determine the zeros of $g(z)$, one can proceed by applying 
Ferrari's transformations and show that the four roots can be found by 
solving the two quadratic equations
\begin{eqnarray}
z^{2} - H z + W + H D/(4W) & = & 0 \label{eq:z21}\\
z^{2} + H z + W - H D/(4W) & = & 0 \label{eq:z22},
\end{eqnarray}
where $D=b/a>0$, $H=\sqrt{2W}$, and $W$ is a real solution of the
\textit{auxiliary cubic} equation
\begin{equation}
\label{eq:aux_cub}
8 w^{3} + 8 C w - D^{2} = 0,
\end{equation}
where $C=E/a>0$. Recall that any polynomial of odd degree 
admits at least one real solution. Notice that $W=0$ is a solution 
only if $D=0$ (but in this case the solution of Equation~(\ref{eq:gz}) 
is trivial).

The statement made above regarding the roots of Equation~(\ref{eq:z21})
and (\ref{eq:z22}) being equal to the roots of Equation~(\ref{eq:gz})
can be readily proved by multiplying the left-hand sides of 
Equations~(\ref{eq:z21}) and (\ref{eq:z22}) and then adding and
subtracting $C$, which yields $z^{4}+Dz-C+(8W^{3}+8CW-D^{2})/8W$.  
Therefore, in the non-trivial case $W\ne 0 $, the validity of our statement 
is demonstrated.

Let us introduce the discriminant
\begin{equation}
\label{eq:discrim}
\Delta=\left(\frac{D}{4}\right)^{4}+\left(\frac{C}{3}\right)^{3},
\end{equation}
and the two quantities
\begin{eqnarray}
A & = & \left[(D/4)^2+\sqrt{\Delta}\right]^{1/3} \label{eq:A}\\
B & = & \left[(D/4)^2-\sqrt{\Delta}\right]^{1/3} \label{eq:B}.
\end{eqnarray}
By applying the method of Cardano-Tartaglia, one can find the three
solutions of Equation~(\ref{eq:aux_cub})
\begin{eqnarray}
w_{1} & = & (A + B) \label{eq:w1}\\
w_{2} & = &-(A+B)/2 + \sqrt{-3} (A - B)/2 \label{eq:w2}\\
w_{3} & = &-(A+B)/2 - \sqrt{-3} (A - B)/2 \label{eq:w3}.
\end{eqnarray}
Since $C>0$ then $\Delta>0$, hence 
both $A$ and $B$ are real numbers, and
thus $w_{1}$ is a real solution, while $w_{2}$ and $w_{3}$
are complex conjugate solutions. Therefore, we shall
identify $W$, in Equations~(\ref{eq:z21}) and (\ref{eq:z22}), 
with $w_{1}$.

It is easy to show that the real solution, $W$, is positive. In fact,
$A>0$ and $A > |B|$ ($B<0$), from which follows that $W>0$.
It is then straightforward to conclude that the two solutions of 
Equation~(\ref{eq:z21}), $z_{1}$ and $z_{2}$, are complex conjugate,
since its discriminant is negative. 
In addition, since we proved above that the quartic $g(z)$ must have two
real solutions of opposite sign, the roots of Equation~(\ref{eq:z22}),
$z_{3}$ and $z_{4}$, must be real. 
The positive solution of Equation~(\ref{eq:gz}) is therefore 
\begin{equation}
\label{eq:z4}
z_{4}=\frac{1}{2}\left(\sqrt{HD/W - H^{2}}-H\right).
\end{equation}
Alternatively, one can directly prove $z_{4}$ in Equation~(\ref{eq:z4}) 
to be positive by showing that the inequality 
$HD/W - H^{2}>H^{2}$ is always true.
We recall that $D=b/a$ and $H=\sqrt{2W}$ and $W$ is
given by Equation~(\ref{eq:w1}). The inequality holds if
$W^{3}< 2(D/4)^{2}$.
From Equation~(\ref{eq:discrim}), we know that 
$\Delta> (D/4)^{4}$ and $B<0$, and thus $A> -B >0$. 
Therefore, we have
$W^{3}=A^{3}+B^{3}+3AB(A+B)=2(D/4)^{2}+3AB(A+B)$
with $AB<0$ and $A+B>0$, hence $W^{3}<2(D/4)^{2}$.

%%--------------------------------------------------------------------------
\section{Tests of the Radiation Flux-Limited Diffusion Solver}
\label{sec:diff_test}
%%--------------------------------------------------------------------------

In this Appendix, we present some tests of the radiation flux-limited 
diffusion solver discussed in Section~\ref{sec:RDS}.
As in the previous Appendices, the notations used here have no relation
to those used in other parts of the paper.

The first case we shall consider is that of the 1D heat 
diffusion equation with a constant diffusion coefficient $\mathcal{K}$
\begin{equation}
\label{eq:Uzz}
\ddt{U}=\mathcal{K}\frac{\partial^{2}{U}}{\partial z^{2}},
\end{equation}
and initial condition $U(z,0)=U_{0}\delta (z-z_{0})$, whose solution is
the well-known heat kernel
\begin{equation}
\label{eq:Uzzsol}
U(z,t)=\frac{U_{0}}{\sqrt{4\pi\mathcal{K}t}}e^{-(z-z_{0})^{2}/(4\mathcal{K}t)}
\end{equation}
%%%%%%
\begin{figure}[]
\centering%
\resizebox{\figlen}{!}{%
\includegraphics[clip]{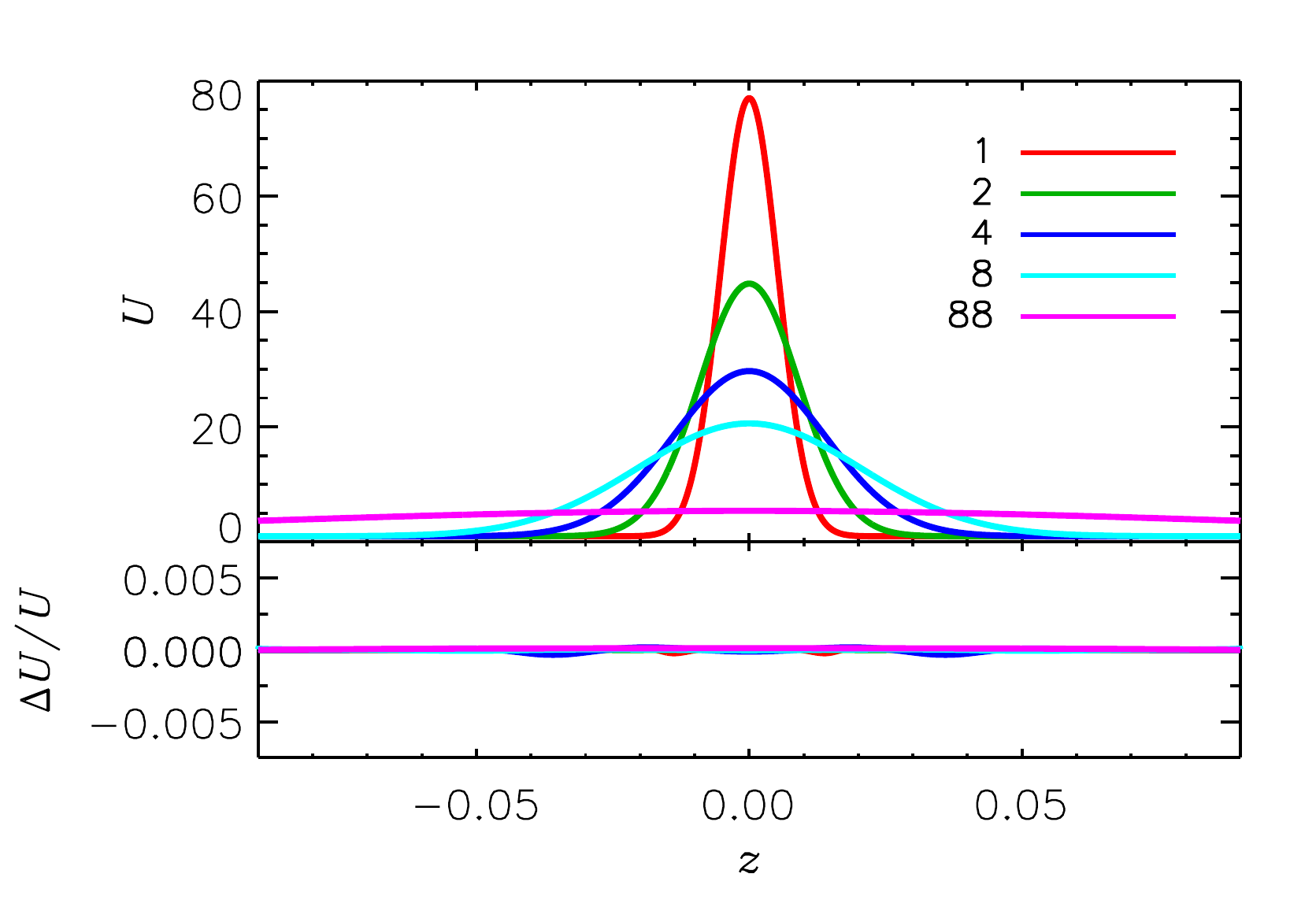}%
\includegraphics[clip]{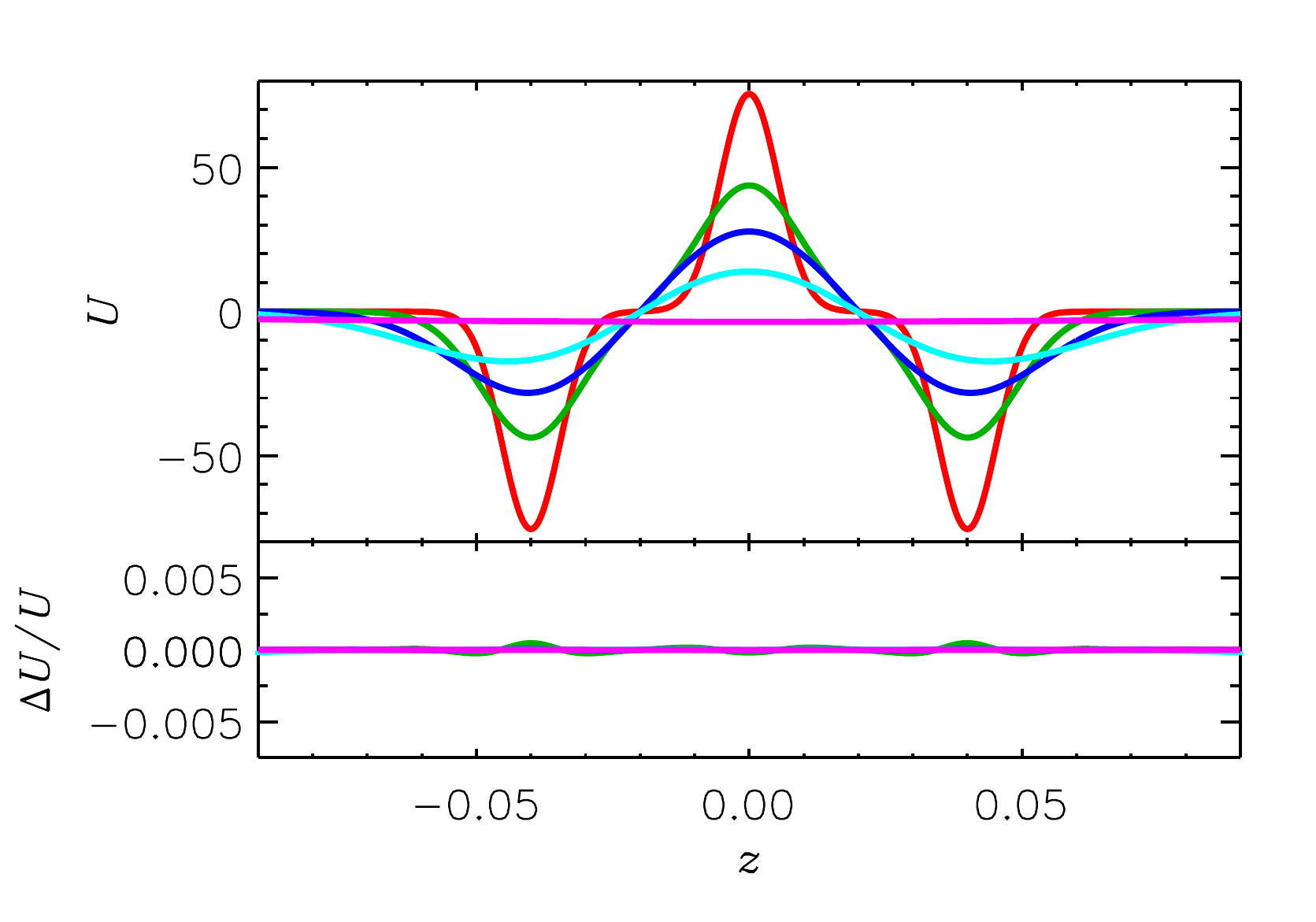}}
\caption{%
             Left:
             diffusion of an initial ``pulse'' represented by the Dirac $\delta$-function,
             $\delta(z)$. The top part of the panel shows numerical and analytical
             solutions of Equation~(\ref{eq:Uzz}), respectively represented 
             (but not distinguishable) by thick and thin lines, 
             at various times in arbitrary units as indicated.
             The difference between numerical and analytical solution, normalized 
             to the analytical solution, is plotted in the bottom part of the panel.
             Right:
             as in the left panel, but for a ``three-pulse'' initial condition proportional
             to $\delta(z+0.04)-\delta(z)+\delta(z-0.04)$.
             }
\label{fig:zp}
\end{figure}
%%%%%%%
The numerical solution of Equation~(\ref{eq:Uzz}), for the initial condition
$U(z,0)\propto\delta(z)$, is plotted in the left panel
of Figure~\ref{fig:zp} (thick lines) at various times (in arbitrary units), 
as indicated in the legend. The analytic solution (Equation~(\ref{eq:Uzzsol}))
is overlaid (as thin lines) to the numerical solution. 
Here, and in the other tests presented in this Appendix, analytic and 
numerical results are indistinguishable on the plot scale.
At the bottom of the panel, 
we also plot the normalized difference $\Delta U/U$ between the numerical 
and analytical solutions.
Obviously, any linear combination of functions of the type in 
Equation~(\ref{eq:Uzzsol}) is also a solution of Equation~(\ref{eq:Uzz}).
The right panel of Figure~\ref{fig:zp} shows the evolution of an initial 
condition $U(z,0)\propto[\delta(z+0.04)-\delta(z)+\delta(z-0.04)]$.
As in the left panel, we plot the numerical and analytical solution (top), 
as well as the normalized difference  $\Delta U/U$ (bottom).
In both calculations, the time step is set equal to $4000$ in units
of $(\Delta z)^{2}/\mathcal{K}$. 

Let us now consider the more general diffusion equation
\begin{equation}
\label{eq:dUdt}
\ddt{U}=\gdiv{(\mathcal{K}\nabla{U})},
\end{equation}
in which $U=U(t,r,\theta,\phi)$. We aim at deriving
simple analytic solutions to this equation using separation of
variables and making appropriate assumptions on the diffusion 
coefficient, $\mathcal{K}$.

%%%%%%
\begin{figure}[]
\centering%
\resizebox{\figlen}{!}{%
\includegraphics[clip]{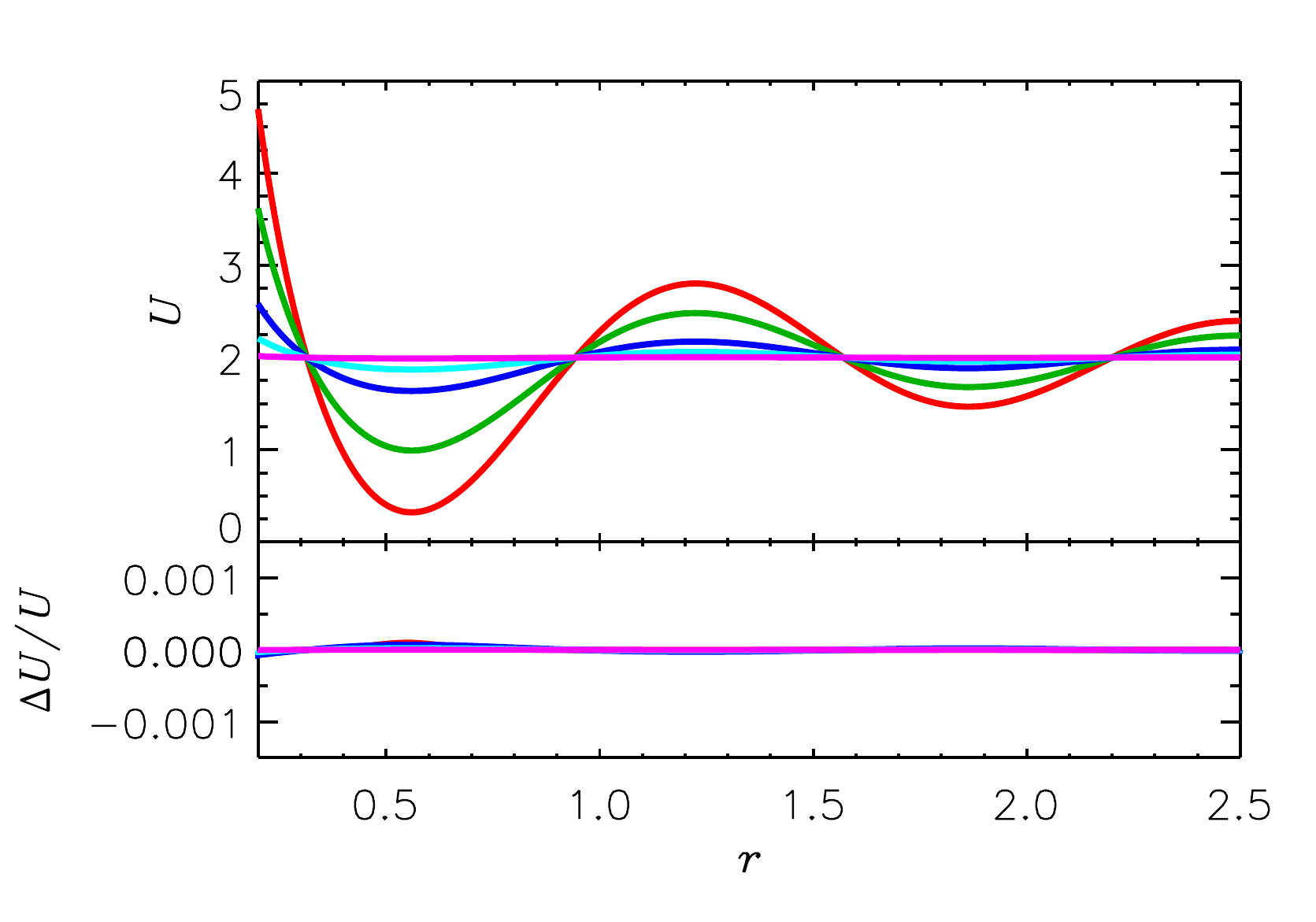}%
\includegraphics[clip]{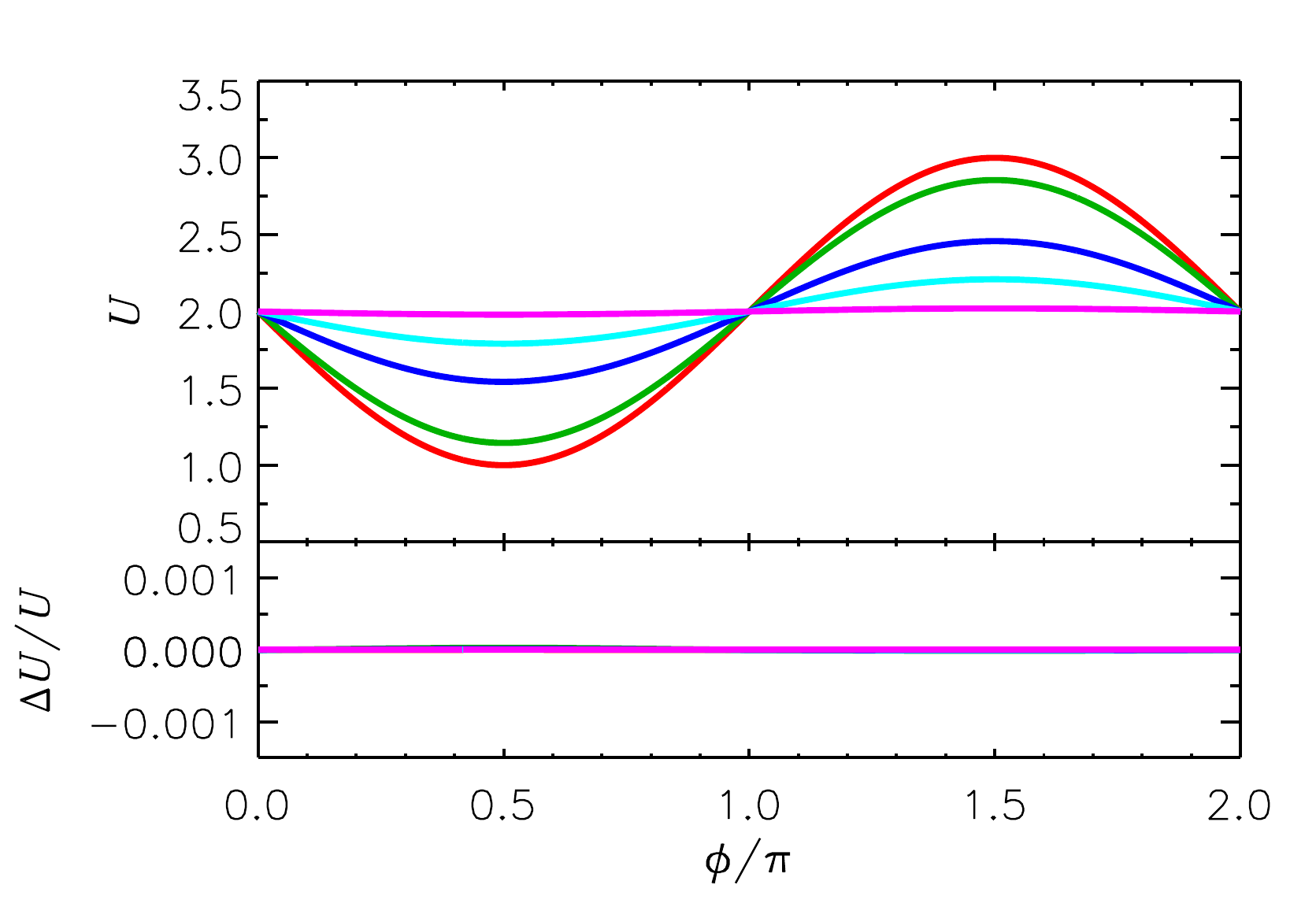}}
\resizebox{\figlen}{!}{%
\includegraphics[clip]{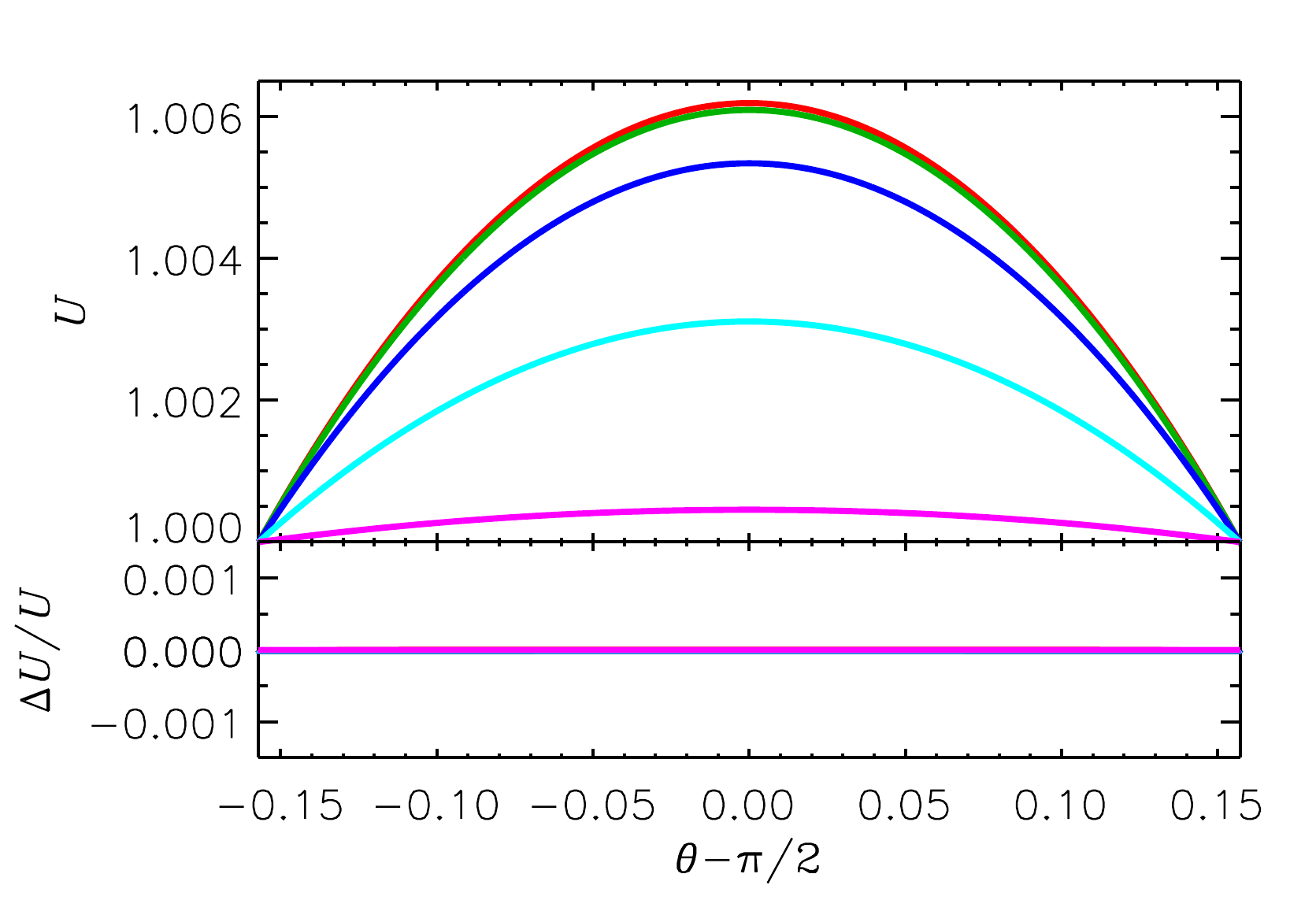}%
\includegraphics[clip]{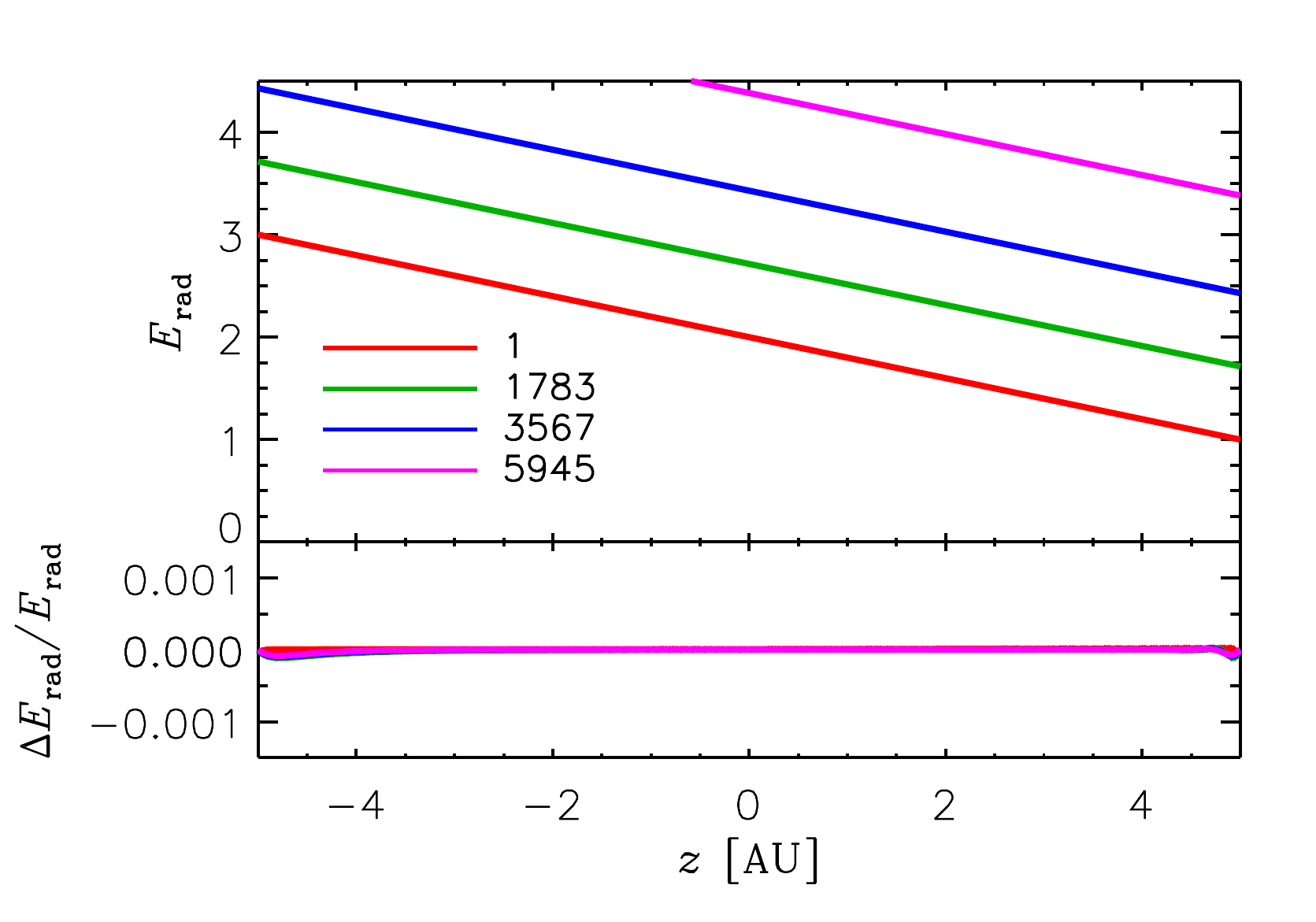}}
\caption{%
             Comparison between numerical and analytical solutions
             of Equation~(\ref{eq:dUdt_r}) (top-left),
             (\ref{eq:dUdt_phi}) (top-right), (\ref{eq:dUdt_rtheta}) (bottom-left),
             and (\ref{eq:streaming_num}) (bottom-right).
             Although not distinguishable, numerical solutions are plotted
             as thick lines and analytical solutions as thin lines.
             In the bottom-left panel, the function $U$ is normalized to its value
             at $\theta_{\mathrm{min}}$, and plotted at a fixed $r$.
             In the bottom-right panel, $E_{\mathrm{rad}}$ is normalized to
             $\bar{E}_{\mathrm{rad}}$ and the labels indicate the time in seconds.
             The difference between numerical and analytical solution, normalized 
             to the analytical solution, is plotted in the bottom part of each panel.             
             }
\label{fig:anasol}
\end{figure}
%%%%%%%
The first assumption we shall make is that $U$ is isotropic, i.e., 
$U=U(t,r)$. Equation~(\ref{eq:dUdt}) is then
\begin{equation}
\label{eq:dUdt_r}
\ddt{U}=\frac{1}{r^{2}}\dd{}{r}\left(r^{2}\mathcal{K}\dd{U}{r}\right).
\end{equation}
If we write $\mathcal{K}=r^{2}_{0}/\tau$, with both $r_{0}$ and
$\tau$ constants and assume the existence of solutions 
of the form $U(t,r)=T(t)R(r)$, Equation~(\ref{eq:dUdt_r}) reduces to
\begin{eqnarray}
\left(\frac{\tau}{T}\right)\frac{dT}{dt} & = & -\xi^{2} \label{eq:dUdt_r_T}\\
\left(\frac{r^{2}_{0}}{R}\right)\frac{d^{2}R}{dr^{2}} %
+  2\left(\frac{r^{2}_{0}}{rR}\right)\frac{dR}{dr} & = & -\xi^{2} \label{eq:dUdt_r_R},
\end{eqnarray}
where $\xi$ is a real number. The solution of Equation~(\ref{eq:dUdt_r_T}) is
$T(t)=T_{0}e^{-\xi^{2}t/\tau}$. Solutions to the second order differential
equation can be found with standard techniques, one of them is
$R(r)=(R_{0}/r)\cos{(\xi r/r_{0})}$. Hence,
$U(t,r)=T_{0}R_{0}e^{-\xi^{2}t/\tau}\cos{(\xi r/r_{0})}/r$ is solution to
Equation~(\ref{eq:dUdt_r}).
The diffusion coefficient need not be constant. In fact, if
$\mathcal{K}=r^{2}/\tau$, a similar approach leads to the solution
$U(t,r)=T_{0}e^{-2t/\tau}(c_{1}r+c_{0})/r^{2}$, where 
$c_{0}$ and $c_{1}$ are integration constants.

The same technique can be applied to look for solutions of
the type $U=U(t,\phi)$, which satisfy
\begin{equation}
\label{eq:dUdt_phi}
\ddt{U}=\frac{1}{r\sin{\theta}}\dd{}{\phi}\left(\frac{\mathcal{K}}{r\sin{\theta}}\dd{U}{\phi}\right).
\end{equation}
If $\mathcal{K}=(r\sin{\theta})^{2}/\tau$ and $U(t,\phi)=T(t)\Phi(\phi)$, 
then one can easily show that
$U(t,\phi)=T_{0}\Phi_{0}e^{-\xi^{2}t/\tau}\cos{(\xi\phi)}$
is a solution of Equation~(\ref{eq:dUdt_phi}).

The diffusion equation
\begin{equation}
\label{eq:dUdt_rtheta}
\ddt{U}=\frac{1}{r^{2}}\dd{}{r}\left(r^{2}\mathcal{K}\dd{U}{r}\right)+
\frac{1}{r\sin{\theta}}\dd{}{\theta}\left(\frac{\sin{\theta}}{r}\mathcal{K}\dd{U}{\theta}\right)
\end{equation}
can also be solved assuming a diffusion coefficient of the form
$\mathcal{K}=r^{2}/(\tau\sin{\theta}$) and expressing the solution as
$U(t,r,\theta)=T(t)R(r)\Theta(\theta)$. A solution for the temporal part
is again $T(t)=T_{0}e^{-\xi^{2}t/\tau}$. The radial and angular 
parts can be found by solving the two differential equations
\begin{eqnarray}
\frac{d}{dr}\left(r^{4}\frac{dR}{dr}\right) + \beta^{2}r^{2}R & = & 0 
                                                                                                      \label{eq:dUdt_rtheta_R}\\
\frac{d^{2}\Theta}{d\theta^{2}} + (\xi^{2}\sin{\theta}-\beta^{2})\Theta
                                                                                          & = & 0
                                                                                                      \label{eq:dUdt_rtheta_Theta},
\end{eqnarray}
where $\beta^{2}$ is a real number, which for simplicity 
we set equal to $2$. 
In such case, as mentioned earlier, a solution to 
Equation~(\ref{eq:dUdt_rtheta_R}) is $(c_{1}r+c_{0})/r^{2}$.
Equation~(\ref{eq:dUdt_rtheta_Theta}) is a linear differential
equation with a transcendental function as coefficient, the solution
of which does not typically have a compact form.
By expanding the coefficient $\xi^{2}\sin{\theta}-2$
in a Taylor series around $\theta=\pi/2$ and expressing the solution 
as $\Theta=\sum_{0}^{\infty}{a_{n}\vartheta^{n}}$, where 
$\vartheta=\pi/2-\theta$, we can determine the coefficients 
$a_{n}$ of the series from Equation~(\ref{eq:dUdt_rtheta_Theta}). 
For symmetry reasons, $\Theta$ is an even function and therefore
the series has only even terms. After some tedious algebra, 
one can write the solution as
\begin{equation}
\label{eq:Theta}
\Theta(\vartheta)=1+(2-\xi^{2})\vartheta^{2}/2+%
                             [(2-\xi^{2})^{2}+\xi^{2}]\vartheta^{4}/24+%
                             [(2-\xi^{2})^{3}+7\xi^{2}(2-\xi^{2})-\xi^{2}]\vartheta^{6}/720+%
                             \mathcal{O}(\vartheta^{8}).
\end{equation}

A comparison between the analytical solutions presented here and
numerical solutions is carried out in Figure~\ref{fig:anasol}.
The top-left, top-right, and bottom-left panels show, respectively, 
comparisons for the solutions of Equation~(\ref{eq:dUdt_r}), 
(\ref{eq:dUdt_phi}), and (\ref{eq:dUdt_rtheta}).
Numerical results are plotted as thick lines, whereas analytical results are
plotted as thin lines.
The normalized difference between numerical and analytical solution
is plotted at the bottom of each panel (see the figure caption for further details). 
The time step of the three calculations is between $100$ and $500$ in units
of $(\Delta S)^{2}/\mathcal{K}$, where $\Delta S$ is $\Delta r$, $r\Delta \theta$,
or $r \sin{\theta}\Delta \phi$ (for values of $r$ and $\sin{\theta}\approx 1$).

The solutions presented above apply to diffusion problems. Let us consider
the equation
\begin{equation}
   \label{eq:streaming_num}
    \ddt{E_{\mathrm{rad}}} = - \gdiv{\gvec{F}}.
\end{equation}
As mentioned in Section~\ref{sec:GT}, in very optically thin media, 
radiation transfer reaches the streaming limit for which the flux-limiter 
in Equation~(\ref{eq:lambda}) is 
$\lambda=\rho\kappa E_{\mathrm{rad}}/|\nabla{E_{\mathrm{rad}}}|\ll 1$
(which requires the condition $|\nabla{E_{\mathrm{rad}}}|\ne 0$)
and the frequency-integrated radiation flux reduces to
$\gvec{F}=-\gvec{n} c E_{\mathrm{rad}}$, where
$\gvec{n}=\nabla{E_{\mathrm{rad}}}/|\nabla{E_{\mathrm{rad}}}|$.
Therefore, in a 1D problem 
(or along the direction $\gvec{n}$), 
Equation~(\ref{eq:streaming_num}) becomes a first-order wave or
transport equation
\begin{equation}
\label{eq:streaming}
\ddt{E_{\mathrm{rad}}}=c\left|\dd{E_{\mathrm{rad}}}{z}\right|,
\end{equation}
in which $z$ is a linear coordinate, e.g., the cylindrical coordinate parallel 
to the disk's axis.
It is trivial to conclude that any function $U=U(z\pm ct)$ is a solution to 
Equation~(\ref{eq:streaming}).
The sign in the argument is equal to the sign
of $\partial E_{\mathrm{rad}}/\partial z $. 
For the sake of simplicity, let us consider the solution 
$E_{\mathrm{rad}}=\bar{E}_{\mathrm{rad}}(ct-z+2)$, where
$\bar{E}_{\mathrm{rad}}$ is a constant, which is an oblique front 
that propagates along the  $z$-direction at speed $c$. 
The numerical solution (thick lines) of
Equation~(\ref{eq:streaming_num}) is compared to the analytic 
solution (thin lines) of Equation~(\ref{eq:streaming}) in the 
bottom-right panel of Figure~\ref{fig:anasol}.
The plot shows the ratio $E_{\mathrm{rad}}/\bar{E}_{\mathrm{rad}}$
versus $z$ at different times, indicated in the legend in seconds, and
the relative difference of numerical and analytic solutions at the bottom.

%%%%%%
\begin{figure}[]
\centering%
\resizebox{\figlen}{!}{%
\includegraphics[clip]{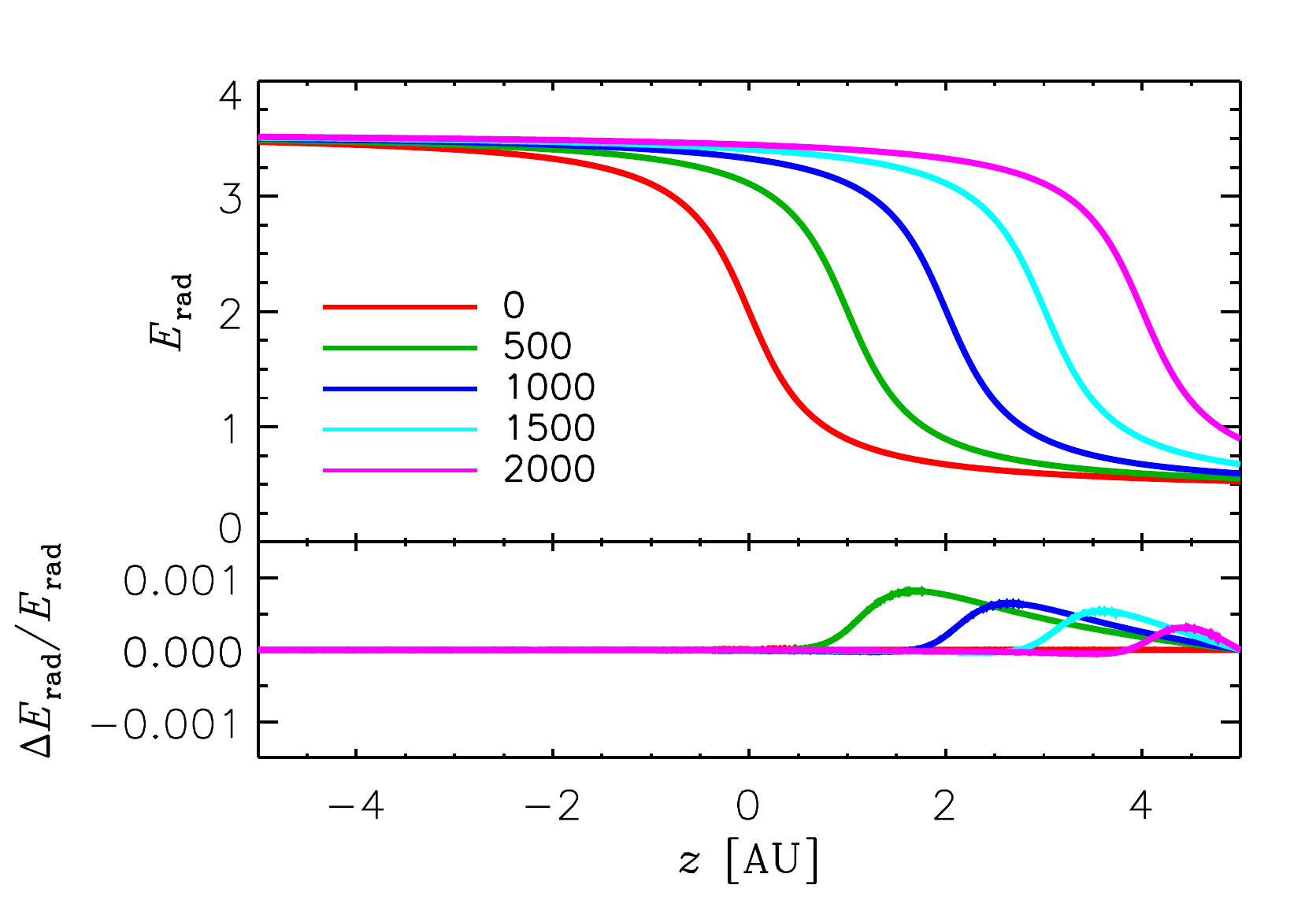}%
\includegraphics[clip]{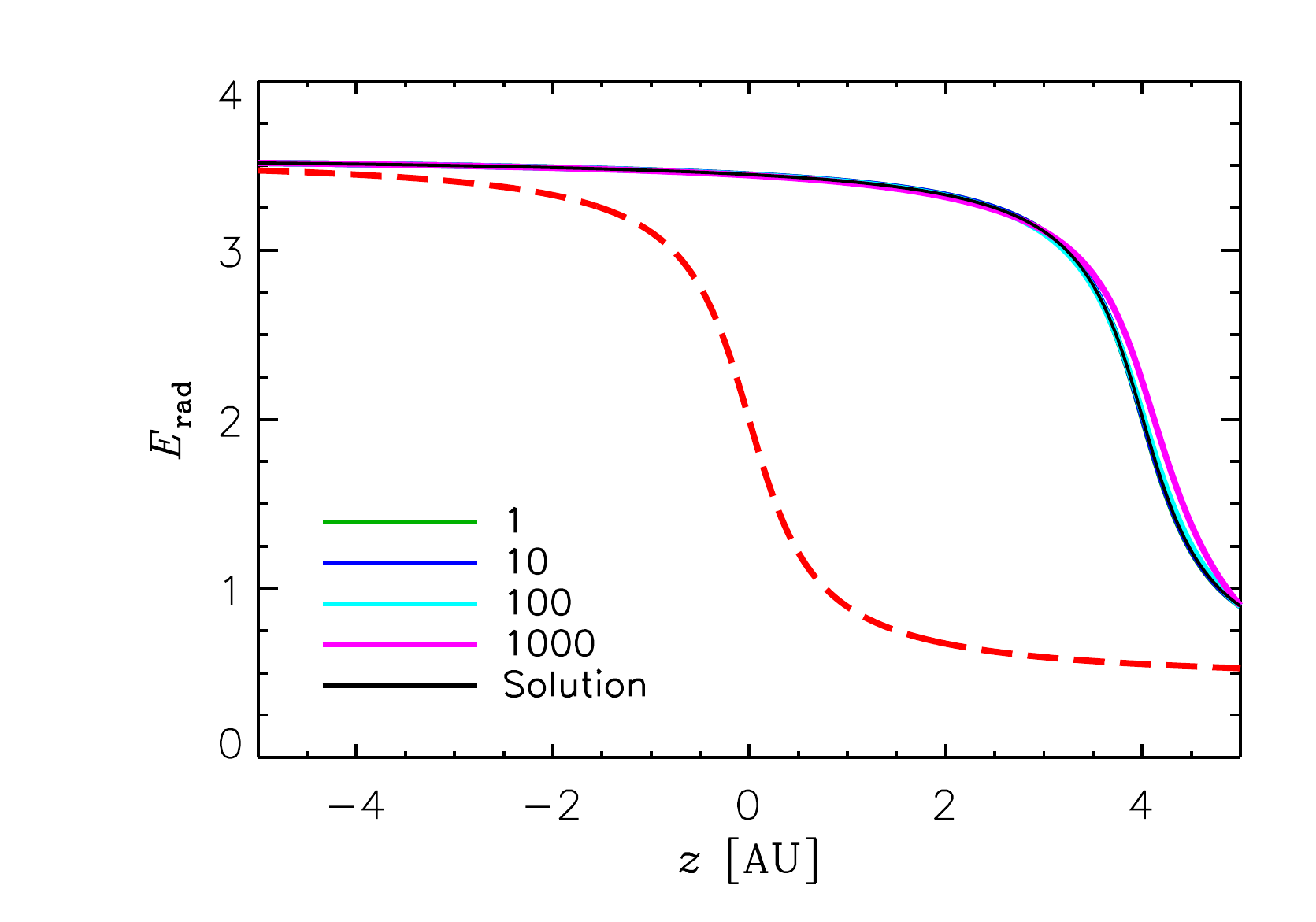}}
\caption{%
             Left: as in the bottom-right panel of Figure~\ref{fig:anasol}, but
             for the propagation of a sharper front and a calculation 
             with time step equal to 
             $(\Delta t)_{\mathrm{CFL}}$, the maximum allowed for stability
             of an explicit numerical solution of Equation~(\ref{eq:streaming}).
             The legend gives the times in seconds.
             Numerical and analytical solutions are plotted, respectively, as
             thick and thin lines.
             Right: comparison of implicit numerical solutions of       
             Equation~(\ref{eq:streaming}) with the analytic solution, 
             at a given time, for various applied time steps, as indicated
             in units of $(\Delta t)_{\mathrm{CFL}}$.
             The dashed line reproduces the front at $t=0$.
             }
\label{fig:stab_tur}
\end{figure}
%%%%%%%
In order to test the stability of the flux-limited approach in the streaming 
limit, when the time step is longer than $(\Delta t)_{\mathrm{CFL}}$ 
(see Section~\ref{sec:NP}), we follow \citet{turner2001}, who simulated
the propagation of a sharp front, represented by a step function, at the 
speed of light. A step function is solution to Equation~(\ref{eq:streaming}),
but $|\nabla{E_{\mathrm{rad}}}|=0$ for $x\ne ct$ so that $\lambda$ in
Equation~(\ref{eq:lambda}) tends to $1/3$ almost anywhere and
Equation~(\ref{eq:streaming_num}) remains a diffusion rather than
a first-order wave equation. Instead, we choose
a propagating front of the type $\arctan{[(ct-z)/10]}$ (plus a constant).
In case of Equation~(\ref{eq:streaming}), 
the Courant-Friedrichs-Lewy limiting time step allowed for stability is
$(\Delta t)_{\mathrm{CFL}}=\Delta z/(2c)$.
In the left panel of Figure~\ref{fig:stab_tur}, we compare the evolution 
of the front, obtained by solving Equation~(\ref{eq:streaming_num}), 
with the analytic solution (top), and show the relative difference
at the bottom, for a calculation with a time step 
$\Delta t=(\Delta t)_{\mathrm{CFL}}$ (see the figure caption for details).
In the right panel, we compare the numerical and analytical solutions 
at a given time, for time steps of various lengths, 
in units of $(\Delta t)_{\mathrm{CFL}}$, as indicated.
Since the aim here is to test the stability of the numerical solution 
rather than the accuracy of the flux-limited diffusion approximation 
of radiative transfer, in the test reported in right panel of 
Figure~\ref{fig:stab_tur} (and in this only!) the ratio 
$\mathcal{R}$ used to construct in the flux-limiter $\lambda$ 
(Equation~(\ref{eq:lambda})) is computed using the analytic solution.

%%%%%%
\begin{figure}[]
\centering%
\resizebox{\figlen}{!}{%
\includegraphics[clip]{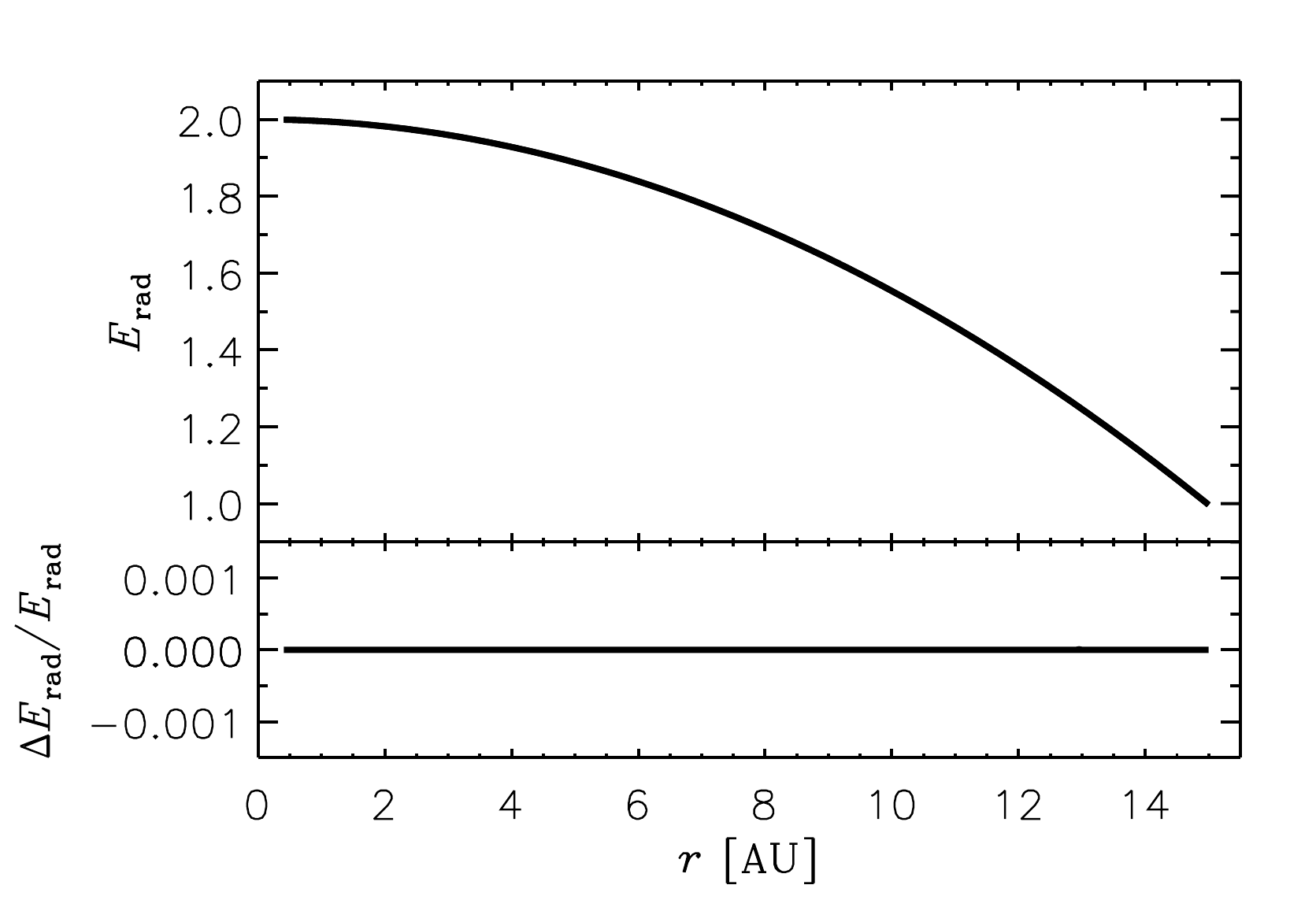}%
\includegraphics[clip]{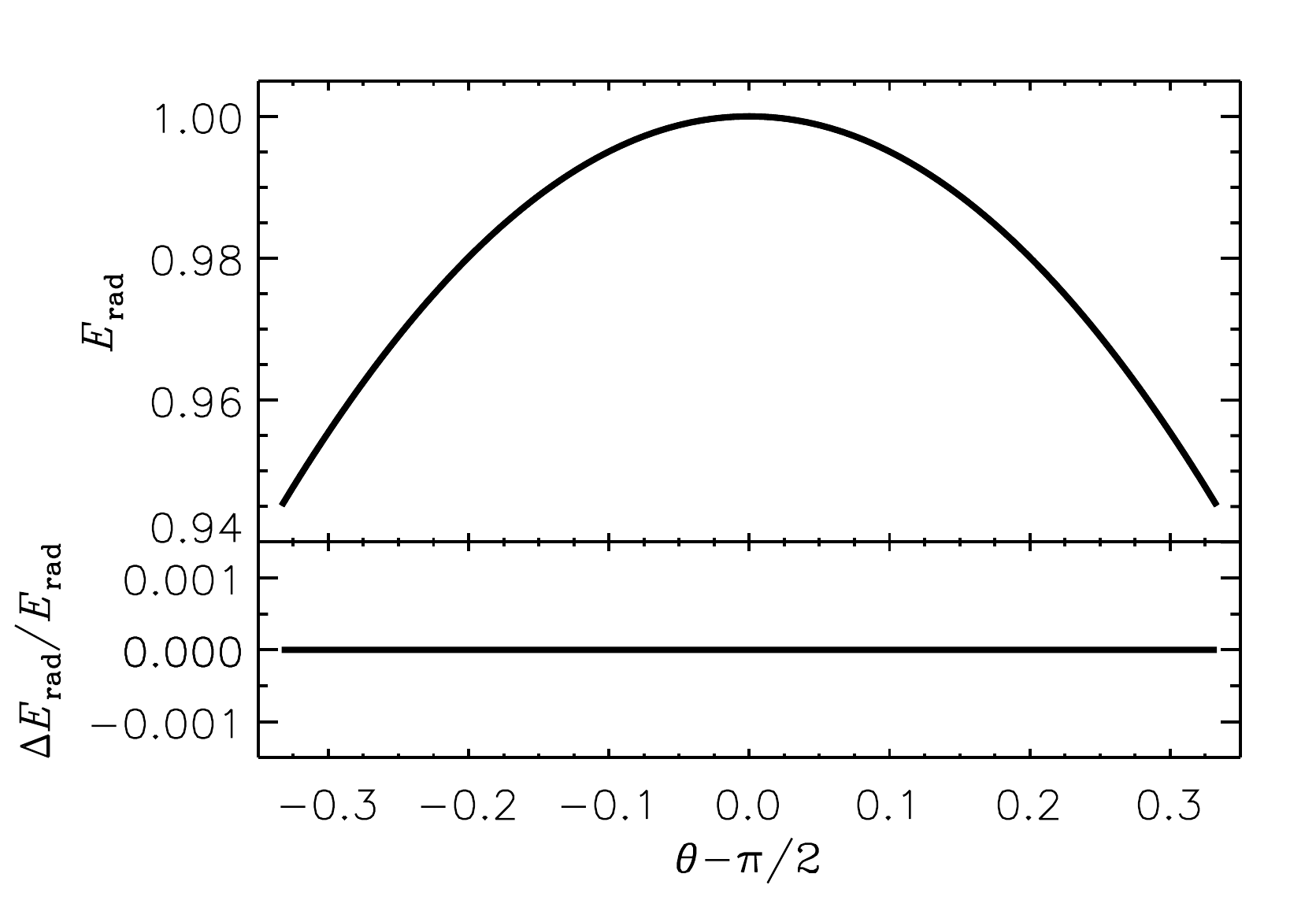}}
\caption{%
            Comparisons between numerical and analytical solutions of the
            radiative diffusion problems, as proposed by \citet{boss2009}
            (see text).
            The upper panels show $E_{\mathrm{rad}}$, divided by the
            constant
            $\bar{E}_{\mathrm{rad}}$, for $\rho=10^{-10}\,\mathrm{g\,cm^{-3}}$
            and $\kappa=1$ and $0.01\,\mathrm{cm^{2}\,g^{-1}}$.
            All calculations are initiated with a uniform radiation energy
            density $E_{\mathrm{rad}}=\bar{E}_{\mathrm{rad}}$. 
            Numerical results are indicated with thick lines and analytical
            results with thin lines (not distinguishable). The lower panels show
            the relative difference between numerical and analytical solutions.
             }
\label{fig:relax_bos}
\end{figure}
%%%%%%%
Stationary solutions to the radiation diffusion problem for a disk
in  spherical polar coordinates were published by \citet{boss2009}, 
who sought special solutions of the equation 
\begin {equation}
\label{eq:bos_stat}
  \frac{1}{3}\frac{c}{\kappa\rho}\nabla^{2}{E_{\mathrm{rad}}} + \varepsilon = 0,
\end{equation}
in which $\kappa$ and $\rho$ are constants.
The quantity $\varepsilon$ is an energy input rate per unit volume. 
In particular, he provided a solution for the isotropic case, 
$E_{\mathrm{rad}}=E_{\mathrm{rad}}(r)$, and a solution 
for the vertically stratified case,
$E_{\mathrm{rad}}=E_{\mathrm{rad}}(\theta)$.
One can easily prove that if 
$\varepsilon=2c\bar{E}_{\mathrm{rad}}/(\kappa\rho \bar{r}^{2})$,
$E_{\mathrm{rad}}=\bar{E}_{\mathrm{rad}}[2-(r/\bar{r})^{2}]$ is a solution
to Equation~(\ref{eq:bos_stat}), with $\bar{E}_{\mathrm{rad}}$ and
$\bar{r}$ constants of integration. Similarly, by setting 
$\varepsilon=-c\bar{E}_{\mathrm{rad}}\cos{(2\theta)}/(3\kappa\rho r^{2} \sin{\theta})$,
then
$E_{\mathrm{rad}}=\bar{E}_{\mathrm{rad}}\sin{\theta}$ is a solution of
Equation~(\ref{eq:bos_stat}).

We compare numerical solutions of Equation~(\ref{eq:bos_stat}) with
the analytic solution of \citet{boss2009} in Figure~\ref{fig:relax_bos}.
We use a density $\rho=10^{-10}\,\mathrm{g\,cm^{-3}}$ and opacities
$\kappa=1$ and $0.01\,\mathrm{cm^{2}\,g^{-1}}$.
Left and right panels refer to comparisons for the radial and the vertical
solutions. 
The upper panels show the ratio $E_{\mathrm{rad}}/\bar{E}_{\mathrm{rad}}$,
while the lower panels show the relative differences of numerical and analytical
solutions.
These calculations, which use the flux-limiter in Equation~(\ref{eq:lambda}),
indicate that $\lambda$ correctly converges to the diffusion limit (i.e., $1/3$)
for such tests.

%%%%%%
\begin{figure}[]
\centering%
\resizebox{\figlen}{!}{%
\includegraphics[clip]{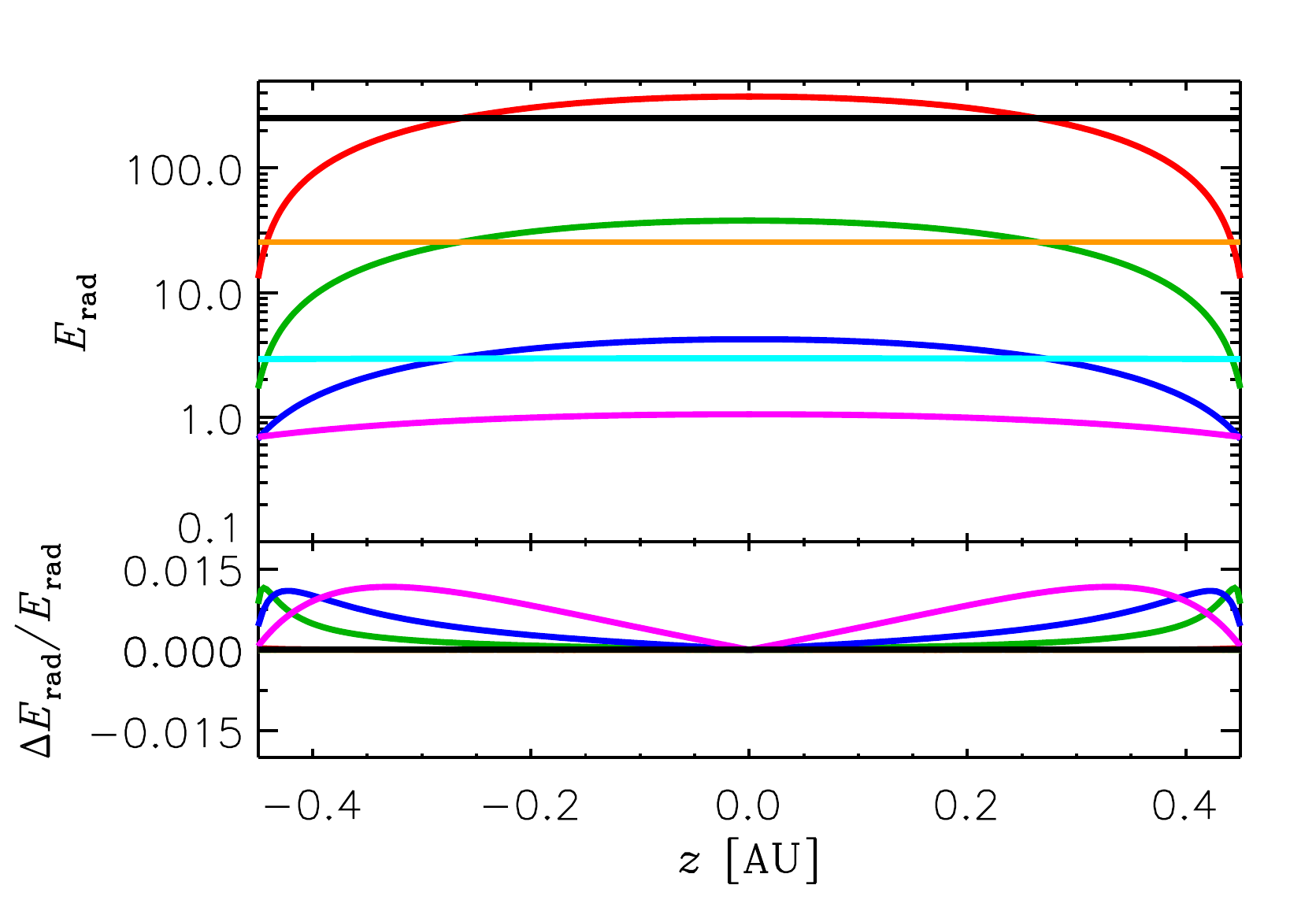}%
\includegraphics[clip]{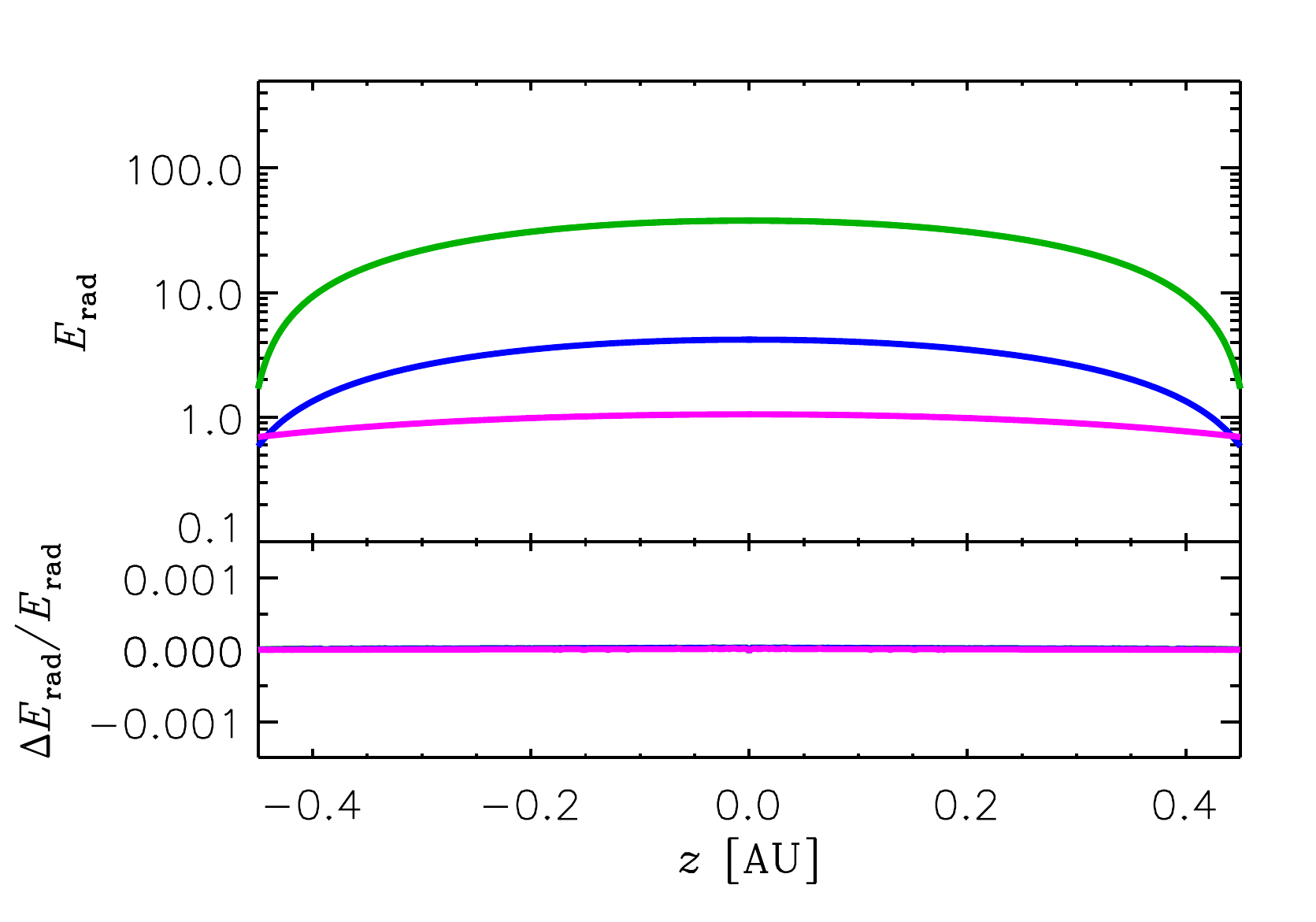}}
\caption{%
             Left:
             relaxation test \citep[see][]{boley2007a} in which a plane-parallel
             disk is heated at a constant rate, and opacity and density are constant.
             In the top of the panel, $E_{\mathrm{rad}}$ calculated numerically and
             divided by $(4\sigma_{\mathrm{SB}}/c)T^{4}_{\mathrm{eff}}$ (thick lines)
             is plotted together with its analytic expression (thin line),
             for $T_{\mathrm{eff}}=100\,\K$
             and $\tau_{M}=1000$, $100$, $10$, $1$ (upper to lower curved line), 
             $0.1$, $0.01$, $0.001$ (lower to upper straight line). 
             The bottom of the panel
             shows, for each value of $\tau_{M}$, the difference between computed
             and analytic result divided by the analytic result.
             Right:
             as in the left panel, but only for $\tau_{M}=100$, $10$, $1$ (upper to
             lower curve) and using the flux-limiter in Equation~(\ref{eq:lambda_win}).
             }
\label{fig:relax_bol}
\end{figure}
%%%%%%%
We also perform the ``relaxation'' test proposed by \citet{boley2007a},
in which a disk with no density stratification is heated at a constant 
rate per unit volume, $\varepsilon$. 
We assume a plane-parallel disk in a cylindrical geometry
$(z,r,\phi)$, with constant properties in $r$ and $\phi$.
For the disk to be stationary,
\begin{equation}
\label{eq:bol_stat}
\frac{1}{3}\frac{c}{\kappa\rho}\frac{d^{2}E_{\mathrm{rad}}}{dz^{2}}+\varepsilon=0,
\end{equation}
where the input energy rate per unit volume is
\begin{equation}
\label{eq:atm_heat}
\varepsilon=\sigma_{\mathrm{SB}}T^{4}_{\mathrm{eff}}%
                   \left(\frac{\kappa\rho}{\tau_{M}}\right).
\end{equation}
In the equations above, both the opacity $\kappa$ and the density 
$\rho$ are constant, and $\tau_{M}\gtrsim 1$ is the optical depth 
at the disk mid-plane (the total optical thickness of the disk is $2\tau_{M}$). 
The radiation flux emitted at the disk surface,
$\sigma_{\mathrm{SB}}T^{4}_{\mathrm{eff}}$,
is expressed through an effective temperature, $T_{\mathrm{eff}}$. 

Equation~(\ref{eq:bol_stat}) is strictly valid in the diffusion limit. 
By integrating, we find that
\begin{equation}
\label{eq:bol_stat_sol}
E_{\mathrm{rad}}=%
\left(\frac{3\sigma_{\mathrm{SB}}}{c}\right)T^{4}_{\mathrm{eff}}%
\left[\tau\left(1-\frac{\tau}{2\tau_{M}}\right)%
+\frac{1}{\sqrt{3}}+\frac{1}{3\tau_{M}}\right],
\end{equation}
in which $z\in[-z_{M},z_{M}]$ is the vertical coordinate ($z=0$ is 
the disk mid-plane), $\tau=\kappa\rho |z_{M}-z|$, and
$\tau_{M}=\kappa\rho z_{M}$. The constants of integration
are chosen so that Equation~(\ref{eq:bol_stat_sol}) agrees
with the derivation of the temperature stratification given
by \citet{hubeny1990}.
If $\tau_{M}\gg1$, the radiation energy density is
$E_{\mathrm{rad}}\sim(3\sigma_{\mathrm{SB}}/c)T^{4}_{\mathrm{eff}}%
(\tau+1/\sqrt{3})$.
In the streaming limit, Equation~(\ref{eq:bol_stat}) reduces to
$dE_{\mathrm{rad}}/dz=\varepsilon/c$. By choosing appropriately 
the constant of integration, the solution can be written as
$E_{\mathrm{rad}}\sim(3\sigma_{\mathrm{SB}}/c)T^{4}_{\mathrm{eff}}%
[(\tau+1)/(3\tau_{M})+1/\sqrt{3}]$.
For $\tau<\tau_{M}\ll 1$,
$E_{\mathrm{rad}}\sim(3\sigma_{\mathrm{SB}}/c)T^{4}_{\mathrm{eff}}%
[1/\sqrt{3}+1/(3\tau_{M})]$, which is the same as the formal solution of
Equation~(\ref{eq:bol_stat_sol}) in the optically thin limit.

Results from comparisons between numerical and analytical solutions
of the relaxation problem are reported in Figure~\ref{fig:relax_bol}. 
The upper part of the left panel shows the ratio
$E_{\mathrm{rad}}/[(4\sigma_{\mathrm{SB}}/c)T^{4}_{\mathrm{eff}}]$
for  $T_{\mathrm{eff}}=100\,\K$ and $\tau_{M}$ ranging from
$0.001$ to $1000$. The lower part of the panel shows the normalized
difference between numerical and analytic solutions. All calculations
are initiated with a uniform radiation field 
$E_{\mathrm{rad}}=(4\sigma_{\mathrm{SB}}/c)T^{4}_{\mathrm{eff}}$.

For these stationary problems, the flux-limiter in Equation~(\ref{eq:lambda})
reproduces reasonably well the transition between diffusion and streaming 
limits. 
It ought to be stressed, however, that the $\approx 1$\% discrepancy, 
observed at low optical 
depths for the cases with $\tau_{M}=1$, $10$, and $100$ (see lower-left 
panel of Figure~\ref{fig:relax_bol}), is not due to inaccuracies or limitations
of the numerical solver, but rather to the way the flux-limiter $\lambda$ in 
Equation~(\ref{eq:lambda}) transitions from the diffusion to the streaming 
limit. In fact, using the flux-limiter 
\citep{levermore1984,castor2007}
\begin{equation}
\label{eq:lambda_win}
\lambda=\frac{1}{\max{(3,\mathcal{R})}},
\end{equation}
which provides a sharper transition (at larger $\mathcal{R}$), 
results in a much better agreement between numerical and analytical 
solutions for those values of $\tau_{M}$, as can be clearly seen in the 
right panel of Figure~\ref{fig:relax_bol}.
\citep[This flux-limiter is not generally recommended, tough, see][]{levermore1984}.

Finally, we comment on the impact of applying a preconditioner
(see Section~\ref{sec:RDS}) to the iterative solvers. In the tests
reported here, for which the accuracy of the global convergence criterion 
(see discussion in Section~\ref{sec:RDS}) can be independently assessed
via comparison with the true solution, the use of the ILU preconditioner 
results in a reduction of the iteration count of factors between $20$ and 
$\gtrsim 200$.
In the actual calculations, where the problem is much larger and rather 
more challenging, the reduction of the iteration count is typically smaller, 
but still substantial (factors $> 10$). More importantly, the ILU
pre-conditioned system (Section~\ref{sec:RDS}) generally converges
in situations where the non-conditioned system fails to converge.

%%--------------------------------------------------------------------------
\section{Piecewise Approximation of the Core's Gravitational Potential}
\label{sec:phi_c}
%%--------------------------------------------------------------------------

The gravitational potential of a point-mass, $\Mc$, the planetary
core, introduces a singularity on the computational grid at the vector 
location of the core, $\gvec{r}_{c}$. 
In disk-planet interaction calculations, this singularity
is typically avoided by regularizing the potential, $\Phi_{c}$, so that
$\Phi_{c}=-G\Mc/\sqrt{R^{2}+\epsilon^{2}}$, where 
$\gvec{R}=\gvec{r}-\gvec{r}_{c}$ and $\epsilon$ is a softening
length.
Clearly, if $\epsilon$ is small enough, for all practical purposes
this potential is very similar to the true potential (almost) 
everywhere on the grid.

More sophisticated approximations, using piecewise 
reconstructions of the true potential, have also been introduced
\citep[e.g.,][]{klahr2006b,kley2009}. These functional forms have 
the property of becoming the true potential, beyond some 
distance from the point-mass. 
In the calculations presented in this paper, we follow this
sophisticated approach and introduce a piecewise reconstruction
of the core's potential that involves fifth-order polynomials
\begin{equation}
   \label{eq:phi_p}
   \Phi_{c}=\left\{%
                 \begin{array}{l l}
                  \left[\frac{1}{10}\left(R/\epsilon\right)^{5} 
                     - \frac{3}{10}\left(R/\epsilon\right)^{4}%
                     +\frac{2}{3}\left(R/\epsilon\right)^{2}%
                      -\frac{14}{10}\right]\left(G \Mc/\epsilon\right) & %
   \mathrm{if}\quad R/\epsilon\le 1\\
                              &   \\
                 \left[-\frac{1}{30}\left(R/\epsilon\right)^{5}%
                       +\frac{3}{10}\left(R/\epsilon\right)^{4} -\left(R/\epsilon\right)^{3}%
                       +\frac{4}{3}\left(R/\epsilon\right)^{2} %
                        -\frac{16}{10}\right]\left(G \Mc/\epsilon\right) %
                 +\frac{1}{15} G \Mc/R & %
   \mathrm{if}\quad 1<R/\epsilon\le 2\\
                              &   \\         
                 -G \Mc/R & \mathrm{otherwise}.
                 \end{array}
                 \right.
\end{equation}
We set the length $\epsilon$ between $1$ and $2$ core 
radii, $R_{c}$, basically on the order of the linear resolution on the finest 
gird level. 
Equation~(\ref{eq:phi_p}) is applied to all grids with the same value 
of $\epsilon$. Therefore, an exact potential is effectively used over
the entire grid domain, on all but the top-most grid level.\\[5mm]

%%++++++++++++++++++++++++++++++++++++++++++++++++++++++++++++++++++++++%%

%%++++++++++++++++++++++++++++++++++++++++++++++++++++++++++++++++++++++%%


\begin{thebibliography}{77}\setlength{\itemsep}{1mm}
\expandafter\ifx\csname natexlab\endcsname\relax\def\natexlab#1{#1}\fi

\bibitem[{{Asplund} {et~al.}(2009){Asplund}, {Grevesse}, {Sauval}, \&
  {Scott}}]{asplund2009}
{Asplund}, M., {Grevesse}, N., {Sauval}, A.~J., \& {Scott}, P. 2009, \araa, 47,
  481

\bibitem[{{Ayliffe} \& {Bate}(2009)}]{ayliffe2009a}
{Ayliffe}, B.~A., \& {Bate}, M.~R. 2009, \mnras, 393, 49

\bibitem[{{Ayliffe} \& {Bate}(2012)}]{ayliffe2012}
{Ayliffe}, B.~A., \& {Bate}, M.~R. 2012, \mnras, 427, 2597

\bibitem[{{Bate} {et~al.}(2003){Bate}, {Lubow}, {Ogilvie}, \&
  {Miller}}]{bate2003}
{Bate}, M.~R., {Lubow}, S.~H., {Ogilvie}, G.~I., \& {Miller}, K.~A. 2003, \mnras, 341, 213

\bibitem[{{Black} \& {Bodenheimer}(1975)}]{black1975}
{Black}, D.~C., \& {Bodenheimer}, P. 1975, \apj, 199, 619

\bibitem[{{Bodenheimer} {et~al.}(2013){Bodenheimer}, {D'Angelo}, {Lissauer},
  {Fortney}, \& {Saumon}}]{bodenheimer2013}
{Bodenheimer}, P., {D'Angelo}, G., {Lissauer}, J.~J., {Fortney}, J.~J., \&
  {Saumon}, D. 2013, \apj, 770, 120

\bibitem[{{Bodenheimer} \& {Pollack}(1986)}]{bodenheimer1986}
{Bodenheimer}, P., \& {Pollack}, J.~B. 1986, Icarus, 67, 391

\bibitem[{{Bohren} \& {Huffman}(1983)}]{bohren1983}
{Bohren}, C.~F., \& {Huffman}, D.~R. 1983, {Absorption and Scattering of Light
  by Small Particles} (New York: Wiley)

\bibitem[{{Boley} {et~al.}(2007{\natexlab{a}}){Boley}, {Durisen}, {Nordlund},
  \& {Lord}}]{boley2007a}
{Boley}, A.~C., {Durisen}, R.~H., {Nordlund}, {\AA}., \& {Lord}, J.
  2007{\natexlab{a}}, \apj, 665, 1254

\bibitem[{{Boley} {et~al.}(2007{\natexlab{b}}){Boley}, {Hartquist}, {Durisen},
  \& {Michael}}]{boley2007b}
{Boley}, A.~C., {Hartquist}, T.~W., {Durisen}, R.~H., \& {Michael}, S.
  2007{\natexlab{b}}, \apjl, 656, L89

\bibitem[{{Boss}(2009)}]{boss2009}
{Boss}, A.~P. 2009, \apj, 694, 107

\bibitem[{Britz {et~al.}(2003)Britz, {\O}sterby, \& Strutwolf}]{britz2003}
Britz, D., {\O}sterby, O., \& Strutwolf, J. 2003, Computational Biology and
  Chemistry, 27, 253

\bibitem[{{Castor}(2007)}]{castor2007}
{Castor}, J.~I. 2007, {Radiation Hydrodynamics} (Cambridge: Cambridge
  University Press)

\bibitem[{Chandrasekhar(1967)}]{chandra1967}
Chandrasekhar, S. 1967, Communications on Pure and Applied Mathematics, 20, 251

\bibitem[{{Cook}(2009)}]{cook2009}
{Cook}, A.~H. 2009, {Interiors of the Planets} (Cambridge: Cambridge
  University Press)

\bibitem[{{D'Alessio} {et~al.}(2001){D'Alessio}, {Calvet}, \&
  {Hartmann}}]{dalessio2001}
{D'Alessio}, P., {Calvet}, N., \& {Hartmann}, L. 2001, \apj, 553, 321

\bibitem[{{D'Angelo} {et~al.}(2005){D'Angelo}, {Bate}, \&
  {Lubow}}]{gennaro2005}
{D'Angelo}, G., {Bate}, M.~R., \& {Lubow}, S.~H. 2005, \mnras, 358, 316

\bibitem[{{D'Angelo} {et~al.}(2002){D'Angelo}, {Henning}, \&
  {Kley}}]{gennaro2002}
{D'Angelo}, G., {Henning}, T., \& {Kley}, W. 2002, \aap, 385, 647

\bibitem[{{D'Angelo} {et~al.}(2003{\natexlab{a}}){D'Angelo}, {Henning}, \&
  {Kley}}]{gennaro2003b}
{D'Angelo}, G., {Henning}, T., \& {Kley}, W. 2003{\natexlab{a}}, \apj, 599, 548

\bibitem[{{D'Angelo} {et~al.}(2003{\natexlab{b}}){D'Angelo}, {Kley}, \&
  {Henning}}]{gennaro2003a}
{D'Angelo}, G., {Kley}, W., \& {Henning}, T. 2003{\natexlab{b}}, \apj,
  586, 540

\bibitem[{{D'Angelo} \& {Lubow}(2008)}]{gennaro2008}
{D'Angelo}, G., \& {Lubow}, S.~H. 2008, \apj, 685, 560

\bibitem[{{D'Angelo} \& {Lubow}(2010)}]{gennaro2010}
{D'Angelo}, G., \& {Lubow}, S.~H. 2010, \apj, 724, 730

\bibitem[{{D'Angelo} \& {Marzari}(2012)}]{gennaro2012}
{D'Angelo}, G., \& {Marzari}, F. 2012, \apj, 757, 50

\bibitem[{{Davis} \& {Rabinowitz}(2007)}]{davis2007}
{Davis}, P.~J., \& {Rabinowitz}, P. 2007, {Methods of Numerical Integration}
  (New York: Dover)

\bibitem[{{Davis}(2005)}]{davis2005}
{Davis}, S.~S. 2005, \apjl, 627, L153

\bibitem[{{de Val-Borro} {et~al.}(2006){de Val-Borro}, {Edgar}, {Artymowicz},
  {Ciecielag}, {Cresswell}, {D'Angelo}, {Delgado-Donate}, {Dirksen}, {Fromang},
  {Gawryszczak}, {Klahr}, {Kley}, {Lyra}, {Masset}, {Mellema}, {Nelson},
  {Paardekooper}, {Peplinski}, {Pierens}, {Plewa}, {Rice}, {Sch{\"a}fer}, \&
  {Speith}}]{devalborro2006}
{de Val-Borro}, M., {Edgar}, R.~G., {Artymowicz}, P., {Ciecielag}, P.,
  {Cresswell}, P., {D'Angelo}, G., {Delgado-Donate}, E.~J., {Dirksen}, G.,
  {Fromang}, S., {Gawryszczak}, A., {Klahr}, H., {Kley}, W., {Lyra}, W.,
  {Masset}, F., {Mellema}, G., {Nelson}, R.~P., {Paardekooper}, S.,
  {Peplinski}, A., {Pierens}, A., {Plewa}, T., {Rice}, K., {Sch{\"a}fer}, C.,
  \& {Speith}, R. 2006, \mnras, 370, 529

\bibitem[{{DeCampli} {et~al.}(1978){DeCampli}, {Cameron}, {Bodenheimer}, \&
  {Black}}]{decampli1978}
{DeCampli}, W.~M., {Cameron}, A.~G.~W., {Bodenheimer}, P., \& {Black}, D.~C.
  1978, \apj, 223, 854

\bibitem[{{Demmel} {et~al.}(2000){Demmel}, {Koev}, \& {Li}}]{demmel2000}
{Demmel}, J., {Koev}, P., \& {Li}, X. 2000, {A Brief Survey of Direct Linear
  Solvers}, ed. Z.~{Bai}, J.~{Demmel}, J.~{Dongarra}, A.~{Ruhe}, \& H.~{van der
  Vorst} (Philadelphia, PA: Society for Industrial and Applied Mathematics), 326--331

\bibitem[{{Draine}(2011)}]{draine2011}
{Draine}, B.~T. 2011, {Physics of the Interstellar and Intergalactic Medium}
  (Princeton, NJ: Princeton University Press)

\bibitem[{{Draine} \& {Lee}(1984)}]{draine1984}
{Draine}, B.~T., \& {Lee}, H.~M. 1984, \apj, 285, 89

\bibitem[{{Ferguson} {et~al.}(2005){Ferguson}, {Alexander}, {Allard}, {Barman},
  {Bodnarik}, {Hauschildt}, {Heffner-Wong}, \& {Tamanai}}]{ferguson2005}
{Ferguson}, J.~W., {Alexander}, D.~R., {Allard}, F., {Barman}, T., {Bodnarik},
  J.~G., {Hauschildt}, P.~H., {Heffner-Wong}, A., \& {Tamanai}, A. 2005, \apj,
  623, 585

\bibitem[{{Freedman} {et~al.}(2008){Freedman}, {Marley}, \&
  {Lodders}}]{freedman2008}
{Freedman}, R.~S., {Marley}, M.~S., \& {Lodders}, K. 2008, \apjs, 174, 504

\bibitem[{{Gray}(1992)}]{gray1992}
{Gray}, D.~F. 1992, {The Observation and Analysis of Stellar Photospheres} 
 (Cambridge: Cambridge University Press)

\bibitem[{{Grevesse} \& {Sauval}(1998)}]{grevesse1998}
{Grevesse}, N., \& {Sauval}, A.~J. 1998, \ssr, 85, 161

\bibitem[{Gutknecht(2006)}]{gutknecht2006}
Gutknecht, M. 2006, {Block Krylov Space Methods for Linear Systems with
  Multiple Right-Hand Sides: An Introduction}
  (Zurich, Switzerland: ETH-Zurich)

\bibitem[{{Hansen} {et~al.}(2004){Hansen}, {Kawaler}, \&
  {Trimble}}]{hansen2004}
{Hansen}, C.~J., {Kawaler}, S.~D., \& {Trimble}, V. 2004, {Stellar Interiors:
  Physical Principles, Structure, and Evolution} (New York: Springer)

\bibitem[{{Hubeny}(1990)}]{hubeny1990}
{Hubeny}, I. 1990, \apj, 351, 632

\bibitem[{{Hubickyj} {et~al.}(2005){Hubickyj}, {Bodenheimer}, \&
  {Lissauer}}]{hubickyj2005}
{Hubickyj}, O., {Bodenheimer}, P., \& {Lissauer}, J.~J. 2005, Icarus, 179, 415

\bibitem[{{Kippenhahn} {et~al.}(2013){Kippenhahn}, {Weigert}, \&
  {Weiss}}]{kippenhahn2013}
{Kippenhahn}, R., {Weigert}, A., \& {Weiss}, A. 2013, {Stellar Structure and
  Evolution} (Berlin: Springer)

\bibitem[{{Kittel}(2004)}]{kittel2004}
{Kittel}, C. 2004, {Elementary Statistical Physics} (New York: Dover)

\bibitem[{{Klahr} \& {Kley}(2006)}]{klahr2006b}
{Klahr}, H., \& {Kley}, W. 2006, \aap, 445, 747

\bibitem[{{Kley} {et~al.}(2009){Kley}, {Bitsch}, \& {Klahr}}]{kley2009}
{Kley}, W., {Bitsch}, B., \& {Klahr}, H. 2009, \aap, 506, 971

\bibitem[{{Kopal}(1978)}]{kopal1978}
{Kopal}, Z., ed. 1978, {Dynamics of Close Binary Systems}
 (Dordrecht: Reidel)
\bibitem[{{Kowalski}(2006)}]{kowalski2006}
{Kowalski}, P.~M. 2006, \apj, 641, 488

\bibitem[{{Landau} \& {Lifshitz}(1976)}]{llvI1976}
{Landau}, L.~D., \& {Lifshitz}, E.~M. 1976, {Course of Theoretical
  Physics: Mechanics} (Oxford: Pergamon Press)

\bibitem[{{Levermore}(1984)}]{levermore1984}
{Levermore}, C.~D. 1984, \jqsrt, 31, 149

\bibitem[{{Levermore} \& {Pomraning}(1981)}]{levermore1981}
{Levermore}, C.~D., \& {Pomraning}, G.~C. 1981, \apj, 248, 321

\bibitem[{{Lissauer}(1987)}]{lissauer1987}
{Lissauer}, J.~J. 1987, Icarus, 69, 249

\bibitem[{{Lissauer} {et~al.}(2009){Lissauer}, {Hubickyj}, {D'Angelo}, \&
  {Bodenheimer}}]{lissauer2009}
{Lissauer}, J.~J., {Hubickyj}, O., {D'Angelo}, G., \& {Bodenheimer}, P. 2009,
  Icarus, 199, 338

\bibitem[{{Lodders}(2010)}]{lodders2010}
{Lodders}, K. 2010, in Principles and Perspectives in Cosmochemistry, ed.
  A.~{Goswami} \& B.~E. {Reddy} (Berlin: Springer), 379

\bibitem[{{Masset} {et~al.}(2006){Masset}, {D'Angelo}, \& {Kley}}]{masset2006a}
{Masset}, F.~S., {D'Angelo}, G., \& {Kley}, W. 2006, \apj, 652, 730

\bibitem[{{Mihalas} \& {Weibel Mihalas}(1999)}]{m&m}
{Mihalas}, D., \& {Weibel Mihalas}, B. 1999, {Foundations of Radiation
  Hydrodynamics} (New York: Dover)

\bibitem[{{Movshovitz} {et~al.}(2010){Movshovitz}, {Bodenheimer}, {Podolak}, \&
  {Lissauer}}]{naor2010}
{Movshovitz}, N., {Bodenheimer}, P., {Podolak}, M., \& {Lissauer}, J.~J. 2010,
  Icarus, 209, 616

\bibitem[{{Murray} \& {Dermott}(2000)}]{murray2000}
{Murray}, C.~D., \& {Dermott}, S.~F. 2000, {Solar System Dynamics} 
  (Cambridge: Cambridge University Press)

\bibitem[{{Nelson} \& {Ruffert}(2013)}]{anelson2013}
{Nelson}, A.~F., \& {Ruffert}, M. 2013, \mnras, 429, 1791

\bibitem[{{Paardekooper} \& {Mellema}(2006)}]{paardekooper2006}
{Paardekooper}, S.-J., \& {Mellema}, G. 2006, \aap, 459, L17

\bibitem[{{Paczy{\'n}ski}(1971)}]{paczynski1971}
{Paczy{\'n}ski}, B. 1971, \araa, 9, 183

\bibitem[{{Pathria} \& {Beale}(2011)}]{pathria2011}
{Pathria}, R.~K., \& {Beale}, P.~D. 2011, Statistical Mechanics 
  (Amsterdam: Elsevier)

\bibitem[{{Podolak}(2003)}]{podolak2003}
{Podolak}, M. 2003, Icarus, 165, 428

\bibitem[{{Pollack} {et~al.}(1994){Pollack}, {Hollenbach}, {Beckwith},
  {Simonelli}, {Roush}, \& {Fong}}]{pollack1994}
{Pollack}, J.~B., {Hollenbach}, D., {Beckwith}, S., {Simonelli}, D.~P.,
  {Roush}, T., \& {Fong}, W. 1994, \apj, 421, 615

\bibitem[{{Pollack} {et~al.}(1996){Pollack}, {Hubickyj}, {Bodenheimer},
  {Lissauer}, {Podolak}, \& {Greenzweig}}]{pollack1996}
{Pollack}, J.~B., {Hubickyj}, O., {Bodenheimer}, P., {Lissauer}, J.~J.,
  {Podolak}, M., \& {Greenzweig}, Y. 1996, Icarus, 124, 62

\bibitem[{{Pollack} {et~al.}(1985){Pollack}, {McKay}, \&
  {Christofferson}}]{pollack1985}
{Pollack}, J.~B., {McKay}, C.~P., \& {Christofferson}, B.~M. 1985, \icarus, 64,
  471

\bibitem[{{Press} {et~al.}(1992){Press}, {Teukolsky}, {Vetterling}, \&
  {Flannery}}]{pressF1992}
{Press}, W.~H., {Teukolsky}, S.~A., {Vetterling}, W.~T., \& {Flannery}, B.~P.
  1992, {Numerical Recipes in FORTRAN. The Art of Scientific Computing}
  (Cambridge: Cambridge University Press)

\bibitem[{{Rafikov} \& {De Colle}(2006)}]{rafikov2006}
{Rafikov}, R.~R., \& {De Colle}, F. 2006, \apj, 646, 275

\bibitem[{Saad(2003)}]{saad2003}
Saad, Y. 2003, {Iterative Methods for Sparse Linear Systems}
 (Philadelphia, PA: Society for Industrial and Applied Mathematics)

\bibitem[{Schmauch \& Singleton(1964)}]{schmauch1964}
Schmauch, G.~E., \& Singleton, A.~H. 1964, Industrial \& Engineering Chemistry,
  56, 20

\bibitem[{{Shakura} \& {Syunyaev}(1973)}]{S&S1973}
{Shakura}, N.~I., \& {Syunyaev}, R.~A. 1973, \aap, 24, 337

\bibitem[{{Sleijpen} \& {Fokkema}(1993)}]{sleijpen1993}
{Sleijpen}, G., \& {Fokkema}, D. 1993, Electronic Transactions on Numerical
  Analysis, 1, 11

\bibitem[{{Stone} \& {Norman}(1992)}]{stone1992a}
{Stone}, J.~M., \& {Norman}, M.~L. 1992, \apjs, 80, 753

\bibitem[{{Tanigawa} {et~al.}(2012){Tanigawa}, {Ohtsuki}, \&
  {Machida}}]{tanigawa2012}
{Tanigawa}, T., {Ohtsuki}, K., \& {Machida}, M.~N. 2012, \apj, 747, 16

\bibitem[{{Turner} \& {Stone}(2001)}]{turner2001}
{Turner}, N.~J., \& {Stone}, J.~M. 2001, \apjs, 135, 95

\bibitem[{van~der Vorst(2003)}]{vandervorst2003}
van~der Vorst, H. 2003, {Iterative Krylov Methods for Large Linear Systems}
  (Cambridge: Cambridge University Press)

\bibitem[{van~der Vorst \& Vuik(1994)}]{vandervorst1994}
van~der Vorst, H.~A., \& Vuik, C. 1994, Numerical Linear Algebra with
  Applications, 1, 369

\bibitem[{{van Leer}(1977)}]{vanleer1977}
{van Leer}, B. 1977, Journal of Computational Physics, 23, 276

\bibitem[{{Weiss} {et~al.}(2006){Weiss}, {Hillebrandt}, {Thomas}, \&
  {Ritter}}]{cox2006}
{Weiss}, A., {Hillebrandt}, W., {Thomas}, H.-C., \& {Ritter}, H. 2006, {Cox and
  Giuli's Principles of Stellar Structure} (Cambridge:  Cambridge Scientific 
  Publishers Ltd)

\bibitem[{{Wuchterl}(1990)}]{wuchterl1990}
{Wuchterl}, G. 1990, \aap, 238, 83

\bibitem[{{Wuchterl}(1991)}]{wuchterl1991a}
{Wuchterl}, G. 1991, \icarus, 91, 39

\bibitem[{Yorke \& Kaisig(1995)}]{yorke1995}
Yorke, H.~W., \& Kaisig, M. 1995, Computer Physics Communications, 89, 29

\end{thebibliography}
\end{document}